\documentclass[longauth,traditabstract]{aa}
\usepackage{graphicx}
\usepackage{amsmath,amsfonts,amssymb}
\usepackage{txfonts}
\usepackage[breaklinks,colorlinks,citecolor=blue,pdfa=true]{hyperref}
\usepackage{color}
\usepackage{fixltx2e}
\usepackage{natbib}
\usepackage{url}
\usepackage{multirow}
\usepackage{epsf}
\usepackage{epsfig}

\usepackage{ifthen}

\usepackage[T1]{fontenc}
\usepackage{lmodern}
\usepackage{ifxetex,ifluatex}

\usepackage[switch]{lineno}

\bibpunct{(}{)}{;}{a}{}{,}

\def\setsymbol#1#2{\expandafter\def\csname #1\endcsname{#2}}
\def\getsymbol#1{\csname #1\endcsname}

\def\Planck{\textit{Planck}}

\def\alltwentythirteenresultspapers{\nocite{planck2013-p01, planck2013-p02, planck2013-p02a, planck2013-p02d, planck2013-p02b, planck2013-p03, planck2013-p03c, planck2013-p03f, planck2013-p03d, planck2013-p03e, planck2013-p01a, planck2013-p06, planck2013-p03a, planck2013-pip88, planck2013-p08, planck2013-p11, planck2013-p12, planck2013-p13, planck2013-p14, planck2013-p15, planck2013-p05b, planck2013-p17, planck2013-p09, planck2013-p09a, planck2013-p20, planck2013-p19, planck2013-pipaberration, planck2013-p05, planck2013-p05a, planck2013-pip56, planck2013-p06b, planck2013-p01a}}

\def\alltwentyfifteenresultspapers{\nocite{planck2014-a01, planck2014-a03, planck2014-a04, planck2014-a05, planck2014-a06, planck2014-a07, planck2014-a08, planck2014-a09, planck2014-a11, planck2014-a12, planck2014-a13, planck2014-a14, planck2014-a15, planck2014-a16, planck2014-a17, planck2014-a18, planck2014-a19, planck2014-a20, planck2014-a22, planck2014-a24, planck2014-a26, planck2014-a28, planck2014-a29, planck2014-a30, planck2014-a31, planck2014-a35, planck2014-a36, planck2014-a37, planck2014-ES}}

\newbox\tablebox    \newdimen\tablewidth
\def\leaderfil{\leaders\hbox to 5pt{\hss.\hss}\hfil}
\def\endPlancktable{\tablewidth=\columnwidth 
    $$\hss\copy\tablebox\hss$$
    \vskip-\lastskip\vskip -2pt}
\def\endPlancktablewide{\tablewidth=\textwidth 
    $$\hss\copy\tablebox\hss$$
    \vskip-\lastskip\vskip -2pt}
\def\tablenote#1 #2\par{\begingroup \parindent=0.8em
    \abovedisplayshortskip=0pt\belowdisplayshortskip=0pt
    \noindent
    $$\hss\vbox{\hsize\tablewidth \hangindent=\parindent \hangafter=1 \noindent
    \hbox to \parindent{$^#1$\hss}\strut#2\strut\par}\hss$$
    \endgroup}
\def\doubleline{\vskip 3pt\hrule \vskip 1.5pt \hrule \vskip 5pt}

\def\L2{\ifmmode L_2\else $L_2$\fi}

\def\DeltaT{\ifmmode \Delta T\else $\Delta T$\fi}
\def\deltat{\ifmmode \Delta t\else $\Delta t$\fi}
\def\fknee{\ifmmode f_{\rm knee}\else $f_{\rm knee}$\fi}
\def\Fmax{\ifmmode F_{\rm max}\else $F_{\rm max}$\fi}
\def\solar{\ifmmode{\rm M}_{\mathord\odot}\else${\rm M}_{\mathord\odot}$\fi}
\def\Msolar{\ifmmode{\rm M}_{\mathord\odot}\else${\rm M}_{\mathord\odot}$\fi}
\def\Lsolar{\ifmmode{\rm L}_{\mathord\odot}\else${\rm L}_{\mathord\odot}$\fi}
\def\inv{\ifmmode^{-1}\else$^{-1}$\fi}
\def\mo{\ifmmode^{-1}\else$^{-1}$\fi}
\def\sup#1{\ifmmode ^{\rm #1}\else $^{\rm #1}$\fi}
\def\expo#1{\ifmmode \times 10^{#1}\else $\times 10^{#1}$\fi}
\def\,{\thinspace}
\def\lsim{\mathrel{\raise .4ex\hbox{\rlap{$<$}\lower 1.2ex\hbox{$\sim$}}}}
\def\gsim{\mathrel{\raise .4ex\hbox{\rlap{$>$}\lower 1.2ex\hbox{$\sim$}}}}

\def\simprop{\mathrel{\raise .4ex\hbox{\rlap{$\propto$}\lower 1.2ex\hbox{$\sim$}}}}
\def\deg{\ifmmode^\circ\else$^\circ$\fi}
\def\pdeg{\ifmmode $\setbox0=\hbox{$^{\circ}$}\rlap{\hskip.11\wd0 .}$^{\circ}
          \else \setbox0=\hbox{$^{\circ}$}\rlap{\hskip.11\wd0 .}$^{\circ}$\fi}
\def\arcs{\ifmmode {^{\scriptstyle\prime\prime}}
          \else $^{\scriptstyle\prime\prime}$\fi}
\def\arcm{\ifmmode {^{\scriptstyle\prime}}
          \else $^{\scriptstyle\prime}$\fi}
\newdimen\sa  \newdimen\sb
\def\parcs{\sa=.07em \sb=.03em
     \ifmmode \hbox{\rlap{.}}^{\scriptstyle\prime\kern -\sb\prime}\hbox{\kern -\sa}
     \else \rlap{.}$^{\scriptstyle\prime\kern -\sb\prime}$\kern -\sa\fi}
\def\parcm{\sa=.08em \sb=.03em
     \ifmmode \hbox{\rlap{.}\kern\sa}^{\scriptstyle\prime}\hbox{\kern-\sb}
     \else \rlap{.}\kern\sa$^{\scriptstyle\prime}$\kern-\sb\fi}
\def\ra[#1 #2 #3.#4]{#1\sup{h}#2\sup{m}#3\sup{s}\llap.#4}
\def\dec[#1 #2 #3.#4]{#1\deg#2\arcm#3\arcs\llap.#4}
\def\deco[#1 #2 #3]{#1\deg#2\arcm#3\arcs}
\def\rra[#1 #2]{#1\sup{h}#2\sup{m}}

\def\dots{\relax\ifmmode \ldots\else $\ldots$\fi}
\def\WHzsr{\ifmmode $W\,Hz\mo\,sr\mo$\else W\,Hz\mo\,sr\mo\fi}
\def\mHz{\ifmmode $\,mHz$\else \,mHz\fi}
\def\GHz{\ifmmode $\,GHz$\else \,GHz\fi}
\def\mKs{\ifmmode $\,mK\,s$^{1/2}\else \,mK\,s$^{1/2}$\fi}
\def\muKs{\ifmmode \,\mu$K\,s$^{1/2}\else \,$\mu$K\,s$^{1/2}$\fi}
\def\muKRJs{\ifmmode \,\mu$K$_{\rm RJ}$\,s$^{1/2}\else \,$\mu$K$_{\rm RJ}$\,s$^{1/2}$\fi}
\def\muKHz{\ifmmode \,\mu$K\,Hz$^{-1/2}\else \,$\mu$K\,Hz$^{-1/2}$\fi}
\def\MJysr{\ifmmode \,$MJy\,sr\mo$\else \,MJy\,sr\mo\fi}
\def\MJysrmK{\ifmmode \,$MJy\,sr\mo$\,mK$_{\rm CMB}\mo\else \,MJy\,sr\mo\,mK$_{\rm CMB}\mo$\fi}
\def\microns{\ifmmode \,\mu$m$\else \,$\mu$m\fi}

\def\muK{\ifmmode \,\mu$K$\else \,$\mu$\hbox{K}\fi}
\def\microK{\ifmmode \,\mu$K$\else \,$\mu$\hbox{K}\fi}
\def\muW{\ifmmode \,\mu$W$\else \,$\mu$\hbox{W}\fi}
\def\kms{\ifmmode $\,km\,s$^{-1}\else \,km\,s$^{-1}$\fi}
\def\kmsMpc{\ifmmode $\,\kms\,Mpc\mo$\else \,\kms\,Mpc\mo\fi}
\providecommand{\sorthelp}[1]{}

\renewcommand*\vec[1]{\ensuremath{\boldsymbol{#1}}}

\def\eqref#1{(\ref{#1})}

\def\smica{{\tt SMICA}}
\def\nilc{{\tt NILC}}
\def\sevem{{\tt SEVEM}}
\def\commander{\texttt{Commander}}

\def\healpix{\texttt{HEALPix}}

\def\LCDM{$\Lambda$CDM}

\newcommand{\nside}{\ensuremath{N_{\mathrm{side}}}}

\def\eq{\begin{eqnarray}}
\def\qe{\end{eqnarray}}

\def\curl{\mathcal}
\def\({\left(}
\def\){\right)}

\def\and{\quad \mbox{and} \quad}
\def\barQ{\kern2pt\overline{\kern-2pt\curl{Q}}}

\def\barR{\kern2pt\overline{\kern-2pt\curl{R}}}

\def\bargamma{\kern2pt\overline{\kern-2pt\gamma}}

\def\leaderfil{\leaders\hbox to 5pt{\hss.\hss}\hfil}
\def\dd{\mathrm{d}} 

\newcommand{\cs}{Cold Spot}
\newcommand{\Qr}{$Q_\mathrm{r}$}
\newcommand{\Ur}{$U_\mathrm{r}$}
\newcommand{\be}{\begin{equation}}
\newcommand{\ee}{\end{equation}}
\newcommand{\bea}{\begin{eqnarray}}
\newcommand{\eea}{\end{eqnarray}}

\renewcommand{\L}[0]{\mathbf{L}}

\def\inv{^{-1}}

\newcommand{\spinup}{\;\raise1.0pt\hbox{$'$}\hskip-6pt\partial\;}
\newcommand{\spindown}{\;\overline{\raise1.0pt\hbox{$'$}\hskip-6pt\partial}\;}

\newcommand{\pval}{$p$-value}

\newcommand{\Tr}[1]{{\mathrm{Tr}\left[ #1 \right]}}
\graphicspath{{./figures/}}

\defcitealias{planck2013-p09}{PCIS13}

\title{\vglue -10mm\Planck\ 2015 results. XVI. Isotropy and statistics of the CMB}
\author{\small
Planck Collaboration: P.~A.~R.~Ade\inst{90}
\and
N.~Aghanim\inst{60}
\and
Y.~Akrami\inst{65, 104}
\and
P.~K.~Aluri\inst{55}
\and
M.~Arnaud\inst{76}
\and
M.~Ashdown\inst{72, 6}
\and
J.~Aumont\inst{60}
\and
C.~Baccigalupi\inst{89}
\and
A.~J.~Banday\inst{101, 9}~\thanks{Corresponding author: A.~J.~Banday \url{anthony.banday@irap.omp.eu}}
\and
R.~B.~Barreiro\inst{67}
% ...
}
\institute{\small
APC, AstroParticule et Cosmologie, Universit\'{e} Paris Diderot, CNRS/IN2P3, CEA/lrfu, Observatoire de Paris, Sorbonne Paris Cit\'{e}, 10, rue Alice Domon et L\'{e}onie Duquet, 75205 Paris Cedex 13, France\goodbreak
\and
Aalto University Mets\"{a}hovi Radio Observatory and Dept of Radio Science and Engineering, P.O. Box 13000, FI-00076 AALTO, Finland\goodbreak
\and
African Institute for Mathematical Sciences, 6-8 Melrose Road, Muizenberg, Cape Town, South Africa\goodbreak
\and
Agenzia Spaziale Italiana Science Data Center, Via del Politecnico snc, 00133, Roma, Italy\goodbreak
\and
Aix Marseille Universit\'{e}, CNRS, LAM (Laboratoire d'Astrophysique de Marseille) UMR 7326, 13388, Marseille, France\goodbreak
\and
Astrophysics Group, Cavendish Laboratory, University of Cambridge, J J Thomson Avenue, Cambridge CB3 0HE, U.K.\goodbreak
\and
Astrophysics \& Cosmology Research Unit, School of Mathematics, Statistics \& Computer Science, University of KwaZulu-Natal, Westville Campus, Private Bag X54001, Durban 4000, South Africa\goodbreak
\and
CITA, University of Toronto, 60 St. George St., Toronto, ON M5S 3H8, Canada\goodbreak
\and
CNRS, IRAP, 9 Av. colonel Roche, BP 44346, F-31028 Toulouse cedex 4, France\goodbreak
\and
CRANN, Trinity College, Dublin, Ireland\goodbreak
\and
California Institute of Technology, Pasadena, California, U.S.A.\goodbreak
\and
Centre for Theoretical Cosmology, DAMTP, University of Cambridge, Wilberforce Road, Cambridge CB3 0WA, U.K.\goodbreak
\and
Centro de Estudios de F\'{i}sica del Cosmos de Arag\'{o}n (CEFCA), Plaza San Juan, 1, planta 2, E-44001, Teruel, Spain\goodbreak
\and
Computational Cosmology Center, Lawrence Berkeley National Laboratory, Berkeley, California, U.S.A.\goodbreak
\and
Consejo Superior de Investigaciones Cient\'{\i}ficas (CSIC), Madrid, Spain\goodbreak
\and
DSM/Irfu/SPP, CEA-Saclay, F-91191 Gif-sur-Yvette Cedex, France\goodbreak
\and
DTU Space, National Space Institute, Technical University of Denmark, Elektrovej 327, DK-2800 Kgs. Lyngby, Denmark\goodbreak
\and
D\'{e}partement de Physique Th\'{e}orique, Universit\'{e} de Gen\`{e}ve, 24, Quai E. Ansermet,1211 Gen\`{e}ve 4, Switzerland\goodbreak
\and
Departamento de Astrof\'{i}sica, Universidad de La Laguna (ULL), E-38206 La Laguna, Tenerife, Spain\goodbreak
\and
Departamento de F\'{\i}sica, Universidad de Oviedo, Avda. Calvo Sotelo s/n, Oviedo, Spain\goodbreak
\and
Departamento de Matem\'{a}ticas, Estad\'{\i}stica y Computaci\'{o}n, Universidad de Cantabria, Avda. de los Castros s/n, Santander, Spain\goodbreak
\and
Department of Astronomy and Astrophysics, University of Toronto, 50 Saint George Street, Toronto, Ontario, Canada\goodbreak
\and
Department of Astrophysics/IMAPP, Radboud University Nijmegen, P.O. Box 9010, 6500 GL Nijmegen, The Netherlands\goodbreak
\and
Department of Physics \& Astronomy, University of British Columbia, 6224 Agricultural Road, Vancouver, British Columbia, Canada\goodbreak
\and
Department of Physics and Astronomy, Dana and David Dornsife College of Letter, Arts and Sciences, University of Southern California, Los Angeles, CA 90089, U.S.A.\goodbreak
\and
Department of Physics and Astronomy, University College London, London WC1E 6BT, U.K.\goodbreak
\and
Department of Physics, Florida State University, Keen Physics Building, 77 Chieftan Way, Tallahassee, Florida, U.S.A.\goodbreak
\and
Department of Physics, Gustaf H\"{a}llstr\"{o}min katu 2a, University of Helsinki, Helsinki, Finland\goodbreak
\and
Department of Physics, Princeton University, Princeton, New Jersey, U.S.A.\goodbreak
\and
Department of Physics, University of California, Santa Barbara, California, U.S.A.\goodbreak
\and
Department of Physics, University of Illinois at Urbana-Champaign, 1110 West Green Street, Urbana, Illinois, U.S.A.\goodbreak
\and
Dipartimento di Fisica e Astronomia G. Galilei, Universit\`{a} degli Studi di Padova, via Marzolo 8, 35131 Padova, Italy\goodbreak
\and
Dipartimento di Fisica e Scienze della Terra, Universit\`{a} di Ferrara, Via Saragat 1, 44122 Ferrara, Italy\goodbreak
\and
Dipartimento di Fisica, Universit\`{a} La Sapienza, P. le A. Moro 2, Roma, Italy\goodbreak
\and
Dipartimento di Fisica, Universit\`{a} degli Studi di Milano, Via Celoria, 16, Milano, Italy\goodbreak
\and
Dipartimento di Fisica, Universit\`{a} degli Studi di Trieste, via A. Valerio 2, Trieste, Italy\goodbreak
\and
Dipartimento di Matematica, Universit\`{a} di Roma Tor Vergata, Via della Ricerca Scientifica, 1, Roma, Italy\goodbreak
\and
Discovery Center, Niels Bohr Institute, Blegdamsvej 17, Copenhagen, Denmark\goodbreak
\and
European Space Agency, ESAC, Planck Science Office, Camino bajo del Castillo, s/n, Urbanizaci\'{o}n Villafranca del Castillo, Villanueva de la Ca\~{n}ada, Madrid, Spain\goodbreak
\and
European Space Agency, ESTEC, Keplerlaan 1, 2201 AZ Noordwijk, The Netherlands\goodbreak
\and
Facolt\`{a} di Ingegneria, Universit\`{a} degli Studi e-Campus, Via Isimbardi 10, Novedrate (CO), 22060, Italy\goodbreak
\and
Gran Sasso Science Institute, INFN, viale F. Crispi 7, 67100 L'Aquila, Italy\goodbreak
\and
HGSFP and University of Heidelberg, Theoretical Physics Department, Philosophenweg 16, 69120, Heidelberg, Germany\goodbreak
\and
Helsinki Institute of Physics, Gustaf H\"{a}llstr\"{o}min katu 2, University of Helsinki, Helsinki, Finland\goodbreak
\and
INAF - Osservatorio Astronomico di Padova, Vicolo dell'Osservatorio 5, Padova, Italy\goodbreak
\and
INAF - Osservatorio Astronomico di Roma, via di Frascati 33, Monte Porzio Catone, Italy\goodbreak
\and
INAF - Osservatorio Astronomico di Trieste, Via G.B. Tiepolo 11, Trieste, Italy\goodbreak
\and
INAF/IASF Bologna, Via Gobetti 101, Bologna, Italy\goodbreak
\and
INAF/IASF Milano, Via E. Bassini 15, Milano, Italy\goodbreak
\and
INFN, Sezione di Bologna, Via Irnerio 46, I-40126, Bologna, Italy\goodbreak
\and
INFN, Sezione di Roma 1, Universit\`{a} di Roma Sapienza, Piazzale Aldo Moro 2, 00185, Roma, Italy\goodbreak
\and
INFN, Sezione di Roma 2, Universit\`{a} di Roma Tor Vergata, Via della Ricerca Scientifica, 1, Roma, Italy\goodbreak
\and
INFN/National Institute for Nuclear Physics, Via Valerio 2, I-34127 Trieste, Italy\goodbreak
\and
IPAG: Institut de Plan\'{e}tologie et d'Astrophysique de Grenoble, Universit\'{e} Grenoble Alpes, IPAG, F-38000 Grenoble, France, CNRS, IPAG, F-38000 Grenoble, France\goodbreak
\and
IUCAA, Post Bag 4, Ganeshkhind, Pune University Campus, Pune 411 007, India\goodbreak
\and
Imperial College London, Astrophysics group, Blackett Laboratory, Prince Consort Road, London, SW7 2AZ, U.K.\goodbreak
\and
Infrared Processing and Analysis Center, California Institute of Technology, Pasadena, CA 91125, U.S.A.\goodbreak
\and
Institut N\'{e}el, CNRS, Universit\'{e} Joseph Fourier Grenoble I, 25 rue des Martyrs, Grenoble, France\goodbreak
\and
Institut Universitaire de France, 103, bd Saint-Michel, 75005, Paris, France\goodbreak
\and
Institut d'Astrophysique Spatiale, CNRS (UMR8617) Universit\'{e} Paris-Sud 11, B\^{a}timent 121, Orsay, France\goodbreak
\and
Institut d'Astrophysique de Paris, CNRS (UMR7095), 98 bis Boulevard Arago, F-75014, Paris, France\goodbreak
\and
Institut f\"ur Theoretische Teilchenphysik und Kosmologie, RWTH Aachen University, D-52056 Aachen, Germany\goodbreak
\and
Institute for Space Sciences, Bucharest-Magurale, Romania\goodbreak
\and
Institute of Astronomy, University of Cambridge, Madingley Road, Cambridge CB3 0HA, U.K.\goodbreak
\and
Institute of Theoretical Astrophysics, University of Oslo, Blindern, Oslo, Norway\goodbreak
\and
Instituto de Astrof\'{\i}sica de Canarias, C/V\'{\i}a L\'{a}ctea s/n, La Laguna, Tenerife, Spain\goodbreak
\and
Instituto de F\'{\i}sica de Cantabria (CSIC-Universidad de Cantabria), Avda. de los Castros s/n, Santander, Spain\goodbreak
\and
Istituto Nazionale di Fisica Nucleare, Sezione di Padova, via Marzolo 8, I-35131 Padova, Italy\goodbreak
\and
Jet Propulsion Laboratory, California Institute of Technology, 4800 Oak Grove Drive, Pasadena, California, U.S.A.\goodbreak
\and
Jodrell Bank Centre for Astrophysics, Alan Turing Building, School of Physics and Astronomy, The University of Manchester, Oxford Road, Manchester, M13 9PL, U.K.\goodbreak
\and
Kavli Institute for Cosmological Physics, University of Chicago, Chicago, IL 60637, USA\goodbreak
\and
Kavli Institute for Cosmology Cambridge, Madingley Road, Cambridge, CB3 0HA, U.K.\goodbreak
\and
Kazan Federal University, 18 Kremlyovskaya St., Kazan, 420008, Russia\goodbreak
\and
LAL, Universit\'{e} Paris-Sud, CNRS/IN2P3, Orsay, France\goodbreak
\and
LERMA, CNRS, Observatoire de Paris, 61 Avenue de l'Observatoire, Paris, France\goodbreak
\and
Laboratoire AIM, IRFU/Service d'Astrophysique - CEA/DSM - CNRS - Universit\'{e} Paris Diderot, B\^{a}t. 709, CEA-Saclay, F-91191 Gif-sur-Yvette Cedex, France\goodbreak
\and
Laboratoire Traitement et Communication de l'Information, CNRS (UMR 5141) and T\'{e}l\'{e}com ParisTech, 46 rue Barrault F-75634 Paris Cedex 13, France\goodbreak
\and
Laboratoire de Physique Subatomique et Cosmologie, Universit\'{e} Grenoble-Alpes, CNRS/IN2P3, 53, rue des Martyrs, 38026 Grenoble Cedex, France\goodbreak
\and
Laboratoire de Physique Th\'{e}orique, Universit\'{e} Paris-Sud 11 \& CNRS, B\^{a}timent 210, 91405 Orsay, France\goodbreak
\and
Lawrence Berkeley National Laboratory, Berkeley, California, U.S.A.\goodbreak
\and
Lebedev Physical Institute of the Russian Academy of Sciences, Astro Space Centre, 84/32 Profsoyuznaya st., Moscow, GSP-7, 117997, Russia\goodbreak
\and
Leung Center for Cosmology and Particle Astrophysics, National Taiwan University, Taipei 10617, Taiwan\goodbreak
\and
Max-Planck-Institut f\"{u}r Astrophysik, Karl-Schwarzschild-Str. 1, 85741 Garching, Germany\goodbreak
\and
McGill Physics, Ernest Rutherford Physics Building, McGill University, 3600 rue University, Montr\'{e}al, QC, H3A 2T8, Canada\goodbreak
\and
National University of Ireland, Department of Experimental Physics, Maynooth, Co. Kildare, Ireland\goodbreak
\and
Nicolaus Copernicus Astronomical Center, Bartycka 18, 00-716 Warsaw, Poland\goodbreak
\and
Niels Bohr Institute, Blegdamsvej 17, Copenhagen, Denmark\goodbreak
\and
Optical Science Laboratory, University College London, Gower Street, London, U.K.\goodbreak
\and
SISSA, Astrophysics Sector, via Bonomea 265, 34136, Trieste, Italy\goodbreak
\and
School of Physics and Astronomy, Cardiff University, Queens Buildings, The Parade, Cardiff, CF24 3AA, U.K.\goodbreak
\and
School of Physics and Astronomy, University of Nottingham, Nottingham NG7 2RD, U.K.\goodbreak
\and
Simon Fraser University, Department of Physics, 8888 University Drive, Burnaby BC, Canada\goodbreak
\and
Sorbonne Universit\'{e}-UPMC, UMR7095, Institut d'Astrophysique de Paris, 98 bis Boulevard Arago, F-75014, Paris, France\goodbreak
\and
Space Research Institute (IKI), Russian Academy of Sciences, Profsoyuznaya Str, 84/32, Moscow, 117997, Russia\goodbreak
\and
Space Sciences Laboratory, University of California, Berkeley, California, U.S.A.\goodbreak
\and
Special Astrophysical Observatory, Russian Academy of Sciences, Nizhnij Arkhyz, Zelenchukskiy region, Karachai-Cherkessian Republic, 369167, Russia\goodbreak
\and
Stanford University, Dept of Physics, Varian Physics Bldg, 382 Via Pueblo Mall, Stanford, California, U.S.A.\goodbreak
\and
Sub-Department of Astrophysics, University of Oxford, Keble Road, Oxford OX1 3RH, U.K.\goodbreak
\and
Theory Division, PH-TH, CERN, CH-1211, Geneva 23, Switzerland\goodbreak
\and
UPMC Univ Paris 06, UMR7095, 98 bis Boulevard Arago, F-75014, Paris, France\goodbreak
\and
Universit\'{e} de Toulouse, UPS-OMP, IRAP, F-31028 Toulouse cedex 4, France\goodbreak
\and
University of Granada, Departamento de F\'{\i}sica Te\'{o}rica y del Cosmos, Facultad de Ciencias, Granada, Spain\goodbreak
\and
University of Granada, Instituto Carlos I de F\'{\i}sica Te\'{o}rica y Computacional, Granada, Spain\goodbreak
\and
University of Heidelberg, Institute for Theoretical Physics, Philosophenweg 16, 69120, Heidelberg, Germany\goodbreak
% ...
}

\authorrunning{Planck Collaboration}
\titlerunning{Isotropy and statistics of the CMB}

\begin{document}


\abstract{
  We test the statistical isotropy and Gaussianity of the cosmic
  microwave background (CMB) anisotropies using observations made by
  the \Planck\ satellite.  Our results are based mainly on the full
  \Planck\ mission for temperature, but also include some polarization
  measurements.
  In particular, we consider the CMB anisotropy maps derived from the
  multi-frequency \Planck\ data by several component-separation
  methods. For the temperature anisotropies, we find excellent
  agreement between results based on these sky maps over both a very
  large fraction of the sky and a broad range of angular
  scales, 
  establishing that potential foreground residuals do not
  affect our studies.
  Tests of skewness, kurtosis, multi-normality, $N$-point functions,
  and Minkowski functionals indicate consistency with Gaussianity,
  while a power deficit at large angular scales is manifested in
  several ways, for example low map variance.  The results of a peak
  statistics analysis are consistent with the expectations of a
  Gaussian random field.  The ``Cold Spot'' is detected with several
  methods, including map kurtosis, peak statistics, and mean
  temperature profile.  We thoroughly probe the large-scale dipolar power
  asymmetry, detecting it with several independent tests, and address
  the subject of a posteriori correction.  Tests of directionality
  suggest the presence of angular clustering from large to small
  scales, but at a significance that is dependent on the details of
  the approach.  We perform the first examination of polarization
  data, finding the morphology of stacked peaks to be consistent with
  the expectations of statistically isotropic simulations.  Where they
  overlap, these results are consistent with the \Planck\ 2013
  analysis based on the nominal mission data and provide our most
  thorough view of the statistics of the CMB fluctuations to date.
}


\maketitle

\alltwentythirteenresultspapers

\alltwentyfifteenresultspapers

\section{Introduction}
\label{sec:introduction}

This paper, one of a set associated with the 2015 release of data from
the \Planck\footnote{\Planck\ (\url{http://www.esa.int/Planck}) is a
  project of the European Space Agency (ESA) with instruments provided
  by two scientific consortia funded by ESA member states and led by
  Principal Investigators from France and Italy, telescope reflectors
  provided through a collaboration between ESA and a scientific
  consortium led and funded by Denmark, and additional contributions
  from NASA (USA).}  mission \citep{planck2014-a01}, describes a set
of studies undertaken to determine the statistical properties of both
the temperature and polarization anisotropies of the cosmic microwave
background (CMB).

The standard cosmological model is described well by the
Friedmann-Lema\^{\i}tre-Robertson-Walker solution of the
Einstein field equations. This model is characterized by a homogeneous
and isotropic background metric and a scale factor of the expanding
Universe. It is hypothesized that at very early times the Universe
went through a period of accelerated expansion, the so-called
``cosmological inflation,'' driven by a hypothetical scalar field, the
``inflaton.'' During inflation the Universe behaves approximately as a
de Sitter space, providing the conditions by which some of its present
properties can be realized and specifically relaxing the problem of
initial conditions. In particular, the seeds that gave rise to the
present large-scale matter distribution via gravitational instability
originated as quantum fluctuations of the inflaton about its vacuum
state. These fluctuations in the inflaton produce energy density
perturbations that are distributed as a statistically homogeneous and
isotropic Gaussian random field. Linear theory relates those
perturbations to the temperature and polarization anisotropies of the
CMB, implying a distribution for the anisotropies very close to that
of a statistically isotropic Gaussian random field.

The aim of this paper is to use the full mission \Planck\ data to test
the Gaussianity and isotropy of the CMB as measured in both intensity
and, in a more limited capacity, polarization.  Testing these
fundamental properties is crucial for the validation of the standard
cosmological scenario, and has profound implications for our
understanding of the physical nature of the Universe and the initial
conditions of structure formation.  Moreover, the confirmation of the
statistically isotropic and Gaussian nature of the CMB is essential for justifying
the corresponding assumptions usually made when estimating the CMB
power spectra and other quantities to be obtained from the \Planck\
data. Indeed,  the isotropy and Gaussianity of the CMB anisotropies are
implicitly assumed in critical science papers from the 2015 release,
in particular those describing the likelihood and
the derivation of cosmological parameter constraints
\citep{planck2014-a13,planck2014-a15}.
Conversely, if the detection of significant deviations from these
assumptions cannot be traced to known systematic effects or
foreground residuals, the presence of which should be diagnosed by the
statistical tests set forth in this paper, this would necessitate a major
revision of the current methodological approaches adopted in deriving the
mission's many science results.

Well-understood physical processes due to the integrated Sachs-Wolfe
(ISW) effect \citep{planck2013-p12,planck2014-a26} and gravitational
lensing \citep{planck2013-p14,planck2014-a17} lead to secondary
anisotropies that exhibit marked deviation from Gaussianity. In
addition, Doppler boosting, due to our motion with respect to the CMB
rest frame, induces both a dipolar modulation of the temperature
anisotropies and an aberration that corresponds to a change in the
apparent arrival directions of the CMB photons
\citep{challinor2002}. Both of these effects are aligned with the CMB
dipole, and were detected at a statistically significant level on
small angular scales in \citet{planck2013-pipaberration}.  Beyond
these, \citet[hereafter \citetalias{planck2013-p09}]{planck2013-p09}
established that the \Planck\ 2013 data set showed little evidence for
non-Gaussianity, with the exception of a number of CMB temperature
anisotropy anomalies on large angular scales that confirmed earlier
claims based on WMAP data.  Moreover, given that the broader frequency
coverage of the \Planck\ instruments allowed improved component
separation methods to be applied in the derivation of
foreground-cleaned CMB maps, it was generally considered that the case
for anomalous features in the CMB had been strengthened.  Hence, such
anomalies have attracted considerable attention in the community,
since they could be the visible traces of fundamental physical
processes occurring in the early Universe.

However, the literature also supports an ongoing debate about the
significance of these anomalies. The central issue in this discussion
is connected with the role of a~posteriori choices --- whether
interesting features in the data bias the choice of statistical tests,
or if arbitrary choices in the subsequent data analysis enhance the
significance of the features.  Indeed, the WMAP team
\citep{bennett2010} base their rejection of the presence of anomalies
in the CMB on such arguments.  Of course, one should attempt to
correct for any choices that were made in the process of detecting an
anomaly. However, in the absence of an alternative model for
comparison to the standard Gaussian, statistically isotropic one adopted to quantify
significance, this is often simply not possible. In this work, whilst
it is recognized that care must be taken in the assessment of
significance, we proceed on the basis that allowing a posteriori
reasoning permits us to challenge the limits of our existing knowledge
\citep{Pontzen2010}.  That is, by focusing on specific properties of
the observed data that are shown to be empirically interesting, we may
open up new paths to a better theoretical understanding of the
Universe.  We will clearly describe the methodology applied to the
data, and attempt to study possible links among the anomalies in order
to search for a physical interpretation.

The analysis of polarization data introduces a new opportunity to
explore the statistical properties of the CMB sky, including the
possibility of improvement of the significance of detection of
large-scale anomalies.  However, this cannot be fully included in the
current data assessment, since the component-separation products in
polarization are high-pass filtered to remove large angular scales
\citep{planck2014-a11}, owing to the persistence of significant
systematic artefacts originating in the High Frequency Instrument
(HFI) data \citep{planck2014-a08,planck2014-a09}. In addition,
limitations of the simulations with which the data are to be compared
\citep{planck2014-a14}, in particular a significant mismatch in noise
properties, limit the extent to which any polarization results can be
included. Therefore, we only present a stacking analysis of the
polarized data, although this is a significant extension of previous
approaches found in the literature.

With future \Planck\ data releases, it will be important to determine
in more detail whether there are any pecularities in the CMB
polarization, and if so, whether they are related to existing features
in the CMB temperature field.  Conversely, the absence of any
corresponding features in polarization might imply that the the
temperature anomalies (if they are not simply flukes) could be due to
a secondary effect such as the ISW effect, or
alternative scenarios in which the anomalies arise from physical
processes that do not correlate with the temperature, e.g., texture or
defect models. Either one of these possible outcomes could yield a
breakthrough in understanding the nature of the CMB anomalies. Of
course, there also remains the possibility that anomalies may be found
in the polarization data that are unrelated to existing features in
the temperature measurements.

Following the approach established in \citet{planck2013-p09},
throughout this paper we quantify the significance of a test statistic
in terms of the \pval. This is the probability of obtaining a test
statistic at least as extreme as the observed one, under the
assumption that the null hypothesis (i.e., primordial Gaussianity and
isotropy of the CMB) is true. In some tests, where it is clearly
justified to only use a one-tailed probability, the \pval\ is replaced
by the corresponding upper- or lower-tail probability.

This paper covers all relevant aspects related to the phenomenological
study of the statistical isotropy and Gaussian nature of the CMB
measured by the \Planck\ satellite. Specific theoretically-motivated
model constraints on isotropy or non-Gaussianity, as might arise from
non-standard inflationary models, the geometry and topology of the
Universe, and primordial magnetic fields are provided in the companion
papers~\citep{planck2014-a19,planck2014-a24,planck2014-a20,planck2014-a22}.
The paper is organized as follows.  Section~\ref{sec:data} summarizes
the \Planck\ full mission data used for the analyses, and important
limitations of the polarization maps that are
studied. Section~\ref{sec:simulations} describes the characteristics
of the simulations that constitute our reference set of Gaussian sky
maps representative of the null hypothesis. In
Sect.~\ref{sec:non_gaussianity} the null hypothesis is tested with a
number of standard tests that probe different aspects of
non-Gaussianity. Several important anomalous features of the CMB sky,
originally detected with the WMAP data and subsequently confirmed in
\citetalias{planck2013-p09}, are reassessed in Sect.~\ref{sec:anomalies}.
Aspects of the CMB fluctuations specifically related to dipolar asymmetry
are examined
in Sect.~\ref{sec:dipmodsection}.  The sensitivity of the results for a 
number of statistical tests to the sky fraction is examined in 
Sect.~\ref{sec:sky_coverage}.  Section~\ref{sec:polarization_results}
presents tests of the statistical nature of the polarization signal
observed by \Planck\ using a local analysis of stacked patches of the
sky.  Finally, Sect.~\ref{sec:conclusions} provides the main
conclusions of the paper.

\section{Data description}
\label{sec:data}

In this paper, we use data from the \Planck-2015 full mission data
release. This contains approximately 29 months of data for the HFI and 50 months for
the Low Frequency Instrument (LFI).  The release includes sky maps at nine frequencies in
intensity (seven in polarization), with corresponding ``half-mission''
maps that are generated by splitting the full-mission data sets
in various ways.  The maps are provided in {\tt
  HEALPix} format \citep{gorski2005},\footnote{\url{http://healpix.sourceforge.net}}
 with a pixel size defined by the \nside\
parameter.  This set of maps allows a variety of consistency checks to
be made, together with estimates of the instrumental noise
contributions to our analyses and limits on time-varying systematic
artefacts. Full details are provided in a series of companion
papers~\citep{planck2014-a03,planck2014-a04,planck2014-a05,planck2014-a06,planck2014-a07,planck2014-a08,planck2014-a09}.

Our main results are based on estimates of the CMB generated by four
distinct component-separation algorithms --- \commander, \nilc,
\sevem, and \smica\ --- as described in \citet{planck2014-a11}. These
effectively combine the raw \Planck\ frequency maps in such a way as
to minimize foreground residuals from diffuse Galactic emission.  Note
that the additional information in the full mission data set allows us
to improve the reconstruction noise levels by roughly a factor of 2
(in temperature) as compared to the \Planck-2013 nominal mission data
release.  The CMB intensity maps were derived using all channels, from
30 to 857\,GHz, and provided at a common angular resolution of 5\arcm\
FWHM and $\nside=2048$.
The intensity maps are only partially corrected for the second order
temperature quadrupole \citep{kamionkowski2003}. Therefore, where
appropriate, the component-separated maps should be corrected for the
residual contribution \citep{Notari2015}, specifically as described in
\citet{planck2014-a11}.
The polarization solutions include all channels sensitive to
polarization, from 30 to 353\,GHz, at a resolution of 10\arcm\ FWHM
and $\nside=1024$.  Possible residual emission is then mitigated in our
analyses by the use of sky-coverage masks, provided for both intensity
and polarization.

Since in some cases it is important to study the frequency dependence
of the cosmological signal, either to establish its primordial origin
or to test for the frequency dependence associated with specific
effects such as Doppler boosting (see Sect.~\ref{sec:biposh}), we also
consider the foreground-cleaned versions of the 100, 143, and 217\,GHz
sky maps generated by the \sevem\ algorithm \citep{planck2014-a11},
hereafter referred to as {\tt SEVEM-100}, {\tt SEVEM-143}, and {\tt
  SEVEM-217}, respectively.

For the present release, a post-processing high-pass-filtering
has been applied to the CMB polarization maps in order to mitigate
residual large-scale systematic errors in the HFI channels
\citep{planck2014-a08}. The filter results in the elimination of
structure in the maps on angular scales larger than about
10\deg, and a weighted suppression of power down to scales of 5\deg,
below which the maps remain unprocessed.

Lower-resolution versions of these data sets are also used in the
analyses presented in this paper.  The downgrading procedure for maps
is to decompose them into spherical harmonics on the full sky at the
input \healpix\ resolution.  The spherical harmonic coefficients,
$a_{\ell m}$, are then convolved to the new resolution using
\begin{linenomath*}
\begin{equation}
a_{\ell m}^{\rm out} = \frac{b_{\ell}^{\rm out}p_{\ell}^{\rm
    out}}{b_{\ell}^{\rm in}p_{\ell}^{\rm in}} a_{\ell m}^{\rm in},
\end{equation}
\end{linenomath*}
where $b_{\ell}$ is the beam transfer function, $p_{\ell}$ is the
\healpix\ pixel window function, and the ``in'' and ``out''
superscripts denote the input and output resolutions.  They are then
synthesized into a map directly at the output \healpix\ resolution.
Masks are downgraded in a similar way.  The binary mask at the
starting resolution is first downgraded like a temperature map.  The
smooth downgraded mask is then thresholded by setting pixels where the
value is less than~$0.9$ to zero and all others to unity in order to
make a binary mask.  Table~\ref{tab:masks} lists the \nside\ and FWHM
values defining the resolution of these maps, together with the
different masks and their sky coverages that accompany the signal
maps.  In general, we make use of standardized masks that are the
union of those associated with the individual component-separation
methods.

As recommended in \citet{planck2014-a11}, the mask {\tt UT78} is
adopted for all high-resolution analyses of temperature data. {\tt
  UTA76} is an extended version of this mask more suitable for some
non-Gaussianity studies.  The mask preferred for polarization studies,
{\tt UPB77}, is again the union of those determined for each component
separation method, but in addition the polarized point sources
detected at each frequency channel are excluded.  These masks are then
downgraded for lower-resolution studies.  As a consequence of the
common scheme applied in order to generate such low-resolution masks,
they are generally more conservative than the corresponding ones used
in the 2013 analyses.

In what follows, we will undertake analyses of the data at a given
resolution denoted by a specific \nside\ value. Unless otherwise
stated, this implies that the data have been smoothed to a
corresponding FWHM as described above, and a standardized mask
employed. Irrespective of the resolution in question, we will then
often simply refer to the latter as the ``common mask.''

\begin{table}
\begingroup
\newdimen\tblskip \tblskip=5pt
\caption{Standardized data sets used in this paper. The resolutions of
  the sky maps used are defined in terms of the \nside\ parameter and
  corresponding FWHM of the Gaussian beam with which they are
  convolved. The corresponding common masks and the fraction of unmasked
  pixels used for analysis are also specified.
}
\label{tab:masks}
\nointerlineskip
\vskip -3mm
\footnotesize
\setbox\tablebox=\vbox{
   \newdimen\digitwidth
   \setbox0=\hbox{\rm 0}
   \digitwidth=\wd0
   \catcode`*=\active
   \def*{\kern\digitwidth}
   \newdimen\signwidth
   \setbox0=\hbox{+}
   \signwidth=\wd0
   \catcode`!=\active
   \def!{\kern\signwidth}
\halign{\hbox to 0.8in{#\leaderfil}\tabskip 5pt&
\hfil#\hfil\tabskip 5pt&
#\hfil\tabskip 5pt&
\hfil#\hfil\/\tabskip 0pt\cr
\noalign{\doubleline}
\omit& FWHM\hfil&\hfil Mask\hfil&  Unmasked\cr
\omit\nside\ \hfil& [arcmin]&  \omit& pixels [\%]\cr
\noalign{\vskip 3pt\hrule\vskip 3pt}
2048& **5& {\tt UT78}& 77.6\cr
2048& **5& {\tt UTA76}& 76.1\cr
1024& *10& {\tt UT$_\mathrm{1024}$76}& 75.6\cr
*512& *20& {\tt UT$_\mathrm{512}$74}& 73.7\cr
*256& *40& {\tt UT$_\mathrm{256}$73}& 72.5\cr
*128& *80& {\tt UT$_\mathrm{128}$70}& 69.7\cr
**64& 160& {\tt UT$_\mathrm{64}$67}& 67.0\cr
**32& 320& {\tt UT$_\mathrm{32}$64}& 63.8\cr
**16& 640& {\tt UT$_\mathrm{16}$58}& 58.4\cr
1024&  *10& {\tt UPB77}& 77.4\cr
\noalign{\vskip 3pt\hrule\vskip 3pt}
\noalign{\vskip 3pt}}}
\endPlancktable 
\endgroup
\end{table}

\section{Simulations}
\label{sec:simulations}

The results presented in this paper are derived using the extensive
full focal plane (FFP8) simulations described in
\citet{planck2014-a14}. Of most importance are the Monte Carlo (MC)
simulations that provide the reference set of Gaussian sky maps used
for the null tests employed here. They also form the basis of any
debiasing in the analysis of the real data as required by certain
statistical methods.

The simulations include both CMB signal and instrumental noise
realizations that 
capture important characteristics of the
\Planck\ scanning strategy, telescope, detector responses, and data
reduction pipeline over the full mission period. In particular, the
signal realizations include {\tt FEBeCoP} \citep{mitra2010} beam
convolution at each of the \Planck\ frequencies, and are
propagated through the various component-separation pipelines using
the same weights as derived from the \Planck\ full mission data
analysis.

The FFP8 fiducial CMB power spectrum has been adopted from our best
estimate of the cosmological parameters from the first \Planck\ data
release \citep{planck2013-p01}. This corresponds to a cosmology with
baryon density given by
$\omega_{\rm b} = \Omega_{\rm b}h^2 = 0.0222$,
cold dark matter density
$\omega_{\rm c} =\Omega_{\rm c}h^2 = 0.1203$,
neutrino energy density
$\omega_{\nu} = \Omega_{\nu}h^2 = 0.00064$,
density parameter for the cosmological constant
$\Omega_{\Lambda} = 0.6823$,
Hubble parameter
$H_0 = 100 h \,\mathrm{km}\,\mathrm{s}^{-1}\,\mathrm{Mpc}^{-1}$ with $h=0.6712$,
spectral index of the power spectrum of the primordial curvature perturbation
$n_{\rm s} = 0.96$,
and amplitude of the primordial power spectrum (at
$k=0.05\,\mathrm{Mpc}^{-1}$) $A_{\rm s} = 2.09\times 10^{-9}$,
and with the Thomson optical depth through reionization
defined to be $\tau=0.065$.
Each realization of the CMB sky is generated including
lensing, Rayleigh scattering, and Doppler boosting effects, the latter
two of which are frequency-dependent. A second order temperature
quadrupole \citep{kamionkowski2003} is added to each simulation with
an amplitude corresponding to the residual uncorrected contribution
present in the observed data, as described in \citet{planck2014-a14}.

However, the \Planck\ maps were effectively renormalized by
approximately 2\,\% to 3\,\% in power in the time between the generation
of the FFP8 simulations and the final maps.  As discussed in \cite{planck2014-a14},
correction for this calibration effect should have no significant
impact on cosmological parameters. As recommended, in this paper the
CMB component of the simulations is simply rescaled by a factor of
$1.0134$ before analysis.

Of somewhat more importance is an observed noise mismatch between the
simulations and the data.  Whilst this has essentially no impact on
studies of temperature anisotropy, it imposes important limitations on
the statistical studies of polarization sky maps that can be included
here. Conversely, analyses based on 1-point statistics, such as the
variance, and the $N$-point correlation functions have played important
roles in establishing the nature of this mismatch, which seems to be
scale-dependent with an amplitude around 20\,\% at lower resolutions but
falling to a few per cent at higher resolution. As a
consequence, this paper only includes results from a stacking analysis
of the polarized data, in which the stacking of the data themselves
necessarily acts to lower the effect of the noise
mismatch. Polarization studies that do not rely on auto-statistics can
still yield interesting new results, as found in
\citet{planck2014-a15,planck2014-a19,planck2014-a20}.

\section{Tests of non-Gaussianity}
\label{sec:non_gaussianity}

There is no unique signature of non-Gaussianity, but the application
of a variety of tests over a range of angular scales allows us to
probe the data for departures from theoretically motivated Gaussian
statistics. One of the more important tests in the context of
inflationary cosmology is related to the analysis of the
bispectrum. This is discussed thoroughly in \citet{planck2014-a19},
and is therefore not discussed further in this paper. In this section,
we present the results from a variety of statistical tools. Unless
otherwise specified, the analyses are applied to all four component
separation products (\commander, \nilc, \sevem, and \smica) at a given
resolution with the accompanying common mask, and significance levels
are determined by comparison with the corresponding results derived from
the FFP8 simulations, with typically 1000 being used for this
purpose.  Establishing the consistency of the results derived from the
different component-separation techniques is essential in order to be
able to make robust claims about the statistical nature of the
observed temperature fluctuations, and potential deviations from
Gaussianity.

\subsection{One-dimensional moments}
\label{sec:onepdf}

In this section we consider simple tests of Gaussianity based on the
variance, skewness, and kurtosis of the CMB temperature maps.
Previous analyses found an anomalously low variance in the WMAP sky
maps \citep{Monteserin2008,Cruz2011}, which was subsequently confirmed
in an analysis of the \Planck\ 2013 data
\citepalias{planck2013-p09}.

\citet{Cruz2011} developed the unit variance estimator to determine
the variance, $\sigma^2_0$, of the CMB signal on the sky in the
presence of noise.  The normalized CMB map, $u^X$, is given by
\begin{linenomath*}
\begin{equation}\label{normalizedCMBmap}
u^X_i(\sigma^2_{X,0}) = \frac{X_i}{\sqrt{\sigma^2_{X,0}+\sigma^2_{i,\mathrm{noise}}}} \, ,
\end{equation}
\end{linenomath*}
where $X_i$ is the observed temperature at pixel $i$ and
$\sigma^2_{i,\mathrm{noise}}$ is the noise variance for that pixel.
Although this estimator is not optimal, \cite{Cruz2011} and
\cite{Monteserin2008} have demonstrated that it is unbiased and
sufficiently accurate for our purposes.  The noise variance is
estimated from the noise simulations for each component-separation
algorithm.  The CMB variance is then estimated by requiring that the
variance of the normalized map $u^X$ is unity.  The skewness and
kurtosis can then be obtained from the appropriately normalized map.

\begin{figure}
\centerline{
\includegraphics[width=\hsize]{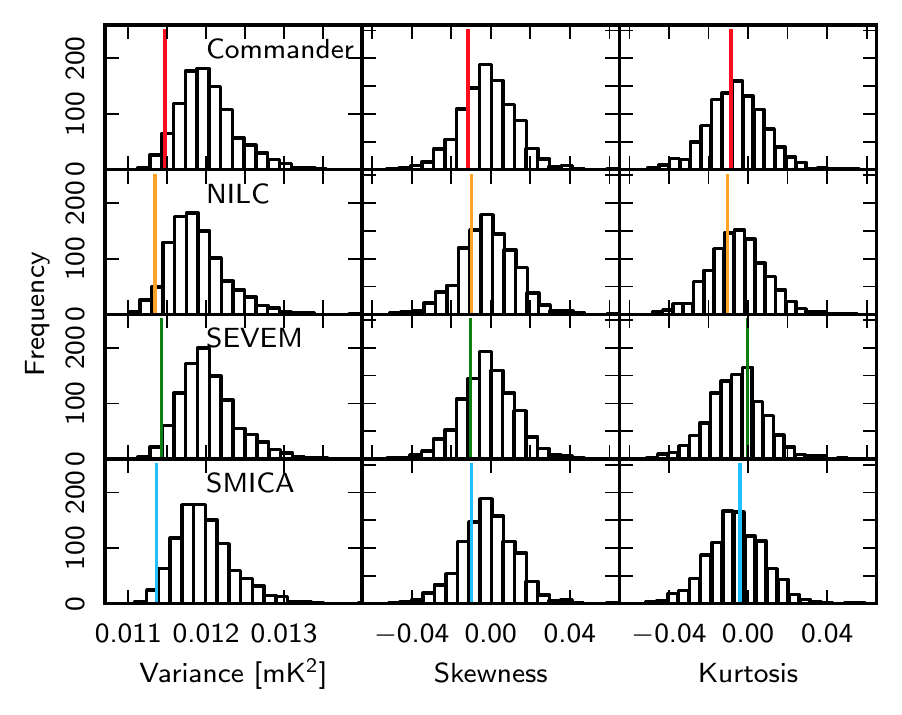}
}
\caption{Variance, skewness, and kurtosis for the four different
  component-separation methods --- \commander\ (red), \nilc\ (orange),
  \sevem\ (green), and \smica\ (blue) --- compared to the distributions
  derived from 1000 Monte Carlo simulations.}
\label{fig:1pdf}
\end{figure}

Figure~\ref{fig:1pdf} presents results for the variance, skewness, and
kurtosis determined from the data at a resolution of 5\arcm,
$N_\mathrm{side}=2048$. Good agreement between the component
separation techniques is found, with small discrepancies likely due to
sensitivity to the noise properties and their variation between
methods.

\begin{table}
\begingroup
\newdimen\tblskip \tblskip=5pt
\caption{Lower-tail probabilities for the variance,
  skewness, and kurtosis of the component-separated maps.
}
\label{tab:onepdf}
\nointerlineskip
\vskip -3mm
\footnotesize
\setbox\tablebox=\vbox{
   \newdimen\digitwidth
   \setbox0=\hbox{\rm 0}
   \digitwidth=\wd0
   \catcode`*=\active
   \def*{\kern\digitwidth}
   \newdimen\signwidth
   \setbox0=\hbox{+}
   \signwidth=\wd0
   \catcode`!=\active
   \def!{\kern\signwidth}
\halign{\hbox to 0.9in{#\leaderfil}\tabskip 4pt&
         \hfil#\hfil&
         \hfil#\hfil&
         \hfil#\hfil\tabskip 0pt\cr   
\noalign{\doubleline\vskip -1pt}
\omit&\multispan3 \hfil Probability [\%] \hfil\cr
\noalign{\vskip -4pt}
\omit&\multispan3\hrulefill\cr
\omit\hfil Method\hfil& Variance& Skewness& Kurtosis\cr   
\noalign{\vskip 4pt\hrule\vskip 6pt}
{\tt Commander}& 3.2& 17.2& 35.3\cr
{\tt NILC}& 3.3& 20.9& 30.9\cr
{\tt SEVEM}& 1.9& 20.5& 56.8\cr
{\tt SMICA}& 1.4& 21.1& 48.2\cr
\noalign{\vskip 4pt\hrule\vskip 6pt}
{\tt SEVEM-100}& 3.4& 13.4& 67.5\cr
{\tt SEVEM-143}& 2.4& 16.9& 61.2\cr
{\tt SEVEM-217}& 3.4& 11.4& 58.3\cr
\noalign{\vskip 3pt\hrule\vskip 4pt}}}
\endPlancktable                    
\endgroup
\end{table}

Table~\ref{tab:onepdf} summarizes the lower-tail probabilities,
defined as the percentage of MC simulations that show a lower
variance, skewness, or kurtosis than the observed map, for these
analyses. The results are in good agreement with
\citetalias{planck2013-p09}; the skewness and
kurtosis are compatible with simulations, but the variance is
marginally lower than in the simulations.

Although the variance is observed to be low, the results could still
be affected by the presence of residual foregrounds at small scales in
these maps, so that the true variance would be lower still. We assess
this by application of the estimator to the cleaned frequency maps
{\tt SEVEM-100}, {\tt SEVEM-143}, and {\tt SEVEM-217}. The results, also
presented in Table~\ref{tab:onepdf}, are similar to those found for
the combined map, although slightly less significant, which is most
likely attributable to higher noise in the cleaned frequency maps.

In conclusion, a simple statistical assessment of the \Planck\ 2015
data using skewness and kurtosis shows no evidence for
non-Gaussianity, although a low variance is found, which we will
readdress in Sect.~\ref{sec:lowvariancemap}.

\subsection{Testing the multi-normality of the CMB}
\label{sec:npdf}
\begin{table}
\begingroup
\newdimen\tblskip \tblskip=5pt
\caption{Lower-tail probabilities for the $N$-pdf $\chi^2$
  statistics derived from the \Planck\ 2015 component-separated maps
  at $N_\mathrm{side}=16$ and 32. }
\label{table:npdf_t_ns}
\nointerlineskip
\vskip -3mm
\footnotesize
\setbox\tablebox=\vbox{
   \newdimen\digitwidth
   \setbox0=\hbox{\rm 0}
   \digitwidth=\wd0
   \catcode`*=\active
   \def*{\kern\digitwidth}
   \newdimen\signwidth
   \setbox0=\hbox{+}
   \signwidth=\wd0
   \catcode`!=\active
   \def!{\kern\signwidth}
\halign{\hbox to 0.7in{#\leaderfil}\tabskip 4pt&
         \hfil#\hfil&
         \hfil#\hfil&
         \hfil#\hfil&
         \hfil#\hfil\tabskip 0pt\cr
\noalign{\doubleline\vskip -1pt}
\omit&\multispan4 \hfil Probability [\%]\hfil\cr
\noalign{\vskip -4pt}
\omit&\multispan4\hrulefill\cr
\omit\hfil$N_{\rm side}$\hfil& {\texttt Comm.}& \nilc& \sevem& \smica\cr  
\noalign{\vskip 4pt\hrule\vskip 6pt}
16& 24.7& 26.2& 25.4& 24.5\cr
32& 11.9& 20.8& 10.6& 10.8\cr
\noalign{\vskip 3pt\hrule\vskip 4pt}}}
\endPlancktable                    
\endgroup
\end{table}
Under the assumption of Gaussianity, the probability density function
(PDF) of the $N$-dimensional pixelized temperature map is given by a
multivariate Gaussian function:
\begin{linenomath*}
\begin{equation}
f\left(\vec{T}\right) = \frac{1}{\left(2\pi\right)^{N_\mathrm{pix}/2}\det{\tens{C}}^{1/2}}
\exp{\left[-\frac{1}{2}\left(\vec{T}\tens{C}^{-1}\vec{T}^\tens{T}\right)\right]},
\label{eq:npdf_npdf}
\end{equation}
\end{linenomath*}
where $\vec{T}$ is a vector formed from the measured temperatures
$T(\vec{x})$ over all positions allowed by the applied mask,
$N_\mathrm{pix}$ is the number of pixels in the vector, and $\tens{C}$
is the covariance of the Gaussian field (of size $N_\mathrm{pix}
\times N_\mathrm{pix}$).

Although the calculation of $\vec{T}\tens{C}^{-1}\vec{T}^\tens{T}$
can be achieved by conjugate gradient methods, the evaluation of
$\det{\tens{C}}$ remains computationally difficult for the full
\Planck\ resolution at {\tt HEALPix} $N_\mathrm{side}=2048$.  At a
lower resolution, the problem is tractable, and the noise level can
also be considered negligible compared to the CMB signal.  That
implies that under the assumption of isotropy the covariance matrix
$\tens{C}$ is fully defined by the \Planck\ angular power spectrum~($C_\ell$):
\begin{linenomath*}
\begin{equation}
\tens{C}_{ij} = \sum_{\ell=2}^{\ell_\mathrm{max}} \frac{2\ell+1}{4\pi} C_\ell b_\ell^2
P_\ell\left(\cos \theta_{ij}\right),
\label{eq:npdf_C}
\end{equation}
\end{linenomath*}
where $\tens{C}_{ij}$ is the covariance between pixels $i$ and $j$,
$\theta_{ij}$ is the angle between them, $P_\ell$ are the Legendre
polynomials, $b_\ell$ is an effective window function describing the
combined effects of the instrumental beam and pixel window at
resolution $N_\mathrm{side}$, and $\ell_\mathrm{max}$ is the maximum
multipole probed.

Under the multivariate Gaussian hypothesis, the argument of the
exponential in Eq.~\eqref{eq:npdf_npdf} should follow a $\chi^2$
distribution with $N_\mathrm{pix}$ degrees of freedom, or,
equivalently (for $N_\mathrm{pix} \gg 1$) a normal distribution
$\mathcal{N}\left(N_\mathrm{pix}, \sqrt{2N_\mathrm{pix}}\right)$.

These $\chi^2$ statistics are computed for the \Planck\ 2015
component-separated CMB maps at $N_\mathrm{side}=16$ and 32, then compared
with the equivalent quantities derived from the corresponding FFP8
simulations.  For those cases in which the covariance matrix is
ill-conditioned, we use a principal component analysis approach
to remove the lowest degenerate eigenvalues of the covariance matrix
\citep[see, e.g.,][]{curto:2011a}. This process is equivalent to
adding uncorrelated regularization noise of amplitude $\approx 1\, \mu$K
to the data before inversion.  The results of the analysis
are presented in Table~\ref{table:npdf_t_ns} and indicate that the
data are consistent with Gaussianity.  We note that the lower-tail
probabilities for the $N$-pdf decrease when the resolution of the data
is increased from $N_\mathrm{side}=16$ to 32. However, this behaviour
is consistent with that seen for simulations, and should not be
considered to be significant.

\subsection{N-point correlation functions}
\label{sec:npoint_correlation}

In this section, we present tests of the non-Gaussianity of the
\Planck\ 2015 temperature CMB maps using real-space $N$-point
correlation functions. While harmonic-space methods are often
preferred over real-space methods for studying primordial
fluctuations, real-space methods have an advantage with respect to
systematic errors and foregrounds, since such effects are usually localized
in real space. It is therefore important to analyse the data in both
spaces in order to highlight different features.

An $N$-point correlation function is defined as the average product of
$N$ temperatures, measured in a fixed relative orientation on the sky,
\begin{linenomath*}
\begin{equation}
C_{N}(\theta_{1}, \ldots, \theta_{2N-3})
= \left\langle T(\vec{\hat{n}}_{1})\cdots
T(\vec{\hat{n}}_{N}) \right\rangle,
\label{eqn:npoint_def}
\end{equation}
\end{linenomath*}
where the unit vectors $\vec{\hat{n}}_1, \ldots,\vec{\hat{n}}_{N}$
span an $N$-point polygon. Under the  assumption of statistical isotropy, these
functions depend only on the shape and size of the $N$-point polygon,
and not on its particular position or orientation on the sky. Hence,
the smallest number of parameters that uniquely determines the shape
and size of the $N$-point polygon is $2N-3$.

The correlation functions are estimated by simple product averages
over all sets of $N$ pixels fulfilling the geometric requirements set
by $\theta_1, \ldots, \theta_{2N-3}$ characterizing the shape and size
of the polygon,
\begin{linenomath*}
\begin{equation}
\hat{C}_{N}(\theta_{1}, \ldots, \theta_{2N-3})
= \frac{\sum_i \left(w_1^i \cdots w_N^i \right) \left( T_1^i \cdots
T_N^i \right) }{\sum_i \left(w_1^i \cdots w_N^i\right)} \ .
\label{eqn:npoint_estim}
\end{equation}
\end{linenomath*}
Pixel weights $w_1^i,\ldots,w_N^i$ can be introduced in order to
reduce noise or mask boundary effects. Here they represent masking by
being set to 1 for included pixels and to 0 for excluded pixels.

The shapes of the polygons selected for the analysis are the
pseudo-collapsed and equilateral configurations for the 3-point
function, and the rhombic configuration for the 4-point function,
composed of two equilateral triangles that share a common side.  We
use the same definition of pseudo-collapsed as in \cite{eriksen2005},
i.e.,~an isosceles triangle where the length of the baseline falls
within the second bin of the separation angles. The length of the
longer edge of the triangle, $\theta$, parameterizes its
size. Analogously, in the case of the equilateral triangle and
rhombus, the size of the polygon is parameterized by the length of the
edge, $\theta$. Note that these functions are chosen for ease
of implementation, not because they are better suited for testing
Gaussianity than other configurations. For a Gaussian field, Wick's
theorem \citep{Wick1950} means that the ensemble average of the
4-point function may be written in terms of the 2-point function. In
the following, all results refer to the connected 4-point function,
i.e., are corrected for this Gaussian contribution.

We use a simple $\chi^2$ statistic to quantify the agreement between
the observed data and simulations, defined by
\begin{linenomath*}
\begin{equation}
\chi^2 = \sum_{i,j=1}^{N_{\rm bin}} \left( \hat{C}_N (\theta_i) -
  \left< C_N (\theta_i) \right> \right) \tens{M}_{ij}^{-1} \left(
  \hat{C}_N(\theta_j) - \left< C_N (\theta_j) \right> \right)
\ .
\label{eqn:chisqr_stat}
\end{equation}
\end{linenomath*}
Here, $\hat{C}_N(\theta_i)$ is the $N$-point correlation function for
the bin with separation angle $\theta_i$, $\left<C_N(\theta_i)\right>
$ is the corresponding average from the MC simulation ensemble, and
$N_\mathrm{bin}$ is the number of bins used for the analysis. If
$\hat{C}^{k}_N(\theta_i)$ is the $k$th simulated $N$-point correlation
function and $N_\mathrm{sim}$ is the number of simulations, then the
covariance matrix $\tens{M}_{ij}$ is given by
\begin{linenomath*}
\begin{multline}
 \tens{M}_{ij} = \frac{1}{N'_\mathrm{sim}}\sum_{k=1}^{N_\mathrm{sim}}
\left( \hat{C}^{(k)}_N(\theta_i) - \bigl<C_N(\theta_i)\bigr>\right) \times \\
\left( \hat{C}^{(k)}_N(\theta_j) -
  \bigl<C_N(\theta_j)\bigr> \right) ,
\label{eqn:cov_mat}
\end{multline}
\end{linenomath*}
where $N'_\mathrm{sim} = N_\mathrm{sim} - 1$.  Following
\citet{hartlap2007}, we then correct for bias in the inverse
covariance matrix by multiplying it by the factor
$(N'_\mathrm{sim}-N_\mathrm{bin}-1)/N'_\mathrm{sim}$.  Below, we quote
the significance level in terms of the fraction of simulations with a
larger $\chi^2$ value than the observed map.

\begin{figure*}[htp!]
\begin{center}
\includegraphics[width=0.47\textwidth]{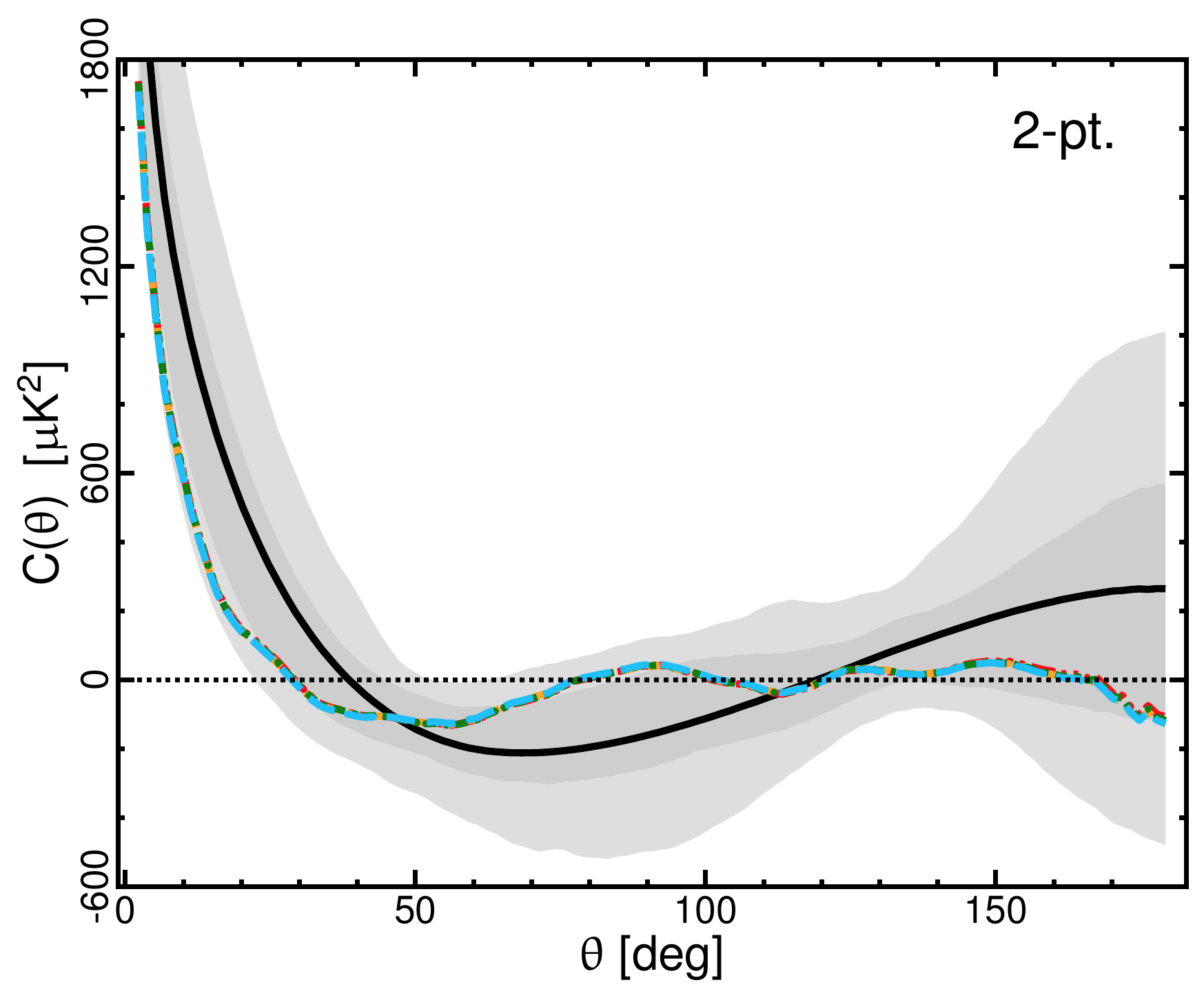}
\includegraphics[width=0.48\textwidth]{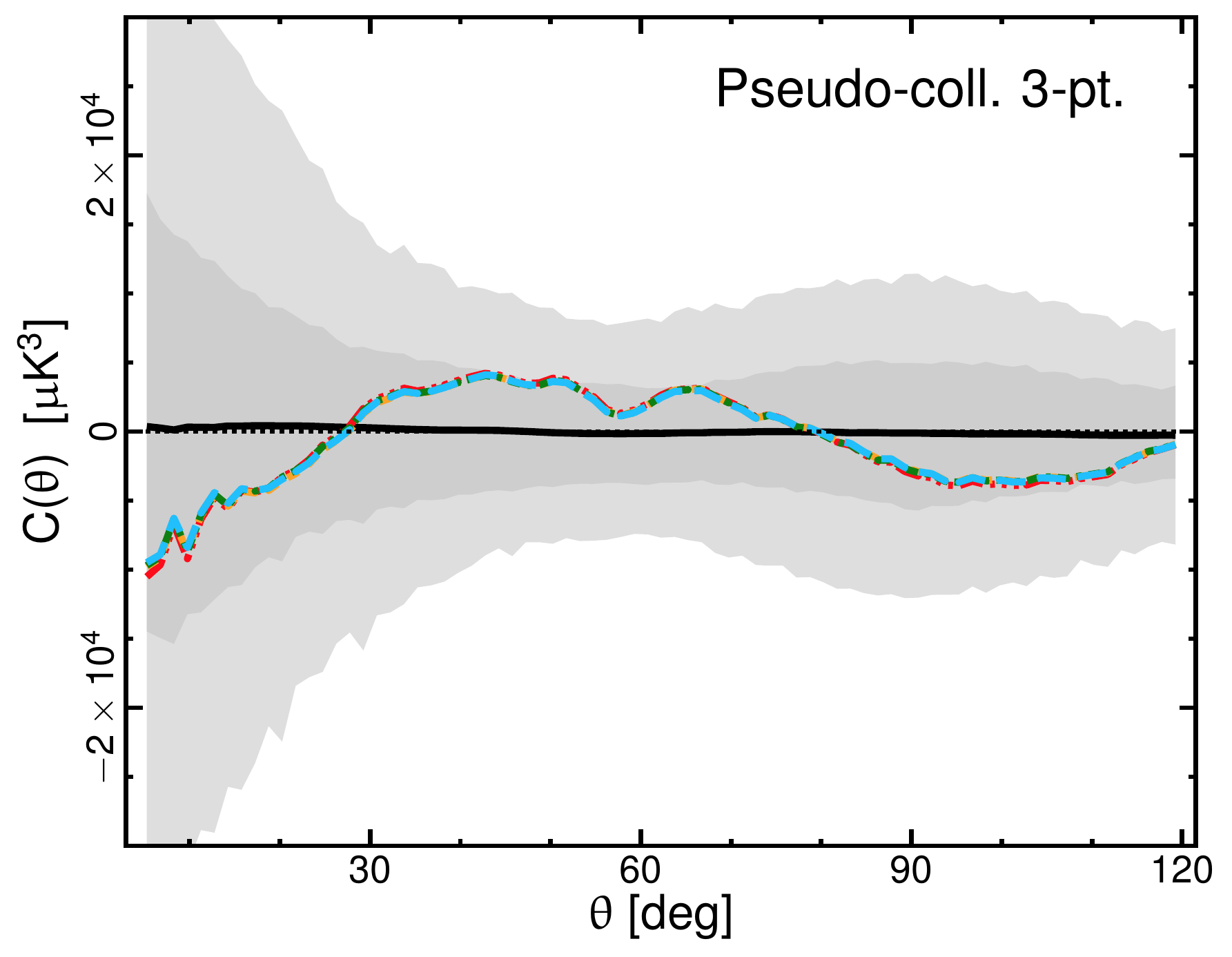}\\
\includegraphics[width=0.48\textwidth]{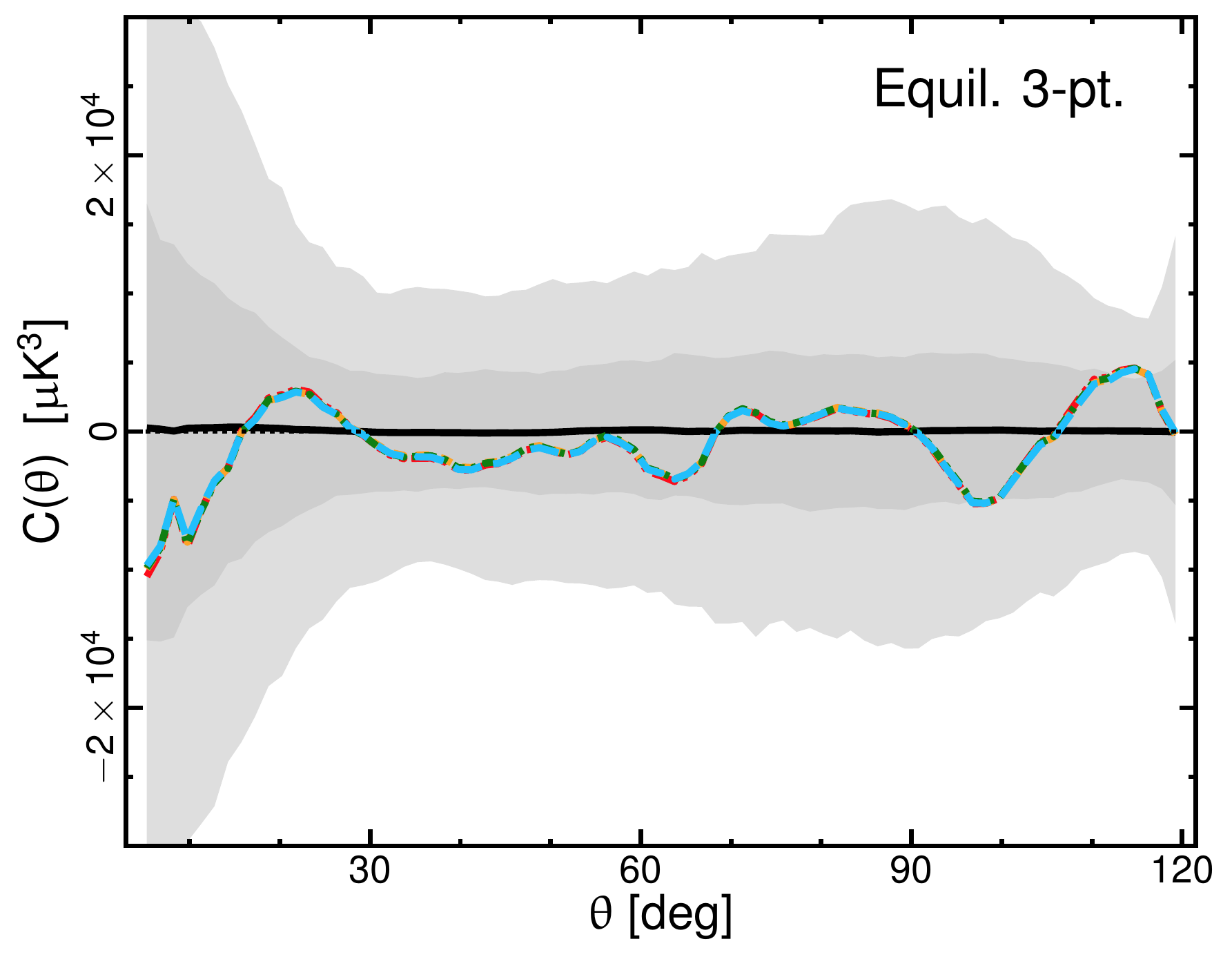}
\includegraphics[width=0.48\textwidth]{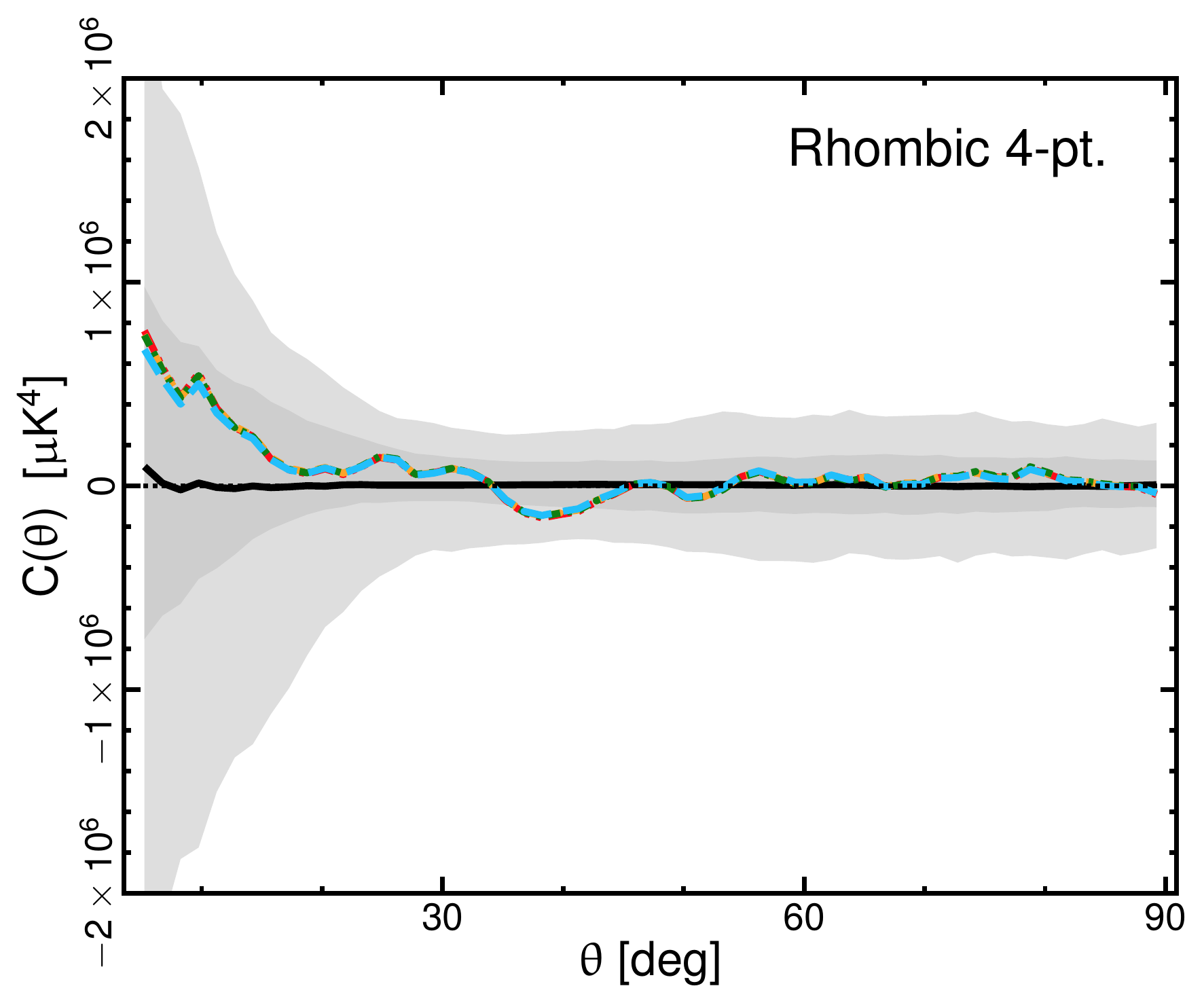}
\caption{$N$-point correlation functions determined from the $\nside=64$
  \Planck\ CMB 2015 temperature maps.  Results are shown for the
  2-point, pseudo-collapsed 3-point (upper left and right panels,
  respectively), equilateral 3-point, and connected rhombic 4-point
  functions (lower left and right panels, respectively).  The red
  dot-dot-dot-dashed, orange dashed, green dot-dashed, and blue long
  dashed lines correspond to the {\tt Commander}, {\tt NILC}, {\tt
    SEVEM}, and {\tt SMICA} maps, respectively. Note that the lines
  lie on top of each other. The black solid line indicates the mean
  determined from 1000 \smica\ simulations. The shaded dark and light
  grey regions indicate the corresponding 68\,\% and 95\,\% confidence
  regions, respectively. See Sect.~\ref{sec:npoint_correlation} for
  the definition of the separation angle~$\theta$.}
\label{fig:npt_data_temp}
\end{center}
\end{figure*}

We analyse the CMB estimates at a resolution of $\nside = 64$, this
being constrained by computational limitations.  The results are
presented in Fig.~\ref{fig:npt_data_temp}, where we compare the
$N$-point functions for the data and the mean values estimated from
1000 MC simulations. The probabilities of obtaining values of the
$\chi^2$ statistic for the \Planck\ fiducial $\Lambda$CDM model at
least as large as the observed values are given in
Table~\ref{tab:prob_chisq_npt_temp}.

\begin{table}[tp]
\begingroup
\newdimen\tblskip \tblskip=5pt
\caption{Probabilities of obtaining values for the $\chi^2$ statistic
  of the $N$-point functions for the
  \Planck\ fiducial $\Lambda$CDM model at least as
  large as the observed values of the statistic for the \Planck\ 2015
  temperature CMB maps with resolution parameter $N_{\rm side}=64$, estimated
  using the {\tt Commander}, {\tt NILC}, {\tt SEVEM},
  and {\tt SMICA} methods.}
\label{tab:prob_chisq_npt_temp}
\nointerlineskip
\vskip -3mm
\footnotesize
\setbox\tablebox=\vbox{
   \newdimen\digitwidth
   \setbox0=\hbox{\rm 0}
   \digitwidth=\wd0
   \catcode`*=\active
   \def*{\kern\digitwidth}
   \newdimen\signwidth
   \setbox0=\hbox{+}
   \signwidth=\wd0
   \catcode`!=\active
   \def!{\kern\signwidth}
\halign{\hbox to 1.6in{#\leaderfil}\tabskip 4pt&
\hfil#\hfil&
\hfil#\hfil&
\hfil#\hfil&
\hfil#\hfil\tabskip 0pt\cr
\noalign{\doubleline\vskip -1pt}
\omit&\multispan4 \hfil Probability [\%]\hfil\cr
\noalign{\vskip -4pt}
\omit&\multispan4\hrulefill\cr
\omit\hfil Function\hfil&{\tt Comm.}&{\tt NILC}&{\tt SEVEM}&{\tt SMICA}\cr
\noalign{\vskip 3pt\hrule\vskip 3pt}
 2-pt.&97.2&98.9&97.4& 98.1\cr
 Pseudo-coll.\ 3-pt.&92.1&94.7&91.8& 92.2\cr
 Equil.\ 3-pt.&74.0&80.4&75.8& 79.0\cr
 Rhombic 4-pt.&64.6& 70.9&65.6& 65.9\cr
 \noalign{\vskip 3pt\hrule\vskip 3pt}}}
\endPlancktable                    
\endgroup
\end{table}

It is worth noting that the values of the $N$-point functions for
different angular separations are strongly correlated, and for this
reason the figures show only one profile of each function in
multi-dimensional space. Since the estimated probabilities take into
account the correlations, they provide more reliable information on
the goodness of fit between the data and a given model than simple
inspection of the figures.

The results show excellent consistency between the CMB maps estimated
using the different component-separation methods. No statistically
significant deviations of the CMB maps from Gaussianity are
found. Indeed, the slight preference for super-Gaussianity of the
equilateral 3-point and 4-point functions observed for the 2013 data
is now less pronounced. That may be caused by differences between the
masks used for the analysis.  Interestingly, the 2-point function
shows clear evidence of a lack of structure for large separation
angles. Such behaviour was originally noted for the WMAP first-year
data by \citet{bennett2003a}, and has subsequently been discussed at
length in the literature \citep{efstathiou2004,Copi2007,copi2013}.  We
will return to this issue in Sect.~\ref{sec:n_point_asymmetry}.

\subsection{Minkowski functionals}
\label{sec:Minkowski_functionals}

\begin{table}[h!tb]
\begingroup
\newdimen\tblskip \tblskip=5pt
\caption{
  Probability $P\left( \chi^2>\chi^2_{\rm Planck}\right)$ as a function of resolution
  for the unnormalized, classical Minkowski functionals.
}
\label{mf:nside_t}
\nointerlineskip
\vskip -3mm
\footnotesize  
\setbox\tablebox=\vbox{
   \newdimen\digitwidth
   \setbox0=\hbox{\rm 0}
   \digitwidth=\wd0
   \catcode`*=\active
   \def*{\kern\digitwidth}
   \newdimen\signwidth
   \setbox0=\hbox{+}
   \signwidth=\wd0
   \catcode`!=\active
   \def!{\kern\signwidth}
\halign{\hbox to 0.8in{#\leaderfil}\tabskip 4pt&
\hfil#\hfil&
\hfil#\hfil&
\hfil#\hfil&
\hfil#\hfil\tabskip 0pt\cr
\noalign{\doubleline}
\noalign{\vskip -1pt}
\omit&\multispan4 \hfil Probability [\%]\hfil\cr
\noalign{\vskip -4pt}
\omit&\multispan4\hrulefill\cr
\omit\hfil $N_{\rm side}$\hfil& {\tt Comm.}& {\tt NILC}& {\tt SEVEM}& {\tt SMICA}\cr
\noalign{\vskip 3pt\hrule\vskip 3pt}
1024& 91.4& 90.7& 95.5& 95.8\cr
*512& 95.4& 90.9& 62.6& 92.6\cr
*256& 55.8& 34.5& 55.9& 55.9\cr
*128& 43.6& 56.4& 19.9& 19.2\cr
**64& 59.3& 37.8& 22.7& 80.0\cr
**32& 62.0& 16.2& 29.9& 67.0\cr
**16& 43.4& 45.8& 47.7& 31.0\cr
\noalign{\vskip 3pt\hrule\vskip 3pt}}} %
\endPlancktable     
\endgroup %
\end{table}

\begin{table}[h!tb]
\begingroup
\newdimen\tblskip \tblskip=5pt
\caption{ Probability $P\left( \chi^2>\chi^2_{\rm Planck}\right)$ as a function of
  resolution determined using normalized MFs.}
\label{tab:mfs_smoothscales}
\nointerlineskip
\vskip -3mm
\footnotesize  
\setbox\tablebox=\vbox{
   \newdimen\digitwidth
   \setbox0=\hbox{\rm 0}
   \digitwidth=\wd0
   \catcode`*=\active
   \def*{\kern\digitwidth}
   \newdimen\signwidth
   \setbox0=\hbox{+}
   \signwidth=\wd0
   \catcode`!=\active
   \def!{\kern\signwidth}
\halign{\hbox to 0.8in{#\leaderfil}\tabskip 4pt&
\hfil#\hfil&
\hfil#\hfil&
\hfil#\hfil&
\hfil#\hfil\/\tabskip=0pt\cr
\noalign{\doubleline}
\noalign{\vskip -1pt}
\omit&\multispan4\hfil Probability [\%]\hfil\cr
\noalign{\vskip -4pt}
\omit&\multispan4\hrulefill\cr
\omit\hfil $N_{\rm side}$\hfil& {\tt Comm.}& {\tt NILC}& {\tt SEVEM}& {\tt SMICA}\cr
\noalign{\vskip 3pt\hrule\vskip 3pt}
2048& 97.2& 77.7& 99.0& 93.0\cr 
1024& 93.1& 98.0& 90.2& 92.6\cr 
*512& 53.7& 36.7& 30.4& 77.6\cr 
*256& 89.0& 85.9& 96.8& 58.1\cr 
*128& 93.0& 63.5& 94.1& 37.1\cr 
**64& 37.1& 70.4& 54.1& 62.5\cr 
**32& 28.9& 77.4& 75.5& 46.7\cr 
**16& 33.1& 39.4& 44.1& 38.8\cr 
\noalign{\vskip 3pt\hrule\vskip 3pt}}} %
\endPlancktable     
\endgroup %
\end{table}

\begin{table}[h!tb]
\begingroup
\newdimen\tblskip \tblskip=5pt
\caption{Probability $P\left( \chi^2>\chi^2_{\rm Planck}\right)$ as a function of needlet
  scale.
}
\label{tab:mfs_needlet}
\nointerlineskip
\vskip -3mm
\footnotesize  
\setbox\tablebox=\vbox{
   \newdimen\digitwidth
   \setbox0=\hbox{\rm 0}
   \digitwidth=\wd0
   \catcode`*=\active
   \def*{\kern\digitwidth}
   \newdimen\signwidth
   \setbox0=\hbox{+}
   \signwidth=\wd0
   \catcode`!=\active
   \def!{\kern\signwidth}
\halign{\hbox to 1.2in{#\leaderfil}\tabskip 4pt&
\hfil#\hfil&
\hfil#\hfil&
\hfil#\hfil&
\hfil#\hfil\/\tabskip=0pt\cr
\noalign{\doubleline}
\noalign{\vskip -1pt}
\omit&\multispan4 \hfil Probability [\%] \hfil\cr
\noalign{\vskip -4pt}
\omit&\multispan4\hrulefill\cr
\omit\hfil Needlet scale ($\ell$ range)\hfil& {\tt Comm.}& {\tt NILC}& {\tt SEVEM}& {\tt SMICA}\cr
\noalign{\vskip 3pt\hrule\vskip 3pt}
    3 (4,16)& 32.1& 36.1& 40.4& 39.8\cr
    4 (8,32)& 84.0& 57.9& 79.4& 59.4\cr
    5 (16,64)& 23.8& 11.2& 29.1& 43.8\cr
    6 (32,128)& 14.8& 38.9& 33.5& 34.1\cr
    7 (64,256)& 11.9& *7.5& 15.4& *1.1\cr
    8 (128,512)& 46.1& 55.2& 67.7& 52.2\cr
\noalign{\vskip 3pt\hrule\vskip 3pt}}} %
\endPlancktable
\endgroup %
\end{table}

The Minkowski functionals (hereafter MFs) describe the morphology of
fields in any dimension and have long been used as estimators of
non-Gaussianity and anisotropy in the CMB \citep[see
e.g.,][]{mecke1994,schmalzing1997,schmalzing1998,komatsu2003,eriksen2004c,curto2007,troia2007,spergel2007,curto2008,hikage2008,komatsu2009,planck2013-p09}. They
are additive for disjoint regions of the sky and invariant under
rotations and translations. In the literature, the
contours are traditionally defined by a threshold $\nu$, usually given in units of
the sky standard deviation ($\sigma_0$).

We compute MFs for the regions colder and hotter than a given
threshold $\nu$. Thus, the three MFs, namely the area
$V_0(\nu)=A(\nu)$, the perimeter $V_1(\nu)=C(\nu)$, and the genus
$V_2(\nu)=G(\nu)$, are defined respectively as
\begin{linenomath*}
\begin{equation}
V_0(\nu) = A(\nu) = \frac{N_\nu}{N_{\rm pix}},
\end{equation}
\end{linenomath*}
\begin{linenomath*}
\begin{equation}
V_1(\nu)= C(\nu) = \frac{1}{4A_{\rm tot}}\sum_{i}S_i,
\end{equation}
\end{linenomath*}
\begin{linenomath*}
\begin{equation}
V_2(\nu)=G(\nu) = \frac{1}{2\pi A_{\rm tot}}\big( N_{\rm hot} - N_{\rm
  cold}\big),
\end{equation}
\end{linenomath*}
where $N_{\nu}$ is the number of pixels where $\Delta T / \sigma_0 >
\nu$, $N_{\rm pix}$ is the total number of available pixels, $A_{\rm
  tot}$ is the total area of the available sky, $N_{\rm hot}$ is the
number of compact hot spots, $N_{\rm cold}$ is the number of compact
cold spots, and $S_i$ is the contour length of each hot spot.

For a Gaussian random field in pixel space, the MFs can be written in
terms of two functions: $A_k$, which depends only on the power spectrum,
and $v_k$, which is a function only of the threshold $\nu$ \citep[see,
e.g.,][]{vanmarcke1983,pogosyan2009a,gay2012a,matsubara2010,fantaye2014a}.
The analytical expressions are
\begin{linenomath*}
\begin{equation}
V_{k}(\nu)=A_{k}v_{k}(\nu),
\end{equation}
\end{linenomath*}
with
\begin{linenomath*}
\begin{eqnarray}
v_{k}(\nu) &= &\exp(-\nu^{2}/2) H_{k-1}(\nu), \quad k \leq 2, \label{eq:nuk1} \\
v_{3}(\nu) &= & \frac{e^{-\nu^{2}}}{\mathrm{erfc}\left(\nu /\sqrt{2}\right)}, \label{eq:nuk2}
\end{eqnarray}
\end{linenomath*}
and
\begin{linenomath*}
\begin{equation}
H_{n}(\nu)={\rm e}^{\nu^{2}/2}\left(-\frac{\rm d}{{\rm d}\nu}\right)^{n} {\rm e}^{-\nu^{2}/2}.
\end{equation}
\end{linenomath*}
The amplitude $A_k$ depends only on the shape of the power spectrum
$C_{\ell}$ through the
rms of the field  $\sigma_{0}$ and its first derivative $\sigma_{1}$:
\begin{linenomath*}
\begin{eqnarray}
A_{k} &= &\frac{1}{(2\pi)^{(k+1)/2}}\frac{\omega_{2}}{\omega_{2-k}\omega_{k}}\left( \frac{\sigma_{1}}{\sqrt{2}\sigma_{0}}\right)^{k},\quad k \leq 2, \\
A_{3} &= &\frac{2}{\pi}\left( \frac{\sigma_{1}}{\sqrt{2}\sigma_{0}}\right)^{2},
 \end{eqnarray}
\end{linenomath*}
where $\omega_{k}\equiv\pi^{k/2}/\Gamma(k/2+1)$.

Since this factorization is still valid in the weakly non-Gaussian
case, we can use the normalized MFs, $v_k$, to focus on
deviations from Gaussianity, with a reduced sensitivity to cosmic
variance.

Apart from the characterization of the MFs using full-resolution
temperature sky maps, we also consider results at different angular
scales. In this paper, two different approaches are considered to
study these degrees of freedom: in real space via a standard Gaussian
smoothing and degradation of the maps; and, for the first time, in
harmonic space by using needlets. Such a complete investigation
provides an insight regarding the harmonic and spatial nature of
possible non-Gaussian features detected with the MFs.

First, we apply scale-dependent analyses in real space by considering
the sky maps at different resolutions. The three classical MFs ---
area, contour length, and genus --- are evaluated over the threshold
range $-3 \le \nu \le 3$ in $\sigma$ units, with a step of 0.5. This
provides a total of 39 different statistics. The values of these
statistics for the \Planck\ data are all within the 95\,\% confidence
region when compared with Gaussian simulations for all of the
resolutions considered. A $\chi^2$ value is computed for each
component-separation method by combining the 39 statistics and taking
into account their correlations \citep[see
e.g.,][]{curto2007,curto2008}. The corresponding covariance matrix is
computed using 1000 simulations. The $p$-value of this $\chi^2$
test is presented in Table~\ref{mf:nside_t} for each component
separation technique and for map resolutions between $\nside =
1024$ and $\nside = 16$. We find no significant deviations from
Gaussianity for any of the resolutions considered.

Then we consider the four normalized functionals described above. For
every scale we used 26 thresholds ranging between $-3.5$ and $3.5$ in
$\sigma$ units, except for $\theta=640'$ where 13 thresholds between
$-3.0$ and $3.0$ in $\sigma$ units were more appropriate.
Table~\ref{tab:mfs_smoothscales} indicates that no significant
deviation from Gaussianity is found.

Third, we tested MFs on needlet components. The needlet components of
the CMB field as defined by \citet{marinucci08a} and \citet{baldi09a} are given
by:
\begin{linenomath*}
\begin{eqnarray}
\beta _{j}(\vec{\hat n})&=&\sum_{\ell =B^{j-1}}^{B^{j+1}}b^{2}\(\frac{\ell }{B^{j}}%
\)\sum_{m}a_{\ell m}Y_{\ell m}(\vec{\hat n})  \nonumber\\
&=&\sum_{\ell =B^{j-1}}^{B^{j+1}}b^{2}\(\frac{\ell }{%
B^{j}}\)T_{\ell }(\vec{\hat n})\text{ .}  \label{needfield}
\end{eqnarray}
\end{linenomath*}
Here, $T_{\ell }(\vec{\hat n})$ denotes the component at multipole
$\ell $ of the CMB map $T(\vec{\hat n})$, i.e.,
\begin{linenomath*}
\begin{equation}
T(\vec{\hat n})=\sum_{\ell }T_{\ell }(\vec{\hat n})\text{ ,}
\end{equation}
\end{linenomath*}
where $\vec{\hat n} \in S^{2}$ denotes the pointing direction, $B$ is
a fixed parameter (usually taken to be between $1$ and $2$) and $b(.)$
is a smooth function such that $\sum_{j}b^{2}({\ell }/{B^{j}})=1$
for all $\ell$. \citet{fantaye2014a} show in a rigorous
way that a general analytical expression for MFs at a given needlet
scale $j$, which deals with an arbitrary mask and takes into account
the spherical geometry of the sky, can be written as
\begin{linenomath*}
\begin{eqnarray}
V_k^j = \sum_{i=0}^k  t_{(2-i)} A_i^j v_i,
 \end{eqnarray}
\end{linenomath*}
where $t_0=2$, $t_1=0$, and $t_2=4\pi$ are respectively the Euler-Poincar\'{e}
characteristic, boundary length, and area of the full sphere. The quantities $v_k$ are the normalized MFs given in
Eq.~\eqref{eq:nuk1}, while the needlet scale amplitudes $A_k^j$
have a similar form as $A_k$ but with the variances of the map and
its first derivative given by
\begin{linenomath*}
\begin{eqnarray}
\sigma_0^2 &=& \sum_{\ell }b^{4}\(\frac{\ell }{B^{j}}\)C_{\ell }%
\frac{2\ell +1}{4\pi } , \\
\sigma_1^2 &=& \sum_{\ell }b^{4}\(\frac{\ell }{B^{j}}\)C_{\ell }%
\frac{2\ell +1}{4\pi }\frac{\ell (\ell +1)}{2} .
 \end{eqnarray}
\end{linenomath*}

Implementing the MFs in needlet space has several advantages: the
needlet filter is localized in pixel space, hence the needlet
component maps are minimally affected by masked regions, especially at
high-frequency $j;$ and the double-localization properties of needlets
(in real and harmonic space) allow a much more precise,
scale-by-scale, interpetation of any possible anomalies. While the
behaviour of standard all-scale MFs is contaminated by the large
cosmic variance of the low multipoles, this is no longer the case for
MFs evaluated at the highest needlet scales; in such circumstances,
the variance of normalized components may be shown to decrease
steadily, entailing a much greater detection power in the presence of
anomalies. Finally, and most importantly, the needlet MFs are more
sensitive to the shape of the power spectrum than the corresponding
all-scale MFs.

The needlet parameters we use in this analysis are $B=2$,
$j=3,4,5,6,7,8$.  Since the masks in pixel space are map-resolution
dependent, we also use different masks for each needlet scale. These
new masks are constructed by multiplying the high-resolution common
mask with the upgraded version of the appropriate low-resolution
common mask. For needlet scales $j=2$ and $j=3$, we use the common
mask defined at $N_\mathrm{side}=16$, and upgraded to
$N_\mathrm{side}=2048$. Similarly, for the higher needlet scales,
$j=2^n$, where $n=4,5,6,7,8$, we use upgraded versions of the common
masks defined at $N_\mathrm{side}=2^n$.

\begin{figure*} 
\begin{center}
\includegraphics[width=0.32\textwidth]{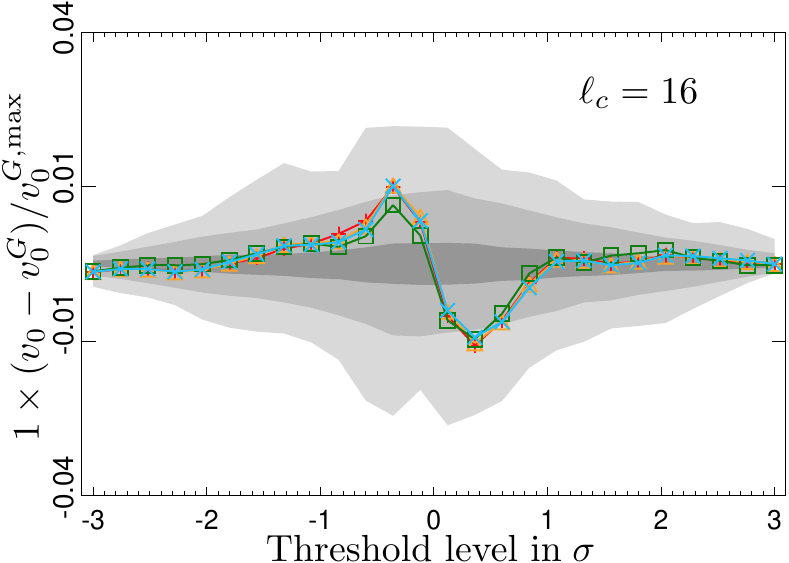}
\includegraphics[width=0.32\textwidth]{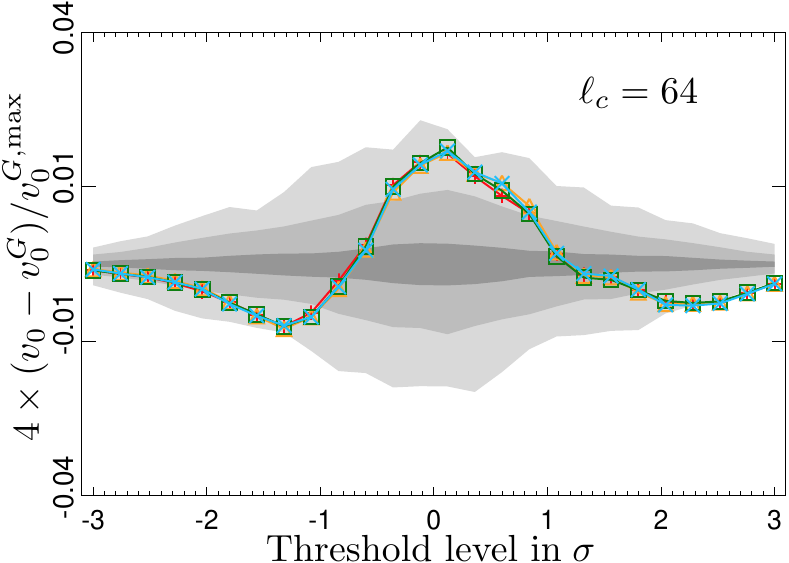}
\includegraphics[width=0.32\textwidth]{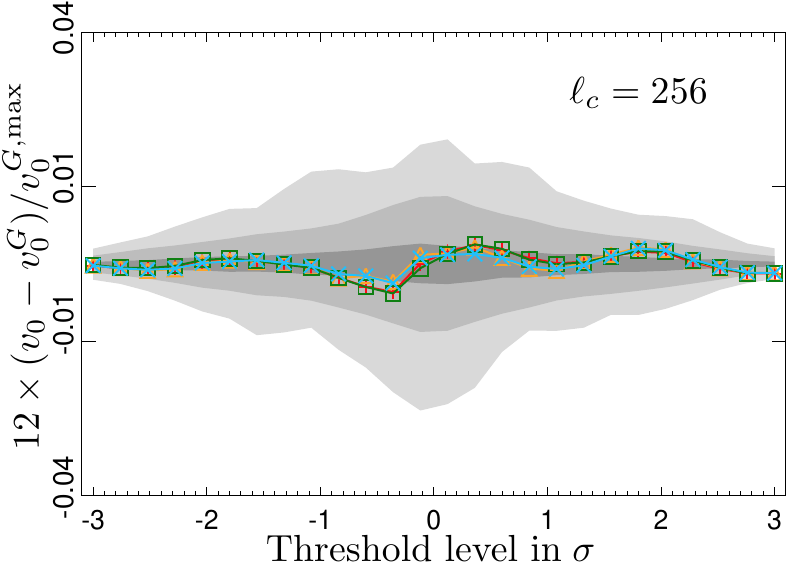} \\
\includegraphics[width=0.32\textwidth]{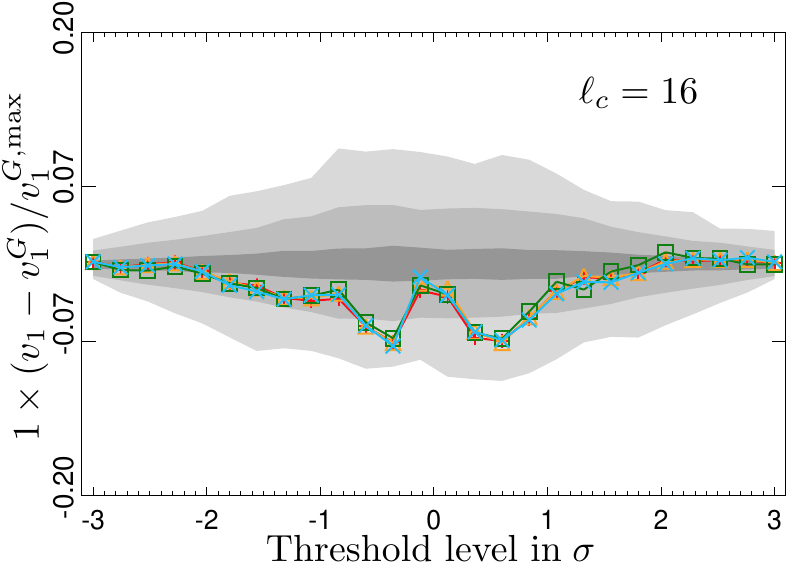}
\includegraphics[width=0.32\textwidth]{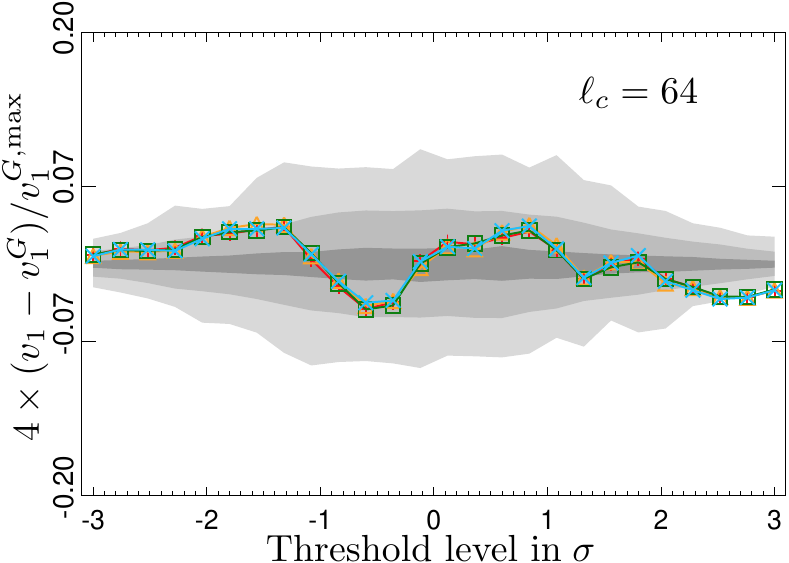}
\includegraphics[width=0.32\textwidth]{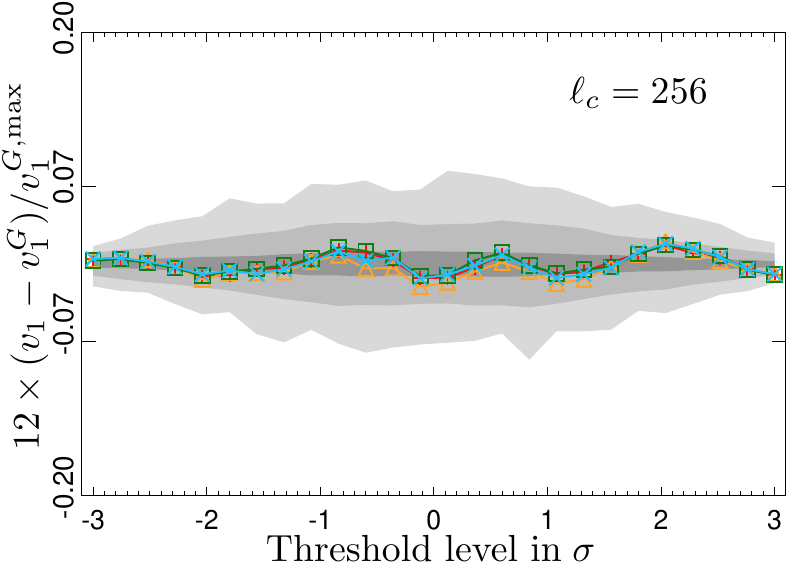}\\
\includegraphics[width=0.32\textwidth]{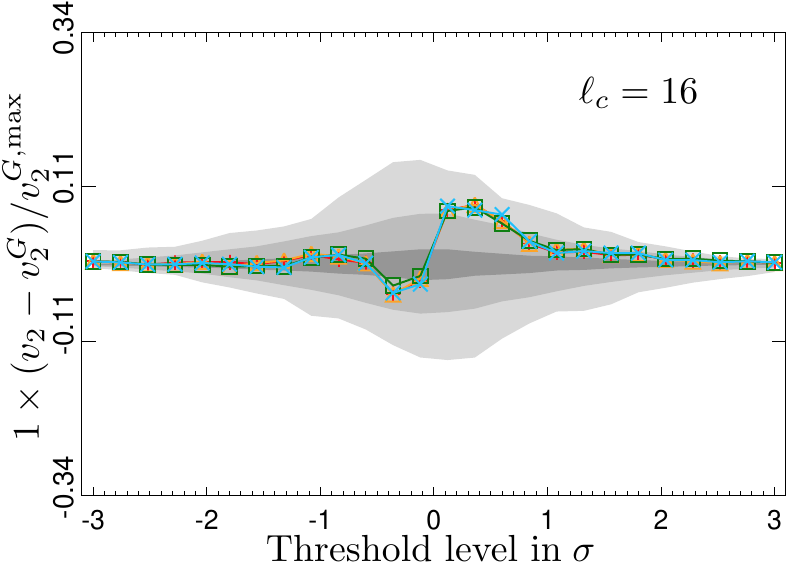}
\includegraphics[width=0.32\textwidth]{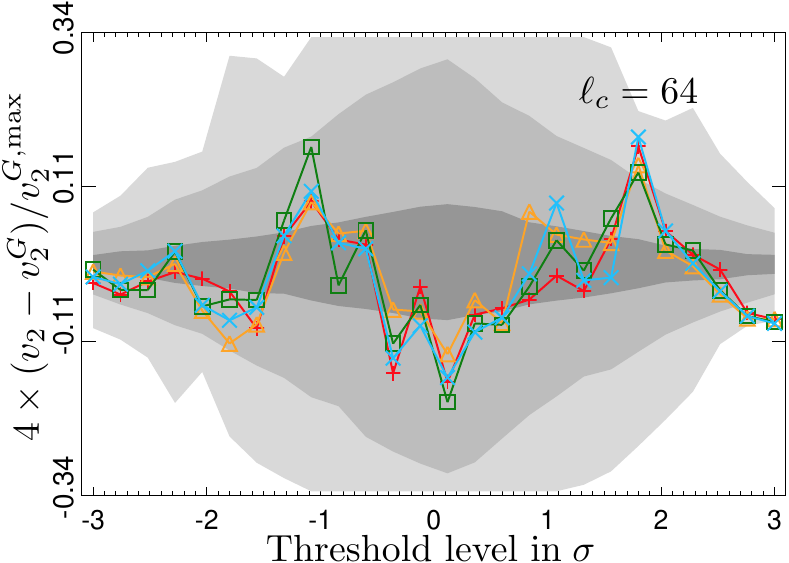}
\includegraphics[width=0.32\textwidth]{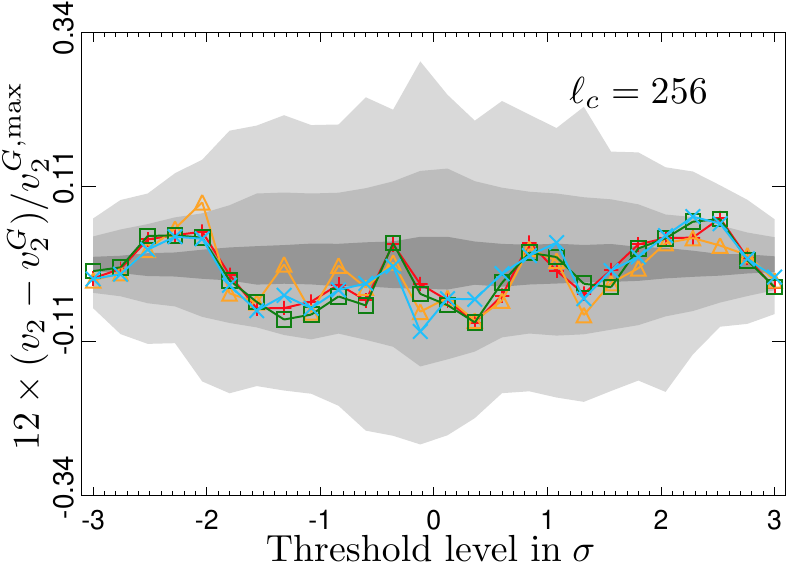}
\caption{Needlet space MFs for \Planck\ 2015 data using the four
  component-separated maps, \commander\ (red), \nilc\ (orange),
  \sevem\ (green), and \smica\ (blue); the grey regions, from dark to
  light, correspond, respectively, to 1, 2, and $3\,\sigma$
  confidence regions estimated from the 1000 FFP8 simulations processed by
  the \commander\ method. The columns from left to right correspond to
  the needlet parameters $j=4, 6,$ and $8$, respectively; the $j$th needlet
  parameter has compact support over multipole ranges $[2^{j-1},
  2^{j+1}]$. The $\ell_c=2^j$ value indicates the central
  multipole of the corresponding needlet map. Note that to have the
  same range at all the needlet scales, the vertical axis has been multiplied by a
  factor that takes into account the steady decrease of the
  variance of the MFs as a function of scale.}
\label{fig:mfneedle1}
\end{center}
\end{figure*}

\begin{figure*} 
\begin{center}
\includegraphics[width=0.32\textwidth]{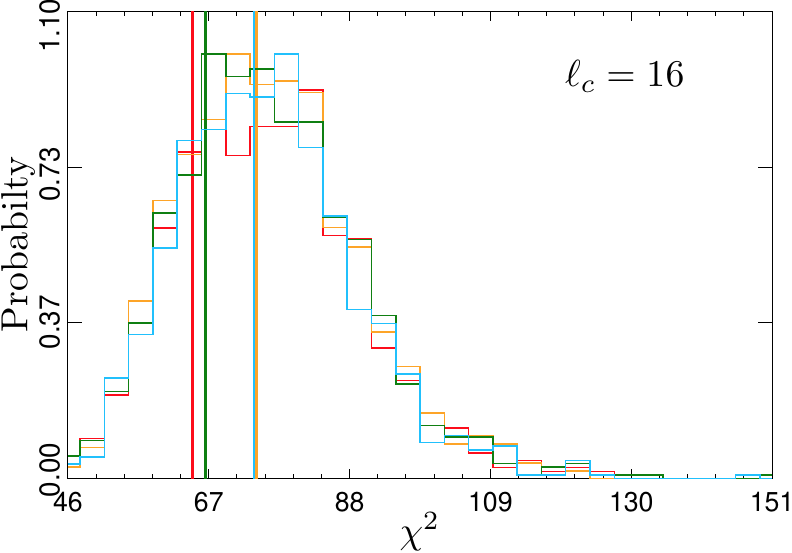}
\includegraphics[width=0.32\textwidth]{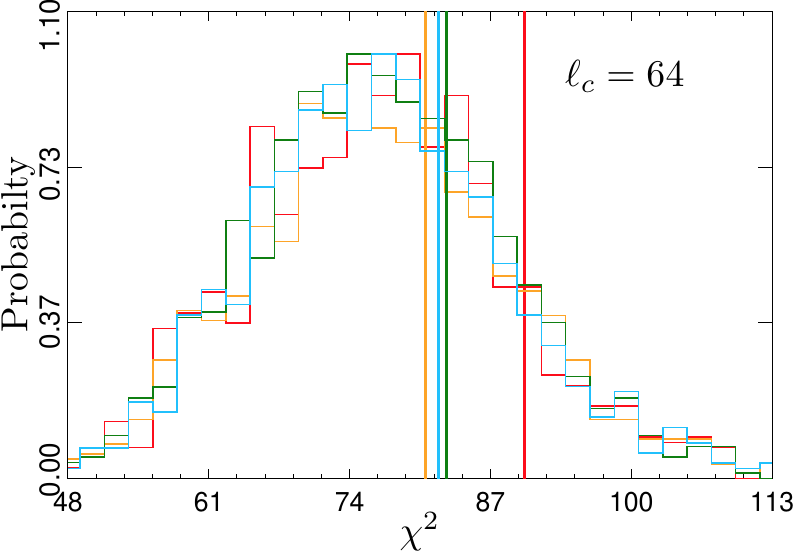}
\includegraphics[width=0.32\textwidth]{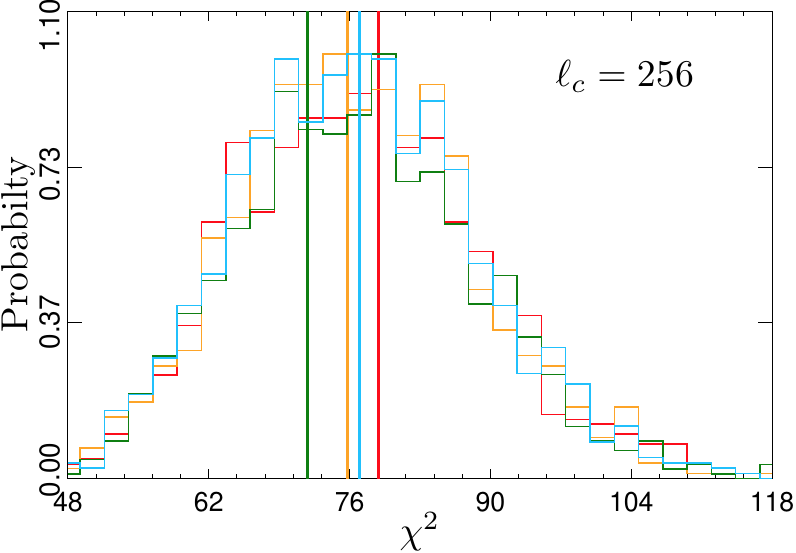}
\caption{Histograms of $\chi^2$ for the \Planck\ 2015
  \commander\ (red), \nilc\ (orange), \sevem\ (green), and \smica\
  (blue) foreground-cleaned maps analysed with the common mask. The
  $\chi^2$ is obtained by combining the three MFs in needlet space
  with an appropiate covariance matrix. The histograms are for the
  FFP8 simulations, while the vertical lines are for the data. The
  figures from left to right are for the needlet scales $j=4, 6,$ and
  $8$, with the central multipoles $\ell_\mathrm{c}=2^j$ shown in each
  panel.}
\label{fig:mfneedle2}
\end{center}
\end{figure*}

The results concerning needlet MFs from the \commander, \nilc, \sevem,
and \smica\ foreground-cleaned temperature maps for needlet scales
$B=2$, $j=4, 6, 8$ are shown in Fig.~\ref{fig:mfneedle1}. All cases
are computed using 26 thresholds ranging between $-3.5$ and $3.5$ in
$\sigma$ units.  The figure shows the fractional
difference between the \Planck\ data and the FFP8 simulations
in area (top panels), boundary
length (middle panels), and genus (bottom panels) for
different needlet scales. The $j$th needlet scale has compact support
over the multipole ranges $[2^{j-1}, 2^{j+1}]$. All the scales we
considered are consistent with the Gaussian FFP8 simulations. This can
be seen in Fig.~\ref{fig:mfneedle2}, where we compare the data and
simulation $\chi^2$ values, which are computed by combining the three
MFs with an appropiate covariance matrix. The vertical lines in these
figures represent the data, while the histogram shows the results for
the 1000 FFP8 simulations. We also show in Table~\ref{tab:mfs_needlet}
the $p$-values for the four component-separation methods, as well
as all needlet scales we considered.  Despite the relatively small
$p$-values for some scales, the \Planck\ temperature maps show no
significant deviation from the Gaussian simulations up to $\ell_{\rm
  max}=512$, which corresponds to the maximum multipole of our highest-frequency needlet map.

\subsection{Multiscale analyses}
\label{sec:multiscale}

Multiscale data analysis is a powerful approach for probing the
fundamental hypotheses of the isotropy and Gaussianity of the CMB. The
exploration of different scales (in an almost independent manner) not
only helps to test the specific predictions of a given scenario for
the origin and evolution of the fluctuations, but also is an important
check on the impact of systematic errors or other contaminants on the
cosmological signal.

There are several ways of performing a multiscale analysis, the
simplest being to smooth/degrade the CMB map to different resolutions.
However, in this section, we will focus on image processing techniques
related to the application of wavelets and more general band-pass
filtering kernels to the original CMB fluctuations. The advantage of
wavelet-like analyses over scale degradation is clear: they allow
the exploration of characteristics of the data that are related to specific
angular scales. Wavelets have already been extensively used in the
study of the Gaussianity and isotropy of the
CMB~\citep[e.g.,][]{mcewen2007,vielva2007b}. Indeed, a wavelet-based
(needlet) analysis of the \Planck\ 2015 data has already been
presented in Sect.~\ref{sec:Minkowski_functionals}.

We recall that in the 2013 analysis, some of the applied estimators
deviated from the null hypothesis. In particular, it was determined
that the cold area of the spherical Mexican hat wavelet
~\citep[SMHW,][]{martinez:2002} coefficients at scales of around
5$\degr$ yielded a \pval\ of 0.3\,\%. In addition, we also found an
excess in the kurtosis of the wavelet coefficients on the same
scales. Previous analyses~\citep[for a review, see][]{vielva2010a} have
suggested that the ``Cold Spot'' (see Sect.~\ref{sec:coldspot}) was
the major contributor to these statistical outliers.

In what follows, we will consider the application of the SMHW,
together with matched filters for a 2D-Gaussian profile (GAUSS), and
for generalized spherical Savitzky-Golay kernels~\citep[SSG,][see
Appendix~\ref{asec:gsgp}]{Savitzky1964}.

The application of a filter $\psi(R,p)$ to a signal on the sky $S(p)$
can be written as
\begin{linenomath*}
\begin{equation}
\omega_S\left(R,p\right) = \sum_{\ell = 0}^{\ell_{\rm max}} \sum_{m = -\ell}^{m = \ell} s_{\ell m}
W^{\psi}_\ell\left(R\right)Y_{\ell m}\left(p\right),
\label{eq:filtering}
\end{equation}
\end{linenomath*}
where $p$ represents a given position/pixel, $R$ parameterizes a
characteristic scale for the filter (e.g., a wavelet scale),
$W^{\psi}_\ell\left(R\right)$ is the window function associated with
the filter $\psi(R,p)$, $\ell_{\rm max}$ is the maximum multipole
allowed by the corresponding {\tt HEALPix} pixelization, and $Y_{\ell
  m}\left(p\right)$ is the spherical harmonic basis. Here, $s_{\ell
  m}$, the spherical harmonic coefficients of the analysed map, are
given by
\begin{linenomath*}
\begin{equation}
s_{\ell m} = \int \dd\Omega\, Y^*_{\ell m}\left(p\right) S\left(p\right),
\label{eq:cs_def2}
\end{equation}
\end{linenomath*}
where $\mathrm{d}\Omega = \mathrm{d}\theta\sin{\theta}\mathrm{d}\phi$
and the asterisk denotes complex conjugation.  Note that the filtered
map (or the wavelet coefficient map, if $\psi(R,p)$ is a continous
wavelet) conserves the statistical properties of the original map,
since the convolution is a linear operation. In particular, if $S(p)$
is a Gaussian and statistically isotropic random signal,
$\omega_S\left(R,p\right)$ is also Gaussian and statistically
isotropic.

In the present work, the signal $S(p)$ corresponds to a
temperature map $T(p)$.  Several statistics can then be
computed from the derived filtered map as a function of the filter
scale, in particular, the first moments (the dispersion $\sigma_R$,
the skewness $S_R$, and the kurtosis $K_R$), the total area
above/below a given threshold, and the peak distribution.  These
statistics are compared to the corresponding results determined from
the FFP8 simulations to establish the degree of compatibility with the
null hypothesis.

\subsubsection{First-order moments of the multiscale maps}

\begin{figure} \begin{center}
    \includegraphics[width=9cm]{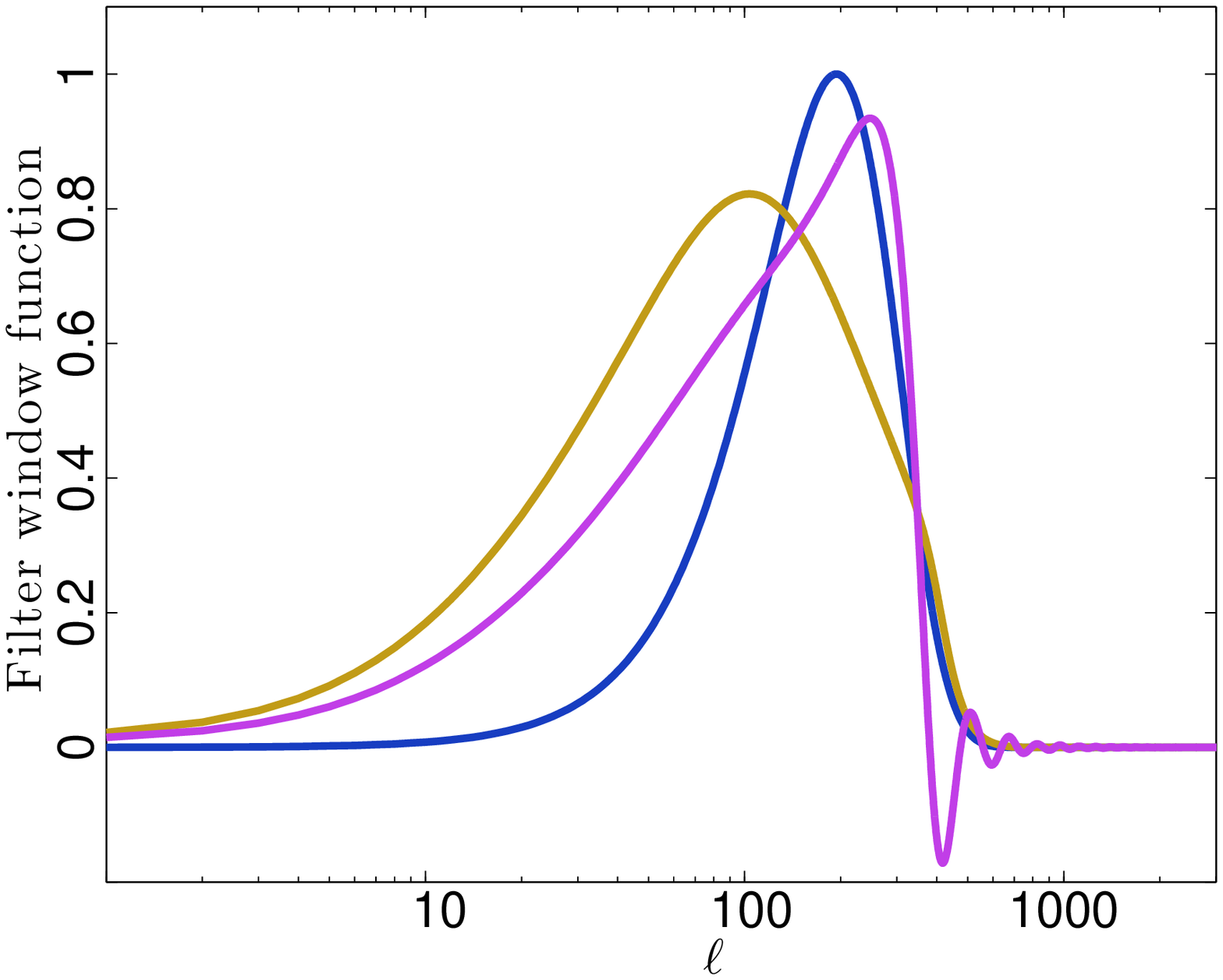}
    \includegraphics[width=9cm]{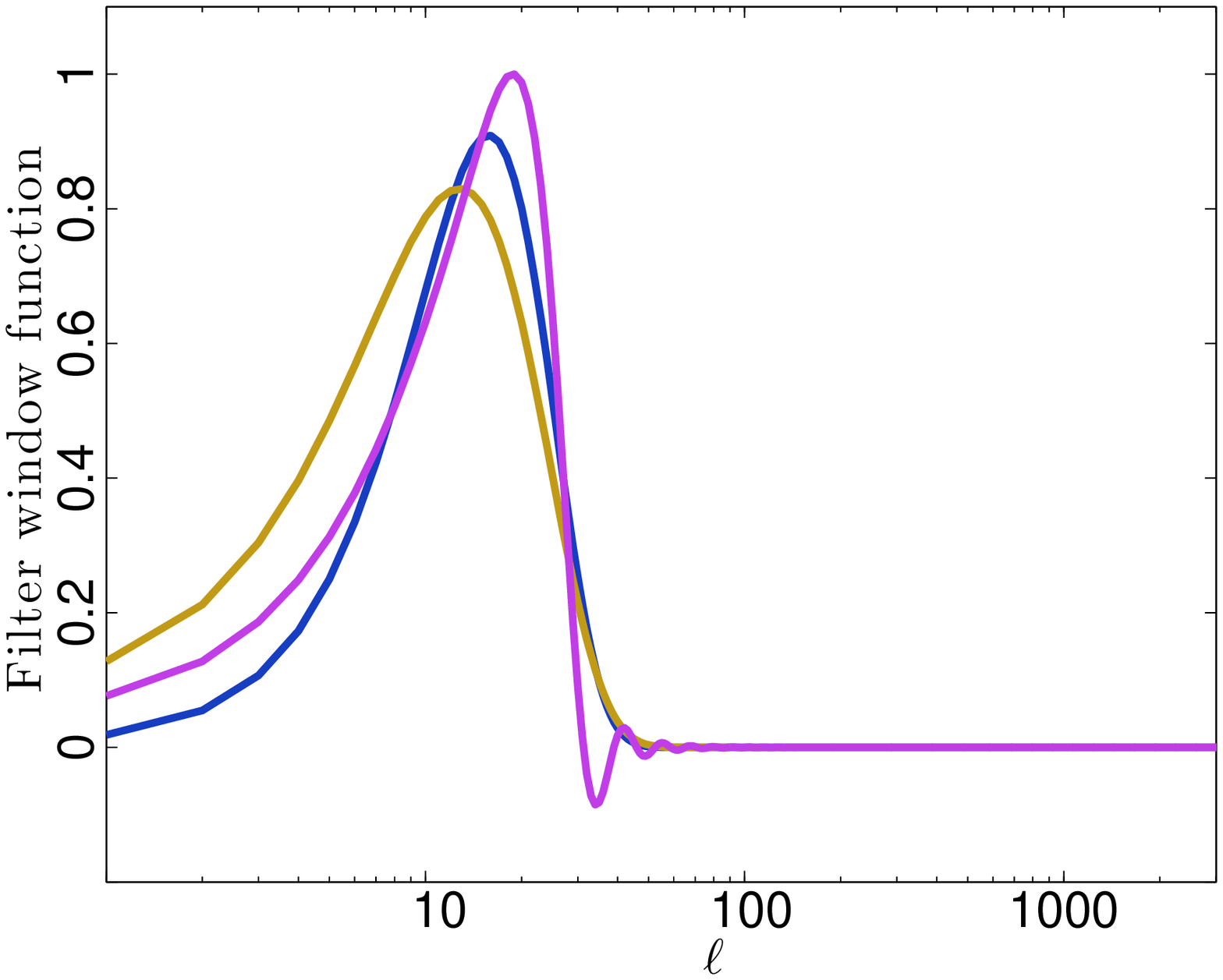}
    \caption{Comparison of the window functions (normalized to have
      equal area) for the SMHW (blue), GAUSS (yellow), and SSG84 (magenta)
      filters. The scales shown are 25\arcm\ (top) and 250\arcm\
      (bottom). } \label{fig:smhw_gauss_ssg84_comparison} \end{center}
\end{figure}

For the three filters considered (SMHW, GAUSS, and SSG84\footnote{The
  digits 8 and 4 denote the order of the spherical Savitzky-Golay
  kernel and the smoothing weight, described in
  Appendix~\ref{asec:gsgp}.}) the variance,
skewness, and kurtosis are computed at 18
scales, $R \mathrm{(arcmin)}= \lbrace$2, 4, 7, 14, 25, 50, 75, 100,
150, 200, 250, 300, 400, 500, 600, 750, 900, 1050$\rbrace$. These
scales are chosen to be consistent with previous analyses.  They cover
a wide angular range, and are selected so that the intervals between
them increase with scale.  Notice that, for a given scale, the three
filters do not cover exactly the same multipole range, since that
depends on the specific filter definition. This can be seen in
Fig.~\ref{fig:smhw_gauss_ssg84_comparison}: the SMHW is the
narrowest filter, followed by SSG84, then GAUSS. The three filters
have an equivalent effective $\ell_\mathrm{max}$, but differ
in the effective $\ell_\mathrm{min}$. Overall, the differences
between the filters become smaller with increasing effective scale.
In this paper, we refer to both the scale, $R$, and FWHM as parameters
defining the size of the filters. For the SMHW, these are related by
$\mathrm{FWHM} = R\sqrt{8\ln 2}$, whereas for the GAUSS and SSG84
filters, the scale is defined to be half the FWHM.  The latter
definition is appropriate for filters that include pre-whitening since
it is simple yet matches the $\ell$-space bandwidth reasonably well.

Following the procedure explained in~\citetalias{planck2013-p09},
after convolution with a given filter, the common mask is extended to
omit pixels from the analysis that could be contaminated by the
mask. These pixels introduce an extra correlation between the data and
the simulations, degrading the statistical power of the comparison with
the null hypothesis \citep[see, e.g.,][]{vielva2004}. For a
given scale $R$, the exclusion mask is defined by extending an
auxiliary mask by a distance $2R$ from its border, where the auxiliary
mask is that part of the common mask related to residual diffuse
Galacic emission (i.e., the auxiliary mask does not mask
point sources).

The following figures represent the upper-tail probability (UTP) for a
given statistic, i.e., the fraction of simulations that yield a value equal
to or greater than that obtained for the data. In fact, as explained
in~\citetalias{planck2013-p09}, if a given UTP is larger than 0.5, a
new quantity is defined as mUTP = $1 - \mathrm{UTP}$. Therefore, mUTP is
constrained to lie between $1/N$ and 0.5, where $N$ is the number of
simulations used for each statistic.

\begin{figure*}[h!]
\begin{center}
    \includegraphics[width=5.9cm]{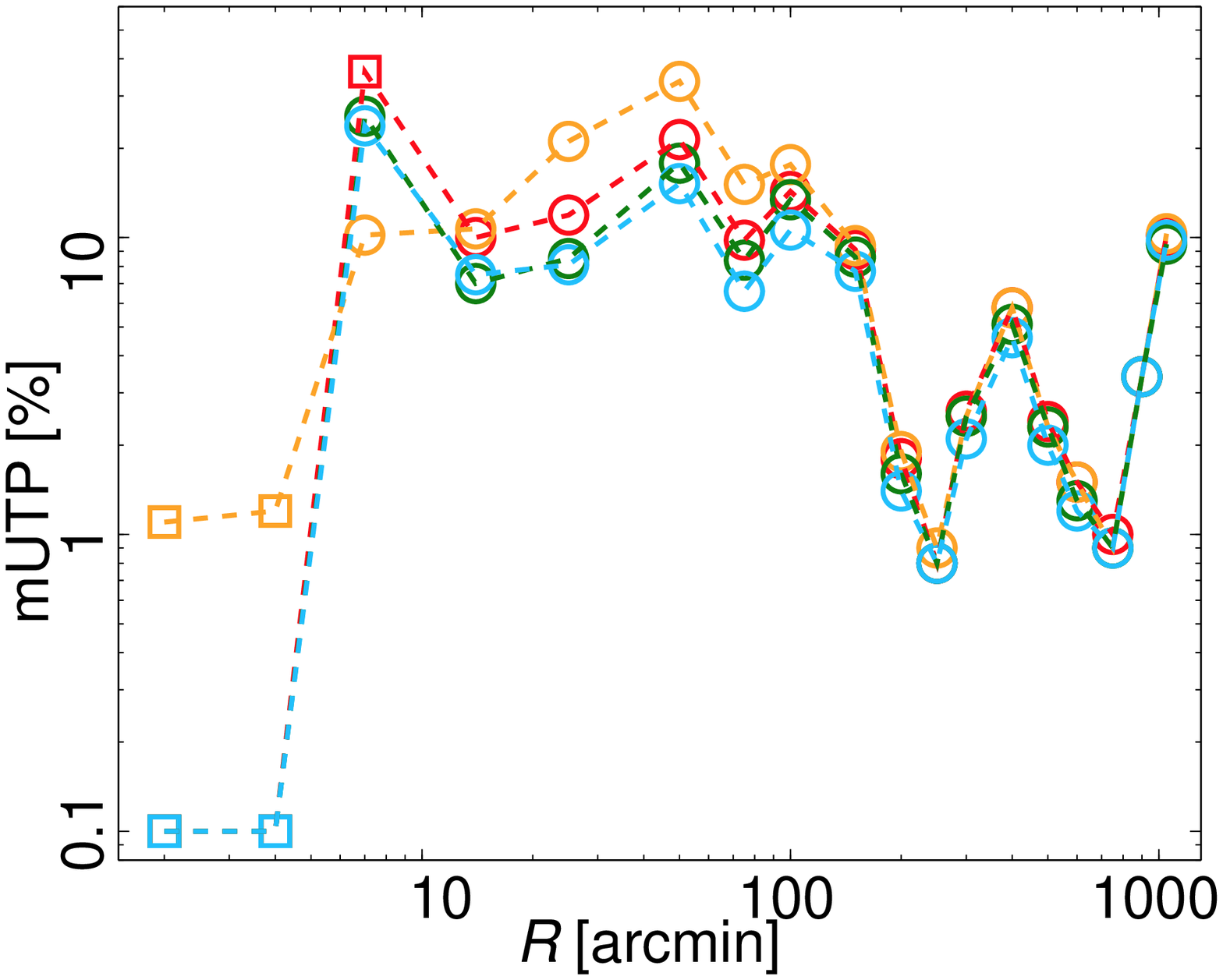}
    \includegraphics[width=5.9cm]{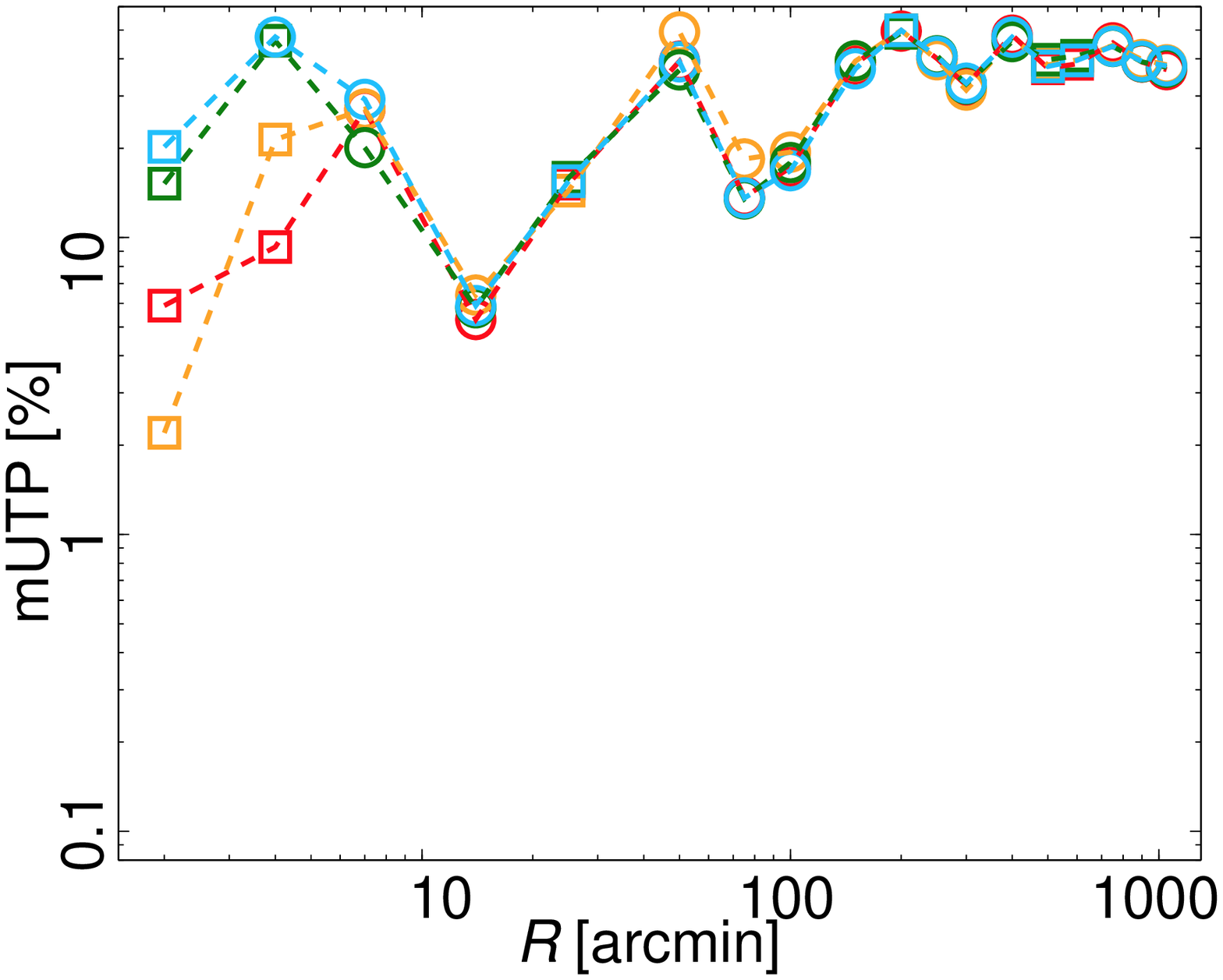}
    \includegraphics[width=5.9cm]{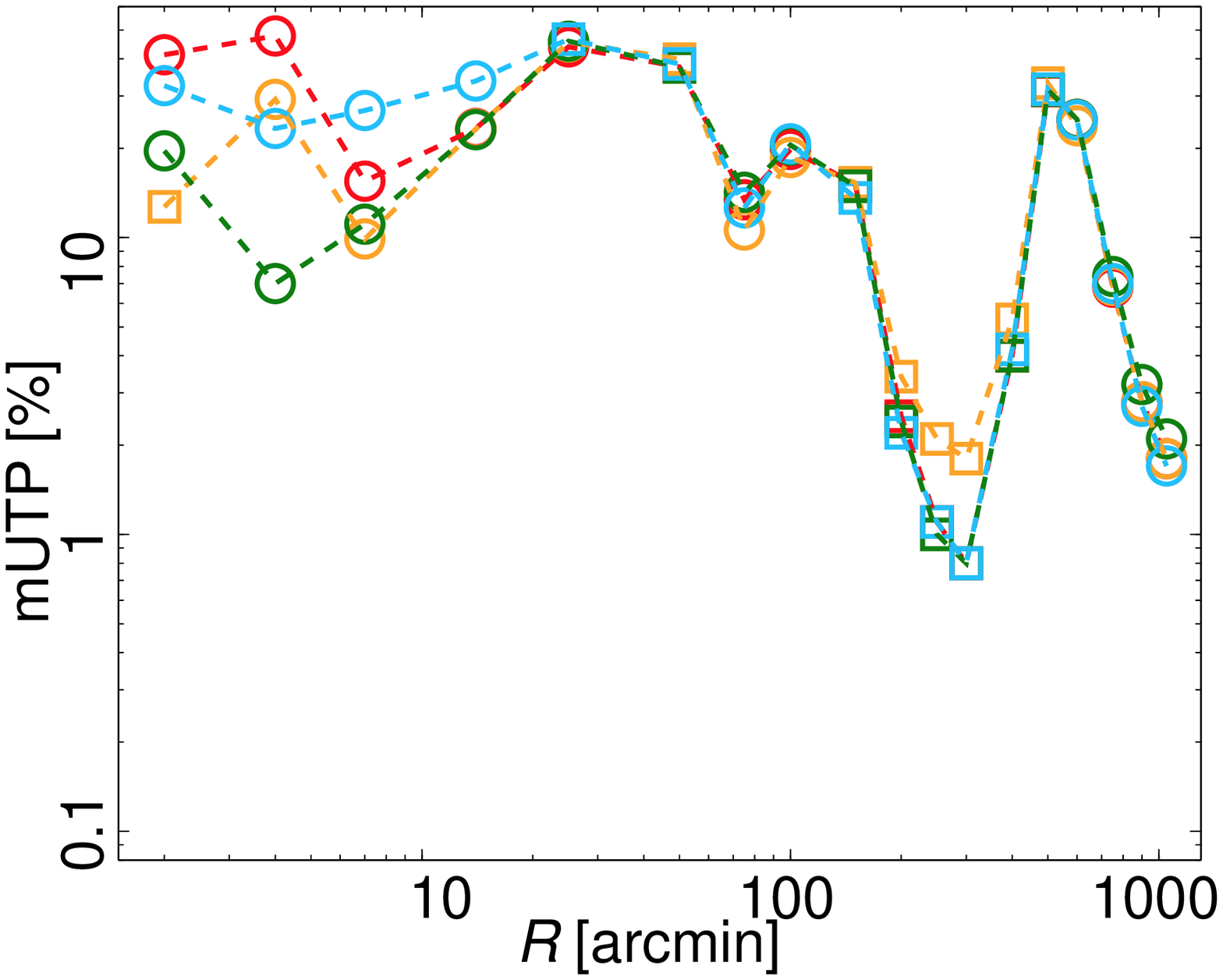}\\
    \includegraphics[width=5.9cm]{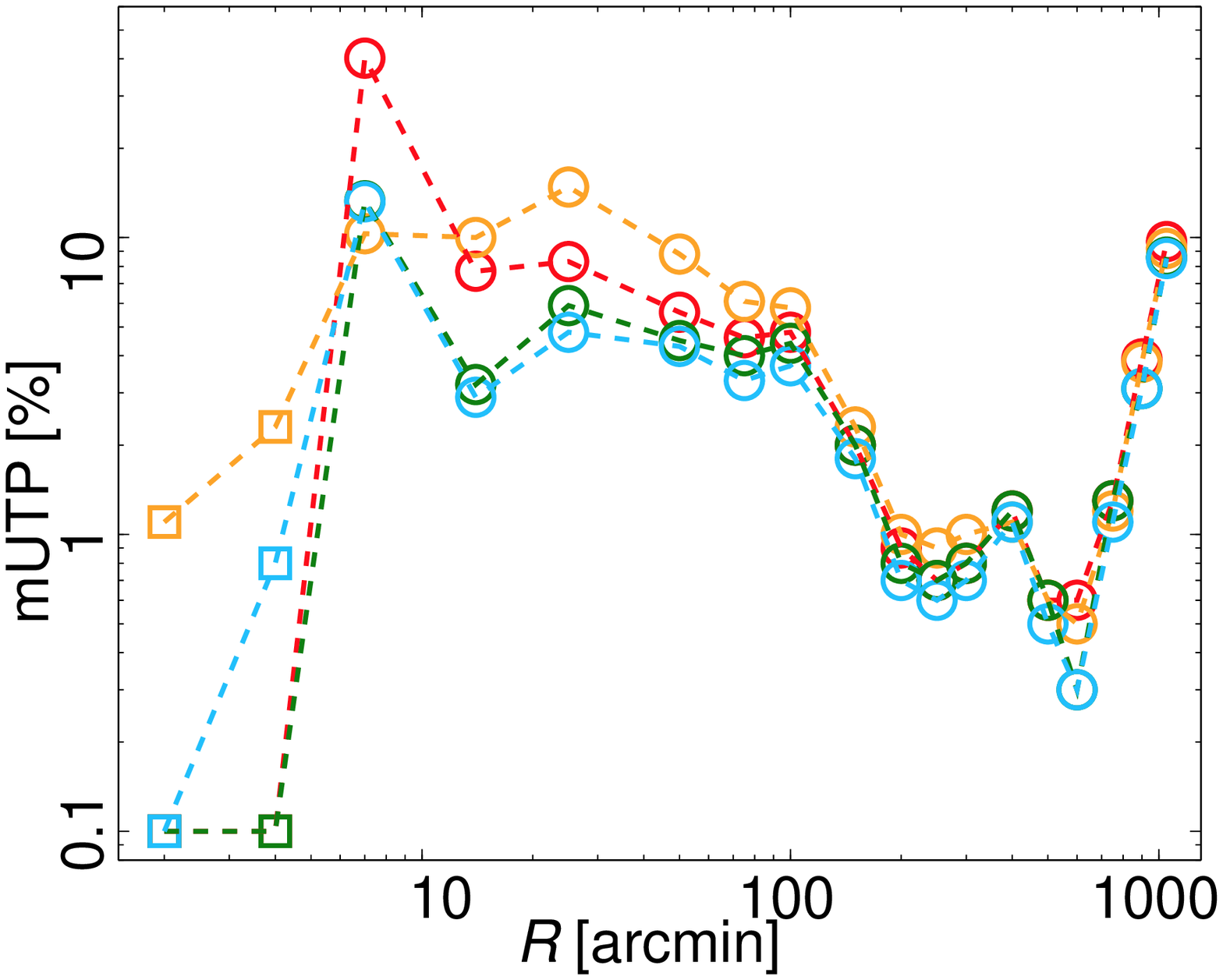}
    \includegraphics[width=5.9cm]{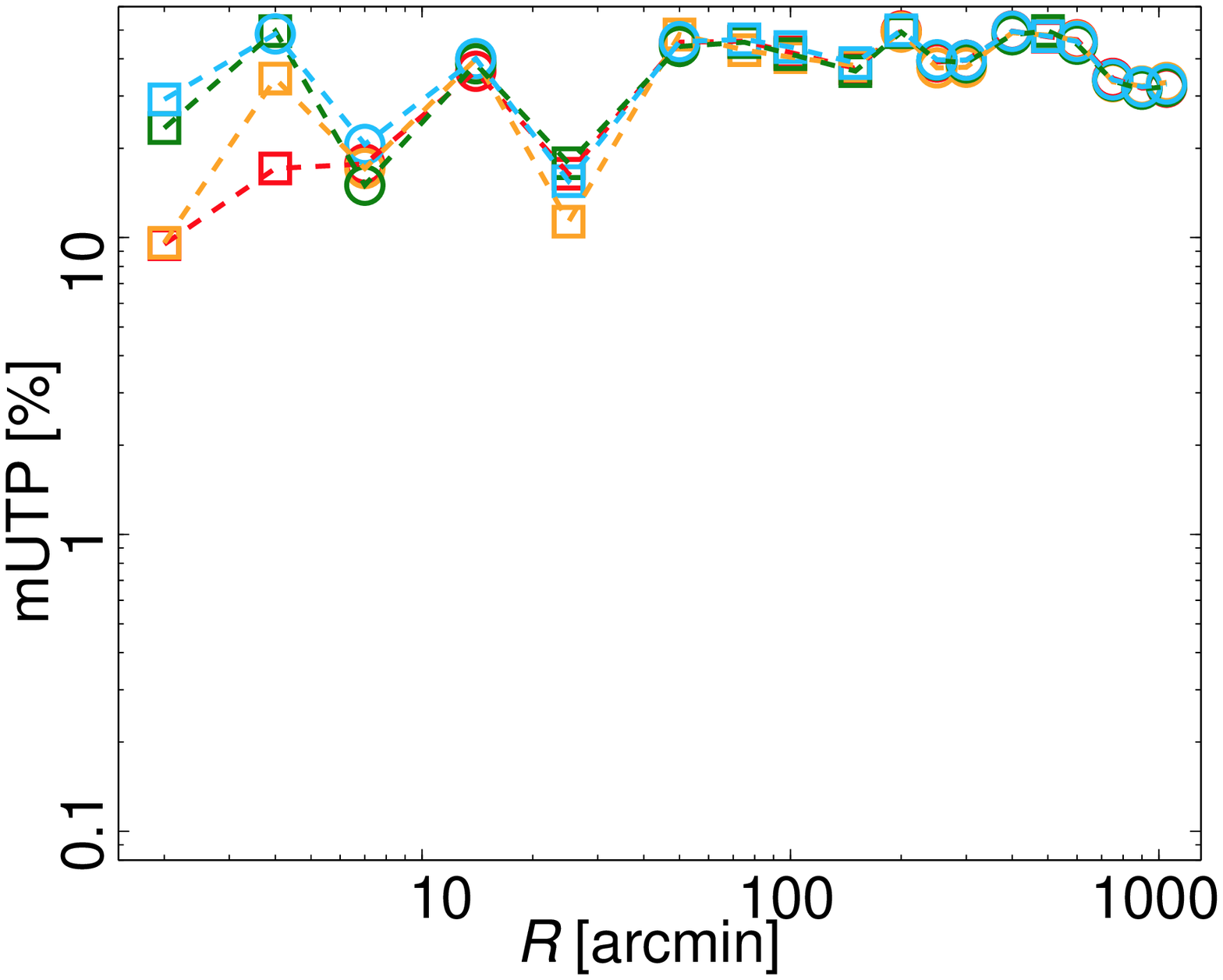}
    \includegraphics[width=5.9cm]{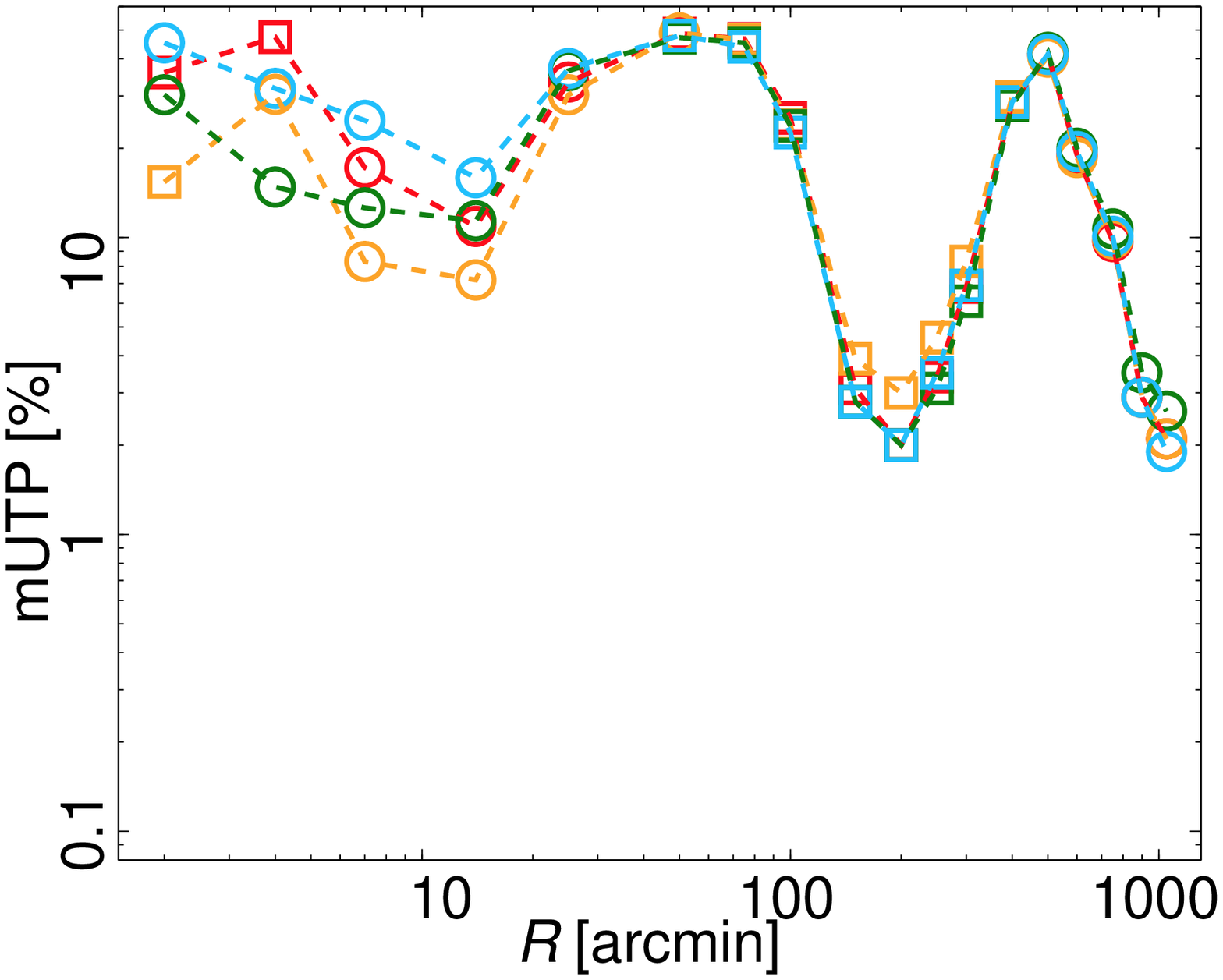}\\
    \includegraphics[width=5.9cm]{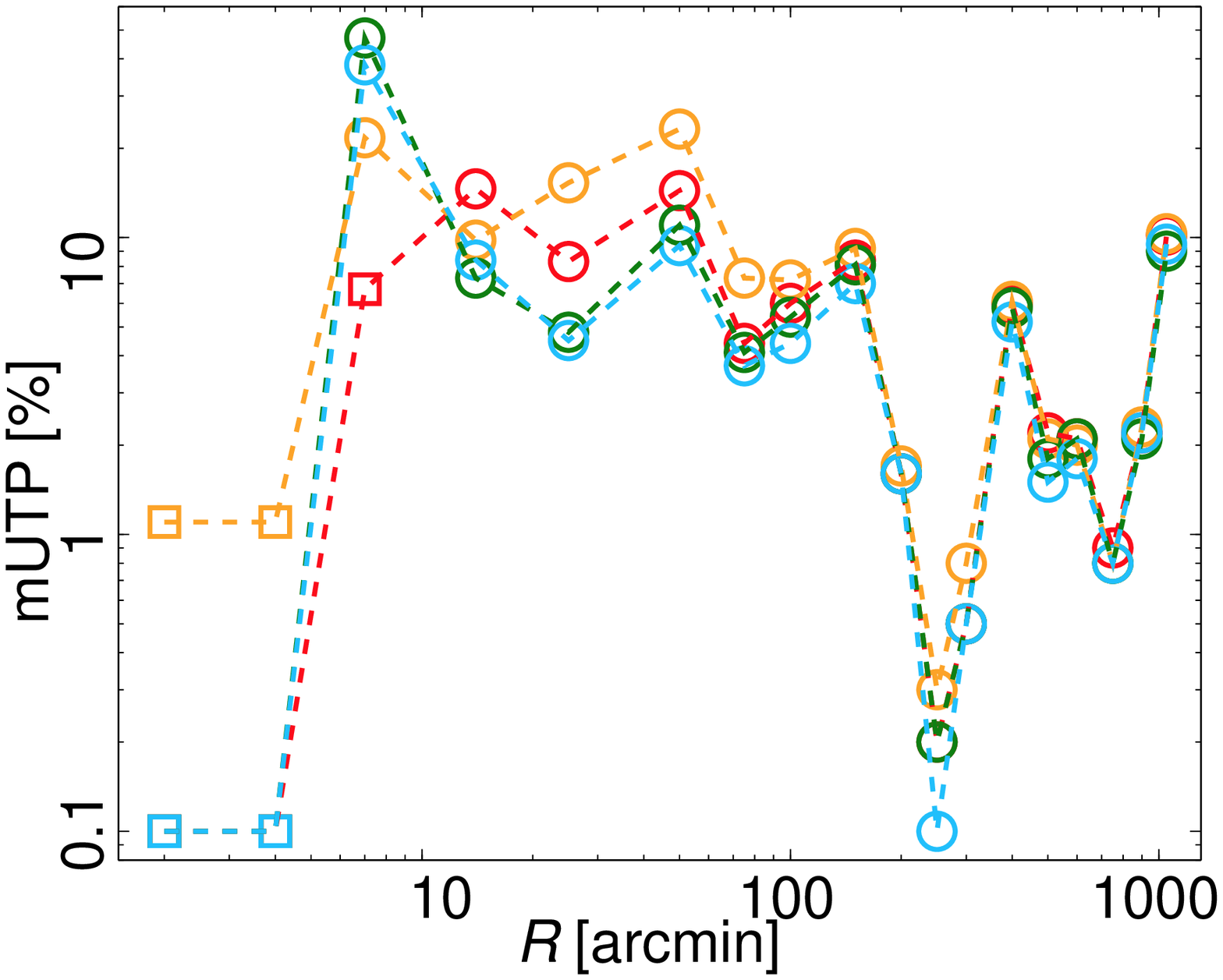}
    \includegraphics[width=5.9cm]{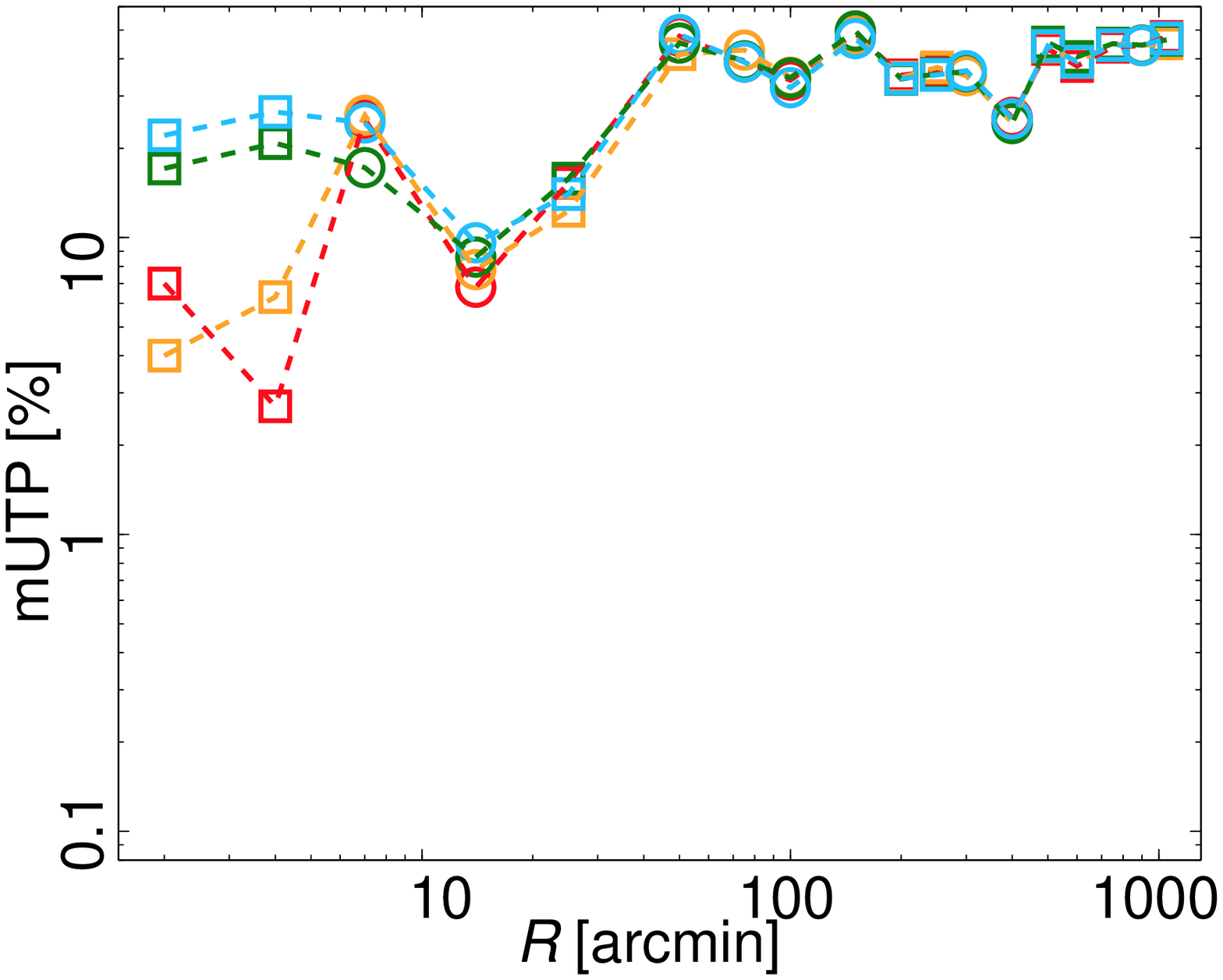}
    \includegraphics[width=5.9cm]{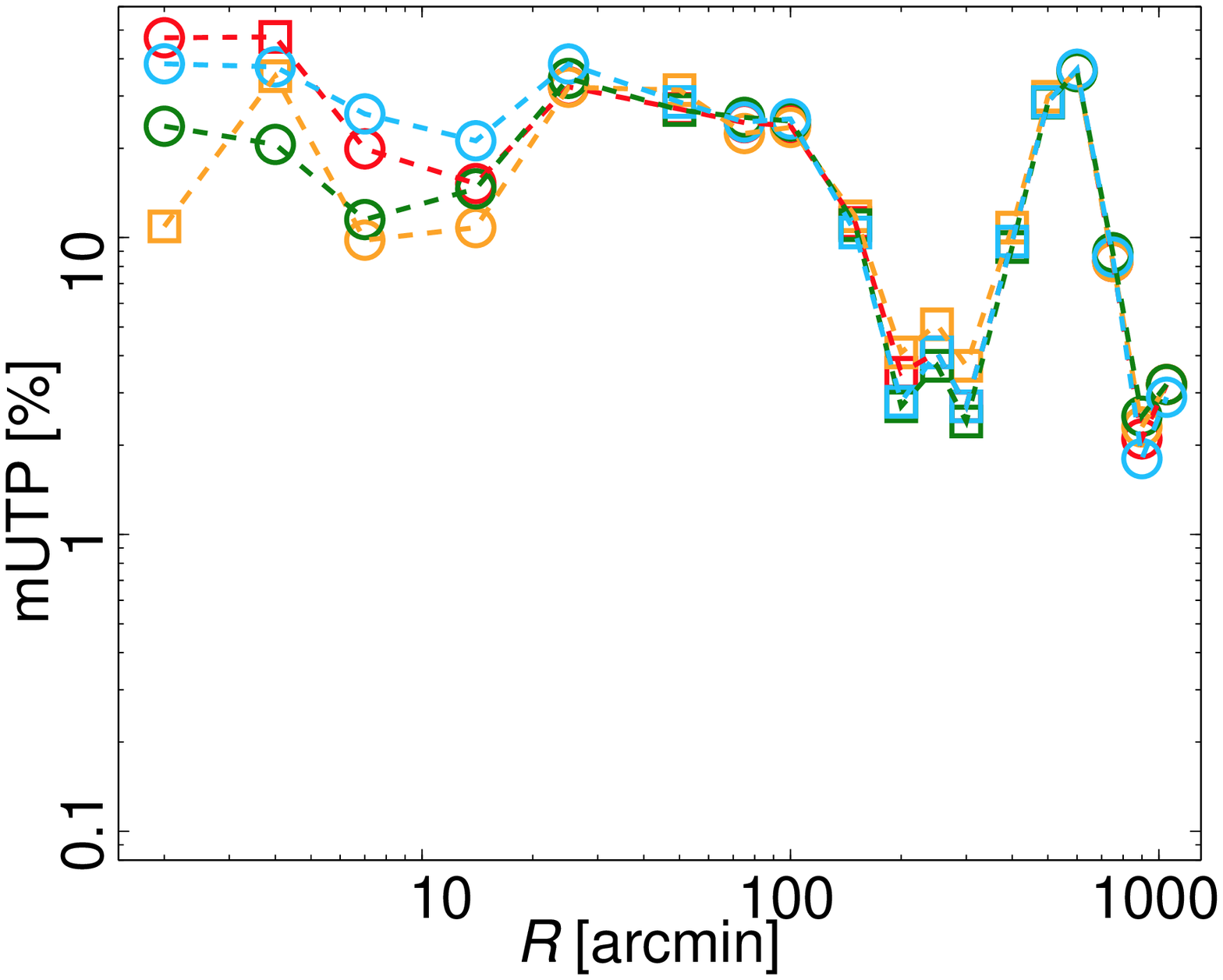}

    \caption{Modified upper tail probabilities (mUTP) obtained from
      the analyses of the filter coefficients as a function of the
      filter scale $R$ for the \commander\ (red), \nilc\ (orange),
      \sevem\ (green), and \smica\ (blue) sky maps. From left to
      right, the panels correspond to standard deviation, skewness, and
      kurtosis results, when determined using the SMHW (top), GAUSS
      (middle), and SSG84 (bottom) filters. The squares represent
      UTP values above 0.5, whereas circles represent UTP values below 0.5.  }
\label{fig:T_stat}
\end{center}
\end{figure*}

Fig.~\ref{fig:T_stat} presents the comparison of the CMB temperature
maps with the corresponding simulations for the SMHW, GAUSS, and SSG84
filters.  The full mission \Planck\ data confirm the results already
obtained with the 2013 release for temperature. In particular, for the
SMHW, we find (i)~an excess of kurtosis ($\approx 0.8\,\%$) at
scales of around 300\arcm; (ii)~that the dispersion of the wavelet
coefficients at these scales and at around 700\arcm\ is relatively
low ($\approx 1\,\%$); and (iii)~that the dispersion of the wavelet
coefficients at scales below 5\arcm\ is significantly high ($\lesssim
0.1\,\%$).

The excess of kurtosis has been previously associated with the
``\cs''\ \citep[e.g.,][]{vielva2004}, and the low value of the
standard deviation of the coefficients on large scales could be
related to the low variance discussed in
Sect.~\ref{sec:lowvariancemap}. Regarding the large dispersion of the
coefficients on the smallest scales, this can be understood either by
the presence of residual foreground contributions (extragalactic point
sources) or by incomplete characterization of the true instrumental
noise properties by the FFP8 simulations. We explore these
possibilities with two additional tests undertaken with the SMHW.

\begin{figure*}[h!]
\begin{center}
    \includegraphics[width=5.9cm]{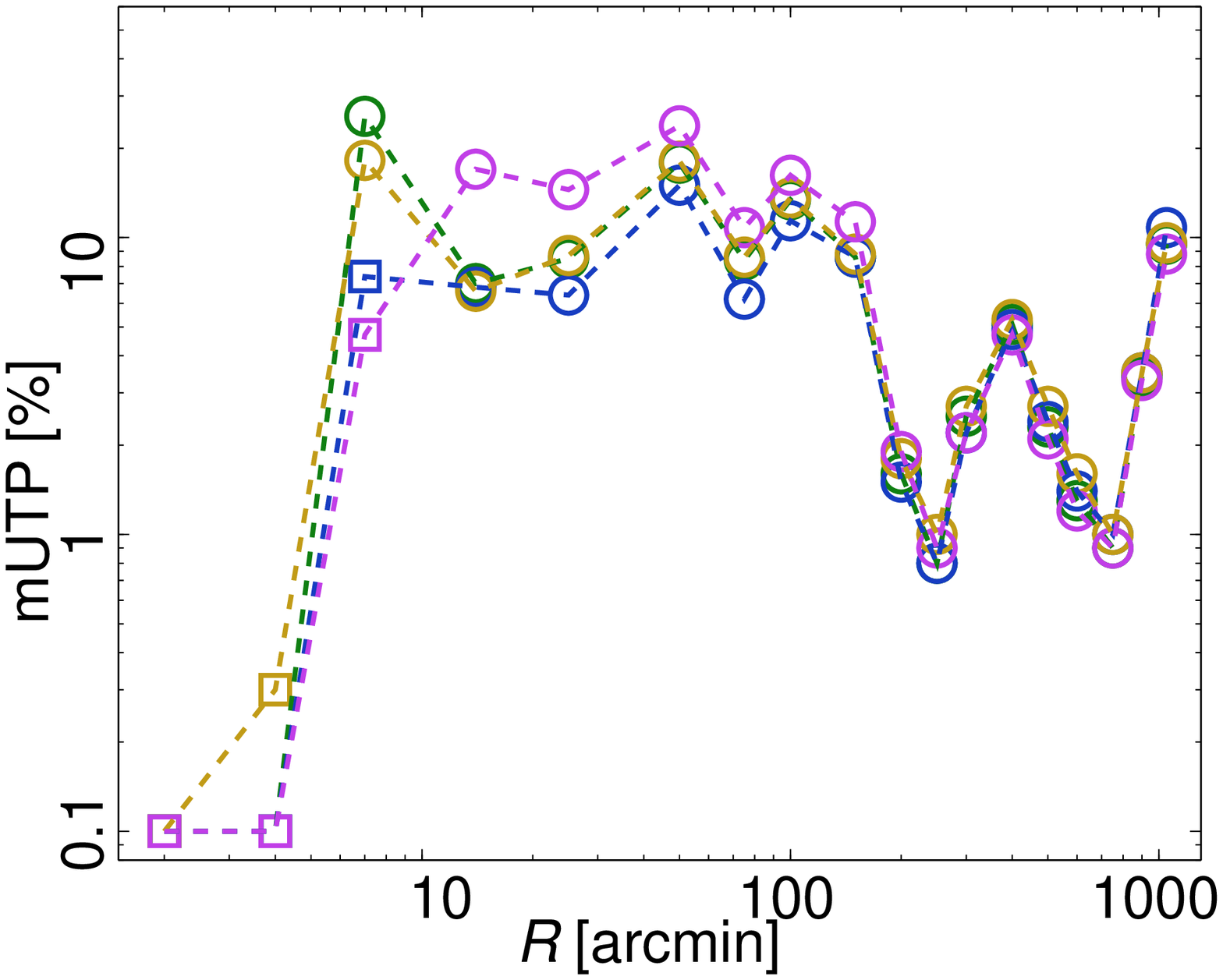}
    \includegraphics[width=5.9cm]{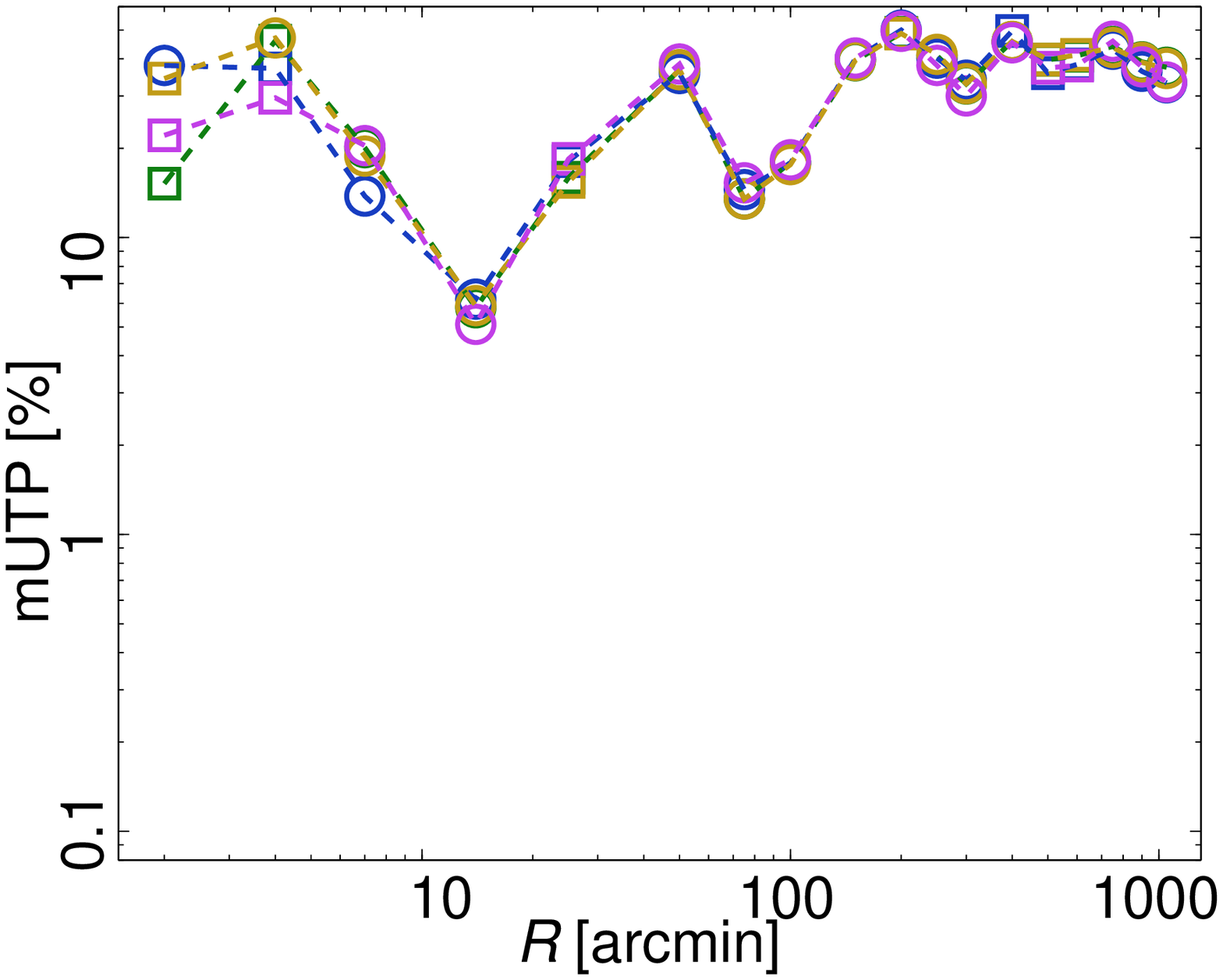}
    \includegraphics[width=5.9cm]{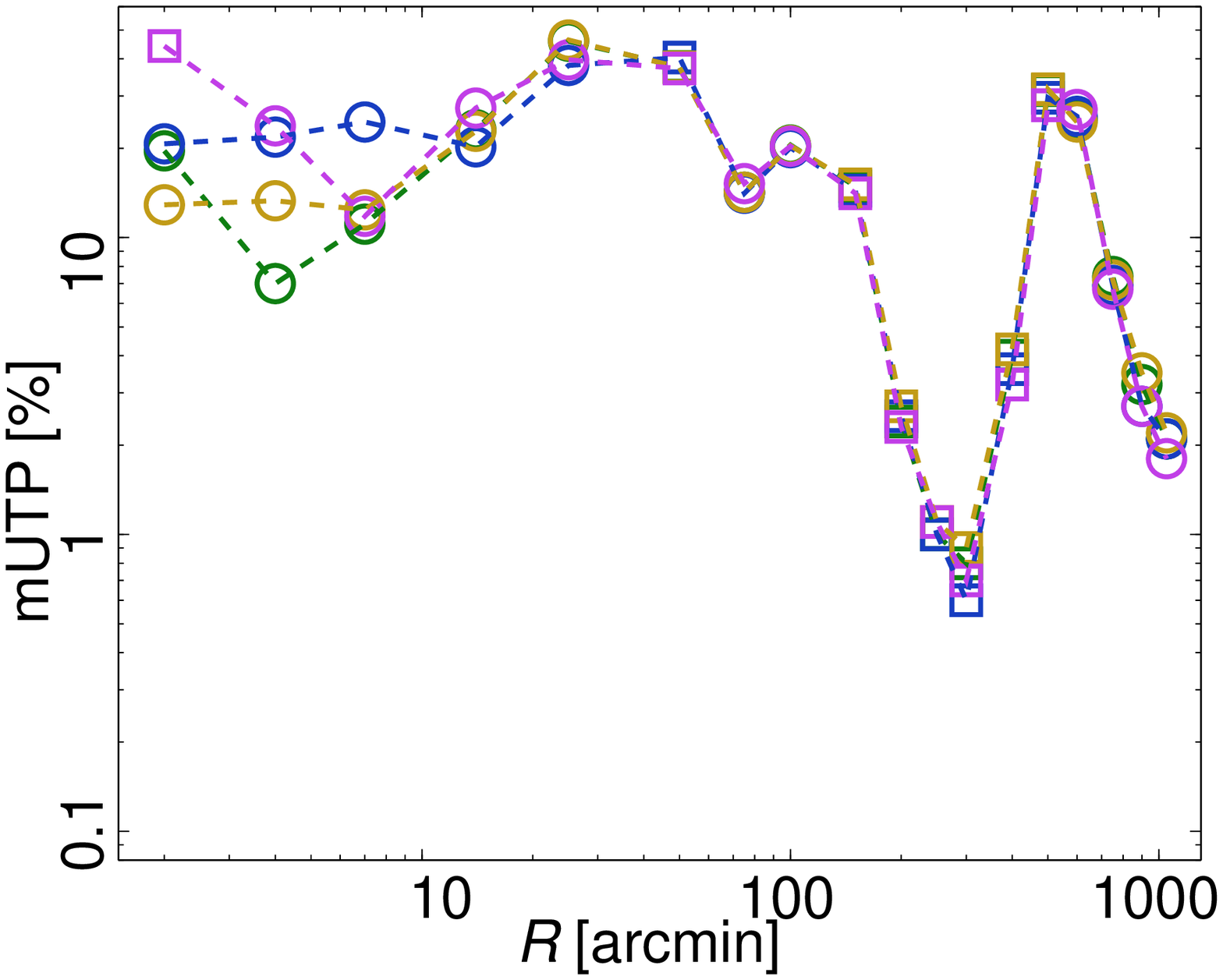}
    \caption{Modified upper tail probabilities (mUTP) obtained from
      the analyses of the SMHW coefficients as a function of the
      wavelet scale $R$ for the {\tt SEVEM-100} (blue), {\tt
        SEVEM-143} (yellow), {\tt SEVEM-217} (magenta), and \sevem\
      (green) maps. From left to right, the panels correspond to the
      standard deviation, skewness, and kurtosis. }
\label{fig:T_stat_SMHW_freq_clean}
\end{center}
\end{figure*}

\begin{figure*}[h!]
\begin{center}
    \includegraphics[width=5.9cm]{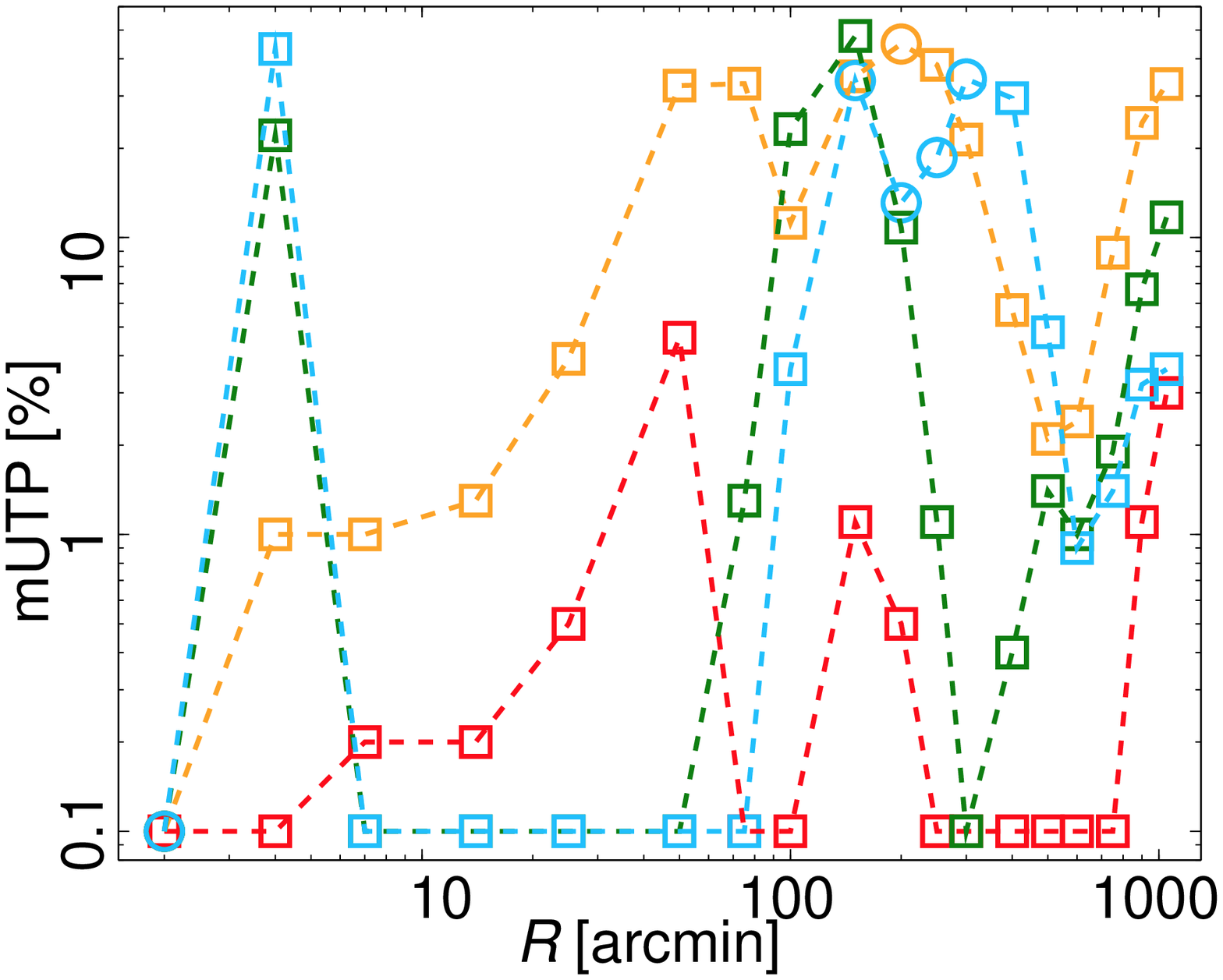}
    \includegraphics[width=5.9cm]{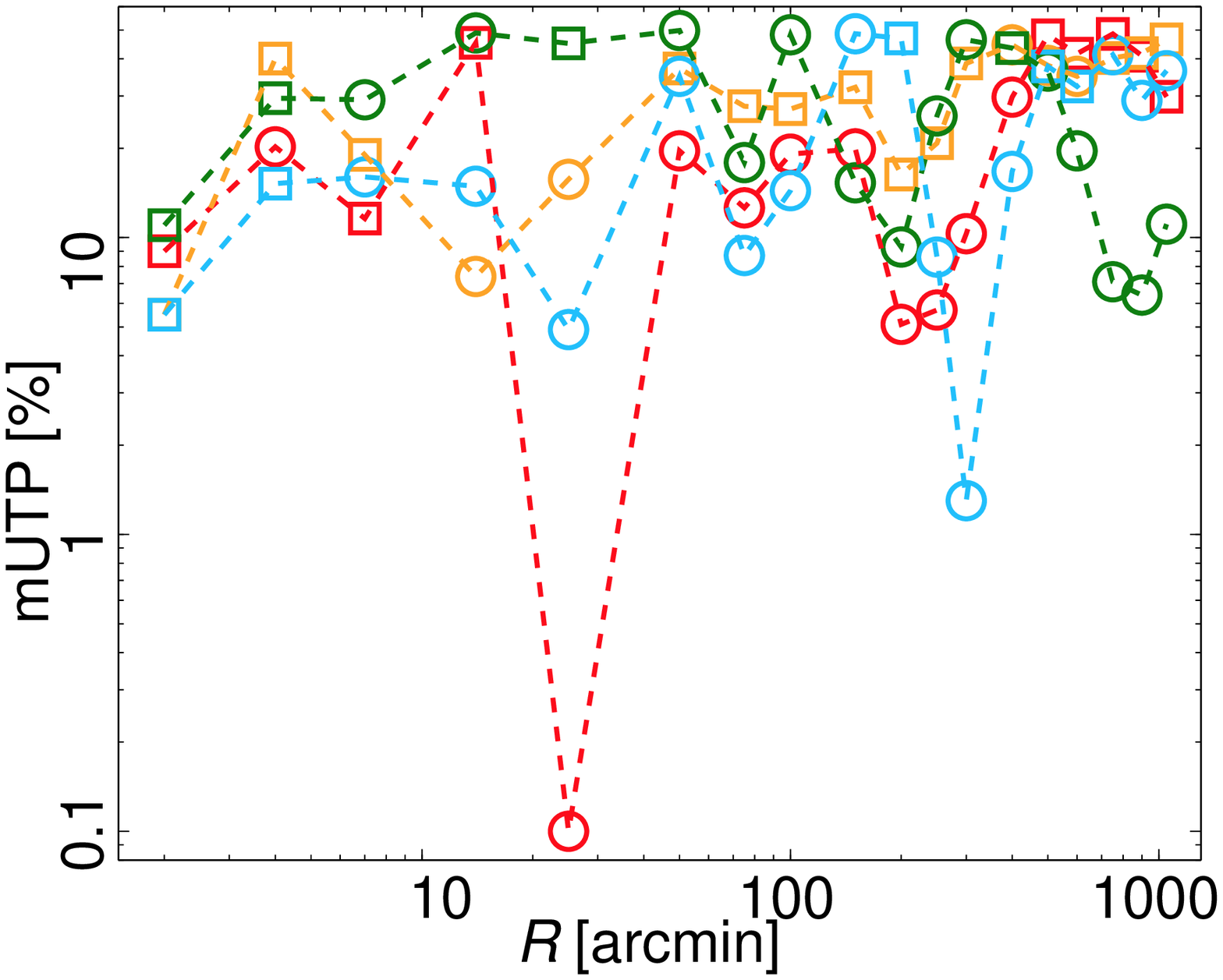}
    \includegraphics[width=5.9cm]{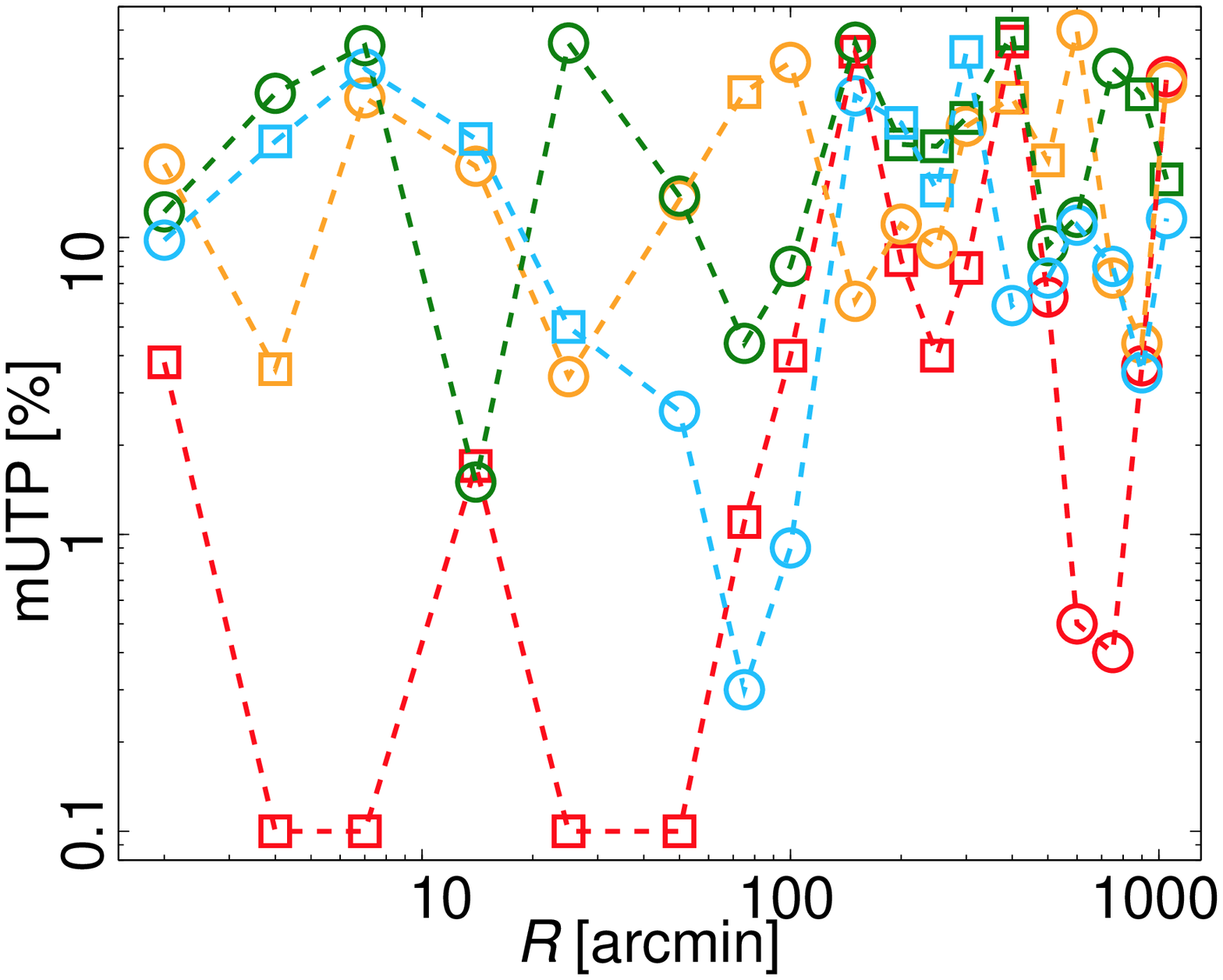}
    \caption{Modified upper tail probabilities (mUTP) obtained
      from the analyses of the SMHW coefficients as a function of the
      wavelet scale $R$ for the \commander\ (red), \nilc\ (orange),
      \sevem\ (green), and \smica\ (blue) half-ring half-difference
      noise estimates. From left to right, the panels correspond to the
      standard deviation, skewness, and kurtosis.
}
\label{fig:T_stat_SMHW_noise}
\end{center}
\end{figure*}

Figure~\ref{fig:T_stat_SMHW_freq_clean} shows the significance of the
statistics derived from the {\tt SEVEM-100}, {\tt SEVEM-143}, and {\tt
  SEVEM-217} maps.  The three cleaned maps yield very consistent
values of the mUTP for the standard deviation, skewness, and kurtosis
of the wavelet coefficients, with only small differences seen at small
scales.  This frequency-independence of the results argues against the
foreground residuals hypothesis.  Figure~\ref{fig:T_stat_SMHW_noise}
presents the same statistics as applied to an estimator of the noise
properties of the CMB maps. This is derived from the half-difference
of the half-ring data sets, which provides the best estimate of the
noise properties of the full mission data set. However, since there is
still a known mismatch in noise properties, any conclusions will be
more qualitative than quantitative.  Nevertheless, the noise study
reveals that, at the smallest scales, there are some discrepancies
with the FFP8 simulations, and in particular the estimated dispersion
of the SMHW noise coefficients is higher than predicted.

\subsubsection{The area above/below a threshold}
\label{sec:area_cs}
In the context of the study of the \cs, the area above/below a
given threshold, as a function of the SMHW wavelet scale, has been
demonstrated to provide a useful and robust statistic
\citep[e.g.,][]{cruz2005}, since it is rather independent of any
masking required. Our previous
analysis \citepalias{planck2013-p09} confirmed that the CMB
temperature fluctuations exhibit an anomalously large cold area on
scales of around $10\degr$, which can be mostly associated with the
\cs. Here, we extend the analysis by including
results derived using the GAUSS and SSG84 filters.

At a given scale $R$ and threshold $\nu$, the cold ($A_R^{-\nu}$) and
hot ($A_R^{+\nu}$) areas of a filtered map are defined as
\begin{linenomath*}
\begin{eqnarray}
A_R^{-\nu} & \equiv & \# \lbrace \omega_S\left(R,p\right) < -\nu \rbrace \, ,\\
A_R^{+\nu} & \equiv & \# \lbrace \omega_S\left(R,p\right) < +\nu \rbrace \, ,
\end{eqnarray}
\end{linenomath*}
where the operator $\#$ represents the number of
pixels $p$ in which the condition defined between the braces is
satisfied.

Table~\ref{tb:cs_areas_all_t} summarizes the results for the hot and
cold areas determined for the four CMB temperature maps analysed with
the common mask (and its associated exclusion masks).  The results are
similar to those obtained in 2013, with some small differences on
those scales related to the \cs\ (between 200\arcm\ and 400\arcm).
Specifically, the cold area is slightly less significant for smaller
values of $R$, whereas the anomalous behaviour remains for larger
filter scales.  The three filters are in reasonable agreement, but, as
expected from Fig.~\ref{fig:T_stat}, the SMHW
yields higher significance levels than the SSG84 and GAUSS
filters. However, it is worth recalling that, for a given scale, the
three filters are not probing exactly the same multipole range and
therefore some differences should be expected.

\begin{figure}
\begin{center}
\includegraphics[width=0.48\textwidth]{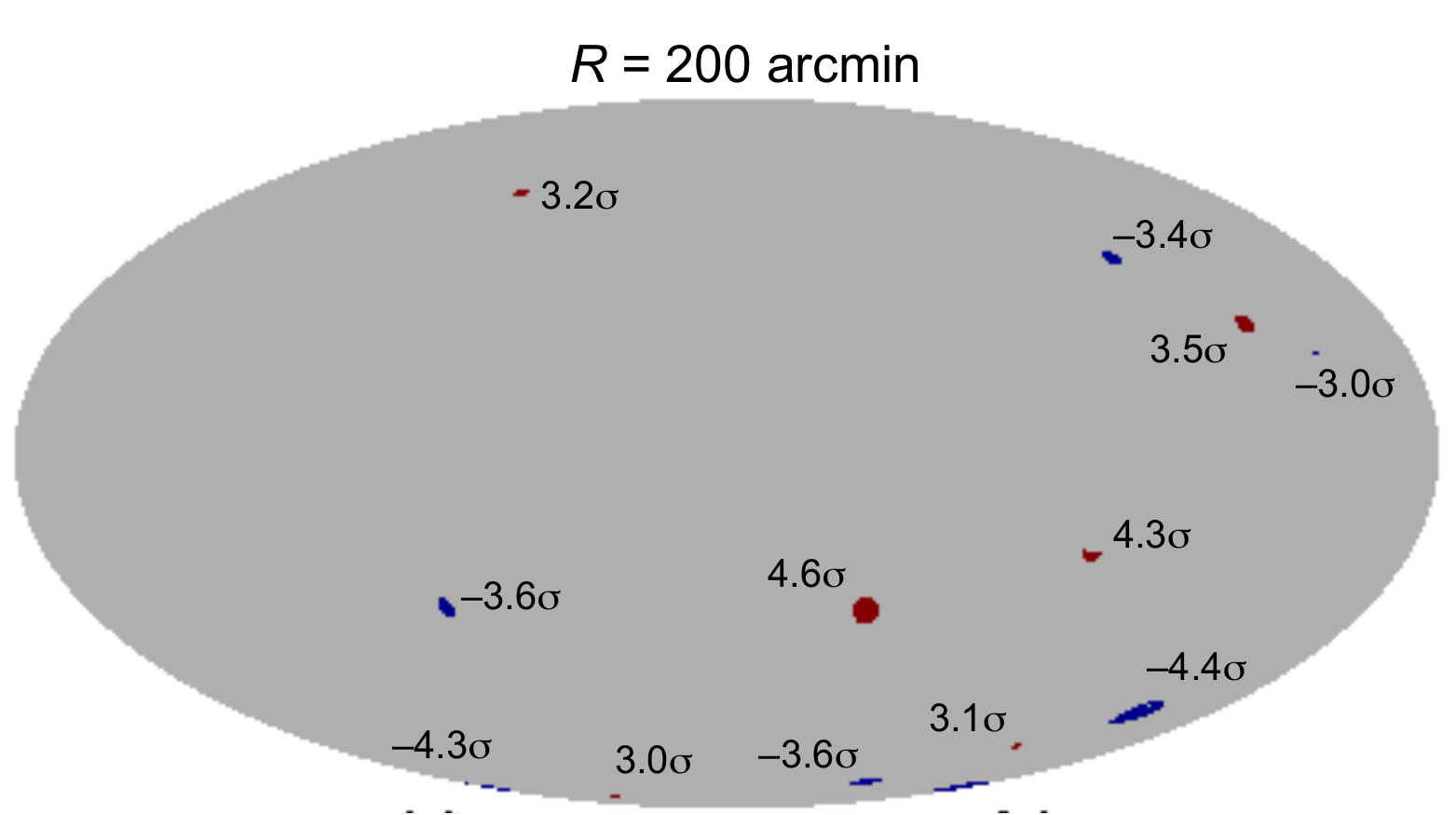}
\includegraphics[width=0.48\textwidth]{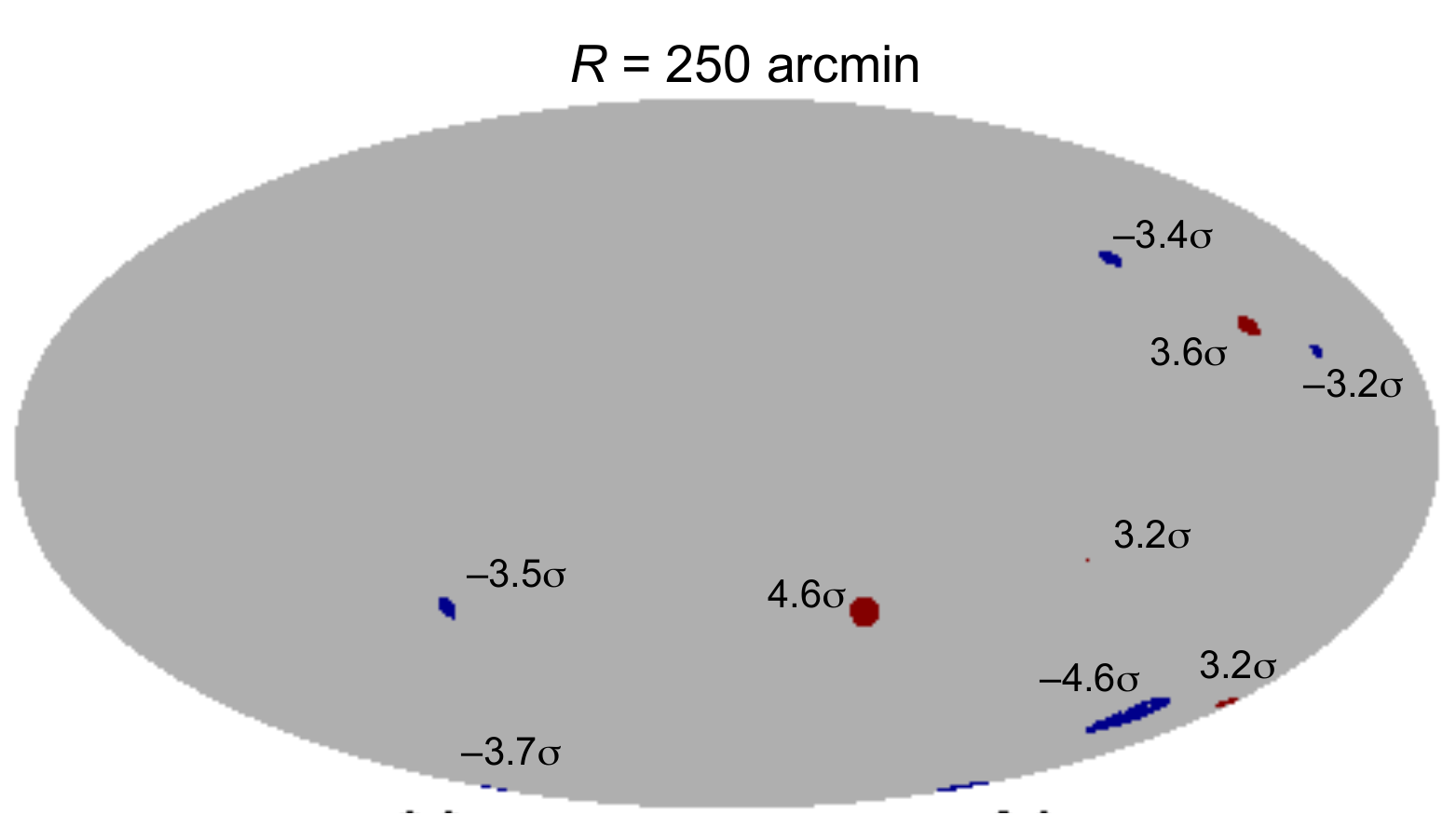}
\includegraphics[width=0.48\textwidth]{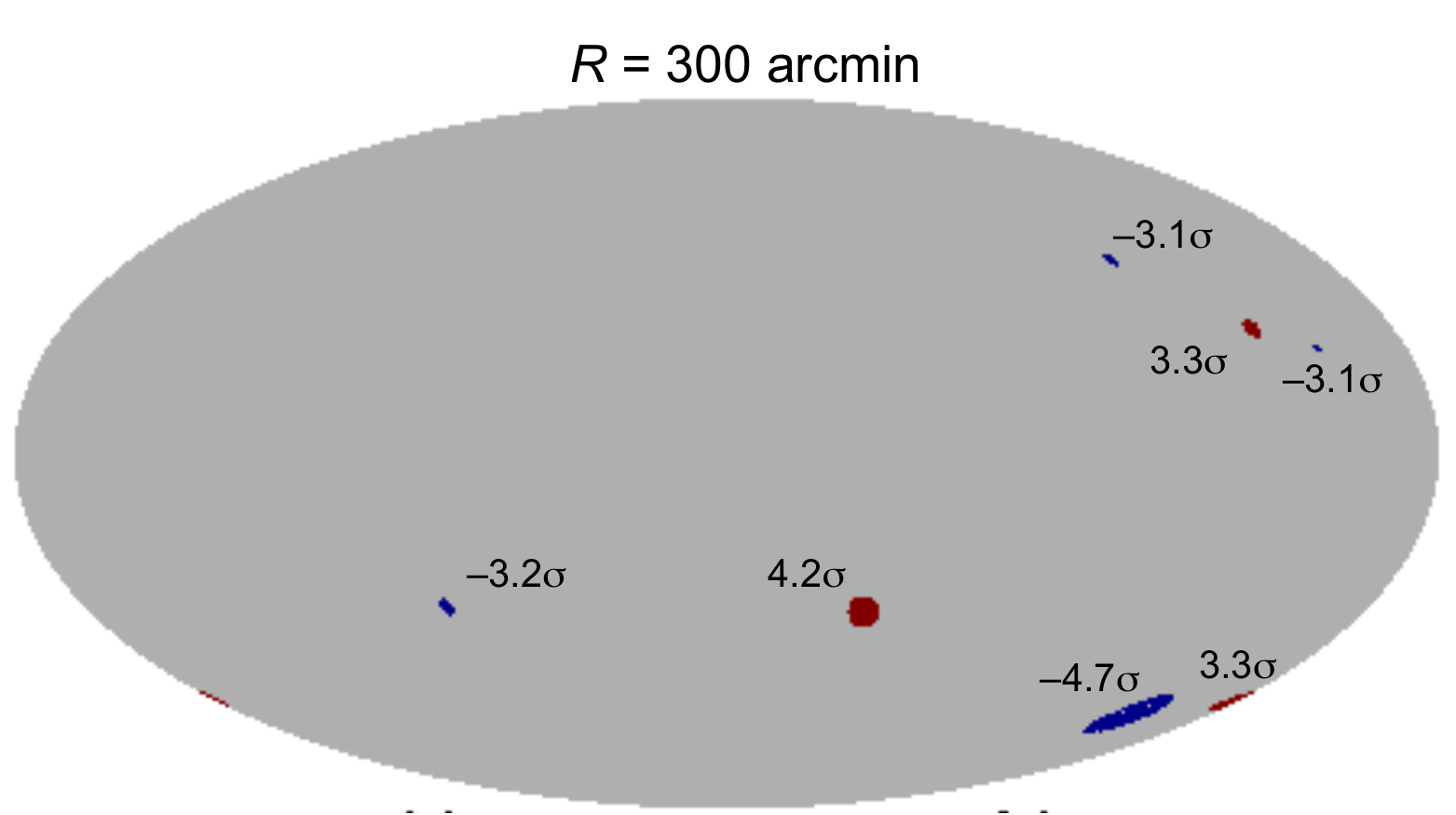}
\includegraphics[width=0.48\textwidth]{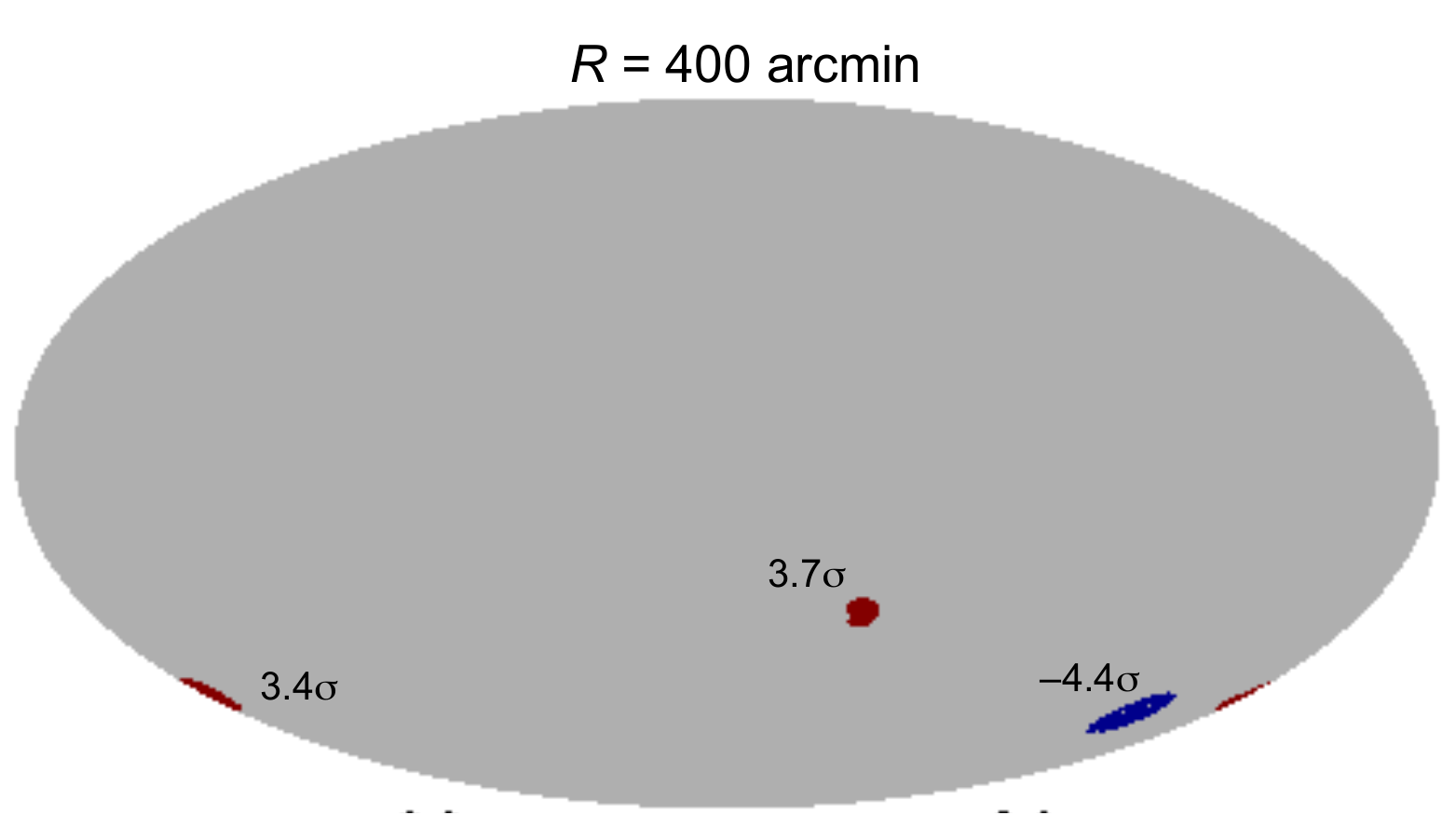}
\caption{\label{fig:cs_area} Cold and hot areas for thresholds $\nu >
  3.0$ as determined from the {\tt SEVEM} temperature
  map. From top to bottom, the maps are for SMHW scales of $R =
  200$\arcm, $R = 250$\arcm, $R = 300$\arcm, and $R = 400$\arcm.}
\end{center}
\end{figure}

In Fig.~\ref{fig:cs_area} we plot the areas for thresholds $\nu >
3.0$, where the threshold is defined in units of $\sigma_R$, as
determined from the \sevem\ temperature map. The results for
\commander, \nilc, and \smica\ are in good agreement with these.  The
panels refer to SMHW scales of $R = 200\arcm$, 250\arcm, 300\arcm, and
400\arcm. The most extreme value (in terms of $\sigma_R$) for each
area is indicated.

The coldest area corresponds to the \cs\ with the minimum value of
the wavelet coefficient at the position $(209\deg, -57\deg)$ in
Galactic coordinates. The hottest area has already been identified in
the WMAP data~\citep[e.g.,][]{vielva2007}.  The results are
insensitive to the choice of CMB temperature map that is adopted. It
is clear that the southern Galactic hemisphere yields more anomalous
signatures than the northern one. These results confirm the importance
of the \cs\ as the most extreme feature in the analysed sky. More
insights about its nature are provided in Sect.~\ref{sec:coldspot}.

\begin{table}[tb] \begingroup \newdimen\tblskip \tblskip=5pt
  \caption{Modified upper tail probability
    (mUTP ) for the cold (top) and hot (bottom) areas.
    Results are given for the $\nu > 4\,\sigma_R$ threshold of the SMHW,
    GAUSS, and SSG84 coefficients. The four most significant scales
    related to the \cs\ feature are shown. An ellipsis (\ldots)
    indicates that no area above that threshold was found in the
    data.}
\label{tb:cs_areas_all_t}
\nointerlineskip
\vskip -3mm
\footnotesize
\setbox\tablebox=\vbox{
   \newdimen\digitwidth
   \setbox0=\hbox{\rm 0}
   \digitwidth=\wd0
   \catcode`*=\active
   \def*{\kern\digitwidth}
   \newdimen\signwidth
   \setbox0=\hbox{+}
   \signwidth=\wd0
   \catcode`!=\active
   \def!{\kern\signwidth}
\halign{
\hbox to 0.7in{#\leaderfil}\tabskip 4pt&
\hfil#\hfil\tabskip=1pt&
\hbox to 0.4in{\hfill#\hfill}&
\hfill#\hfill&
\hfill#\hfill&
\hfill#\hfill\/\tabskip=0pt\cr
\noalign{\doubleline}
\noalign{\vskip -1pt}
\omit&\omit&\multispan4 \hfil Probability [\%] \hfil\cr
\noalign{\vskip -6pt}
\omit&\omit&\multispan4\hrulefill\cr
\omit Area \hfil& \hfil Scale\hfil\tabskip=2pt& \hfil {\tt Comm.} \tabskip=10pt\hfil& \hfil {\tt NILC} \hfil& \hfil {\tt SEVEM} \hfil& \hfil {\tt SMICA}\hfil\cr
\omit& \hfil [arcmin] \hfil\tabskip=2pt& \omit& \omit& \omit& \omit\cr
\noalign{\vskip 4pt\hrule\vskip 3pt}
\multispan6 \hfil SMHW \hfil\cr
\noalign{\vskip 3pt}
\omit& 200& *3.8& *5.1& *3.7& *3.8\cr
Cold& 250& *1.4& *2.4& *1.4& *1.4\cr
\omit& 300& *0.4& *1.5& *0.4& *0.4\cr
\omit& 400& *0.9& *0.9& *0.9& *0.9\cr
\omit&\omit&\omit&\omit&\omit&\omit\cr
\omit& 200& *2.0& *2.6& *1.7& *1.5\cr
Hot& 250& *2.4& *3.0& *2.1& *2.0\cr
\omit& 300& *4.2& *5.0& *4.1& *3.9\cr
\omit& 400& \dots& \dots& \dots& \dots\cr
\noalign{\vskip 3pt}
\multispan6 \hfil GAUSS \hfil\cr
\noalign{\vskip 3pt}
\omit& 200& *1.7& *2.7& *1.7& *1.7\cr
Cold& 250& *1.2& *1.2& *1.2& *1.2\cr
\omit& 300& *1.6& *1.8& *1.2& *1.8\cr
\omit& 400& \dots& \dots& \dots& \dots\cr
\omit&\omit&\omit&\omit&\omit&\omit\cr
\omit& 200& *2.9& *3.5& *2.8& *2.6\cr
Hot& 250& *5.7& *6.4& *5.6& *5.4\cr
\omit& 300& \dots& \dots& \dots& \dots\cr
\omit& 400&  \dots& \dots& \dots& \dots\cr
\noalign{\vskip 3pt}
\multispan6 \hfil SSG84\hfil\cr
\noalign{\vskip 3pt}
\omit& 200& *9.4& 11.0& *9.4& *9.0\cr
Cold& 250& 12.3& 13.4& 10.8& 12.3\cr
\omit& 300& *1.4& *2.6& *1.4& *1.5\cr
\omit& 400& *0.9& *1.9& *0.8& *0.9\cr
\omit&\omit&\omit&\omit&\omit&\omit\cr
\omit& 200& *1.1& *1.8& *1.0& *0.9\cr
Hot& 250& *4.8& *5.1& *4.5& *4.3\cr
\omit& 300& \dots& \dots& \dots& \dots\cr
\omit& 400&  \dots& \dots& \dots& \dots\cr
\noalign{\vskip 5pt\hrule\vskip 3pt}}}
\endPlancktable                    
\endgroup
\end{table}

\subsubsection{Peak statistics}
\label{sec:peaks}

The statistical properties of local extrema (both minima and maxima,
which we refer to collectively as ``peaks'') provide an alternative
approach to search for evidence of non-Gaussianity in the data. Such
peaks, defined as pixels whose amplitudes are higher or lower than the
corresponding values for all of their nearest neighbours, trace
topological properties of the data.  Peak locations and amplitudes,
and various derived quantities, such as their correlation functions,
have previously been used to characterize the WMAP sky maps by
\citet{Larson:2004vm, Larson:2005vb} and \citet{Hou:2009au}.

The statistical properties of peaks for a statistically isotropic
Gaussian random field were derived by \citet{BE1987}.  The integrated
number density of peaks, $n_{\text{pk}}$ (composed of maxima and
minima with corresponding densities $n_{\text{max}}$ and
$n_{\text{min}}$), with amplitudes $x$ above a certain threshold $\nu
= x/\sigma$ is given by
\begin{linenomath*}
\begin{eqnarray}\label{eq:peaks:be}
  \frac{n_{\text{max}}+n_{\text{min}}}{n_{\text{pk}}}\left(\frac{x}{\sigma} > \nu\right) &=&
    \sqrt{\frac{3}{2\pi}}\, \gamma^2\, \nu \exp\left(-\frac{\nu^2}{2}\right)\\
    &&\!\! +
    \frac{1}{2}\, \text{\rm erfc} \left[ \frac{\nu}{\sqrt{2-\frac{4}{3}\,\gamma^2}} \right], \nonumber
\end{eqnarray}
\end{linenomath*}
where $\sigma$ is the rms fluctuation amplitude measured on the
sky, and $\gamma$ is the spectral shape parameter of the underlying
field. Uncharacteristically cold and hot spots are then manifested as
extreme outliers in the peak values, and can constitute evidence for
non-Gaussianity or deviation from isotropy.

\begin{figure}
  \centering
  \includegraphics[width=88mm]{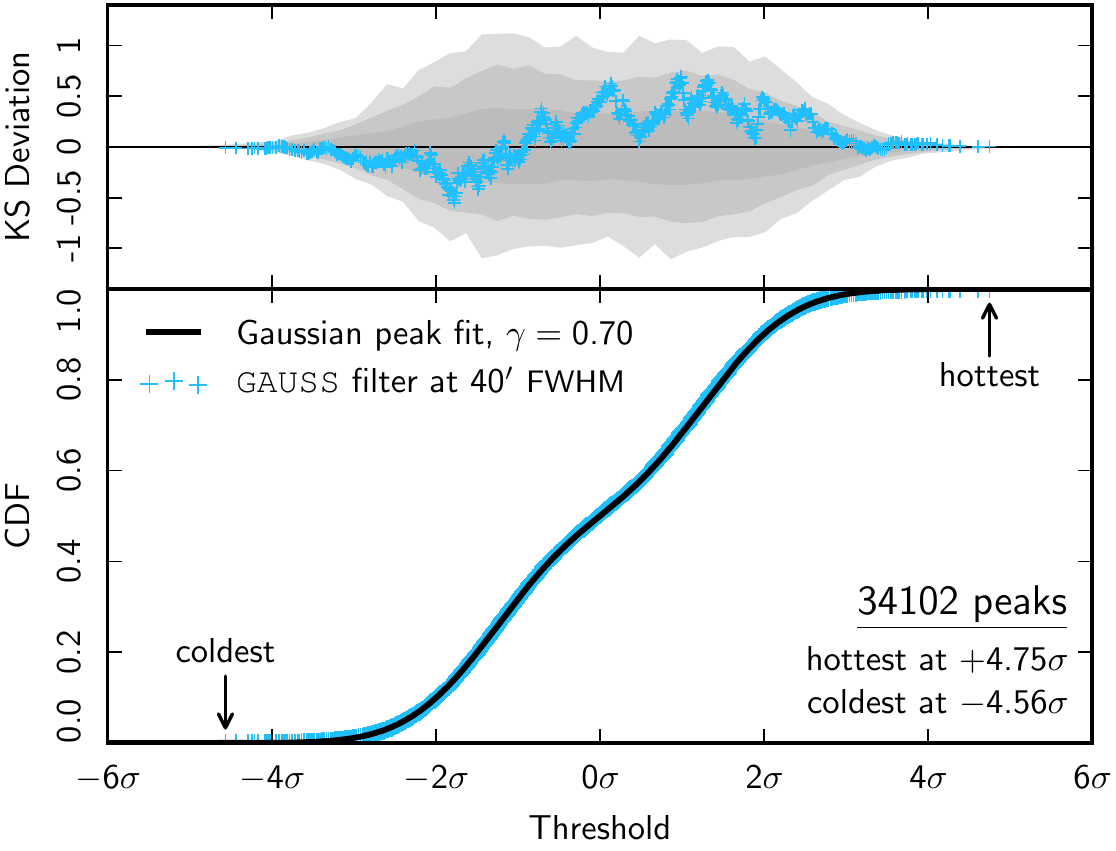}\\
  \includegraphics[width=88mm]{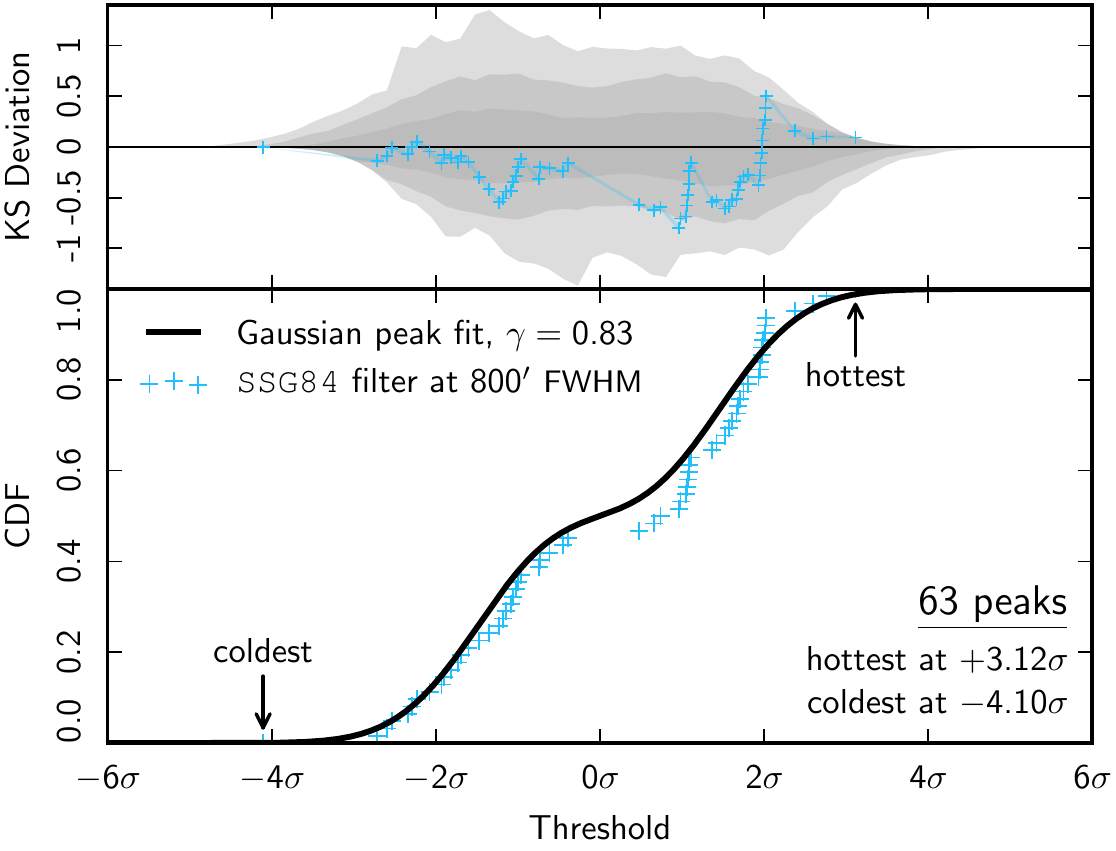}\\
  \caption{ Cumulative density function of the peak distribution for
    the \smica\ CMB temperature map. The top row shows the peak CDF
    for maps filtered with a GAUSS kernel of $40\arcm$ FWHM.  The
    bottom row shows the corresponding peak CDF for an {\tt SSG84}
    kernel of $800\arcm$ FWHM. The spectral shape parameter $\gamma$
    (see Eq.~\ref{eq:peaks:be}) is the best-fit value for the
    simulated ensemble, as indicated by the cyan circle in
    Fig.~\ref{fig:peaks:params}. Similar results are obtained for the
    other component-separation methods.  }
  \label{fig:peaks:stat}
\end{figure}

\begin{figure}
  \centering
  \includegraphics[width=88mm]{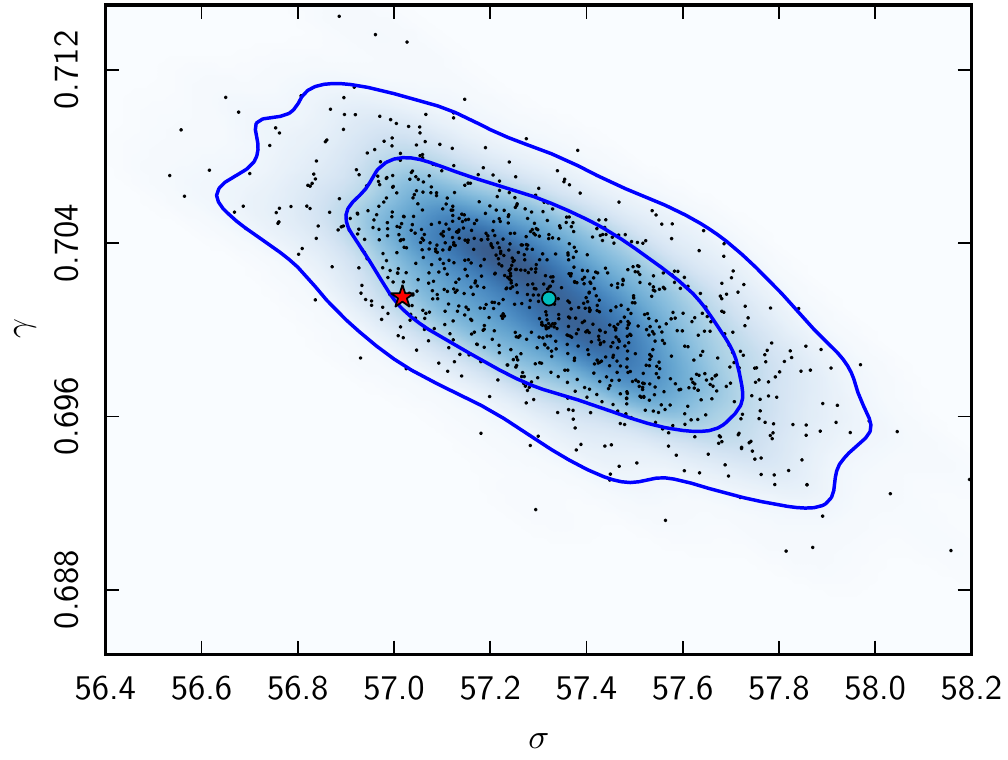}\\
  \includegraphics[width=88mm]{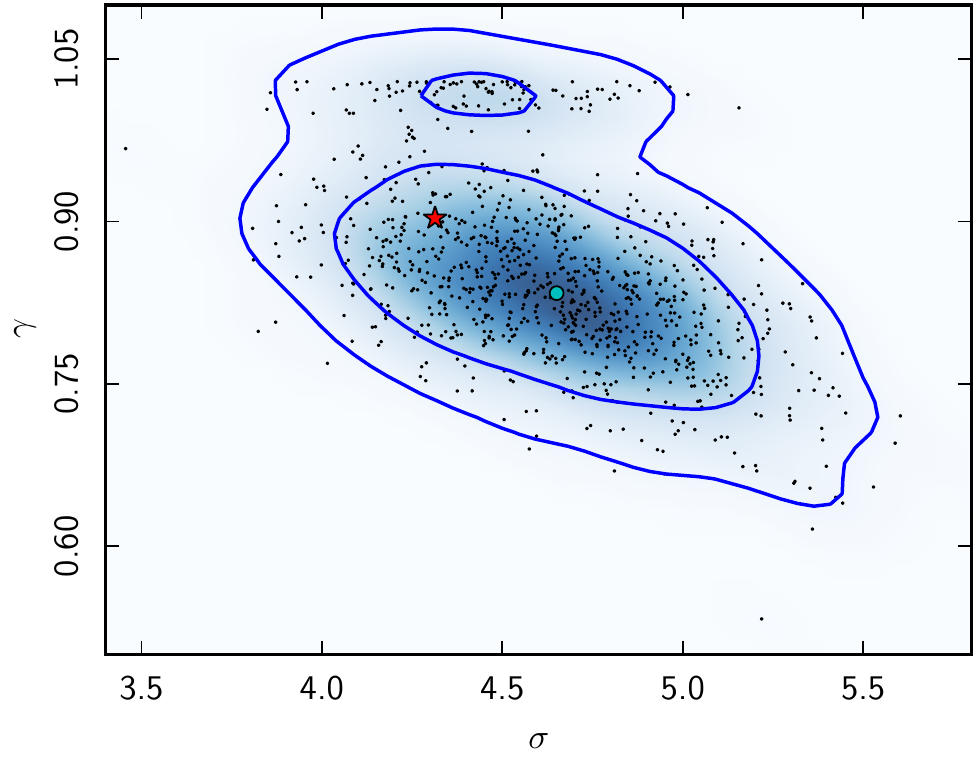}\\
  \caption{ Distribution of best-fit Gaussian peak CDF spectral shape
    parameters, $\sigma$ and $\gamma$ (as defined in
    Eq.~\ref{eq:peaks:be}), recovered from 1000 simulations,
    as indicated by the black dots and the smoothed density map and
    compared to those derived for the observed sky (shown by the red
    star).  The blue contours enclose 68\,\% and 95\,\% of the
    parameter distribution, and the cyan circle represents the
    best-fit parameters for the median peak CDF determined from
    simulations.  The upper panel shows the peak CDF parameters for
    the \smica\ map filtered with a GAUSS kernel of $40\arcm$ FWHM.
    The lower panel shows the corresponding peak CDF for an {\tt
      SSG84} kernel of $800\arcm$ FWHM. Similar results are obtained
    for the other component-separation methods.  }
  \label{fig:peaks:params}
\end{figure}

\begin{figure}
  \centering
  \includegraphics[width=88mm]{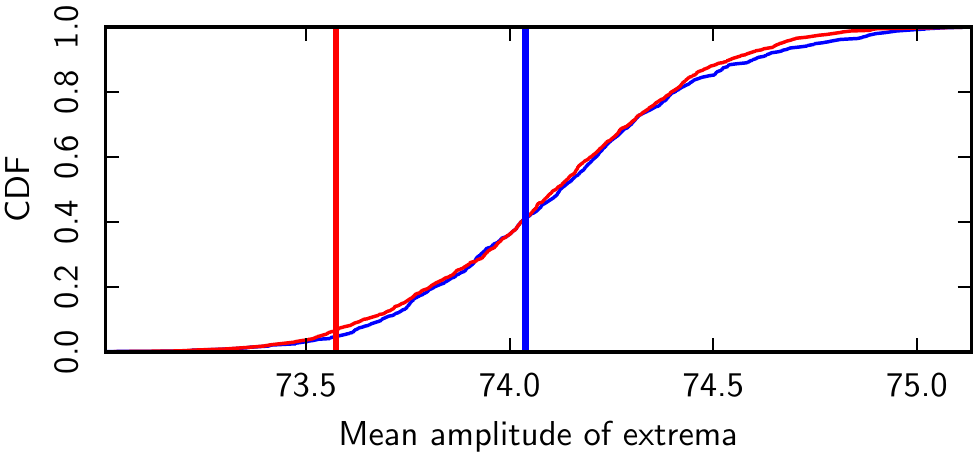}\\
  \includegraphics[width=88mm]{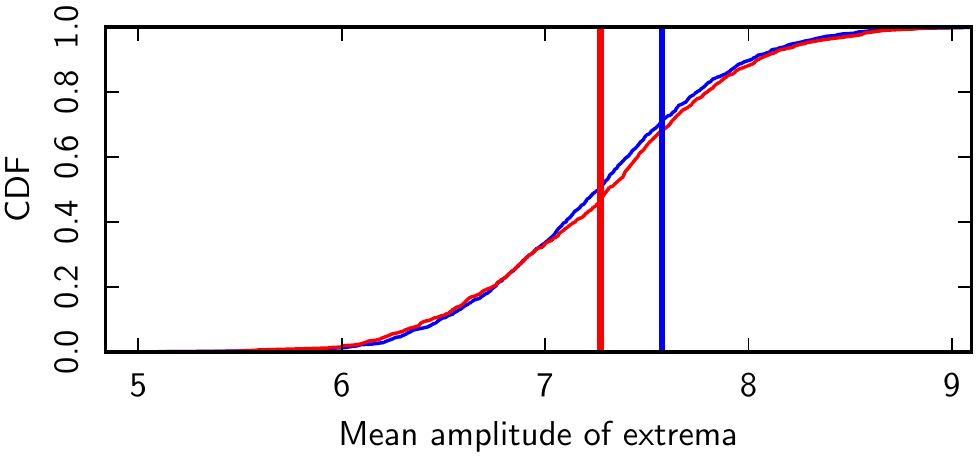}\\
  \caption{ Cumulative density function of the mean amplitude of all extrema,
    maxima (red) and minima (blue),
    derived from simulations, compared to the equivalent values observed for the \smica\
    CMB temperature map.  The upper panel shows the peak mean amplitudes for maps filtered
    with a GAUSS kernel of $40\arcm$ FWHM.  The lower panel shows the
    corresponding peak CDF for an {\tt SSG84} kernel of $800\arcm$
    FWHM. Similar results are obtained for the other component
    separation methods. Since the filter kernel normalization is free,
    and the pre-whitened map to which the filter is applied is
    dimensionless, the plots are essentially in arbitrary units.
 }
  \label{fig:peaks:means}
\end{figure}

Here, we consider the peak statistics of the \Planck\
component-separated temperature maps at $\nside = 2048$.  The maps are
pre-whitened as described in Appendix~\ref{asec:gsgp}.  This step
allows the construction of an estimator that is nearly optimal with
respect to the fiducial CMB properties.  After application of the
common mask, weighted convolutions of the data are performed with
either SSG or GAUSS kernels of variable scale.  In order to avoid
potential contamination by boundary effects, the mask is extended by
rejecting pixels with an effective convolution weight that differs
from unity by more than 12\,\%.  Peaks are extracted from the filtered
map (removing any that are adjacent to masked pixels), their positions
and values are recorded for further analysis, and their cumulative
density function (CDF) is constructed by sorting peak values.
Table~\ref{tab:pkcounts} presents peak counts for the
component-separated sky maps for several different kernels and
representative filtering scales, together with the number of peaks
that are common to all maps. There is excellent agreement between the
various CMB estimates.
All statistical inference is then performed by comparison of the peak
distributions derived from the data with equivalently processed
simulations.  As an internal consistency check, the properties of the
FFP8 simulations are found to be in agreement with the predictions of
Eq.~\eqref{eq:peaks:be}.

\begin{table}[tp]
\begingroup
\newdimen\tblskip \tblskip=5pt
\caption{Peak counts in maps filtered to different scales.}
\label{tab:pkcounts}
\nointerlineskip
\vskip -3mm
\footnotesize
\setbox\tablebox=\vbox{
   \newdimen\digitwidth
   \setbox0=\hbox{\rm 0}
   \digitwidth=\wd0
   \catcode`*=\active
   \def*{\kern\digitwidth}
   \newdimen\signwidth
   \setbox0=\hbox{+}
   \signwidth=\wd0
   \catcode`!=\active
   \def!{\kern\signwidth}
\halign{
\hbox to 0.80in{#\leaderfil}\tabskip 4pt&
\hfil#\hfil&
\hfil#\hfil&
\hfil#\hfil&
\hfil#\hfil&
\hfil#\hfil\tabskip 0pt\cr
\noalign{\doubleline\vskip -1pt}
\omit&\multispan4 \hfil Number of minima/maxima\hfil\cr
\noalign{\vskip -4pt}
\omit&\multispan4\hrulefill\cr
\omit \hfil Filter Scale\hfil& {\tt Comm.}& \nilc& \sevem& \smica& Match\cr
\omit\hfil [arcmin]\hfil&\omit& \omit& \omit& \omit& \omit\cr
\noalign{\vskip 4pt\hrule\vskip 3pt}
\multispan6\hfil SMHW\hfil\cr
\noalign{\vskip 1pt}
200& 176/187& 170/178& 173/182& 169/182& 161/169\cr
250& 105/105& 104/103& 107/123& 105/107& *97/*99\cr
300& *70/*70& *71/*70& *70/*72& *68/*71& *66/*66\cr
400& *43/*32& *46/*32& *44/*31& *43/*33& *37/*30\cr
\noalign{\vskip 4pt}
\multispan6\hfil GAUSS\hfil\cr
\noalign{\vskip 1pt}
200& 152/170& 152/166& 157/179& 156/165& 142/155\cr
250& 102/*93& 104/*95& 108/*99& *99/101& *92/*85\cr
300& *60/*63& *57/*62& *63/*64& *56/*62& *50/*53\cr
400& *33/*28& *29/*29& *31/*33& *29/*28& *24/*27\cr
\noalign{\vskip 4pt}
\multispan6\hfil SSG84\hfil\cr
\noalign{\vskip 1pt}
200& 180/187& 178/183& 180/185& 183/183& 167/175\cr
250& 131/119& 118/114& 122/123& 121/110& 109/103\cr
300& *68/*69& *73/*68& *73/*73& *70/*68& *56/*61\cr
400& *29/*35& *29/*36& *29/*32& *30/*38& *27/*27\cr
\noalign{\vskip 3pt\hrule\vskip 3pt}}}
\endPlancktable                    
\endgroup
\end{table}

Figure~\ref{fig:peaks:stat} presents the distributions of peaks for
the \smica\ CMB map filtered with two representative kernels on scales
of $40\arcm$ and $800\arcm$ FWHM. The lower panels show the empirical
peak CDFs as a function of peak value $x$, defined for a set of $n$
peaks at values $\{X_i\}$ as
\begin{linenomath*}
\begin{equation}\label{eq:peaks:cdf}
F_n(x) = \frac{1}{n} \sum\limits_{i=1}^{n} I_{X_i \le x}, \hspace{1em}
I_{X_i \le x} \equiv \left\{ \begin{array}{cl}1,& \text{if }X_i \le x\\ 0,& \text{otherwise}\end{array} \right. .
\end{equation}
\end{linenomath*}
For plotting purposes alone, the horizontal axis is scaled in units of
$\sigma$ defined by Eq.~\eqref{eq:peaks:be} and derived from the
underlying median CDF, $\bar{F}(x)$, of the simulations.  The upper
panels show the difference between the observed and median simulated
CDF values, $\sqrt{n}\, [F_n(x) - \bar{F}(x)]$, with the grey bands
representing the 68.3\,\%, 95.4\,\%, and 99.7\,\% regions of the
simulated CDF distributions.  The maximal value of this difference
defines a Kolmogorov-Smirnov (KS) deviation estimator:
\begin{linenomath*}
\begin{equation}\label{eq:peaks:ks}
K_n \equiv \sqrt{n}\, \mathop{\text{sup}}\limits_{x} \left|F_n(x) - \bar{F}(x)\right|.
\end{equation}
\end{linenomath*}
This forms the basis of a standard KS test of consistency between the
two distributions. Although the KS deviation has a known limiting
distribution, we also derive its CDF directly from the simulations.

The temperature peak distributions in Fig.~\ref{fig:peaks:stat} are
consistent with Gaussian peak statistics, apart from a single
anomalously cold peak on scales around $800\arcm$ FWHM. This
corresponds to the previously reported \cs.  Although this
exercise confirms that the \cs\ is a rare cold feature, as already
noted by~\citet{cruz2005} and confirmed in this paper, the most
peculiar characteristic of the \cs\ is not its coldness, but
rather its size.  A more detailed analysis of its nature is presented
in Sect.~\ref{sec:coldspot}.

The probability that the observed sky exceeds the value of the KS
deviation for the adopted fiducial cosmology can be determined by
counting the number of simulations with $K_{n'} >
K_{n}^{(\text{sky})}$.  The $p$-values for the KS test
comparing the CDF of the observed sky with the median peak CDF derived
from simulations for several different kernels and representative
scales are summarized in Table~\ref{tab:kstest}.  The similarly
derived $p$-values for the total peak counts are summarized in
Table~\ref{tab:counts:p}. Most of the results indicate that the two
distributions are highly consistent, with the exception of results for
the SSG84 filter on scales of about $500\arcm$ FWHM.  This deviation
appears to be related to a hemispherical asymmetry in the peak CDFs,
and will be discussed further in Sect.~\ref{sec:kstest}.

\begin{figure*}
  \centering
  \includegraphics[width=16cm]{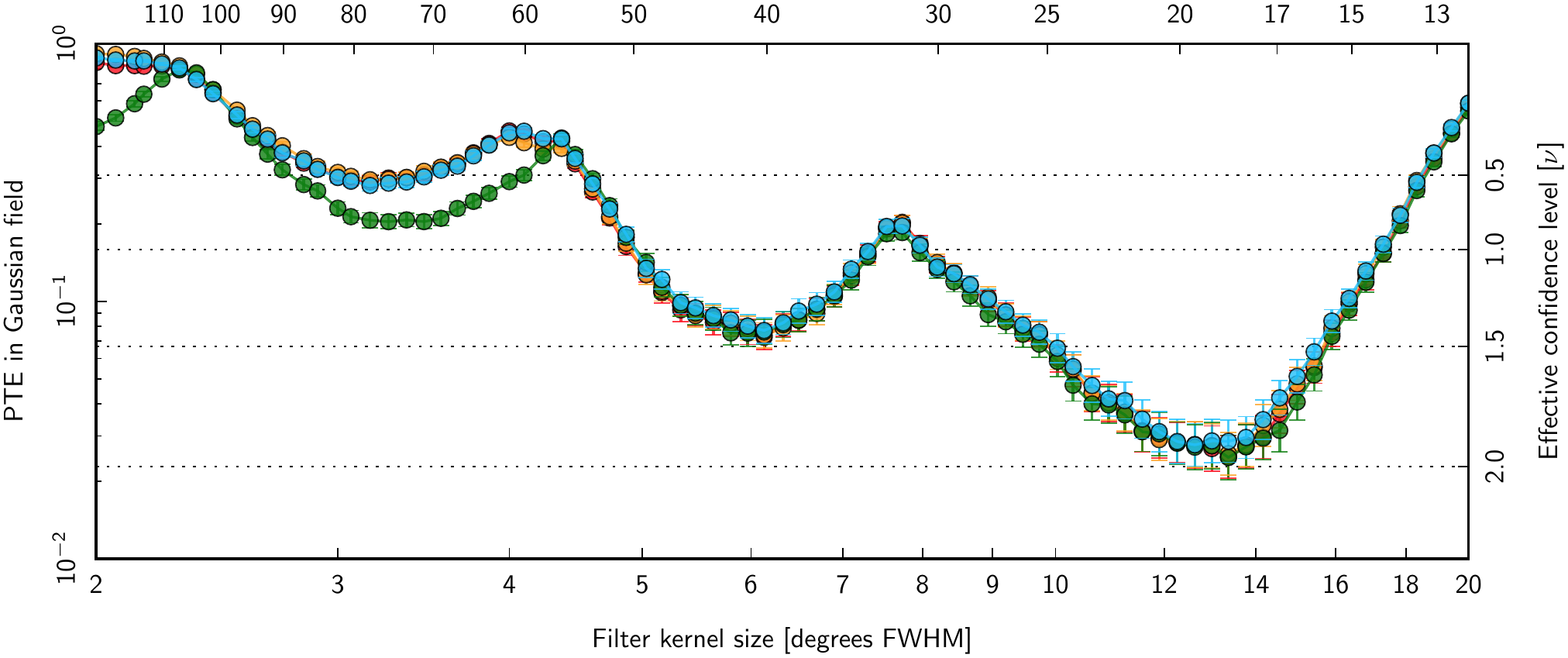}\\
  \caption{Fraction of the Gaussian random field realizations in which the coldest peak is as cold as or colder than that observed, as a function
    of SMHW filter scale for \commander\ (red), \nilc\ (orange), \sevem\
    (green), and \smica\ (blue).
}
  \label{fig:peaks:sign}
\end{figure*}

\begin{table}[tp]
\begingroup
\newdimen\tblskip \tblskip=5pt
\caption{Modified upper tail probability (mUTP) for the
  KS test, comparing the data with the
  median peak CDF derived from simulations.
}
\label{tab:kstest}
\nointerlineskip
\vskip -3mm
\footnotesize
\setbox\tablebox=\vbox{
   \newdimen\digitwidth
   \setbox0=\hbox{\rm 0}
   \digitwidth=\wd0
   \catcode`*=\active
   \def*{\kern\digitwidth}
   \newdimen\signwidth
   \setbox0=\hbox{+}
   \signwidth=\wd0
   \catcode`!=\active
   \def!{\kern\signwidth}
\halign{
\hbox to 0.95in{#\leaderfil}\tabskip 4pt&
\hfil#\hfil&
\hfil#\hfil&
\hfil#\hfil&
\hfil#\hfil\tabskip 0pt\cr
\noalign{\doubleline\vskip -1pt}
\omit&\multispan4 \hfil Probability [\%]\hfil\cr
\noalign{\vskip -4pt}
\omit&\multispan4\hrulefill\cr
\omit\hfil Filter Scale\hfil& {\tt Comm.} & \nilc & \sevem & \smica \cr
\omit\hfil [arcmin] \hfil&\omit& \omit& \omit& \omit\cr
\noalign{\vskip 4pt\hrule\vskip 3pt}
\multispan5\hfil SMHW\hfil\cr
\noalign{\vskip 1pt}
200& 22.0& 42.8& 45.9& 40.5\cr
250& 11.3& 17.6& *3.1& 11.4\cr
300& 49.4& 38.5& 38.4& 32.1\cr
400& 32.6& 24.7& 35.3& 24.7\cr
\noalign{\vskip 3pt}
\multispan5\hfil GAUSS\hfil\cr
\noalign{\vskip 1pt}
200& 41.3& 46.6& 14.4& 47.2\cr
250& 43.7& 34.8& *7.6& 48.4\cr
300& 46.3& *9.9& 28.0& *7.7\cr
400& 30.7& *5.6& 35.8& *6.6\cr
\noalign{\vskip 3pt}
\multispan5\hfil SSG84\hfil\cr
\noalign{\vskip 1pt}
200& 37.1& 36.7& 24.0& 37.5\cr
250& *0.5& *1.7& *0.8& *5.4\cr
300& 17.5& 12.2& *0.3& *9.3\cr
400& 47.4& 44.6& 47.5& 47.8\cr
\noalign{\vskip 3pt\hrule\vskip 3pt}}}
\endPlancktable                    
\endgroup
\end{table}

\begin{table}[tp]
\begingroup
\newdimen\tblskip \tblskip=5pt
\caption{Modified upper tail probability (mUTP) for the
  total peak count, comparing the data with the
  peak count CDF derived from simulations.
}
\label{tab:counts:p}
\nointerlineskip
\vskip -3mm
\footnotesize
\setbox\tablebox=\vbox{
   \newdimen\digitwidth
   \setbox0=\hbox{\rm 0}
   \digitwidth=\wd0
   \catcode`*=\active
   \def*{\kern\digitwidth}
   \newdimen\signwidth
   \setbox0=\hbox{+}
   \signwidth=\wd0
   \catcode`!=\active
   \def!{\kern\signwidth}
\halign{
\hbox to 0.95in{#\leaderfil}\tabskip 4pt&
\hfil#\hfil&
\hfil#\hfil&
\hfil#\hfil&
\hfil#\hfil\tabskip 0pt\cr
\noalign{\doubleline\vskip -1pt}
\omit&\multispan4 \hfil Probability [\%]\hfil\cr
\noalign{\vskip -4pt}
\omit&\multispan4\hrulefill\cr
\omit\hfil Filter Scale \hfil & {\tt Comm.} & \nilc & \sevem & \smica \cr
\omit\hfil [arcmin] \hfil&\omit& \omit& \omit& \omit\cr
\noalign{\vskip 4pt\hrule\vskip 3pt}
\multispan5\hfil SMHW\hfil\cr
\noalign{\vskip 1pt}
200& *6.1& 36.9& 16.2& 27.2\cr
250& 32.9& 47.5& *1.0& 25.6\cr
300& 48.8& 51.7& 44.7& 44.3\cr
400& 33.8& 16.2& 34.6& 26.4\cr
\noalign{\vskip 3pt}
\multispan5\hfil GAUSS\hfil\cr
\noalign{\vskip 1pt}
200& *7.1& 11.2& *0.7& *8.7\cr
250& 18.2& 11.2& *2.1& *8.1\cr
300& 29.0& 12.8& 48.2& 10.0\cr
400& 11.9& *3.0& 26.6& *2.8\cr
\noalign{\vskip 3pt}
\multispan5\hfil SSG84\hfil\cr
\noalign{\vskip 1pt}
200& *0.2& *3.0& *1.0& *1.7\cr
250& *0.1& *1.7& *0.1& *2.1\cr
300& *9.3& 22.6& 50.7& 12.2\cr
400& *0.3& *0.4& *0.1& *2.3\cr
\noalign{\vskip 3pt\hrule\vskip 3pt}}}
\endPlancktable                    
\endgroup
\end{table}

One can also test whether the observed values of the parameters, $\sigma$
and $\gamma$ as defined in Eq.~\eqref{eq:peaks:be}, are consistent with
the simulation ensemble, under the assumption that the peak
distributions in the \Planck\ data are described by a Gaussian peak
CDF. Figure~\ref{fig:peaks:params} demonstrates the consistency of the
best-fit values of these parameters, corresponding to the peak
distributions in Fig.~\ref{fig:peaks:stat}, with equivalent values
derived from the simulations.

Inspired by the analysis of the WMAP first-year data in
\citet{Larson:2004vm} which found fewer extreme peaks than expected,
we additionally evaluate whether the distributions of maxima and
minima are separately consistent with simulations.  The mean of all
maxima, and the negative of the mean of all minima, are calculated for
the filtered map, and the observed values are compared to the
simulated distributions in Fig.~\ref{fig:peaks:means}.  The observed
minima/maxima means are found to be in good agreement with the
fiducial values.

The probability that the coldest peak seen on the sky is consistent
with the adopted fiducial cosmology is evaluated as a function of both
filter shape and size by counting the number of simulations with
$x_{\text{coldest}} < x_{\text{coldest}}^{(\text{sky})}$.  The results
obtained for the SMHW filter are summarized in
Fig.~\ref{fig:peaks:sign}. Consistent behaviour is seen when the GAUSS
and SSG84 filters are applied. The error bars represent the sampling
uncertainty due to the finite number of realizations, and are
determined using a bootstrap method. As the filters overlap
substantially, different points are highly correlated. The \Planck\
CMB maps are consistent with the expectations of a statistically
isotropic Gaussian model.  The most significant deviation, found at an
effective filter bandwidth given by $\ell=20$, is attributable to a
single region on the sky --- the \cs.

\subsubsection{Peak locations as a function of scale}
\label{sec:peaks_tree}

\begin{figure*}
  \centering
  \includegraphics[width=18cm]{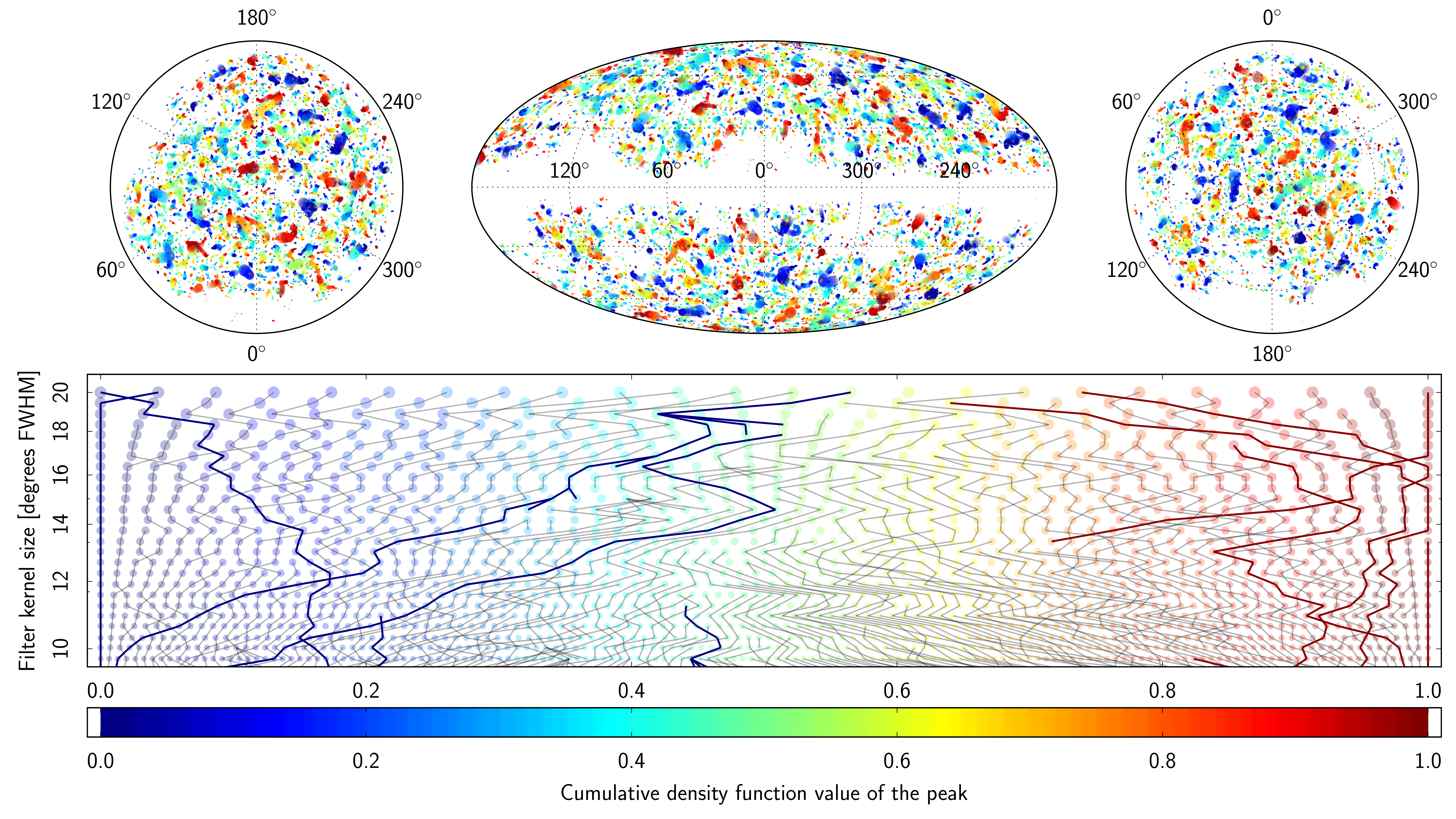}
  \caption{Peak positions and CDF rank summarized for all filtering
    scales.  The three sky-view panels in the top row show a Lambert
    projection of the north pole, the usual full sky Mollweide
    projection, and a Lambert projection of the south pole. The lower
    panel shows the peak heights (in percentile of the peak
    distribution on the horizontal axis) as a function of filter scale (on
    the vertical axis, in logarithmic scale), truncated to larger scales
    for clarity. Circles represent peaks (nodes of the graph) coloured
    according to their percentile level, and scaled according to
    kernel size. Black lines represent edges connecting peaks at
    different scales (according to a minimal distance measure). The
    components connected to the coldest and hottest peaks at any scale
    are highlighted by thicker edges, and are navy blue and dark red
    in colour. Note that there are thick lines that do not touch the 0
    and 1 percentiles in the plot view. Those edges are connected to
    extreme percentile values, but at scales smaller than those shown
    in the plot. The \cs\ is represented by the connected nodes
    that have the smallest percentiles except for the coarsest scale
    in the plot view.}
  \label{fig:peaks:tree}
\end{figure*}

The application of a filter kernel of variable size to a map extends
it into what can be considered a ``multiscale space,'' such that
features on different scales are represented by a one-parameter family
of smoothed maps.  This technique is often used for feature detection
and mathematical morphology analysis. Here, we introduce a
morphological description of temperature maps based on the peak
connectedness graph in multiscale space, and apply this technique to a
statistical analysis of the \Planck\ CMB data.  Like most
morphological analyses, it is equally applicable to intrinsically
non-Gaussian maps, but here we focus on the Gaussian random field
statistics and attempt to understand what features of the CMB
temperature map are responsible for the \cs.

To construct a multiscale representation, we trace the location of the
peaks in the smoothed, whitened CMB map as the smoothing scale is
varied. As the smoothing scale increases, peaks merge and the total
peak count decreases. Linking closest peak neighbours in
position-scale space, from the finest to the coarsest resolution,
produces an acyclic graph that encapsulates the peak ``merger tree''
history as the scale is varied. A summary of all the peak positions and
CDF ranks for the SSG84 filter kernel on scales ranging from
$120\arcm$ to $1200\arcm$ FWHM is shown in
Fig.~\ref{fig:peaks:tree}. The peaks are represented by discs of
varying size (reflecting the filter scale) and colour (reflecting the
peak temperature rank), with peaks at all scales projected onto a
single map. The lower panel shows the peak linkage graph on the
coarser scales; for the statistical analysis 81 filter scales are
used, log-spaced from $120\arcm$ to $1200\arcm$. Peaks of the same
type (i.e., maxima to maxima and minima to minima) are linked to the
closest peak on the coarser scale according to a distance measure,
$\dd s^2 +\dd f^2$, where $\dd s^2$ is the metric on the unit sphere, and
$\dd f^2$ is the difference of peak temperature ranks (but only if that
distance is within a predetermined fraction of the filter scale FWHM).

\begin{figure}
  \centering
  \includegraphics[width=8cm]{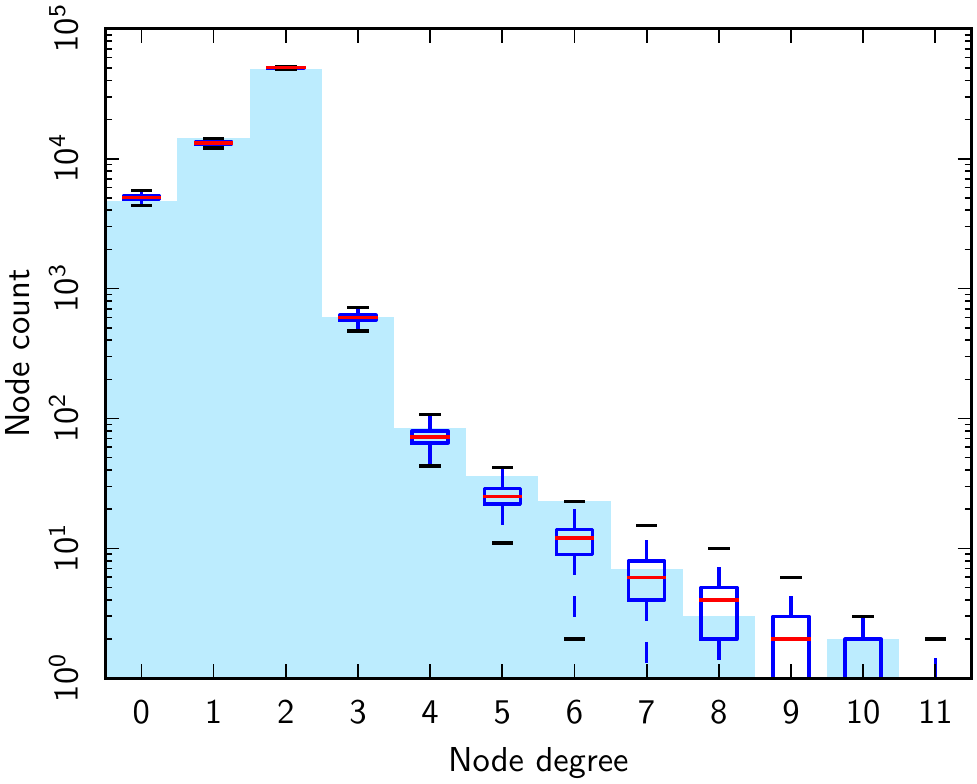}
  \caption{Distribution of node degrees in the multiscale peak linkage
    graph determined for the \smica\ map (cyan), compared with the median
    (red line), first to third quartile (blue box), and 95\,\%
    (whiskers) derived from 1000 FFP8 simulations.  }
  \label{fig:peaks:degree}
\end{figure}

The resulting peak linkage graph is then analysed for
connectedness. The simplest quantifiable measure is the node-degree
distribution, shown in Fig.~\ref{fig:peaks:degree} for \smica. The
node-degree distribution is highly peaked at 2; this population
corresponds to a single peak being traced across multiple
scales. Pre-whitening effectively decorrelates the Gaussian map across
different scales, so that the resulting node distribution shows a
sizeable population of degree 0 and 1 nodes. When compared to the
linkage graphs derived from the simulation ensemble, the node-degree
distribution of the peak linkage graph derived from \Planck\ CMB data
is consistent, with a slight excess in node counts of degrees 5 and 6.

\section{Anomalies in the microwave sky}
\label{sec:anomalies}

The previous section established the lack of evidence for significant
non-Gaussianity in the \Planck\ data. Here we consider several
important anomalies that were originally detected in the WMAP sky
maps, and later confirmed in the analyses described in
\citetalias{planck2013-p09}. Many of these are connected to evidence
for a violation of isotropy, or to a preferred direction, in the CMB.
Tests that involve dipolar power asymmetry, either directly or via
measures of directionality, are collected together in
Sect.~\ref{sec:dipmodsection}.  In this section we consider
only those anomalies not directly related to dipolar power asymmetry.

The microwave sky is intrinsically
statistically anisotropic due to our motion with respect to the CMB
rest frame. The resulting Doppler boosting effect, introduced in
Sect.~\ref{sec:introduction}, was detected in the 2013 \Planck\ data
\citep{planck2013-pipaberration}. For completeness,
Appendix~\ref{sec:dboost} repeats the analysis with the \Planck\ full
mission data set. However, since the effects of Doppler boosting are
now included in the simulations used for that analysis, this
constitutes a consistency check for this release. More importantly,
since both the data and simulations now include the effect, it is not
necessary to consider deboosted data in many of the studies reported
here, unlike in \citetalias{planck2013-p09} (although one exception in
Sect.~\ref{sec:biposh} makes use of unboosted simulations to search
for the frequency-dependent signature of the effect in the {\tt
  SEVEM-100}, {\tt SEVEM-143}, and {\tt SEVEM-217} sky maps).

Before presenting our results, we return to the issue of a posteriori
correction, which particle physicists refer to as correcting for the
``look-elsewhere effect'' (LEE).  Since there are many tests that can
be performed on the data to look for a violation of statistical
isotropy, we expect some to indicate detections at, for example,
roughly $3\,\sigma$ levels, since even a statistically isotropic CMB
sky is a realization of an underlying statistical process
corresponding to many independent random variables.  However, in the
absence of an existing theoretical framework (i.e., a physical model)
to predict such anomalies, it is difficult to interpret their
significance. It is then necessary, and equally challenging, to
address the question of how often such detections would be found for
statistically isotropic Gaussian skies. Unfortunately, it is not
always clear how to answer this question.

There will always be a degree of subjectivity when deciding exactly
how to assess the significance of these types of features in the data.
As an example, one could argue that the large-scale dipole modulation
signal we see is coming specifically from super-Hubble modes, in which
case performing a LEE correction for dipole modulation that could have
been seen on small scales ($\ell \gtrsim 100$) would not make sense.
Models for such a super-Hubble modulation exist and an example was
examined in \citet{planck2014-a24}, the conclusion being that the
model could only explain part of the dipole modulation and that the
allowed part was perfectly consistent with cosmic variance.

In this paper, we adopt a pragmatic approach. When there is a clear
mechanism for doing so, we attempt to correct for the ``multiplicity
of tests,'' or the possible ways in which an anomalous signal might
have been detected but was not, as a consequence of any a posteriori
(data-driven) choices made in searching for it.  In such cases, a strong
dependence of the significance on the correction would indicate that we
should be cautious about the uncorrected result.  When such an obvious
correction is not possible, we clearly describe the methodology applied
to the data and its limitations. With this approach, we also recognize that
any statistical assessment is partially subjective, including those that
purport to correct for the LEE.

Although many of the observed effects described in this and the next
section may elude theoretical prediction today, we continue to
highlight them since there is a real possibility that the significance
of one or more might increase at a later date, perhaps when
polarization data are included in the analysis, and lead to new
insights into early Universe physics. Alternatively, such observations
may directly motivate the construction of models that can make
predictions for features that can be sought in new data sets. This is
particularly the case for anomalies on the largest angular scales,
which may have a specific connection to inflation.

\subsection{Variance, skewness, kurtosis}
\label{sec:lowvariancemap}

Previous analyses of the WMAP data \citep{Monteserin2008, Cruz2011,
  Gruppuso2013} have reported that the variance of the CMB sky is lower
than that of simulations based on the $\Lambda$CDM
model. \citetalias{planck2013-p09} confirmed this, and proposed a
possible explanation of the apparent incompatibility of the observed
variance with a fiducial cosmological model that has been determined
from the same data set.  Specifically, whilst the map-based variance
is dominated by contributions from large angular scales on the sky,
the cosmological parameter fits are relatively insensitive to these
low-order $\ell$-modes, and are instead largely dominated by scales
corresponding to $\ell > 50$. Therefore the variance of the map
appears to be anomalous, since there is a dearth of large-angular-scale 
power compared to the model.

\begin{figure} \centering
  \includegraphics[width=\hsize]{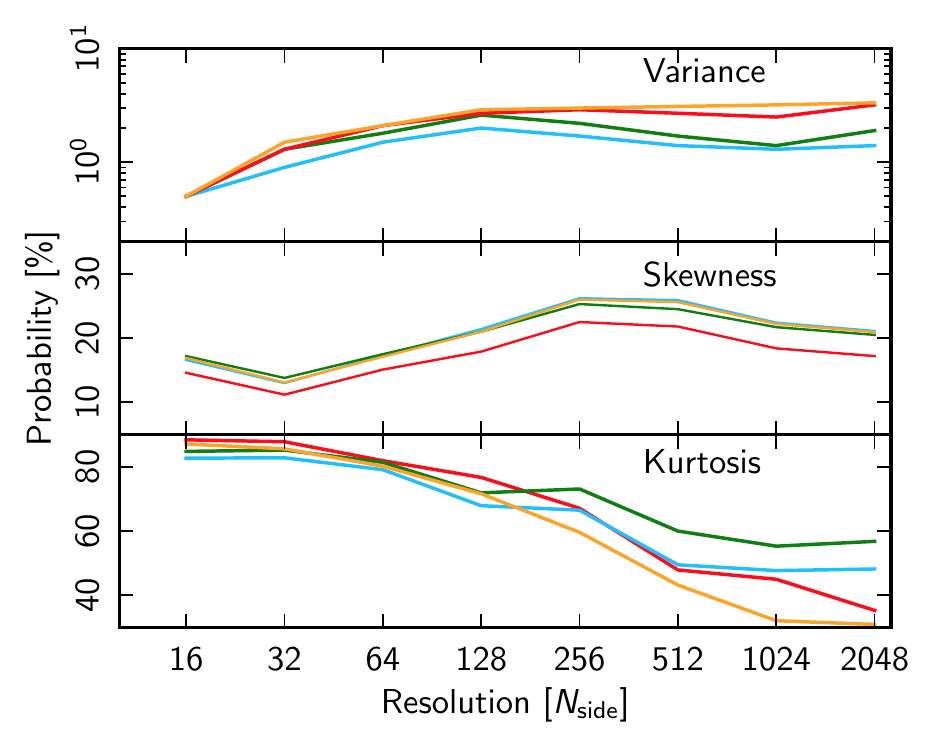} \caption{Lower
    tail probability of the variance (top panel),
    skewness (centre panel), and kurtosis (bottom panel) obtained at
    different resolutions from the {\tt Commander} (red), {\tt NILC}
    (orange), {\tt SEVEM} (green), and {\tt SMICA} (blue) sky maps.}
\label{fig:lowvariance}
\end{figure}

In Sect.~\ref{sec:onepdf}, we again confirmed the presence of low
variance in the data. Here, we extend the analysis to investigate
which angular scales are responsible for the low variance by applying
the unit variance estimator to lower resolution component-separated
maps, specifically those from $N_{\rm side}=1024$ to $N_{\rm
  side}=16$, with the corresponding common masks, and then comparing
the results with those determined from 1000 MC simulations.  The
results are shown in Fig.~\ref{fig:lowvariance}.

All of the component-separation methods that we consider yield very
consistent results which indicate an increasingly anomalous low
variance at lower resolutions, with the lower-tail probability
reaching a minimum value of 0.5\,\% at $N_{\rm side}=16$.  We then
consider the impact of a possible look-elsewhere effect by evaluating
the minimum lower tail probability of each simulation irrespective of
the \nside\ resolution at which it occurs. By comparing the
distribution of these values with that of the data, we infer that the
probability is slightly weakened to a value of about 1\,\%.  These
results are compatible with a lack of power on large angular
scales. Since the variance estimator is heavily weighted towards
low-$\ell$ modes, this has an increasing impact on the observed
variance when going from high to low resolution sky maps. Conversely,
the skewness and kurtosis are consistent with the simulations,
although there is some indication of a weak scale-dependence (albeit
at low significance).

We also investigate the stability of the results at $N_{\rm side}= 16$
with respect to the possible presence of residual foregrounds by
considering two additional masks obtained by extending the edge of the
$N_{\rm side} = 16$ common mask by 5\degr\ and 9\degr, reducing the usable
sky fraction from 58\,\% to 48\,\% and 40\,\%, respectively.  We then
re-apply the unit variance estimator to the low resolution
component-separated maps with these masks and determine the variance,
skewness, and kurtosis values (see Table~\ref{Table:variance16}).

\begin{table} \begingroup \newdimen\tblskip \tblskip=5pt
  \caption{Lower-tail probability for the variance,
    skewness, and kurtosis of the low resolution $\nside = 16$ component-separated
    maps obtained with the common mask and two extended versions thereof.
    }
\label{Table:variance16}
\nointerlineskip
\vskip -3mm
\footnotesize
\setbox\tablebox=\vbox{
   \newdimen\digitwidth
   \setbox0=\hbox{\rm 0}
   \digitwidth=\wd0
   \catcode`*=\active
   \def*{\kern\digitwidth}
   \newdimen\signwidth
   \setbox0=\hbox{+}
   \signwidth=\wd0
   \catcode`!=\active
   \def!{\kern\signwidth}
\halign{\hbox to 1.0in{#\leaderfil}\tabskip 4pt&
         \hfil#\hfil&
         \hfil#\hfil&
         \hfil#\hfil\tabskip 0pt\cr                           
\noalign{\doubleline\vskip -1pt}
\omit&\multispan3 \hfil Probability [\%]\hfil\cr
\noalign{\vskip -4pt}
\omit&\multispan3\hrulefill\cr
\omit\hfil Method\hfil& Variance& Skewness& Kurtosis\cr 
\noalign{\vskip 3pt\hrule\vskip 3pt}
\multispan4 \hfil Common mask ($f_{\rm sky}=58\,\%$) \hfil \cr
\noalign{\vskip 3pt}
{\tt Commander}& *0.5 & 14.6 & 88.4\cr
{\tt NILC}& *0.5 & 16.9 & 87.1\cr
{\tt SEVEM}& *0.5 & 17.2 & 84.8\cr
{\tt SMICA}& *0.5 & 16.6 & 82.7\cr
\noalign{\vskip 3pt}
\multispan4 \hfil $f_{\rm sky}=48\,\%$ \hfil \cr
\noalign{\vskip 3pt}
{\tt Commander}& *0.1 & 29.4 & 65.0\cr
{\tt NILC}& *0.1  & 29.6 & 60.8\cr
{\tt SEVEM}& *0.1 & 29.4 & 62.4\cr
{\tt SMICA}& *0.1 & 29.4 & 57.3\cr
\noalign{\vskip 3pt}
\multispan4 \hfil $f_{\rm sky}=40\,\%$ \hfil \cr
\noalign{\vskip 3pt}
{\tt Commander} & *0.4 & 35.2 & 32.4\cr
{\tt NILC} & *0.4 & 34.4 & 28.7\cr
{\tt SEVEM} & *0.4 & 34.3 & 30.2\cr
{\tt SMICA} & *0.4 & 33.8 & 25.5\cr
\noalign{\vskip 4pt\hrule\vskip 3pt}}}
\endPlancktable                    
\endgroup
\end{table}

The results from 48\,\% of the sky reveal that only 1 simulation in
1000 is found to be more anomalous (i.e., exhibit lower variance) than
the observed map.  In addition, both the skewness and kurtosis become
more compatible with the $\Lambda$CDM model.  With the more aggressive
mask, the lower-tail probability slightly increases again. However,
given the limited number of pixels involved in the analysis, this
shift may be related to the effects of sample variance.

Overall, our results may be explained by the presence of a
low-variance anomaly in the primordial CMB signal --- the stability of
the low-variance significance argues against foreground contamination
being responsible for the lack of observed power. This is reinforced
by the decrease in variance when regions close to the common mask
borders, where foreground residuals are most likely to be observed,
are omitted from the analysis.

\subsection{$N$-point correlation function anomalies}
\label{sec:n_point_asymmetry}

\subsubsection{Lack of large-angle correlations}
We first reassess the lack of correlation seen in the 2-point angular
correlation function at large angular separations as reported in
Sect.~\ref{sec:npoint_correlation}, and previously noted for both WMAP
and the 2013 \Planck\ temperature maps
\citep{bennett2003a,copi2013}. We attempt to quantify this lack of
structure using the statistic proposed by \citet{spergel2003}:
\begin{linenomath*}
\begin{equation}
S(x) =  \int_{-1}^x \left[ \hat{C}_2(\theta) \right]^2 \dd(\cos \theta)\ ,
\label{eqn:s_stat}
\end{equation}
\end{linenomath*}
where $\hat{C}_2(\theta)$ is our estimate of the 2-point correlation
function.  Generally, the upper limit on the integral has been taken
to correspond to a separation angle of 60\deg, possibly \citep[as
noted by][]{copi2009} motivated by the COBE-DMR 4-year results
\citep{hinshaw1996}. Inspection of the top panel of
Fig.~\ref{fig:npt_data_temp} suggests that the \Planck\ 2-point
function lies close to zero between $80\degr$ and $170\degr$, but
for consistency with previous work we compute the statistic $S_{1/2}$,
for $x \equiv \cos 60\degr = {1 \over 2}$. The results are presented
in Table~\ref{tab:prob_stat_2pt_temp}. We find that the data indeed
show a lack of correlations on large angular scales, with a
significance consistent with that found by \citet{copi2013} (although
note that the sense of the $p$-values differs between the papers).

\begin{table}[tp]
\begingroup
\newdimen\tblskip \tblskip=5pt
\caption{Probabilities of obtaining values for the $S_{1/2}$ and
  $\chi_0^2$ statistics for the \Planck\ fiducial $\Lambda$CDM model at least as
  large as the observed values of the statistic for the \Planck\ 2015
  temperature CMB maps with resolution parameter $N_{\rm side}=64$, estimated
  using the \commander, \nilc, \sevem,
  and \smica\ maps. We show also the corresponding estimation of the global $p$-value for
  the $S(x)$ statistic.}
\label{tab:prob_stat_2pt_temp}
\nointerlineskip
\vskip -3mm
\footnotesize
\setbox\tablebox=\vbox{
   \newdimen\digitwidth
   \setbox0=\hbox{\rm 0}
   \digitwidth=\wd0
   \catcode`*=\active
   \def*{\kern\digitwidth}
   \newdimen\signwidth
   \setbox0=\hbox{+}
   \signwidth=\wd0
   \catcode`!=\active
   \def!{\kern\signwidth}
\halign{\hbox to 1.6in{#\leaderfil}\tabskip 4pt&
\hfil#\hfil&
\hfil#\hfil&
\hfil#\hfil&
\hfil#\hfil\tabskip 0pt\cr
\noalign{\doubleline\vskip -1pt}
\omit&\multispan4\hfil Probability [\%]\hfil\cr
\noalign{\vskip -4pt}
\omit&\multispan4\hrulefill\cr
\omit\hfil Statistic \hfil& {\tt Comm.} & {\tt NILC} & {\tt SEVEM} & {\tt SMICA}\cr
\noalign{\vskip 3pt\hrule\vskip 3pt}
$S_{1/2}$&  99.5 &  99.6 &  99.5 &  99.6 \cr
$S(x)$ (global) & 97.7 &   97.8 &   97.8 &   97.9 \cr
$\chi_0^2(\theta > 60\degr)$ &  98.1 &  98.8 &  98.1 &  98.4 \cr
 \noalign{\vskip 3pt\hrule\vskip 3pt}}}
\endPlancktable                    
\endgroup
\end{table}

Possible criticisms of the $S_{1/2}$ statistic include that it has
been designed a posteriori to test for a lack of large-angle
correlations, and that it does not account for the high degree of
correlation between bins at different angular scales. We can address these concerns, at least
in part, by considering a modified version of
the commonly used and well understood $\chi^2$ statistic used in
previous studies.  In order to test the same hypothesis as the
$S_{1/2}$ statistic --- that there are no correlations on scales larger
than some angular cut-off --- we do not subtract an averaged 2-point
function when computing the $\chi^2$, i.e.,~we use a statistic defined
as
\begin{linenomath*}
\begin{equation}
\chi_0^2(\theta_{\rm min},\theta_{\rm max}) = \sum_{i,j=i_{\rm
    min}}^{i_{\rm max}} \hat{C}_2 (\theta_i) \tens{M}_{ij}^{-1} \hat{C}_2(\theta_j) \ ,
\label{eqn:chisqr_stat_zero}
\end{equation}
\end{linenomath*}
where $i_{\rm min}$, $i_{\rm max}$ denote the index of the bins
corresponding to the minimum and maximum value of the separation
angles $\theta_{\rm min}$ and $\theta_{\rm max}$, respectively.  In
this analysis, we adopt $\theta_{\rm min}=60\degr$ and $\theta_{\rm
  max} = 180\degr$. $\tens{M}_{ij}$ is the covariance matrix given by
Eq.~\eqref{eqn:cov_mat}, estimated using MC simulations corresponding
to the fiducial $\Lambda$CDM model.  The results are shown in
Table~\ref{tab:prob_stat_2pt_temp}.  The significance level of the
anomaly is slightly smaller for the $\chi_0^2$ statistic than that
derived with $S_{1/2}$. We note that this statistic is closely related
to the $A(x)$ measure proposed by \citet{hajian2007}.

A further potential criticism of the $S_{1/2}$ statistic relates to
the a posteriori choice of 60\degr\ for the separation angle that
delineates the interesting region of behaviour of the correlation
function. We therefore consider the generalized statistic $S(x)$ and
compute its value for all values of $x$, both for the data and for the
simulations.  Then, for each value of $x$, we determine the number of
simulations with a higher value of $S(x)$, and hence infer the most
significant value of the statistic and the separation angle that it
corresponds to.  However, since such an analysis is sensitive to the
LEE, we define a global statistic to evaluate the true significance of
the result. Specifically, we repeat the procedure for each simulation,
and search for the largest probability irrespective of the value of
$x$ at which it occurs. The fraction of these probabilities higher
than the maximum probability found for the data defines a global $p$-value.
As seen in Table~\ref{tab:prob_stat_2pt_temp}, this
corresponds to values of order 98\,\% for all of the CMB
estimates.

The previous analyses essentially test how consistent the observed
2-point correlation function data is with a lack of correlations on
large angular scales, in particular for separation angles $\theta >
60\degr$.  A conventional $\chi^2$ statistic allows us to test the
consistency of this quantity with the predictions of the $\Lambda$CDM
model. In this case, the statistic is defined as in
Eq.~\eqref{eqn:chisqr_stat}, except that we constrain the computations
to those bins that correspond to the intervals defined by
$\theta<60\degr$ and $\theta>60\degr$. The results of these studies
are shown in Table~\ref{tab:prob_chisq_2pt_temp}.

\begin{table}[tp]
\begingroup
\newdimen\tblskip \tblskip=5pt
\caption{Probabilities of obtaining values for the $\chi^2$ statistic
  for the \Planck\ fiducial $\Lambda$CDM model at least as
  large as the observed values of the statistic for the \Planck\ 2015
  temperature CMB maps with resolution parameter $N_{\rm side}=64$, estimated
  using the \commander, \nilc, \sevem,  and \smica\ maps.}
\label{tab:prob_chisq_2pt_temp}
\nointerlineskip
\vskip -3mm
\footnotesize
\setbox\tablebox=\vbox{
   \newdimen\digitwidth
   \setbox0=\hbox{\rm 0}
   \digitwidth=\wd0
   \catcode`*=\active
   \def*{\kern\digitwidth}
   \newdimen\signwidth
   \setbox0=\hbox{+}
   \signwidth=\wd0
   \catcode`!=\active
   \def!{\kern\signwidth}
\halign{\hbox to 1.6in{#\leaderfil}\tabskip 4pt&
\hfil#\hfil&
\hfil#\hfil&
\hfil#\hfil&
\hfil#\hfil\tabskip 0pt\cr
\noalign{\doubleline\vskip -1pt}
\omit&\multispan4\hfil Probability [\%]\hfil\cr
\noalign{\vskip -4pt}
\omit&\multispan4\hrulefill\cr
\omit\hfil Statistic \hfil& {\tt Comm.} & {\tt NILC} & {\tt SEVEM} & {\tt SMICA}\cr
\noalign{\vskip 3pt\hrule\vskip 3pt}
$\chi^2(\theta < 60\degr)$&  91.5 &  93.3 &  91.6 &  91.7\cr
$\chi^2(\theta > 60\degr)$&  96.8 &  98.3 &  96.9 &  98.1\cr
 \noalign{\vskip 3pt\hrule\vskip 3pt}}}
\endPlancktable                    
\endgroup
\end{table}

The analysis for $\theta<60\degr$ indicates that the
observed 2-point function is a good match to the mean 2-point function
predicted by the $\Lambda$CDM model.  Moreover, for $\theta>60\degr$
the results suggest that the problem is that the fit of the data to
the model is too good, and this is even more pronounced for an
analysis in the full separation angle range.

Overall, the tests indicate an unusually good fit of the observed
2-point function both to zero and to the predictions of the $\Lambda$CDM
model for angles above $60\degr$. This problem may be related to the
fact that the theoretical variance for the best-fit model is larger
than the observed value at large scales, so that the simulations based on
this model that have been used in all of the statistical tests may
overestimate the variance of the 2-point function.

\subsubsection{Hemispherical asymmetry}
We now turn to a reassessment of the asymmetry between the real-space
$N$-point correlation functions computed on hemispheres reported
previously for the WMAP \citep{eriksen2005} and \Planck\ 2013
temperature maps \citepalias{planck2013-p09}.  We initially focus the
analysis on the hemispheres determined in the ecliptic coordinate
frame for which the largest asymmetry was observed. However, we also
carry out the corresponding calculations in other relevant reference
frames, such as those defined by the Doppler boost (DB, see
Sect.~\ref{sec:biposh}, Appendix~\ref{sec:dboost}, and
\citealt{planck2013-pipaberration}) and the dipole modulation (DM, see
Sect.~\ref{sec:dipmod}) directions.  We use the same configurations of
the $N$-point functions as described in
Sect.~\ref{sec:npoint_correlation}. However, here the functions are
not averaged over the full sky and depend on a choice of specific
direction, so they constitute tools for studying statistical
isotropy rather than non-Gaussianity \citep{ferreira1997}.

\begin{figure*}[htp!]
\begin{center}
\includegraphics[width=0.47\textwidth]{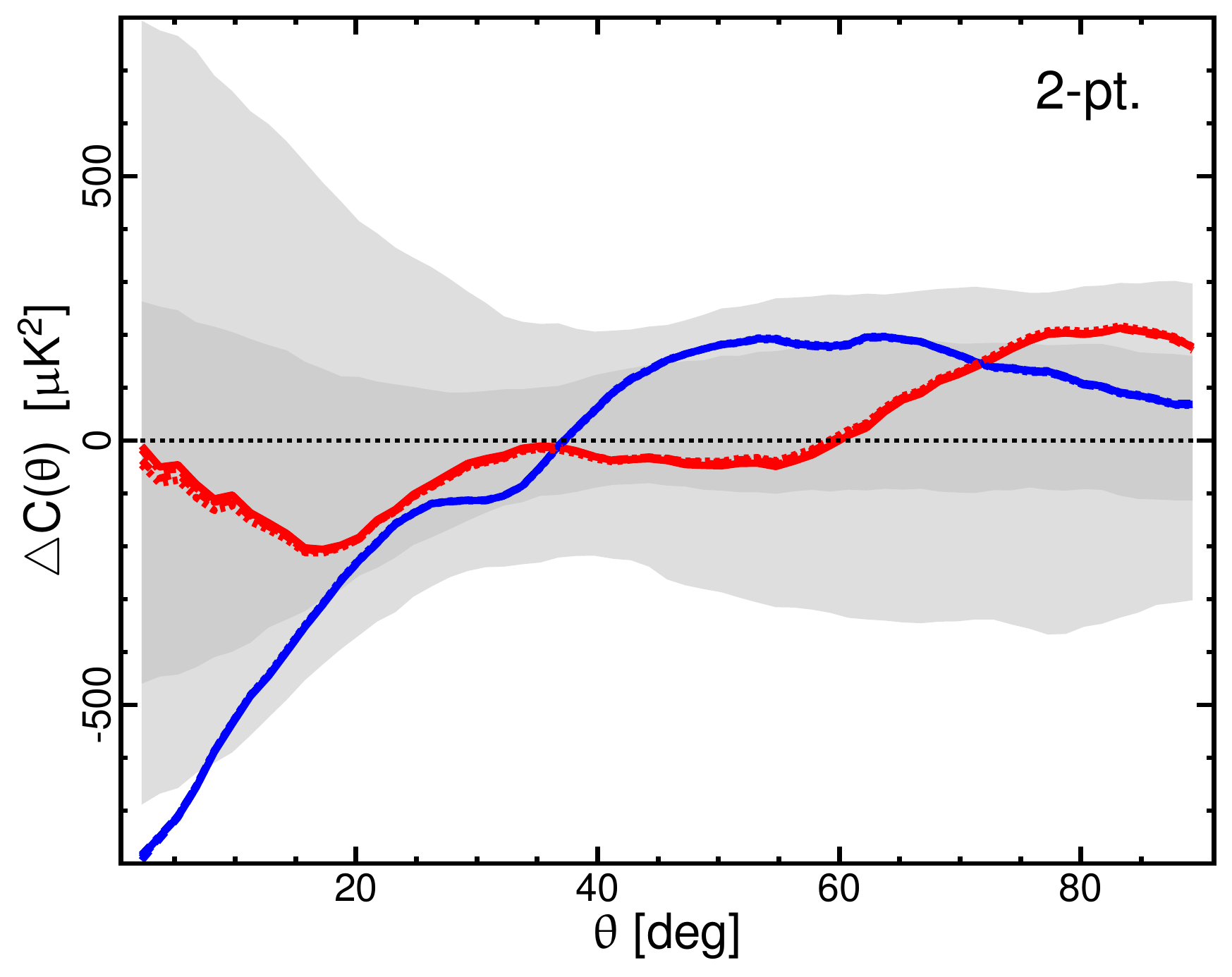}
\includegraphics[width=0.48\textwidth]{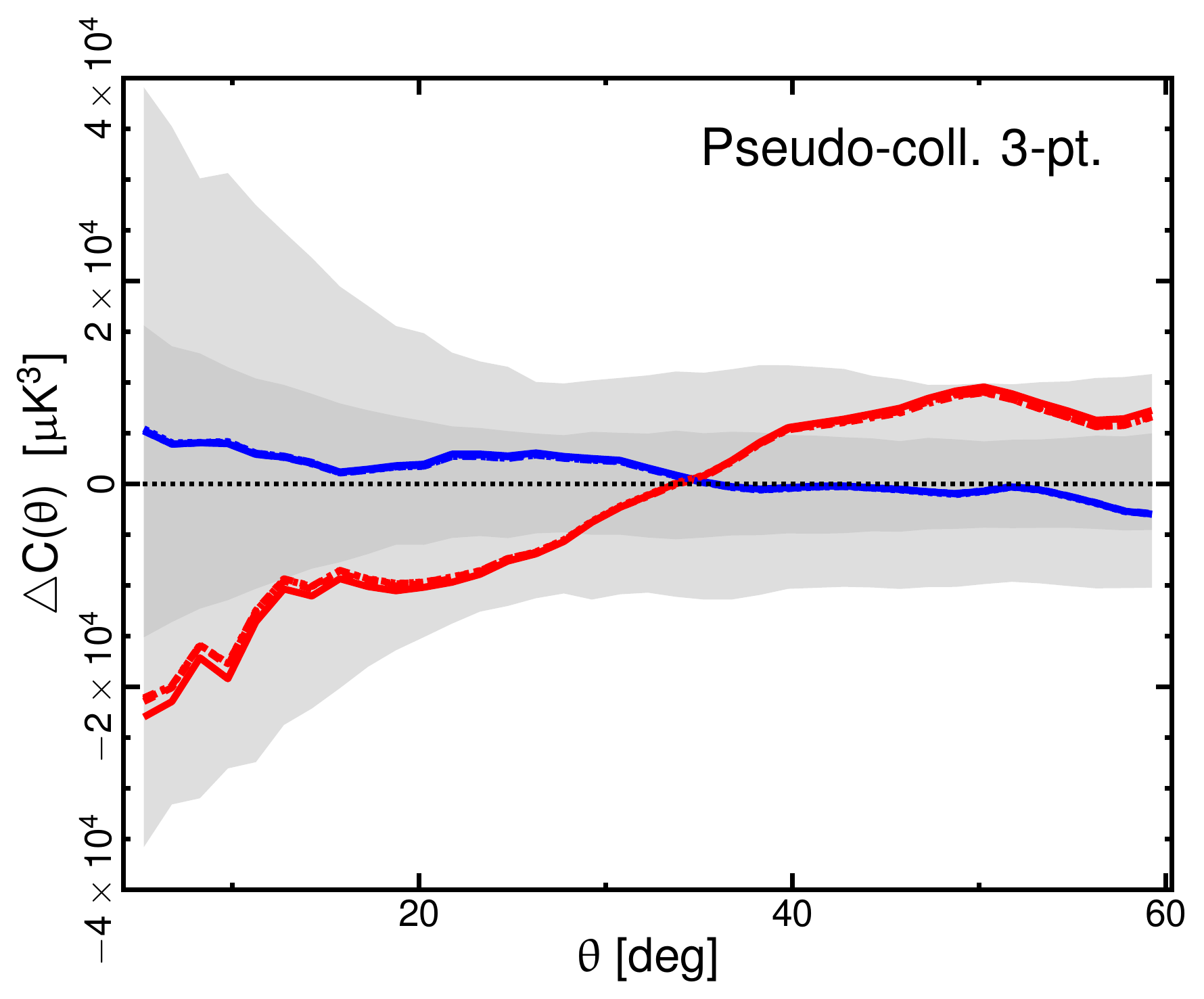}\\
\includegraphics[width=0.48\textwidth]{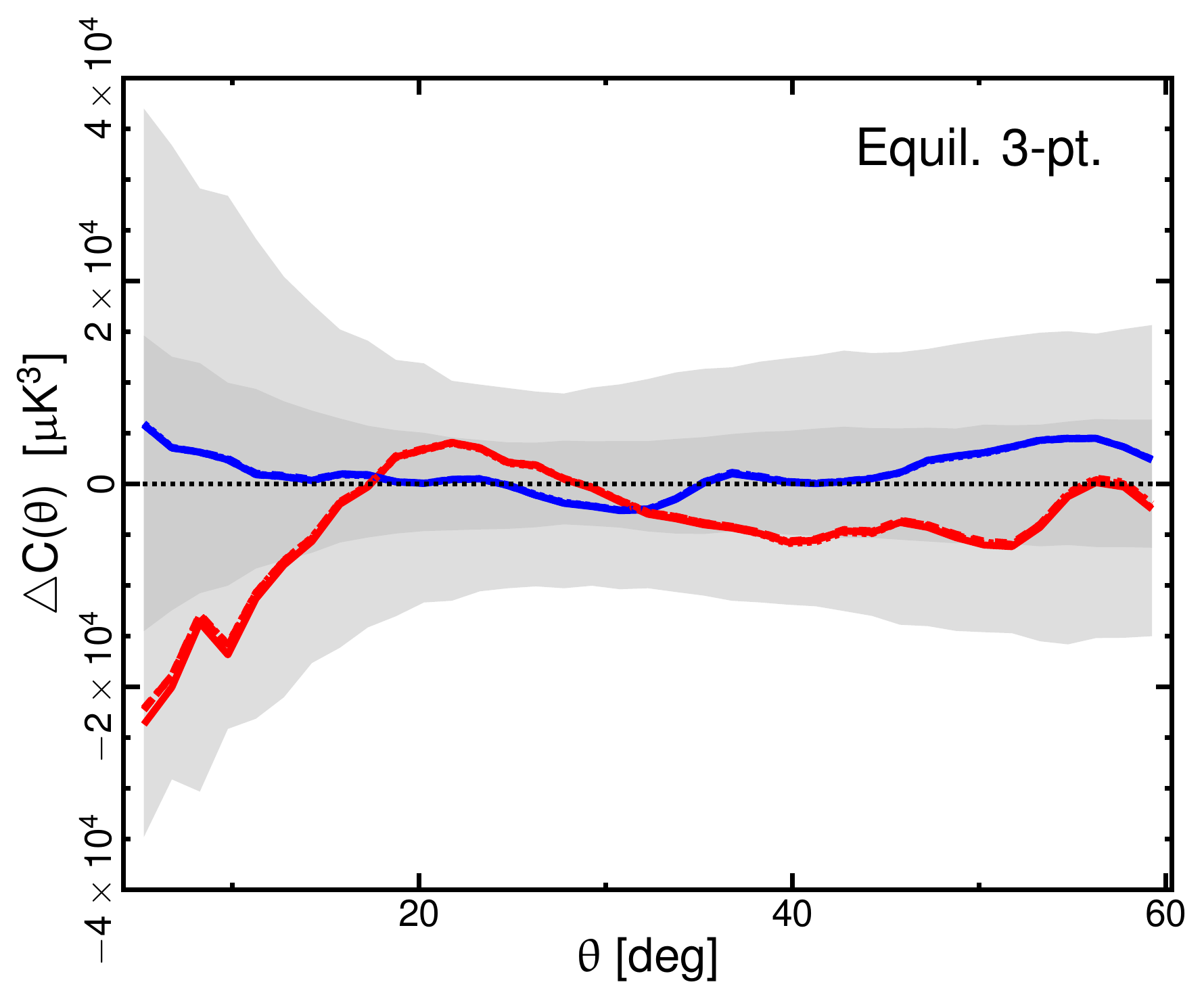}
\includegraphics[width=0.48\textwidth]{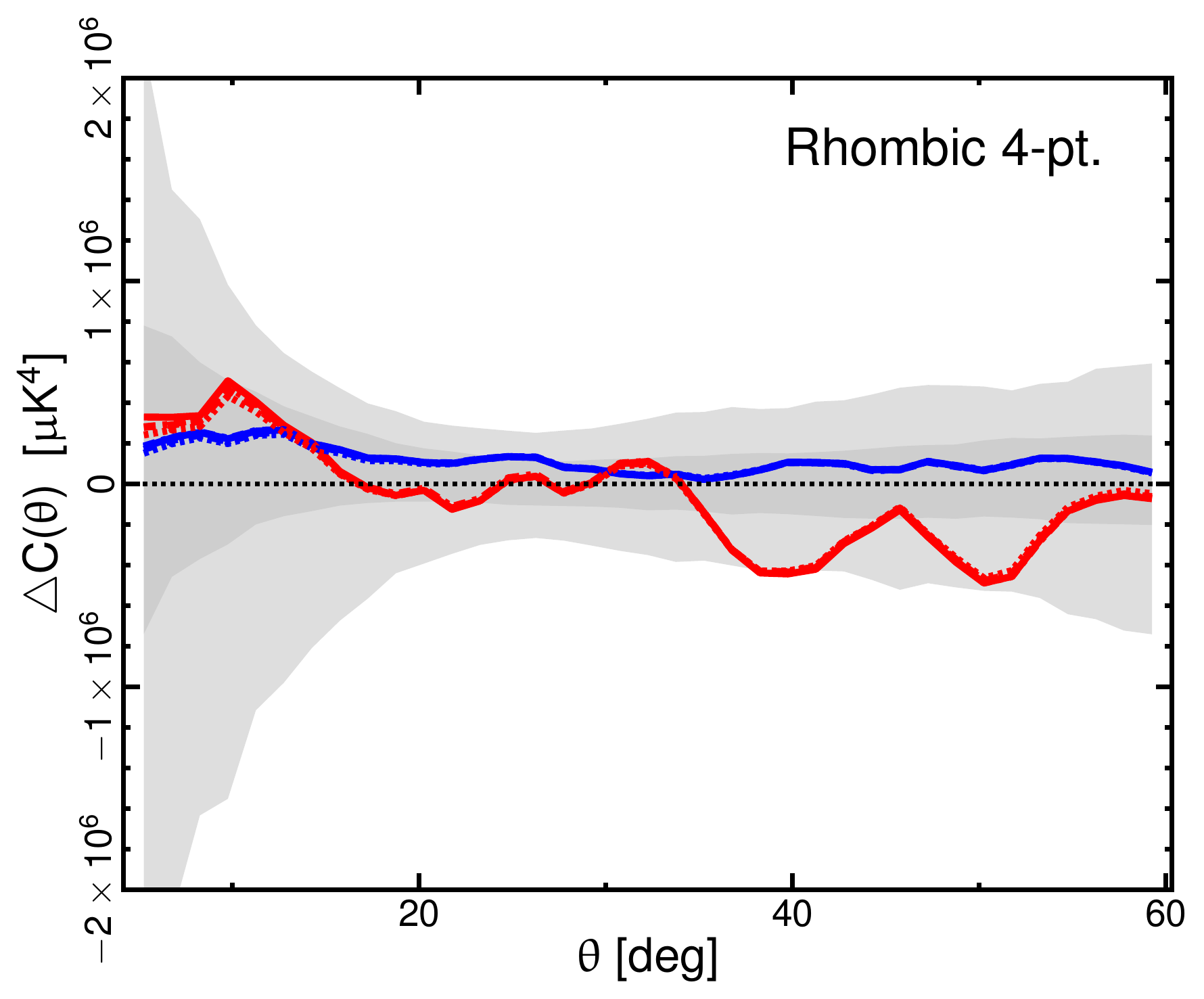}
\caption{Difference of the $N$-point correlation functions
  determined from the $N_{\rm side}=64$ \Planck\ CMB 2015 temperature
  estimates and the corresponding means estimated from 1000
  simulations.
Results are shown for the 2-point,
  pseudo-collapsed 3-point (upper left and right panels, respectively), equilateral 3-point, and connected rhombic 4-point
  functions (lower left and right panels, respectively). Correlation functions are shown for the
  analysis performed on northern (blue) and southern (red) hemispheres
  determined in the ecliptic coordinate frame. The solid, dashed,
  dot-dashed, and dotted lines correspond to the \commander, \nilc,
  \sevem,\ and \smica\ maps, respectively. Note that the lines lie on
  top of each other. The shaded
  dark and light grey regions indicate, for reference, the 68\,\% and 95\,\% confidence
  regions, respectively, determined from the \smica\ simulations.
}
\label{fig:npt_data_temp_ecl}
\end{center}
\end{figure*}

As in Sect.~\ref{sec:npoint_correlation}, we analyse the CMB estimates
at a resolution of $\nside = 64$ and quantify their agreement with the
fiducial cosmological model using a $\chi^2$ statistic.  The results
determined from the \Planck\ 2015 temperature data for the ecliptic
hemispheres are shown in Fig.~\ref{fig:npt_data_temp_ecl}.  If we
consider that the $\chi^2$ statistic itself can act as a measure of
fluctuation level, then asymmetry between the two measured hemispheres
can be quantified by the ratio of the corresponding $\chi^2$ values.
The probabilities of obtaining values of the $\chi^2$ statistic or
ratio for the \Planck\ fiducial $\Lambda$CDM model at least as large
as the observed values are given in
Table~\ref{tab:prob_chisq_npt_temp_ecl}.  Since we do not have any
predictions concerning the behaviour of a given hemisphere, in the
case of the $\chi^2$ ratios we provide the complementary probabilities
of the 2-tailed statistic.

The significance levels of the 3- and 4-point functions in the
northern hemisphere are nominally very high, exceeding 99.9\,\% for
the pseudo-collapsed 3-point function. However, proper interpretation
requires that one recognize that the analysis is affected by a
posteriori choices for the smoothing scale and reference frame
defining the hemispheres. This typically leads to an overestimation of
the significance of the results.  Accounting for such effects requires
the repetition of the analysis for all possible reference directions
and also for data at other resolutions.  Unfortunately, because of
computational limitations, such an extended analysis is not possible
for these higher-order statistics.  Nevertheless, the observed
properties of the \Planck\ data are consistent with a clear lack of
fluctuations in a direction towards the north ecliptic pole.  However,
the $\chi^2$-ratio statistic indicates a slightly smaller significance
level for the asymmetry, not exceeding 99\,\% for any of the $N$-point
functions.

\begin{figure*}[htp!]
\begin{center}
\includegraphics[width=0.47\textwidth]{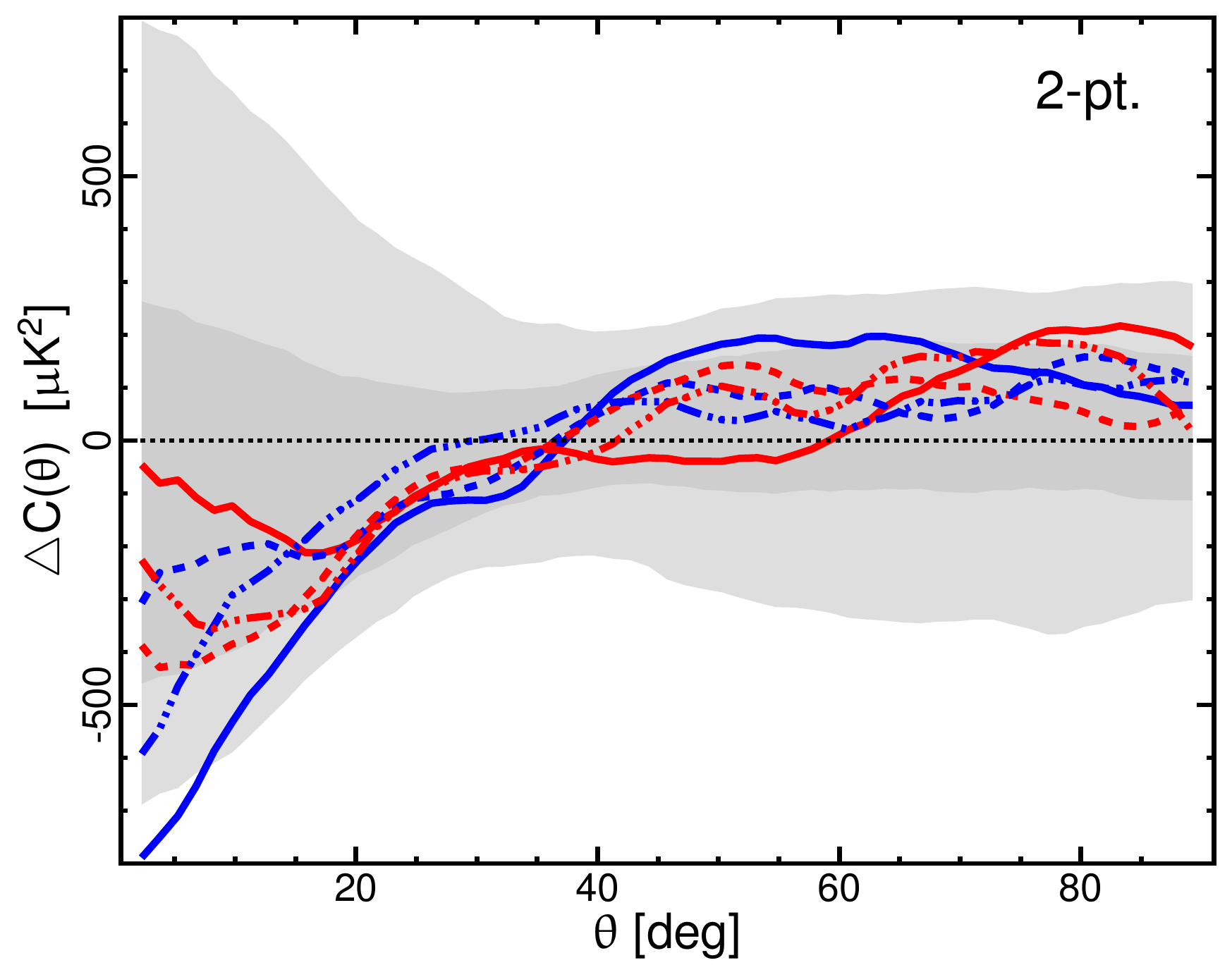}
\includegraphics[width=0.48\textwidth]{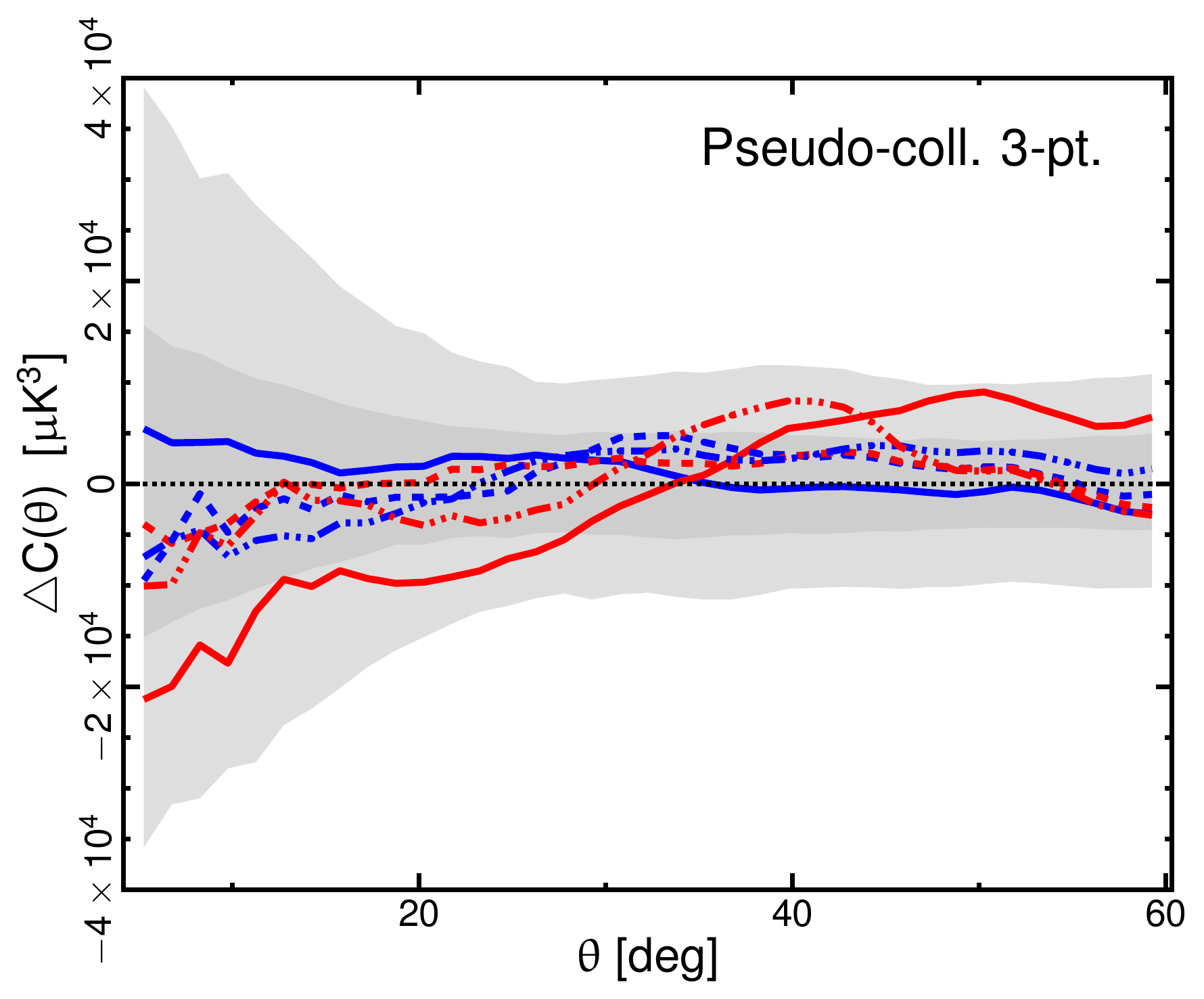}\\
\includegraphics[width=0.48\textwidth]{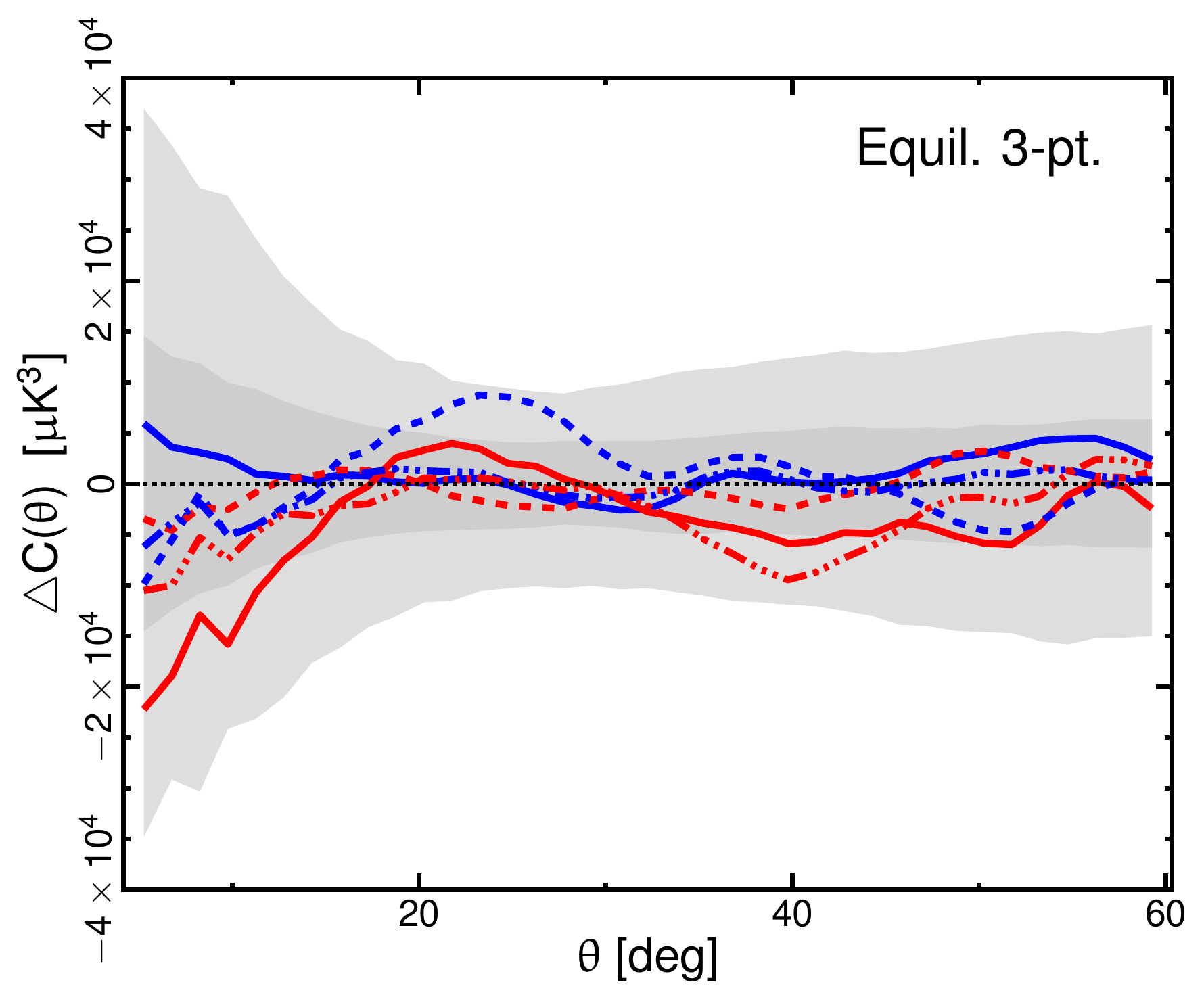}
\includegraphics[width=0.48\textwidth]{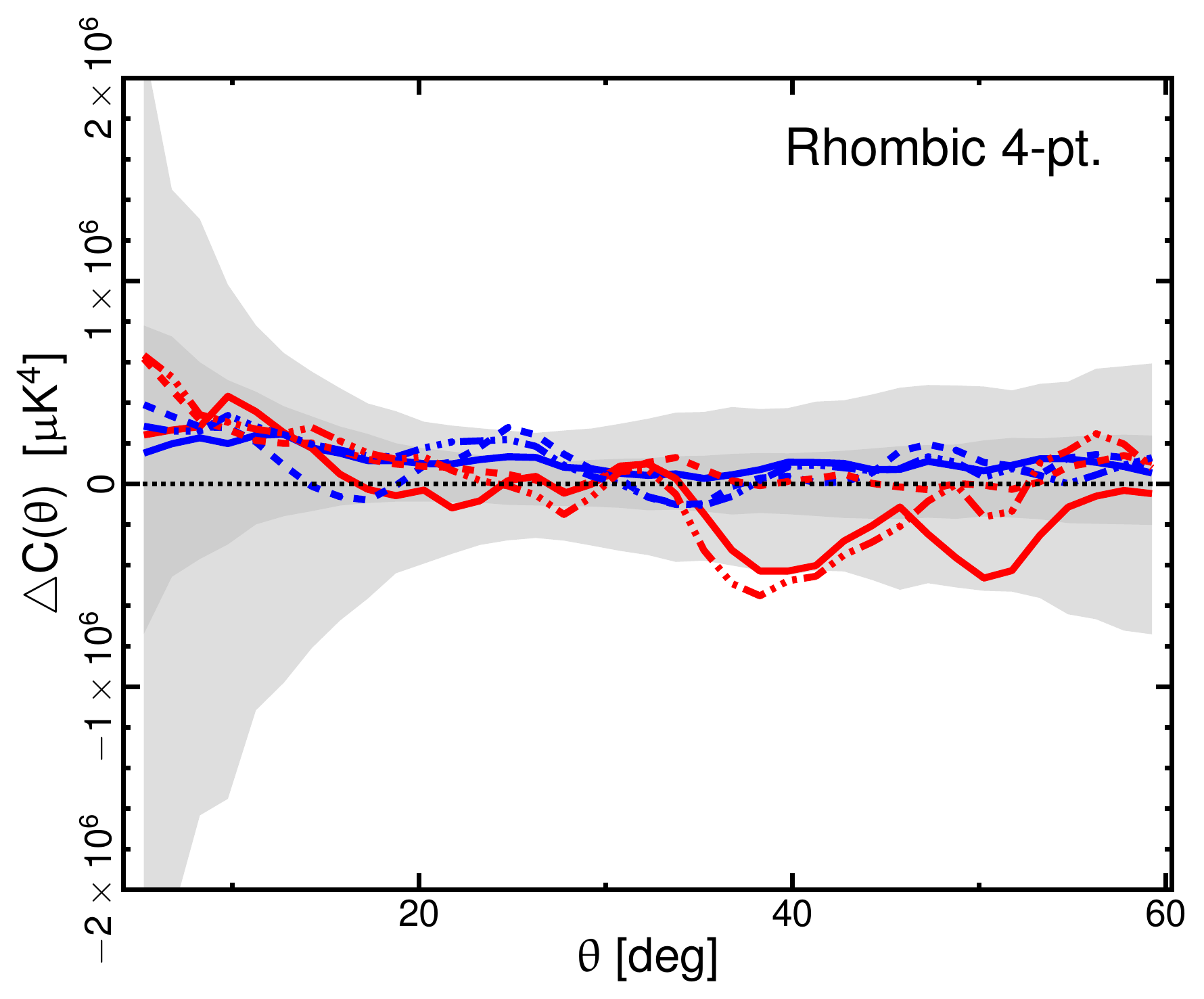}
\caption{Difference of the $N$-point correlation functions
  determined from the $N_{\rm side}=64$ \Planck\ \smica\ CMB 2015 temperature
  estimates and the corresponding means estimated from 1000
  simulations.
Results are shown for the 2-point,
  pseudo-collapsed 3-point (upper left and right panels, respectively), equilateral 3-point, and connected rhombic 4-point
  functions (lower left and right panels, respectively). Correlation functions are shown for the
  analysis performed on negative (blue) and positive (red) hemispheres
  determined in the ecliptic (solid lines), Doppler boost (DB, dashed
  lines), and dipole modulation (DM, dot-dashed lines) coordinate
  frames. The shaded dark and light grey regions indicate the 68\,\% and
  95\,\% confidence regions, respectively.
}
\label{fig:npt_data_temp_asym}
\end{center}
\end{figure*}

The results for the $N$-point correlation functions determined in the
DB and DM reference frames for the \smica\ map are shown in
Fig.~\ref{fig:npt_data_temp_asym} and the probabilities are presented
in Table~\ref{tab:prob_chisq_npt_temp_asym}. Note that the positive
hemisphere for the ecliptic reference frame corresponds to the
southern hemisphere in the previous table. Whilst the largest
asymmetry is seen in ecliptic coordinates, a substantial asymmetry is
present also for the DM direction. This can be explained by the fact
that the DM direction is more closely aligned with the south ecliptic
pole (with a separation of around 47$\degr$) than the DB direction
is. For the DB direction we do not find any significant asymmetry.
The equivalent results for \commander, \nilc, and \sevem\ are
consistent with those shown here.

\begin{table}[tp]
\begingroup
\newdimen\tblskip \tblskip=5pt
\caption{Probabilities of obtaining values for the $\chi^2$ statistic and ratio
  of $\chi^2$ of the $N$-point functions shown in
  Fig.~\ref{fig:npt_data_temp_ecl} for the
  \Planck\ fiducial $\Lambda$CDM model at least as
  large as the observed values of the statistic for the \Planck\ 2015 CMB
  maps estimated on northern and southern ecliptic  hemispheres.
}
\label{tab:prob_chisq_npt_temp_ecl}
\nointerlineskip
\vskip -3mm
\footnotesize
\setbox\tablebox=\vbox{
   \newdimen\digitwidth
   \setbox0=\hbox{\rm 0}
   \digitwidth=\wd0
   \catcode`*=\active
   \def*{\kern\digitwidth}
   \newdimen\signwidth
   \setbox0=\hbox{>}
   \signwidth=\wd0
   \catcode`!=\active
   \def!{\kern\signwidth}
\halign{\hbox to 1.6in{#\leaderfil}\tabskip 4pt&
\hfil#\hfil\tabskip 7pt&
\hfil#\hfil\tabskip 7pt&
\hfil#\hfil\tabskip 7pt&
\hfil#\hfil\/\tabskip 0pt\cr
\noalign{\doubleline}
\noalign{\vskip -1pt}
\omit&\multispan4 \hfil Probability [\%]\hfil\cr
\noalign{\vskip -4pt}
\omit&\multispan4\hrulefill\cr
\omit \hfil Hemisphere\hfil& {\tt Comm.}& {\tt NILC}& {\tt SEVEM}& {\tt SMICA}\cr
\noalign{\vskip 3pt\hrule\vskip 3pt}
\multispan5 \hfil 2-point function \hfil\cr
 Northern&  !89.7&  !90.6&  !89.8&  !88.0\cr
 Southern&  !80.5&  !82.7&  !82.9&  !77.6\cr
 $\chi^2$-ratio&  !22.6& !21.0&  !19.7&  !22.3\cr
\multispan5 \hfil Pseudo-collapsed 3-point function\hfil\cr
 Northern&  >99.9& >99.9& >99.9& !99.7\cr
 Southern&  !35.1&  !34.9&  !35.8&  !31.4\cr
$\chi^2$-ratio&  !98.8&  !98.5&  !98.5&  !98.4\cr
\multispan5 \hfil Equilateral 3-point function\hfil\cr
 Northern&  !98.6&  !98.6&  !98.8&  !98.4\cr
 Southern&  !45.7&  !45.7&  !47.8&  !42.6\cr
 $\chi^2$-ratio&  !86.6&  !86.7&  !86.6&  !86.7\cr
\multispan5 \hfil Rhombic 4-point function\hfil\cr
 Northern&  !99.7&  !99.7&  !99.7&  !99.6\cr
 Southern&  !22.8&  !22.5&  !23.2&  !20.1\cr
 $\chi^2$-ratio&  !97.3&  !97.1&  !97.2&  !97.0\cr
\noalign{\vskip 3pt\hrule\vskip 3pt}}}
\endPlancktable                    
\endgroup
\end{table}

\begin{table}[tp]
\begingroup
\newdimen\tblskip \tblskip=5pt
\caption{Probabilities of obtaining values for the $\chi^2$ statistic and ratio
  of $\chi^2$ of the $N$-point functions shown in Fig.~\ref{fig:npt_data_temp_asym} for the
  \Planck\ fiducial $\Lambda$CDM model at least as
  large as the observed values of the statistic for the
  \smica\ map on hemispheres defined by the ecliptic
  (first column), Doppler boost (DB, second column), and dipolar modulation (DM, third
  column) reference frames.}
\label{tab:prob_chisq_npt_temp_asym}
\nointerlineskip
\vskip -3mm
\footnotesize
\setbox\tablebox=\vbox{
   \newdimen\digitwidth
   \setbox0=\hbox{\rm 0}
   \digitwidth=\wd0
   \catcode`*=\active
   \def*{\kern\digitwidth}
   \newdimen\signwidth
   \setbox0=\hbox{>}
   \signwidth=\wd0
   \catcode`!=\active
   \def!{\kern\signwidth}
\halign{\hbox to 1.6in{#\leaderfil}\tabskip 4pt&
\hfil#\hfil&
        \hfil#\hfil&
\hfil#\hfil\/\tabskip 0pt\cr
\noalign{\doubleline}
\noalign{\vskip -1pt}
\omit&\multispan3 \hfil Probability [\%]\hfil\cr
\noalign{\vskip -4pt}
\omit&\multispan3\hrulefill\cr
\omit \hfil Hemisphere\hfil& Ecl.& DB& DM\cr
\noalign{\vskip 3pt\hrule\vskip 3pt}
\multispan4 \hfil 2-point function\hfil\cr
 Negative&  !88.0&  !86.9&  !61.8\cr
 Positive&  !77.6& !91.1&  !59.9\cr
 $\chi^2$-ratio&  !22.3&   !*5.1&   !*7.7\cr
\multispan4 \hfil Pseudo-collapsed 3-point function\hfil\cr
 Negative&  !99.7&  !64.1&  !95.9\cr
 Positive&  !31.4&  !79.3&  !48.3\cr
 $\chi^2$-ratio&  !98.4&  !23.3&  !78.6\cr
\multispan4 \hfil Equilateral 3-point function\hfil\cr
 Negative&  !98.4&  !54.8& >99.9\cr
 Positive&  !42.6&  !95.0&  !78.4\cr
 $\chi^2$-ratio&  !86.7&  !67.7&  !88.2\cr
\multispan4 \hfil Rhombic 4-point function\hfil\cr
 Negative&  !99.6&  !46.4&  !97.5\cr
 Positive&  !20.1&  !86.3&  !23.2\cr
 $\chi^2$-ratio&  !97.0&  !57.9&  !92.5\cr
\noalign{\vskip 3pt\hrule\vskip 3pt}}}
\endPlancktable                    
\endgroup
\end{table}

In conclusion, the correlation functions for the \Planck\ 2015
temperature data are consistent with the results presented in
\citetalias{planck2013-p09}. Specifically, we observe that the
northern hemisphere correlation functions are relatively featureless
(both the 3- and 4-point functions lie very close to zero), whereas
the southern hemisphere functions exhibit a level of structure
consistent with Gaussian simulations.

\subsection{Constraints on quadrupolar modulation}
\label{sec:quadrupolar_modulation}

The most natural extension of the class of statistically anisotropic
models that we have considered previously involves the quadrupolar
modulation of an initially statistically isotropic CMB sky map.  No
detection of a corresponding quadrupolar power asymmetry is currently
claimed. An initial BipoSH analysis of the WMAP 7-year data
\citep{bennett2010} found evidence of corresponding non-zero spectra,
$A^{20}_{\ell\ell}$ and $A^{20}_{\ell\ell+2}$, in ecliptic
coordinates. However, \citet{Hanson2010a} demonstrated that the signal
could be attributed to an incomplete treatment of beam asymmetries in
the data, and this was subsequently confirmed in \citet{bennett2012}.
The corresponding analysis of the \Planck\ 2013 data indicated
consistency with statistical isotropy \citep{planck2013-p09}.

Here, we proceed further and consider the quadrupolar modulation of
the primordial power spectrum as suggested by \citet{Ackerman:2007}:
\begin{linenomath*}
\begin{equation}
P({\vec k})=P(k)\left [1+\sum_{M}g_{2M}\,Y_{2M}(\hat{\vec k})\right].\label{prim_power_glm}
\end{equation}
\end{linenomath*}
Given such a spectrum, the CMB temperature field is expected to
exhibit a correlation between $a_{\ell m}$ and $a^*_{\ell \pm \Delta
  \, m'}$ with $\Delta=0, 2$. Therefore, the BipoSH coefficients
$A^{2M}_{\ell\ell}$ and $A^{2M}_{\ell\ell+2}$ are sensitive to
$g_{2M}$.  In the limit of weak anisotropy, \citet{Kim2013} proposed
an optimal estimator for $g_{2M}$:
\begin{linenomath*}
\begin{eqnarray}
\lefteqn{\hat g_{2M}=\frac{1}{2}\sum_{M'}\left(\tens F^{-1}\right)_{MM'}\sum_{\ell m}\sum_{\ell' m'}\left.\frac{\partial {\tens C}_{\ell m,\ell' m'}}{\partial g_{2M'}}\right|_{g_{2M}=0}}\label{quad_estimator} \\
&\times&\left[(\tens C^{-1} \vec a^*)_{\ell m} (\tens C^{-1} \vec a)_{\ell' m'}\right.\nonumber \\
&&\left.-\langle\,(\tens C^{-1} \vec a^*)_{\ell m} (\tens C^{-1} \vec a)_{\ell' m'}\rangle\right]_{g_{2M}=0},\nonumber
\end{eqnarray}
\end{linenomath*}
where $\vec a$ is the CMB data vector in harmonic space and $\tens C$
is its covariance matrix, and
\begin{linenomath*}
\begin{eqnarray}
\lefteqn{\tens F_{MM'}\equiv\frac{1}{2} \sum_{\ell m}\sum_{\ell' m'}}\nonumber\\
&& \left[
(\tens C^{-1})_{\ell m} \frac{\partial {\tens C}_{\ell m,\ell' m'}}{\partial g_{2M}}(\tens C^{-1})_{\ell'm'}\frac{\partial {\tens C}_{\ell' m',\ell m}}{\partial g_{2M'}}\right]_{g_{2M}=0}.
\end{eqnarray}
\end{linenomath*}
Here, $\langle\,(\tens C^{-1} \vec a^*)_{\ell m} (\tens C^{-1} \vec
a)_{\ell' m'}\rangle_{g_{2M}=0}$ is the mean field in the absence of
the quadrupolar modulation. Observation-specific issues such as
incomplete sky coverage, inhomogeneous noise, and asymmetric beams
will result in a non-zero mean field, which can be estimated for the
\Planck\ data using simulations. Due to the otherwise prohibitive
computational cost, we adopt a diagonal approximation for the inverse
of the covariance matrix:
\begin{linenomath*}
\begin{eqnarray}
(\tens C^{-1})_{\ell m,\ell' m'}\approx 1/(C_\ell + N_\ell)\,\delta_{\ell \ell'} \delta_{mm'}\label{Cov_diag},
\end{eqnarray}
\end{linenomath*}
where $C_\ell$ and $N_\ell$ are the signal and noise power spectra
respectively. Uncertainties are computed by applying the estimator to
simulations.

\begin{table*}[htb!]               
\begingroup
\newdimen\tblskip \tblskip=5pt
\caption{
  Constraints on the quadrupolar modulation, determined from the
  \texttt{Commander}, \texttt{NILC}, \texttt{SEVEM}, and
  \texttt{SMICA} foreground-cleaned maps. The first three columns
  correspond to the five independent parts of the
  quadrupolar modulation, which we have chosen to present using a
  complex notation for $g_{2M}$.
  The quoted error bars are at the $68\,\%$ confidence level. The
  quadrupolar modulation amplitude
  is given in the fourth column, while the mean and standard deviation
  of $g_2$, estimated from simulations, are provided in the fifth column.
}
\label{tab:quad_mod}                            
\nointerlineskip
\vskip -3mm
\footnotesize
\setbox\tablebox=\vbox{
   \newdimen\digitwidth
   \setbox0=\hbox{\rm 0}
   \digitwidth=\wd0
   \catcode`*=\active
   \def*{\kern\digitwidth}
   \newdimen\signwidth
   \setbox0=\hbox{+}
   \signwidth=\wd0
   \catcode`!=\active
   \def!{\kern\signwidth}
\halign{\hbox to 1in{#\leaderfil}\tabskip 1em&
\hfil#\hfil&\tabskip 0.8em&
\hfil#\hfil&
\hfil#\hfil\tabskip 0pt\cr
\noalign{\doubleline}
\omit&\multispan3\hfil \mbox{$g_{2M}\times 10^2$}\hfil &\multispan2\hfil \mbox{$g_{2}\times 10^2$}\hfil\cr
\noalign{\vskip -4pt}
\omit&\multispan3\hrulefill &\multispan2\hrulefill\cr
\omit\hfil Method \hfil& $M=0$& $M=1$& $M=2$ &\hskip25pt Data & Simulation \cr
\noalign{\vskip 4pt\hrule\vskip 4pt}
\texttt{Commander}& $1.31 \pm 1.22$& $(0.43 \pm 0.86) + i\,(-0.01\pm 0.68)$& $(1.08 \pm 0.89) +i\,(-0.38 \pm 0.86)$&\hskip25pt$0.97$ & $1.12\pm 0.37$\cr
\noalign{\vskip 1pt}
\texttt{NILC}&$0.88 \pm 1.21$& $(0.37 \pm 0.85) + i\,(!0.33\pm 0.67)$& $(0.87 \pm 0.88) +i\,(-0.26 \pm 0.86)$&\hskip25pt$0.77$ & $1.11\pm 0.37$\cr
\noalign{\vskip 1pt}
\texttt{SEVEM}&$0.85 \pm 1.22 $& $(0.35 \pm 0.85) + i\,(!0.34\pm 0.67)$& $(1.00 \pm 0.88) +i\,(-0.25 \pm 0.86)$&\hskip25pt$0.81$ & $1.11\pm 0.37$\cr
\noalign{\vskip 1pt}
\texttt{SMICA}&$1.10 \pm 1.10 $& $(0.46 \pm 0.81) + i\,(!0.26\pm 0.64)$& $(0.93 \pm 0.83) +i\,(-0.26 \pm 0.82)$&\hskip25pt$0.85$ & $1.05\pm 0.34$\cr
\noalign{\vskip 3pt\hrule\vskip 4pt}}}
\endPlancktable                    
\endgroup
\end{table*}                        

Table~\ref{tab:quad_mod} presents results from an analysis of the
\Planck\ data using the extended common mask, UTA76, and limiting the
range of multipoles to $2\le \ell \le 1200$. When including data at
higher $\ell$-values, simulations show evidence for large statistical
uncertainties in the recovered $g_{2M}$ values that are a consequence
of the many holes in the mask related to point sources. Therefore,
imposing this limit $\ell \le 1200$ does not significantly affect the
constraining power of the analysis.  We then estimate the amplitude of
the quadrupolar modulation using the relation $g_2=\left(1/5\sum_M
  |g_{2M}|^2\right)^{1/2}$.  Due to the nature of the estimator, which
is necessarily positive, the estimation is biased.  For an unbiased
assessment, we estimate the mean and standard deviation of $g_2$ from
simulations.  We find no evidence for quadrupolar modulation of the
primordial power spectrum.  However, the derived limits allow us to
impose tight constraints on statistically anisotropic inflationary
models, such as those including vector fields during inflation.  A
companion paper, \citet{planck2014-a24}, contains a more complete
discussion on the theoretical implications of this constraint.

\subsection{Point-parity asymmetry}
\label{sec:pointparity}

The CMB anisotropy field defined on the sky, $T(\vec{\hat{n}})$, may
be divided into symmetric, $T^+(\vec{\hat{n}})$, and antisymmetric,
$T^-(\vec{\hat{n}})$, functions with respect to the centre of the
sphere, as previously described in \citetalias{planck2013-p09}. These
functions have even and odd parity, and thus correspond to spherical
harmonics with even and odd $\ell$-modes, respectively. On the very
large scales corresponding to the Sachs-Wolfe plateau of the
temperature power spectrum ($2 \leq \ell \leq 30$), the Universe
should be parity neutral with no particular parity preference
exhibited by the CMB fluctuations. However, an odd point-parity
preference has previously been observed in the WMAP data releases
\citep{Land2005, LandOdd, kim2010, kim_anomalous_pa, Gruppuso2010} and
the \Planck\ 2013 results.  Here, we investigate the parity asymmetry
in the 2015 temperature maps at $\nside = 32$. We consider the
following estimator:
\begin{linenomath*}
\begin{equation}\label{eq:pointparity}
R^\mathrm{TT}(\ell_\mathrm{max}) = \frac{D^\mathrm{TT}_+(\ell_\mathrm{max})}{D^\mathrm{TT}_-(\ell_\mathrm{max})},
\end{equation}
\end{linenomath*}
where $D_+(\ell_\mathrm{max})$ and $D_-(\ell_\mathrm{max})$ are given by
\begin{linenomath*}
\begin{equation}\label{eq:TTpointparity}
D^\mathrm{TT}_{+,-} = \frac{1}{\ell_\mathrm{tot}^{+,-}}\sum_{\ell=2,\ell_\mathrm{max}}^{+,-} \frac{\ell(\ell+1)}{2 \pi}C^\mathrm{TT}_{\ell},
\end{equation}
\end{linenomath*}
$\ell_\mathrm{tot}^{+,-}$ is the total number of even ($+$) or odd ($-$)
multipoles included in the sum up to $\ell_\mathrm{max}$, and
$D^\mathrm{TT}_{\ell}$ is the temperature angular power spectrum
computed using a QML estimator \citep{Gruppuso2010}. The
$\ell(\ell+1)/(2 \pi)$ factor in Eq.~\eqref{eq:TTpointparity}
effectively flattens the spectrum across the $\ell$-range of the
Sachs-Wolfe plateau (up to $\ell=50$) in a $\Lambda$CDM model.

\begin{figure} \centering
  \includegraphics[width=\hsize]{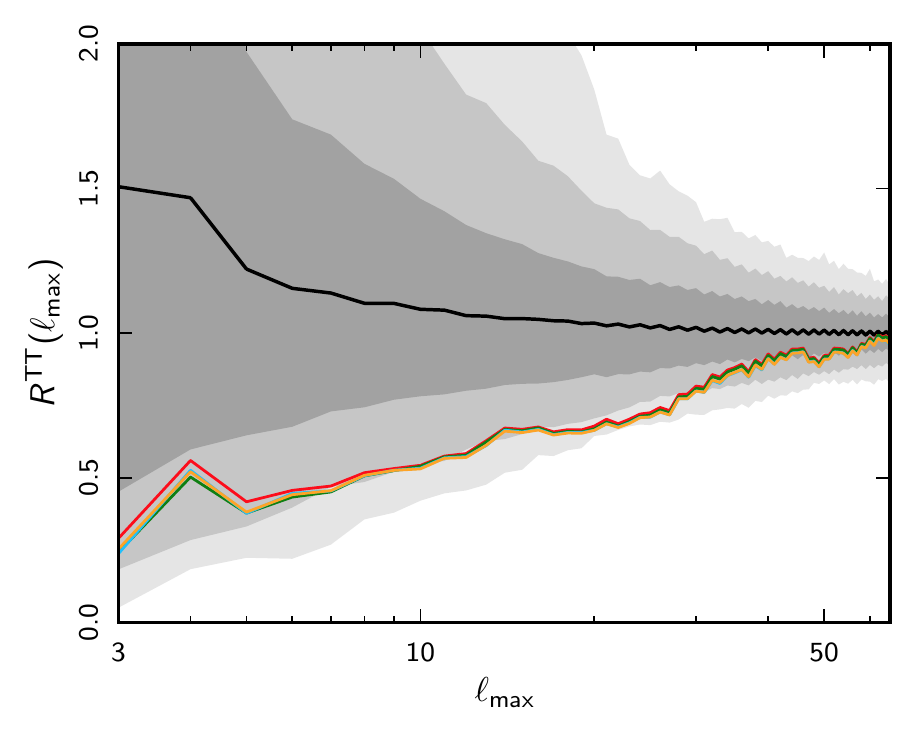}
  \caption{The ratio $R^\mathrm{TT}(\ell_\mathrm{max})$ for \commander\ (red),
    \nilc\ (orange), \sevem\ (green), and \smica\ (blue) determined at $\nside =
    32$.  The shaded grey regions indicate the distribution of the
    statistic derived from the \smica\ MC simulations, with the dark,
    lighter, and light grey bands corresponding to the 1, 2, and 3\,$\sigma$
    confidence levels.}
\label{fig:pointparity}
\end{figure}

Figure~\ref{fig:pointparity} presents the ratio,
$R^\mathrm{TT}(\ell_\mathrm{max})$, for the 2015 component-separated
maps, together with the distribution determined from the \smica\ MC
simulations which serves as a reference for the expected behaviour of
the statistic in a parity-neutral Universe. The distributions for the
other CMB maps are very similar.  The four component-separation
products are in good agreement, indicating an odd-parity preference at
very large scales for the multipole range considered in this test.

\begin{figure} \centering
  \includegraphics[width=\hsize]{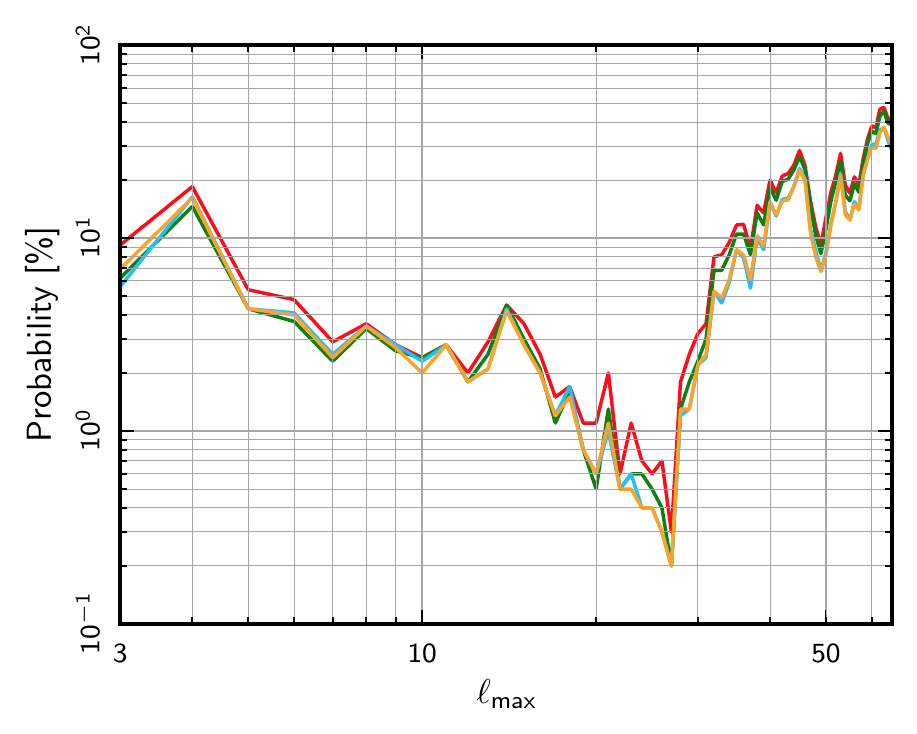}
  \caption{Lower-tail probability of the point-parity
    estimator for \commander\ (red), \nilc\ (orange), \sevem, (green), and
    \smica\ (blue).}
\label{fig:pvalue_pointparity}
\end{figure}

Figure~\ref{fig:pvalue_pointparity} shows the lower-tail probability
for the data as compared to simulations as a function of
$\ell_\mathrm{max}$.  The results are in good agreement with those in
\citetalias{planck2013-p09}. The cleaned CMB maps yield generally
consistent profiles which signify an anomalous odd-parity preference
in the multipole region $\ell_\mathrm{max}=20$--30.  The minimum in
the lower-tail probability occurs at $\ell=28$ corresponding to a
value of 0.2\,\% for \nilc, \sevem, and \smica, and 0.3\,\% for
\commander.\footnote{In the case where we would like to test the
  probability of finding a Universe with either odd or even parity
  preference, the probability would be higher by a factor of about
  two.}

As a first attempt to quantify any a posteriori effects in the
significance levels, we consider how many MC simulations appear in the
lower tail of the MC distribution with a probability equal to, or
lower than, 0.2\,\%, for at least one $\ell_\mathrm{max}$ value over a
specific range.  For $\ell_\mathrm{max}$ in the range 3--50, the total
number of simulated maps with this property is less than 20 over 1000
MC maps, implying that, even considering the LEE, an
odd-parity preference is observed with a lower-tail probability of
less than 2\,\%.

\subsection{Mirror-parity asymmetry}
\label{sec:mirror_parity}

For the \Planck\ 2013 data release, we studied the properties of the
temperature data at a resolution of $\nside = 16$ under reflection with
respect to a plane to search for mirror symmetries. Such a symmetry
might be connected to non-trivial topologies
\citep{Starobinsky:1993yx, Stevens:1993zz, deOliveiraCosta:1995td}.
In \citet{planck2013-p09}, we reported evidence for an antisymmetry
plane, with a perpendicular direction given by
$(l, b)=(264\degr, -17\degr)$, However, the probability of the results
was slightly dependent on the method of foreground cleaning, with a
\pval\ ranging from 0.5\,\% for {\tt Commander-Ruler} to 8.9\,\% for
{\tt SMICA}.  The same direction was also found in the WMAP 7-year
data \citep{Finelli:2011zs}, and is close to that determined for the
dipole modulation in the \Planck\ 2013 data release
\citepalias{planck2013-p09}, suggesting possible connections between
the two directional anomalies.

We now proceed to reanalyse the status of mirror symmetries using the
\Planck\ 2015 full mission temperature data at both $\nside = 16$ and $\nside =
32$.  In order to avoid possible bias introduced by the use of the
Galactic mask\footnote{The Galactic mask induces a preferred direction
  in the analysis of the MC simulation ensemble, which
  affects the significance of the results determined from the
  data. See \citet{Ben-David:2014mea} for a discussion.}  the results
are derived from the full-sky \commander, \nilc, and \smica\ maps
described in Sect.~\ref{sec:data}. For \sevem, a customized map is
first produced by inpainting about 3\,\% of the map along the Galactic
plane using a diffusive inpainting technique. This is then smoothed to
the appropriate lower resolutions for further analysis.  Following
\citet{Finelli:2011zs}, we consider the estimators in the pixel domain
given by:
\begin{linenomath*}
\begin{equation}
S^{\pm} (\vec{\hat{n}}_i) = {1 \over N_\mathrm{pix}} \sum_{j=1}^{N_\mathrm{pix}}
\left[ {1 \over 2} \left( {\delta T \over T} (\vec{\hat{n}}_j) \pm
{\delta T \over T}(\vec{\hat{n}}_k) \right)\right]^2\,,
\label{spm}
\end{equation}
\end{linenomath*}
where the sum is over all $N_{\mathrm {pix}}$ {\tt
  HEALPix} pixels, ${(\delta T / T)} (\vec{\hat{n}}_j)$ is the CMB
temperature anisotropy measured at the pixel defined by the unit
vector $\vec{\hat{n}}_j$, and $\vec{\hat{n}}_k$ is the opposite
direction with respect to the plane defined by $\vec{\hat{n}}_i$,
i.e.,
\begin{linenomath*}
\begin{equation}
\vec{\hat{n}}_k = \vec{\hat{n}}_j - 2\, ( \vec{\hat{n}}_i \cdot \vec{\hat{n}}_j) \vec{\hat{n}}_i \, \,.
\end{equation}
\end{linenomath*}
Note that we expect $S^{+}$ to be small if the points on opposite
sides of the mirror are negatives of each other, and $S^{-}$ to be
small when they are the same.

\begin{figure}[h]
\begin{center}
\includegraphics[scale=0.65]{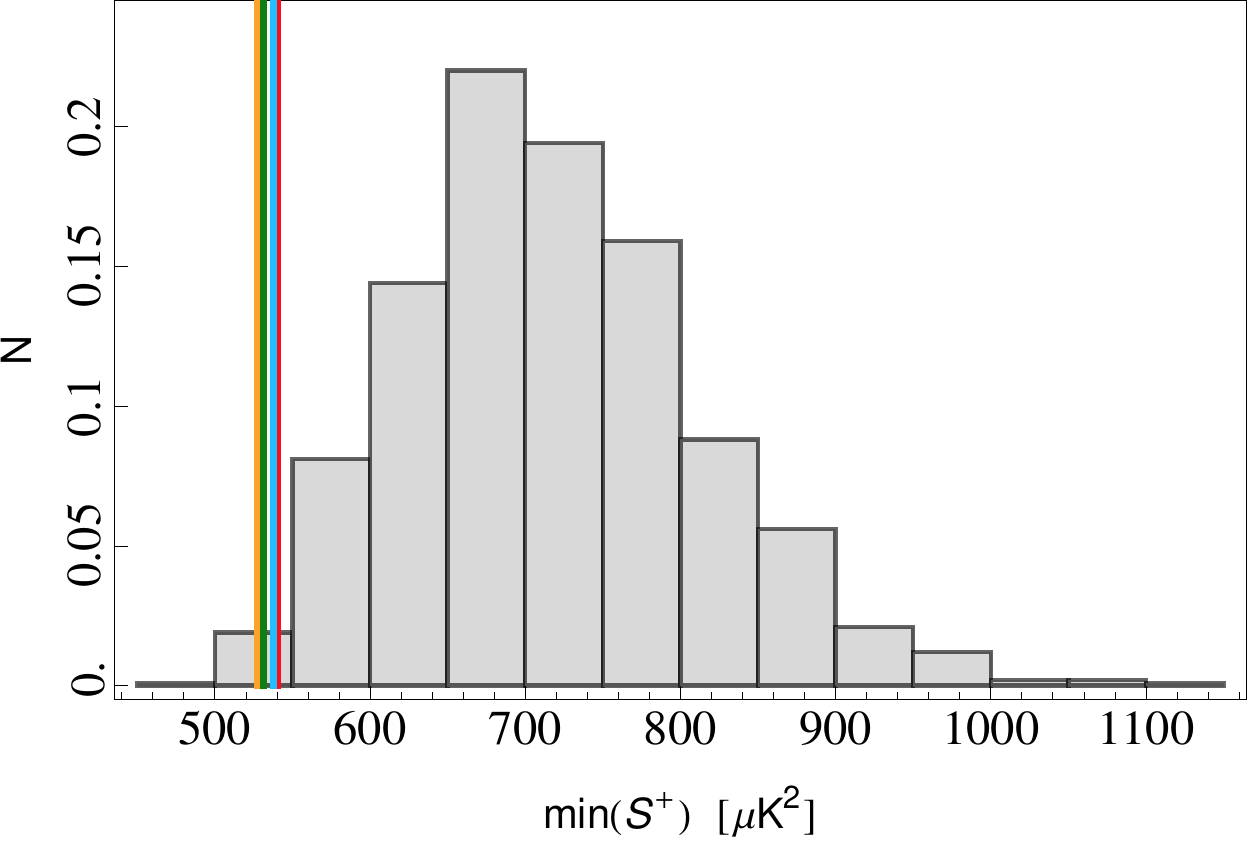}
\includegraphics[scale=0.65]{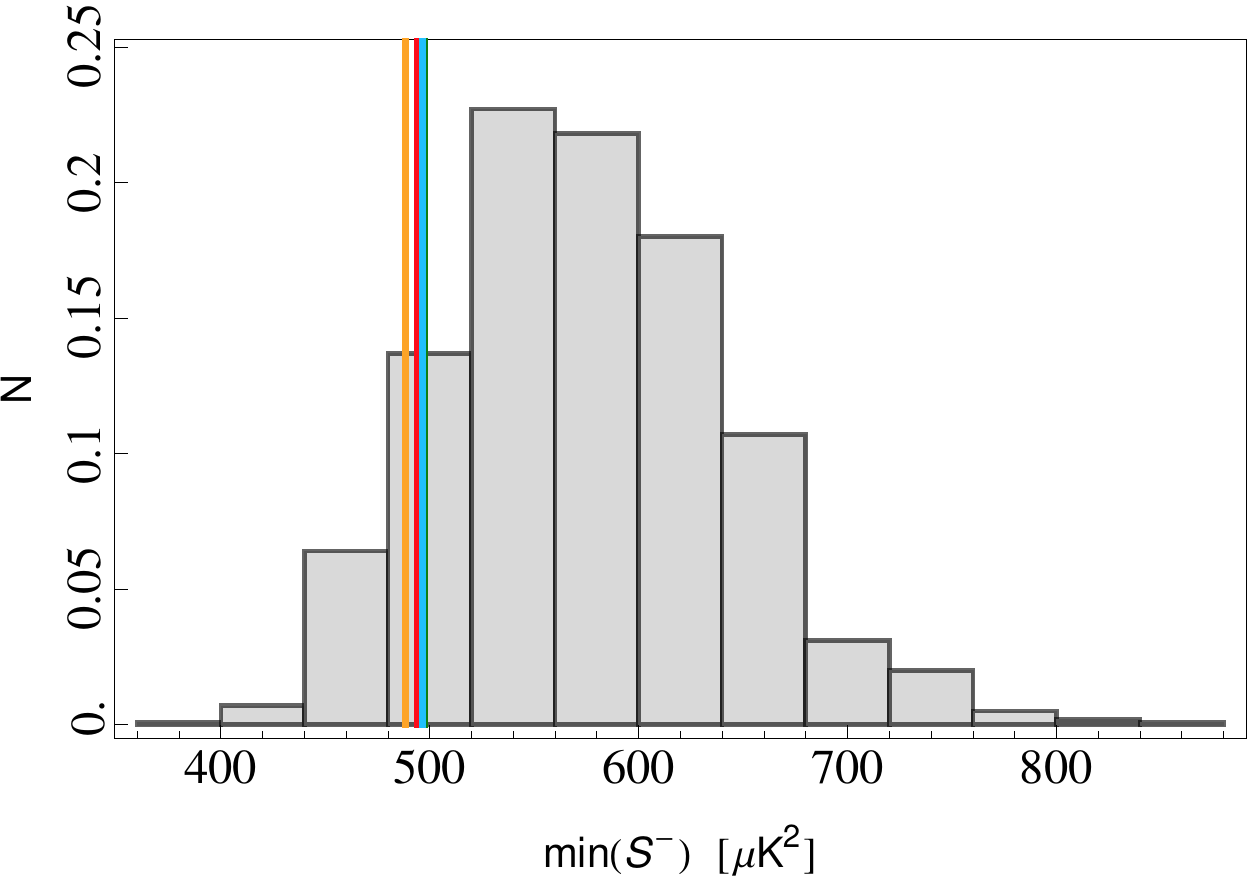}
\caption{Histograms of the $S^+$ (top panel) and $S^-$ (bottom panel) statistic.
  The vertical lines show the
  minimum value for the estimator computed at $N_\mathrm{side}=32$ for
  \commander\ (red), \nilc\ (orange), \sevem\ (green), and \smica\
  (blue) maps. The grey area shows the same quantity computed
  from 1000 simulated \smica\ maps.}
\label{fig:plotmirrorsym}
\end{center}
\end{figure}

We compute these quantities for each of the 3072 (12288) directions
defined at resolution $\nside = 16$ (32), and allow the $j$ and $k$
indices to run over all of the pixels of the low-resolution full-sky
maps.  We perform the same analysis on 1000 FFP8 simulations and
store the minimum value of $S^{\pm}$ for each of these to compute
probabilities. The results are summarized in
Table~\ref{tab:tablevals16and32} and Fig.~\ref{fig:plotmirrorsym}.

We confirm that the full mission \Planck\ temperature  data at
$N_\mathrm{side}=16$ exhibit the most anomalous mirror antisymmetry
in the direction $(l, b)=(264\degr,-17\degr)$, consistent with the
result from the 2013 nominal mission data, with a probability which
ranges from 1.6\,\% for \sevem\ to 2.9\,\% for \commander.  This is
within 40\deg\ of the preferred direction identified by the dipole
modulation analysis in Sect.~\ref{sec:dipmod}.  The corresponding
results at $N_\mathrm{side}=32$ yield approximately the same
direction, $(l, b)=(264\degr,-16\degr)$, with a slightly increased
probability, ranging from 0.8\,\% for \sevem\ to 1.9\,\% for
\commander.

We also note that the CMB pattern exhibits a mirror symmetry in the
direction $(l, b)=(260\deg, 48\deg)$, consistent with that found in
the WMAP 7-year data \citep{Finelli:2011zs}, and close to that
identified by the solar dipole \citep{planck2014-a09}.  However, the
significance of the symmetry pattern is less than in the
antisymmetric case.

This extension of the analysis to higher resolution than in our
previous work shows that the antisymmetry property does not
seem to be confined to the largest angular scales, although we have
not attempted to correct for any a posteriori choices made in the
analysis. The detailed connection of this antisymmetry property to
the low-variance and hemispherical asymmetry observed on these scales
remains an open issue.

\begin{table}[h!tb]
\begingroup
\newdimen\tblskip \tblskip=5pt
\caption{The lower-tail probability for the $S^{\pm}$ statistics of
  the component-separated maps at $N_\mathrm{side}=16$ and $N_\mathrm{side}=32$.
}
\label{tab:tablevals16and32}
\nointerlineskip
\vskip -3mm
\footnotesize  
\setbox\tablebox=\vbox{
   \newdimen\digitwidth
   \setbox0=\hbox{\rm 0}
   \digitwidth=\wd0
   \catcode`*=\active
   \def*{\kern\digitwidth}
   \newdimen\signwidth
   \setbox0=\hbox{+}
   \signwidth=\wd0
   \catcode`!=\active
   \def!{\kern\signwidth}
\halign{\hbox to 0.8in{#\leaderfil}\tabskip 4pt&
\hfil#\hfil\tabskip 8pt&
\hfil#\hfil\/\tabskip 0pt\cr
\noalign{\doubleline}
\noalign{\vskip 1pt}
\omit\hfil& Probability& Direction\cr
\omit\hfil Estimator\hfil& $[\%]$& $(l, b)$ [\deg]\cr
\noalign{\vskip 3pt\hrule\vskip 3pt}
\multispan3 \hfil{$N_\mathrm{side}=16$}\hfil\cr
\noalign{\vskip 3pt\hrule\vskip 3pt}
\multispan3 \hfil{\tt Commander}\hfil\cr
min$(S^+)$&  **2.9& (264.4,$-17.0$)\cr
min$(S^-)$& *12.0& (260.4,$!48.1$)\cr
\noalign{\vskip 1pt}
\multispan3 \hfil{\tt NILC\hfil}\cr
min$(S^+)$&  **2.3& (264.4,$-17.0$)\cr
min$(S^-)$& *16.8& (260.4,$!48.1$)\cr
\noalign{\vskip 1pt}
\multispan3 \hfil{\tt SEVEM}\hfil\cr
min$(S^+)$&  **1.6& (264.4,$-17.0$)\cr
min$(S^-)$& *13.5& (260.4,$!48.1$)\cr
\noalign{\vskip 1pt}
\multispan3 \hfil{\tt SMICA}\hfil\cr
min$(S^+)$&  **2.7& (264.4,$-17.0$)\cr
min$(S^-)$& *19.1& (260.4,$!48.1$)\cr
\noalign{\vskip 3pt\hrule\vskip 3pt} %
\multispan3 \hfil{$N_\mathrm{side}=32$}\hfil\cr
\noalign{\vskip 3pt\hrule\vskip 3pt}
\multispan3 \hfil{\tt Commander}\hfil\cr
min$(S^+)$&  **1.9& (264.4,$-15.7$)\cr
min$(S^-)$& *10.0& (265.3,$!46.2$)\cr
\noalign{\vskip 1pt}
\multispan3 \hfil{\tt NILC\hfil}\cr
min$(S^+)$&  **1.2& (264.4,$-15.7$)\cr
min$(S^-)$& *10.3& (265.3,$!46.2$)\cr
\noalign{\vskip 1pt}
\multispan3 \hfil{\tt SEVEM}\hfil\cr
min$(S^+)$& **0.8& (264.4,$-15.7$)\cr
min$(S^-)$& *11.1&  (265.3,$!46.2$)\cr
\noalign{\vskip 1pt}
\multispan3 \hfil{\tt SMICA}\hfil\cr
min$(S^+)$&  **1.7& (264.4,$-15.7$)\cr
min$(S^-)$& *11.6& (265.3,$!46.2$)\cr
\noalign{\vskip 3pt\hrule\vskip 3pt}}}
\endPlancktable
\endgroup %
\end{table}

\subsection{Local peak statistics}
\label{sec:kstest}

\begin{figure}
  \centering
  \includegraphics[width=88mm]{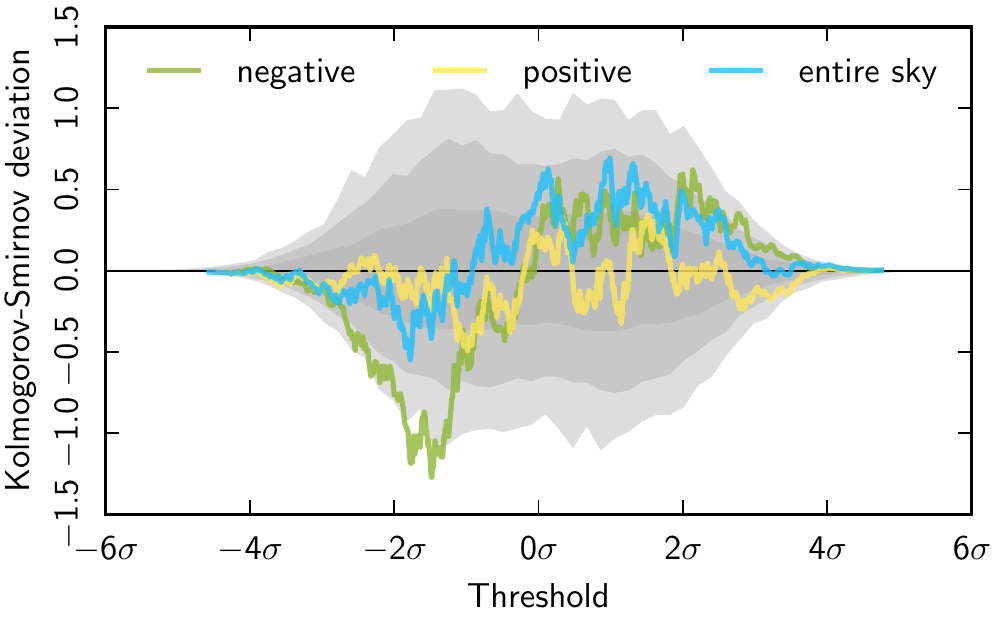}\\
  \includegraphics[width=88mm]{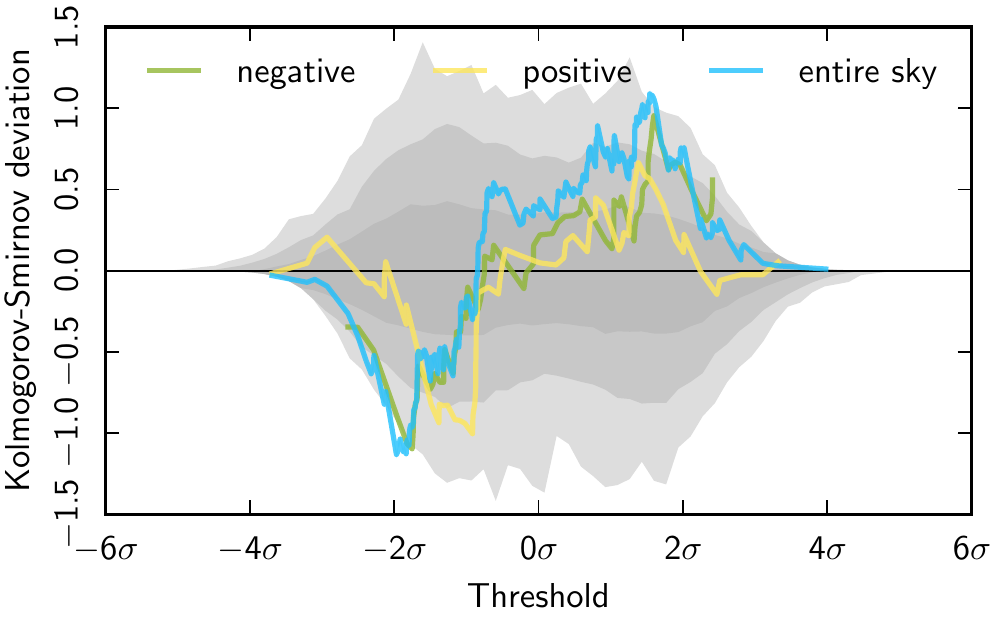}\\
  \includegraphics[width=88mm]{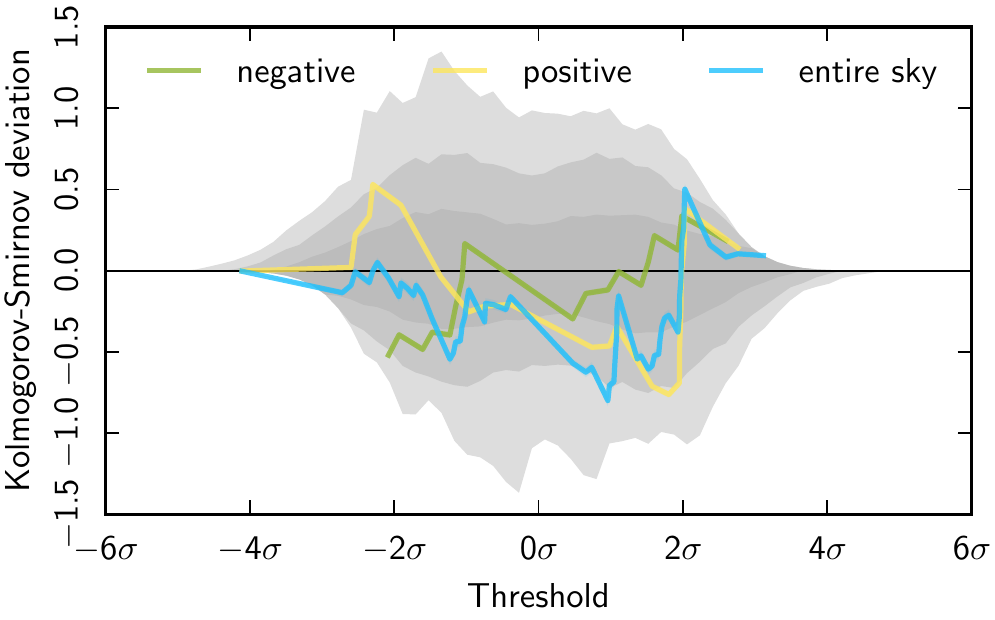}\\
  \caption{ KS-deviation of the peak distribution for $70\deg$ radius
    discs centred on the positive and negative asymmetry directions
    determined from the \smica\ CMB temperature map in
    Sect.~\ref{sec:dipmod}.  From top to bottom, the plots correspond
    to maps filtered with a GAUSS kernel of $40\arcm$ FWHM, an
    SSG84 filter of $500\arcm$ FWHM, and an SSG84 filter of $800\arcm$
    FWHM, respectively.  }
  \label{fig:peaks:asym}
\end{figure}

Local extrema or peaks, as introduced in Sect.~\ref{sec:peaks}, can be
employed to search for localized anomalies on the CMB sky by examining
how their statistical properties vary in patches as a function of
location.

Initially, we consider a further test for asymmetry by examining the
differences in the peak distribution when divided according to
orientation with respect to a previously specified asymmetry
direction.  In particular, we select the peaks both in a disc of
radius $70\deg$ centred on $(l, b) = (225\deg, -18\deg)$ (the positive
direction of the dipole defined in Sect.~\ref{sec:dipmod} for \smica)
and in the corresponding antipodal disc, then construct the empirical
peak height CDFs to be compared with the full-sky median FFP8
distribution, as shown in Fig.~\ref{fig:peaks:asym}.  For maps
filtered with a 40\arcm\ FWHM GAUSS filter the distribution of the
peaks for the positive-direction disc is in general agreement with the
full sky result, while that for the negative-direction is marginally
different. Moreover, this pattern of behaviour is seen over a number
of filtering scales, both for the KS deviation from the median
full-sky simulated CDFs, and the spread of extremal values when
comparing positive and negative regions. We also find that the
properties of the negative disc affect the \textit{p}-value results
for a full sky KS test on data filtered with an SSG84 filter of
$500\arcm$ FWHM, as seen in Sect.~\ref{sec:peaks}.

\begin{figure*}[tbh] 
\begin{tabular}{cccc}
  {\tt Commander} & {\tt NILC} & {\tt SEVEM} & {\tt SMICA} \\
  \includegraphics[width=42mm]{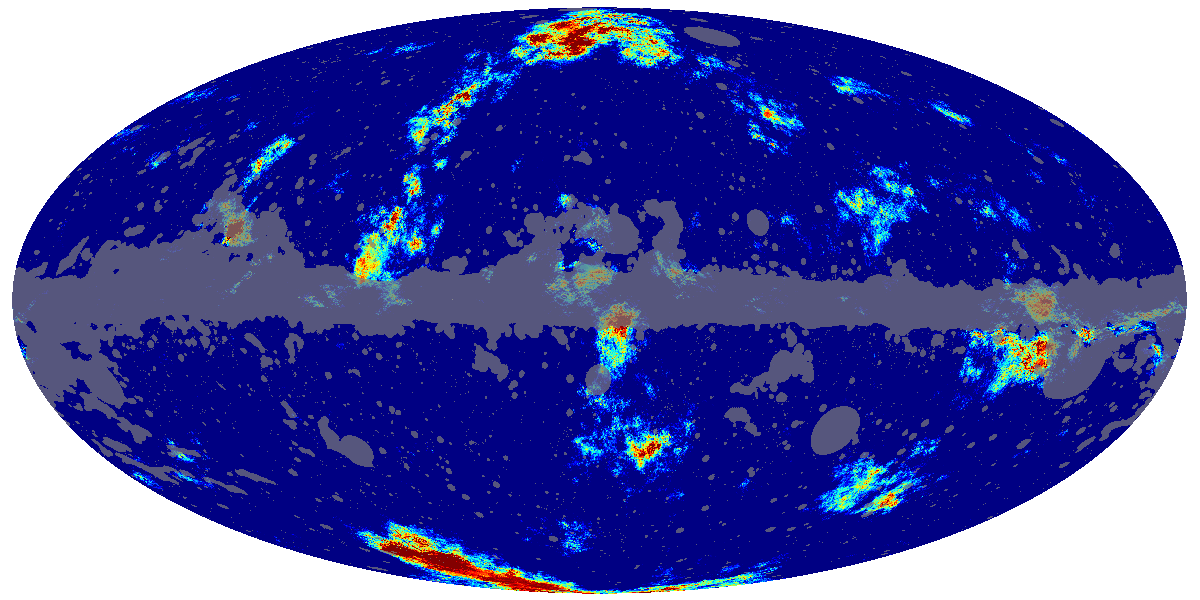} &
  \includegraphics[width=42mm]{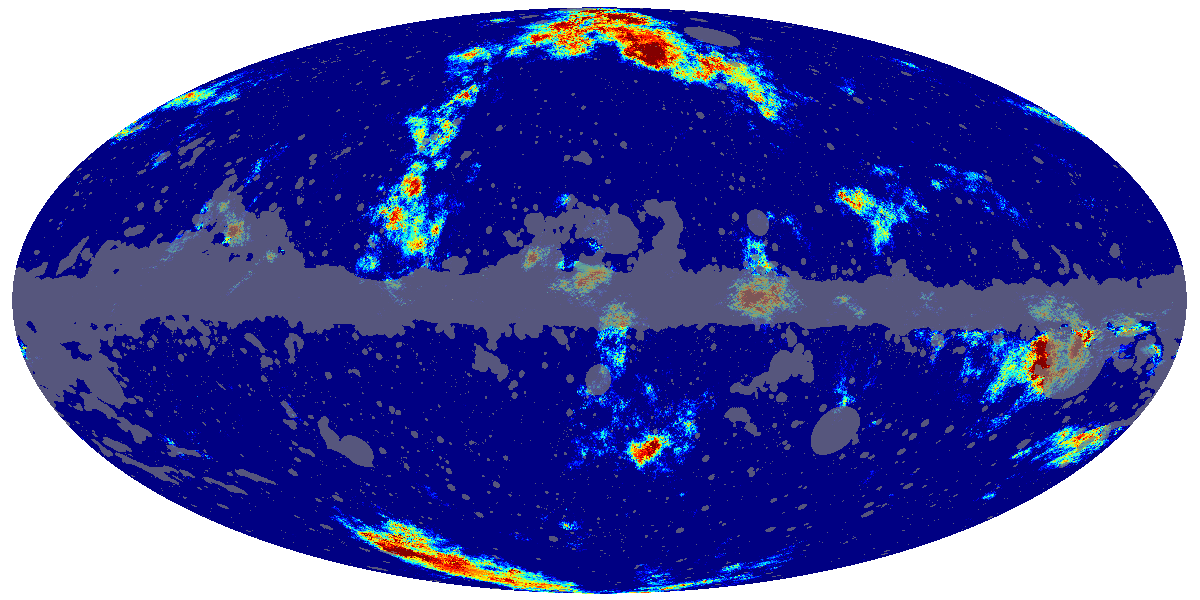} &
  \includegraphics[width=42mm]{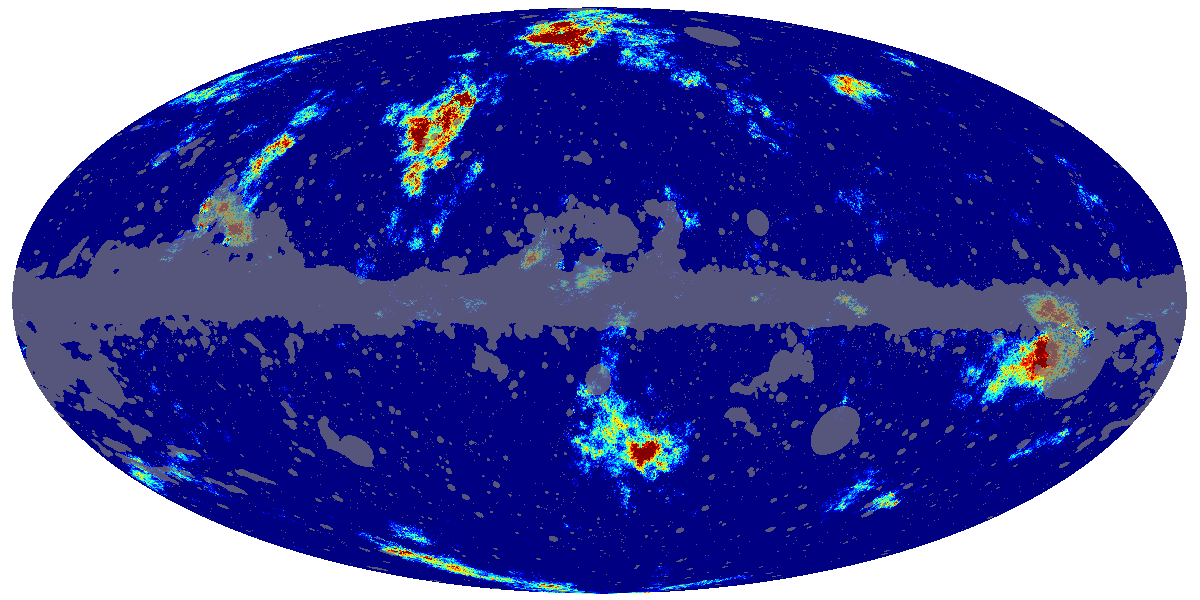} &
  \includegraphics[width=42mm]{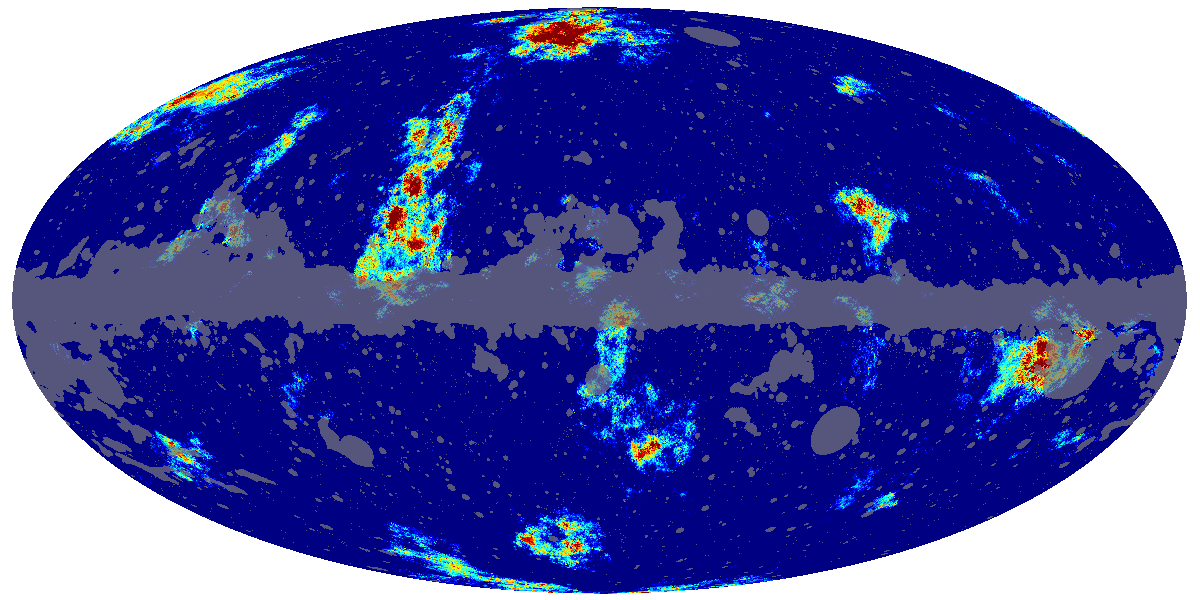} \\
\end{tabular}
\includegraphics[width=\textwidth]{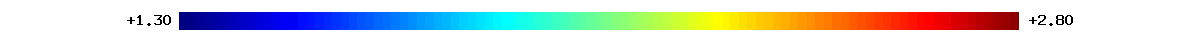}
\centerline{\sf $-\text{log}_{\text{10}} \text{(UTP)}$}
\caption{Map of $-\text{log}_{\text{10}} \text{(UTP)}$  for peak counts in the
 \Planck\ 40\arcm\ GAUSS-filtered temperature data, where each pixel
  encodes the probability determined for a 30\deg\ diameter disc
  centred on it.  }
\label{fig:peaks:counts:30}
\end{figure*}

We can then extend the analysis for the 40\arcm\ GAUSS-filtered data
by considering the variation in the peak statistical properties for a
set of discs, each of which is centred on a pixel defined at $\nside =
256$.  The simplest statistics to consider are the peak number counts.
We therefore consider discs of $30\deg$ diameter and compute the peak
counts for each disc. These are then compared to the corresponding
peak count CDFs determined from simulations, and the upper- and
lower-tail probabilities are assigned by counting the number of
simulations above and below the observed counts at the same
location. These quantities can then be visualized in the form of
$\nside = 256$ sky maps.  The derived $-\text{log}_{\text{10}} \text{(UTP)}$ maps for
each component-separation method are shown in
Fig.~\ref{fig:peaks:counts:30}.  While we find that the total counts
of peaks for the sky coverage defined by the common mask is consistent
with simulations, significant regional variation is seen.  Indeed, the
\textit{p}-value for certain disc locations drops to 0.1\,\% (i.e.,
the sky counts exceed anything seen in simulations). However, one
needs to account for the a posteriori selection of significant regions
in the determination of the true significance.  It should also be
noted that regional variations of the UTP are seen at similar levels
when inspecting the peak-count statistics maps derived for randomly
selected realizations of the simulations. Moreover, the significance
of such peak-counting anomalies is degraded with larger disc
diameters, and becomes insignificant for counts on the full sky. Thus,
no significant anomalies can be claimed for the peak-count statistics
of the \Planck\ data.

\begin{figure*} 
\begin{tabular}{cccc}
  {\tt Commander} & {\tt NILC} & {\tt SEVEM} & {\tt SMICA} \\
  \includegraphics[width=42mm]{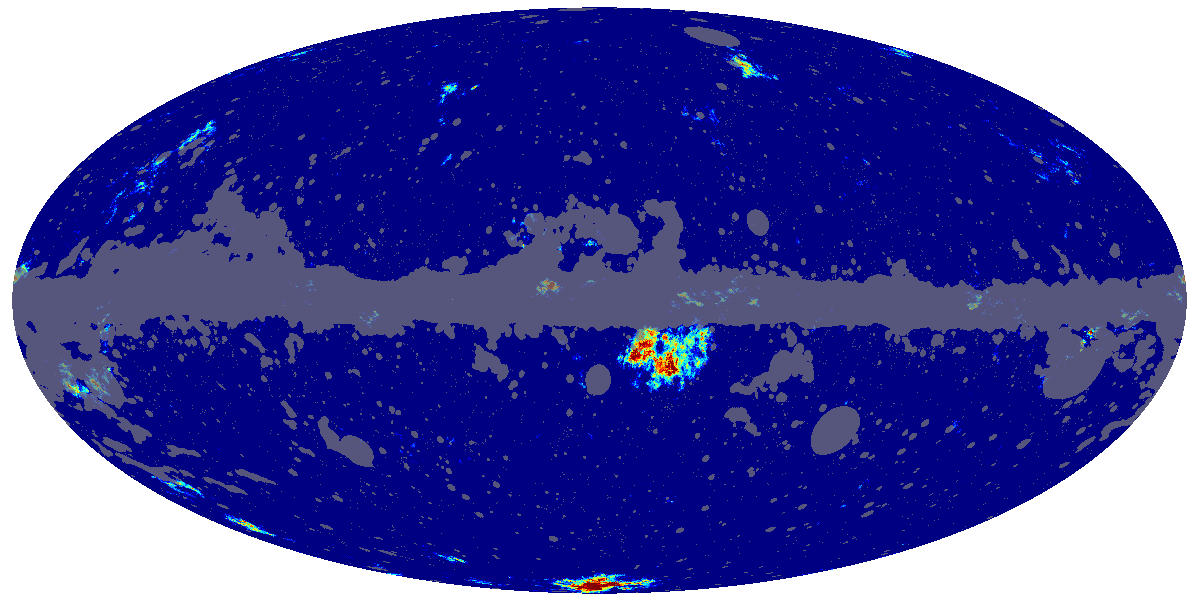} &
  \includegraphics[width=42mm]{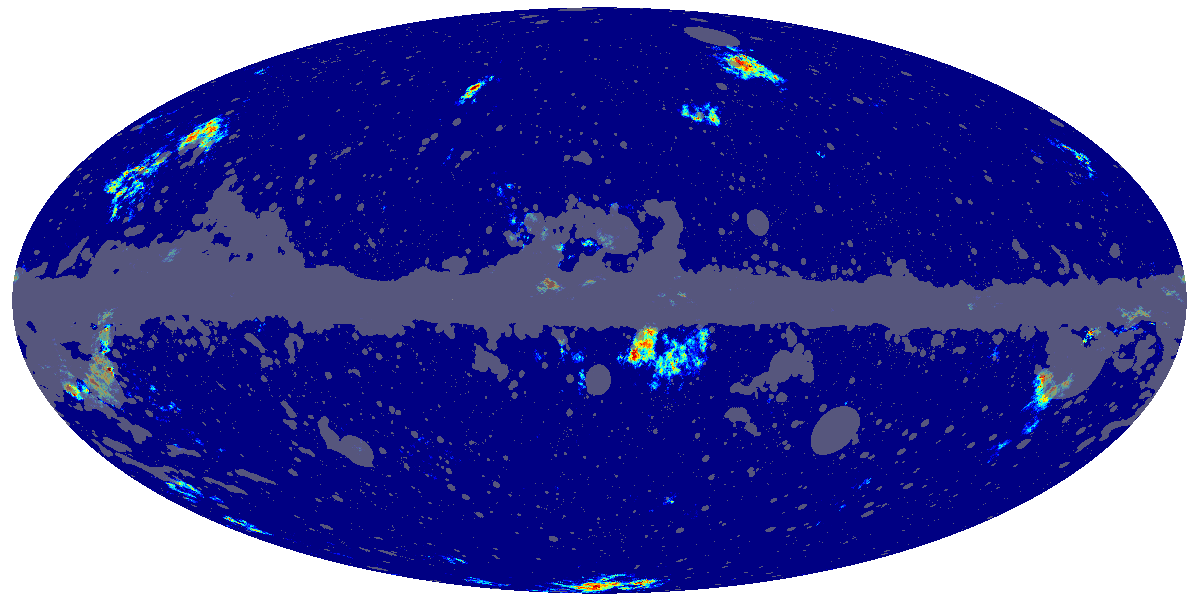} &
  \includegraphics[width=42mm]{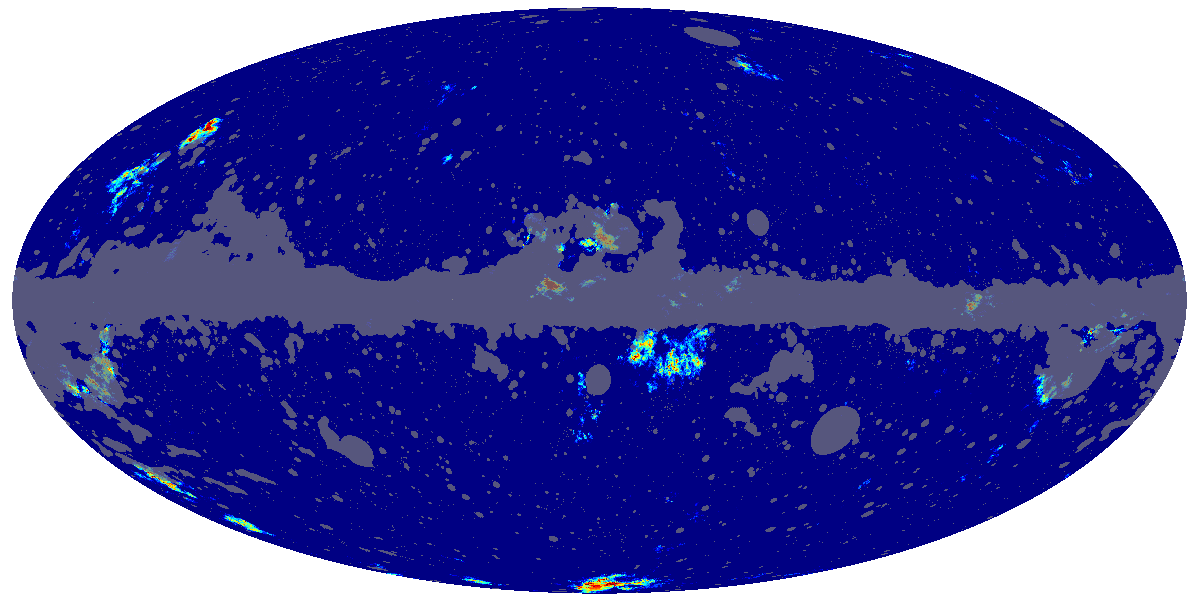} &
  \includegraphics[width=42mm]{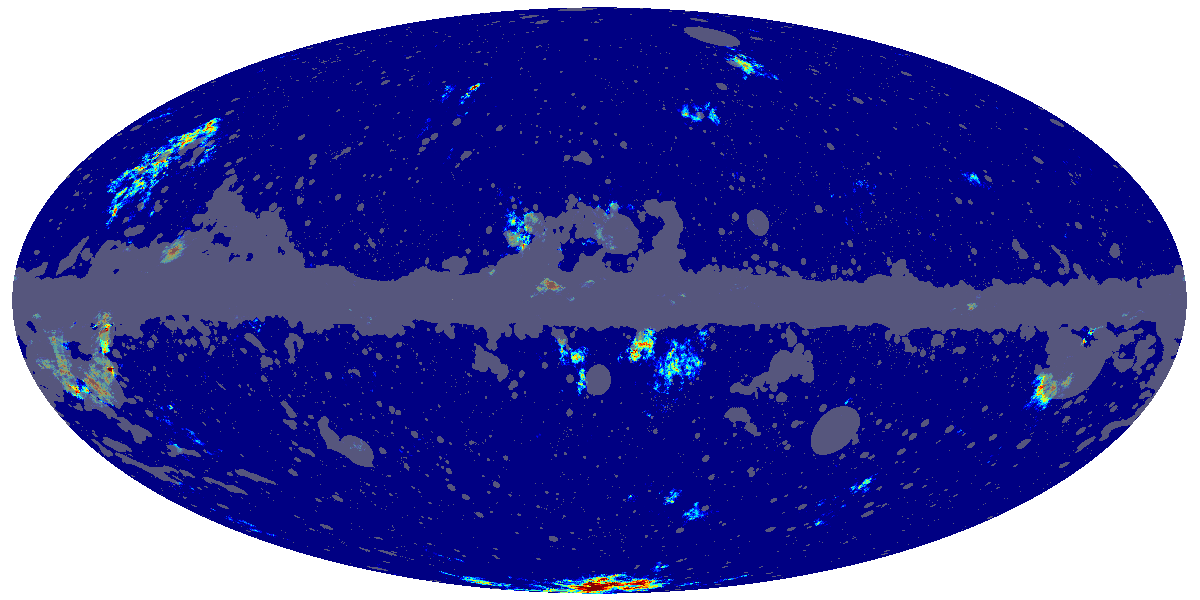} \\
\end{tabular}
\includegraphics[width=\textwidth]{peaks-colorbar-utp}
\centerline{\sf $-\text{log}_{\text{10}} \text{(UTP)}$}
\caption{Map of $-\text{log}_{\text{10}} \text{(UTP)}$ for the two-sample
  KS-deviation where each pixel encodes the
  probability determined for a 30\deg\ diameter disc centred on it, as
  computed from the \Planck\ 40\arcm\ GAUSS-filtered temperature data.
}
\label{fig:peaks:kstest:30}
\end{figure*}

A powerful non-parametric test of statistical isotropy is provided by
the two-sample KS-deviation between the full sky empirical peak height
CDF $F_n(x)$ (see Eq.~\ref{eq:peaks:cdf}) and an empirical peak height
CDF $F_{n'}(x)$ derived from a subsample of
the distribution, again defined by the peaks within discs of $30\deg$
diameter as defined above.  The two-sample KS-deviation
\begin{linenomath*}
\begin{equation}\label{eq:peaks:ks2sample}
K_{nn'} \equiv \sqrt{\frac{nn'}{n+n'}}\, \mathop{\text{sup}}\limits_{x} \left|F_{n'}(x) - F_n(x)\right|
\end{equation}
\end{linenomath*}
for a partial sky region shares samples between the two CDFs, and can
be calculated extremely efficiently using rank statistics according to
\begin{linenomath*}
\begin{equation}\label{eq:peaks:ksrank}
K_{nn'} \equiv \sqrt{\frac{nn'}{n+n'}}\, \max\limits_{i} \left|\frac{{r}'(i) - 1}{n'-1} - \frac{{r}(i) - 1}{n-1}\right|,
\end{equation}
\end{linenomath*}
where $r$ and $r'$ denote the ranks of a value with
index $i$ in the full set of $n$ and restricted set of $n'$ samples,
respectively.  Maps of the upper tail probability are then determined
by comparison with the equivalent quantities computed from
simulations; $-\text{log}_{\text{10}} \text{(UTP)}$ maps are shown in
Fig.~\ref{fig:peaks:kstest:30}. The majority of the selected locations
are consistent with the full-sky distribution, thus indicating the
statistical isotropy of the \Planck\ maps. The most prominent feature
in each of the local KS-deviation maps appears south of the Galactic
centre and may be associated with a cold region crossing the Galactic
plane. However, as with the peak counts, it cannot be interpreted as
statistically anomalous.

\subsection{The Cold Spot}
\label{sec:coldspot}

\newcommand\Angular[1]{\ensuremath{\left\langle#1\right\rangle}}

\begin{figure}
  \centering
\includegraphics[width=8.8cm]{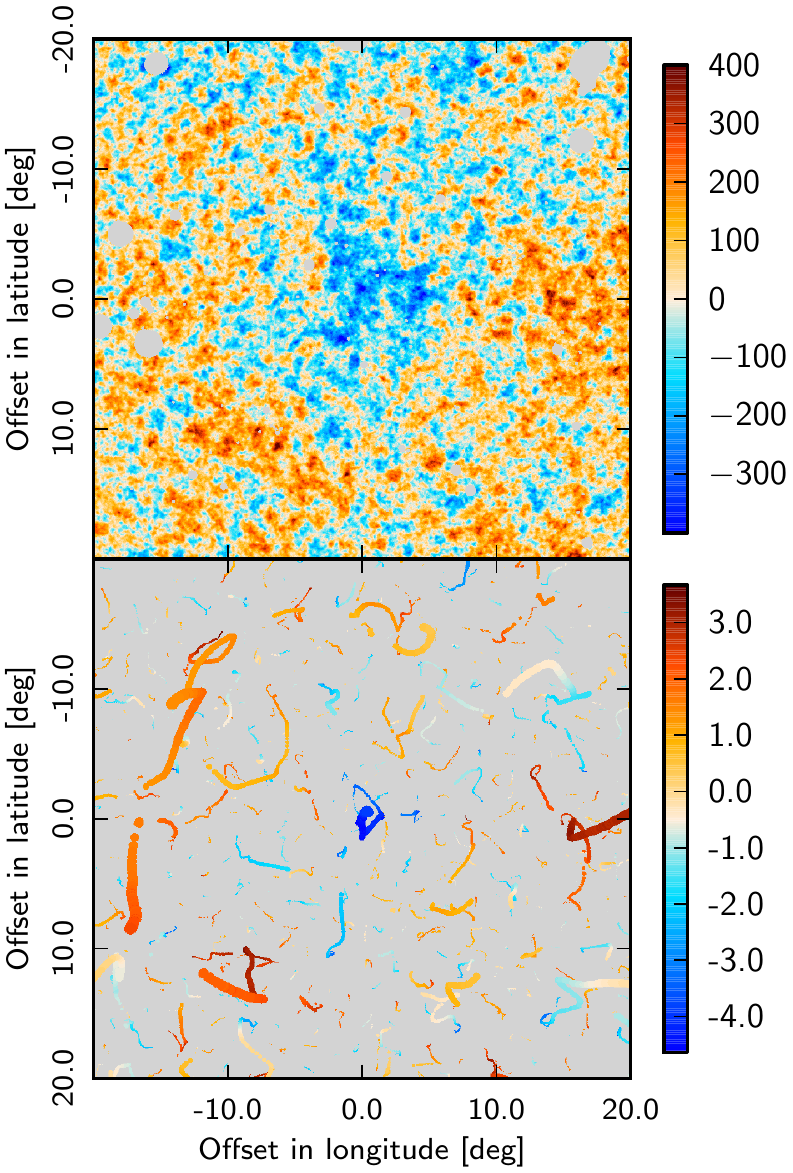}
  \caption{{\it Top:} temperature patch centred on the \cs. {\it Bottom:}
    Peak merger tree within the \cs\ region. The figure shows
    a region centred on the \cs\ location in gnomonic projection, with all
    the peaks in \texttt{SSG84}-filtered maps with FWHM ranging from
    $80\arcm$ to $1200\arcm$ overlaid on the same plot. The size of
    the coloured circles is proportional to the filtering scale. The
    colour corresponds to the peak value, normalized in units of
    $\sigma$ for a given filter scale. In both panels the data are from the \smica\ CMB map at full resolution.}
\label{fig:patch_CS}
\end{figure}

Since its discovery in the WMAP first-year data~\citep{vielva2004},
the \cs, centred at Galactic coordinates $(l, b) = (210\degr,
-57\degr)$ has been one of the most extensively studied large-scale
CMB anomalies.  In the 2013 release~\citep{planck2013-p09}, \Planck\
confirmed the apparently anomalous nature of this feature in
temperature, in terms of the area of the SMHW coefficients on angular
scales of $\approx 10\degr$ on the sky; the 2015 release has also
confirmed this feature (see Sects.~\ref{sec:area_cs}
and~\ref{sec:peaks}). The CMB temperature anisotropies around the \cs\
as observed by \Planck\ are shown in the top panel of
Fig.~\ref{fig:patch_CS}. The peak merger tree within the \cs\ region
is presented in the lower panel of the figure and provides a
multiscale view of its structure (see Sect.~\ref{sec:peaks_tree} for
details).

The robustness of the detection of the anomalies discussed in this
paper is a non-trivial issue. For the particular case of the \cs, this
has been reviewed by \citet{vielva2010a}, and addressed in detail by
\citet{cruz2006}, paying specific attention to the impact of
a~posteriori choices.  In particular, the latter study focused on the
original test that indicated the presence of this feature on the sky,
confirming a significance between 1\,\% and 2\,\%. An alternative
analysis of the significance based on two statistical tests with
different levels of conservativeness was made by \citet{mcewen2005},
providing values of 0.1\,\% and 4.7\,\%, respectively.  The
statistical significance of the \cs\ was questioned by
\citet{zhang2010} who found a low significance after performing a
study based on different kernels.  As discussed in more detail by
\citet{vielva2010a}, this result can also be interpreted as evidence
that not all kernels are necessarily suitable for the detection of
arbitrary non-Gaussian features.

The possibility that the \cs\ arises from instrumental systematics
\citep{vielva2004} or foreground residuals \citep{Liu2005,cruz2006}
has been largely rejected.  However, several non-standard physical
mechanisms have been proposed as possible explanations. These include
the gravitational effect produced by a collapsing cosmic
texture~\citep{cruz2007b}, the linear and nonlinear
ISW effect caused by a void in the large-scale structure
\citep[e.g.,][]{tomita2005,inoue2006,rudnick2007,tomita2008,finelli2014},
a cosmic bubble collision within the eternal inflation
framework~\citep{czech2010,feeney2011,mcewen2012}, and a localized
version of the inhomogeneous reheating scenario within the
inflationary paradigm~\citep{BuenoSanchez2014}.

Since the other scenarios lack additional evidence, the void
hypothesis would seem to be the most plausible, depending on the
sizes, density contrasts, and profiles assumed in the computations,
some of which are not in agreement with either
observation~\citep{cruz2008} or current $N$-body studies
\citep{cai2010,watson2014}. However, \cite{Szapudi2014} have recently
detected a large void in the WISE-2MASS galaxy catalogue aligned with
the \cs, with an estimated radius of around $200 h^{-1}$ Mpc, an
averaged density contrast of $\bar\delta \approx -0.1$, and centred on
a redshift of $z \approx 0.15$.  Large voids with similar
characteristics are not unusual in the standard $\Lambda$CDM model
\citep{Nadathur2014}. In fact, $N$-body simulations predict about 20
such voids in the local Universe ($z<0.5$).  However,
\citet{Zibin2014} and \citet{Nadathur2014} indicate that the expected
signal due to the linear and nonlinear ISW effects caused by this
structure is not large enough to explain the temperature decrement
associated with the \cs.

The new \Planck\ data release allows us to further explore the
statistical nature of the \cs. Two previous studies
\citep{Zhao2013,Gurzadyan2014} have claimed inconsistencies of the
internal properties of the \cs\ with the Gaussian hypothesis, which we
re-address here.  In particular, we consider the small-scale
fluctuations within a disc-like region of radius $\approx 25\degr$.

Several statistical quantities are computed from the full-resolution
temperature maps within the \cs\ region. This is divided into a
central disc of diameter 1\deg\ surrounded by a set of 13 concentric
annuli with central radii spaced in steps of about 2\deg, thus
allowing us to build angular profiles for the mean, variance,
skewness, and kurtosis. These are then compared to specialized CMB
realizations, generated as follows. A set of Gaussian CMB skies is
simulated using the FFP8 reference spectrum, and convolved with a
Gaussian beam of 5\arcm\ FWHM.  As for the FFP8 simulations themselves,
these maps are rescaled, as discussed previously. Only those that
contain a spot as extreme as the \cs\ at a scale $R=300\arcm$  in
SMHW space are retained, and these are rotated such that each simulated
cold spot is relocated to the actual position of the \cs\ (this
ensures that the noise properties are identical for both data and
simulations). This selection criterion corresponds to the
characteristic that originally indicated the presence of the \cs\ in
the observed sky. As a final step, for each remaining CMB simulation a
noise realization is added, consistent with each component-separation
method.

\begin{table}[tp] \begingroup \newdimen\tblskip \tblskip=5pt
  \caption{Probabilities of obtaining values for the $\chi^2$
    statistic of the angular profiles of the estimators shown in
    Fig.~\ref{fig:stats} larger than those determined from the data.}
\label{tab:chi2}
\nointerlineskip
\vskip -3mm
\footnotesize
\setbox\tablebox=\vbox{
   \newdimen\digitwidth
   \setbox0=\hbox{\rm 0}
   \digitwidth=\wd0
   \catcode`*=\active
   \def*{\kern\digitwidth}
   \newdimen\signwidth
   \setbox0=\hbox{+}
   \signwidth=\wd0
   \catcode`!=\active
   \def!{\kern\signwidth}
\halign{\hbox to 1.20in{#\leaderfil}\tabskip 4pt&
\hfil#\hfil\tabskip 8pt&
\hfil#\hfil\tabskip 8pt&
\hfil#\hfil\tabskip 8pt&
\hfil#\hfil\/\tabskip 0pt\cr
\noalign{\doubleline}
\noalign{\vskip -3pt}
\omit&\multispan4 \hfil Probability [\%] \hfil\cr
\noalign{\vskip -6pt}
\omit&\multispan4\hrulefill\cr
\omit\hfil Angular profiles\hfil& {\tt Comm.}& \nilc& \sevem& \smica\cr
\noalign{\vskip 3pt\hrule\vskip 3pt}
 Mean& *0.9& *0.8& *1.0& *0.9\cr
 Variance& 40.0& 40.0& 38.0& 42.0\cr
 Skewness& 79.0& 82.0& 85.0& 80.0\cr
 Kurtosis& 75.0& 56.0& 75.0& 77.0\cr
\noalign{\vskip 3pt\hrule\vskip 3pt}}}
\endPlancktable                    
\endgroup
\end{table}

\begin{figure*}
  \centering
  \includegraphics[scale=0.55]{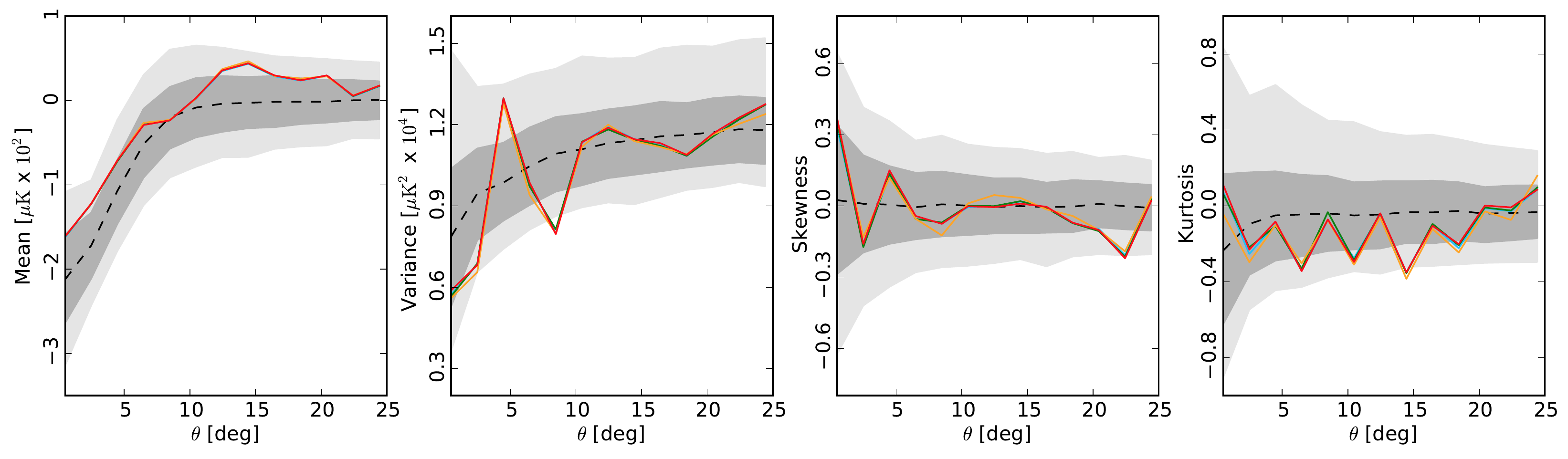}
  \caption{From left to right: the mean, variance, skewness, and kurtosis
    angular profiles computed for rings at radii $\theta$ centred on
    the \cs\ position for \commander\ (red), \nilc\ (orange), \sevem\ (green),
    and \smica\ (blue). The expected value obtained from the
    simulations is denoted by the black dashed line and the grey regions represent
    the $1\,\sigma$ and $2\,\sigma$ intervals.}
\label{fig:stats}
\end{figure*}

The results are presented in Fig.~\ref{fig:stats}.  Focusing on the
profile of the mean value, it is apparent that the largest deviations
from the simulations appear on scales around 15\degr, which
corresponds to a hot ring structure, as seen in
Fig.~\ref{fig:patch_CS} and previously discussed in~\cite{cayon2005}
and~\cite{Nadathur2014}. Notice that on the smallest scales the mean
profile is also somewhat deviant with respect to the simulations, but
this may be connected to selection bias, since we are considering CMB
simulations containing a spot that is at least as cold as the
\cs. However, if we consider the distribution of the profiles
corresponding to the coldest spots instead of the spots as extreme as
the \cs\ (removing the bias at the smallest scales) then the results
do not change substantially (see below).

In order to quantify possible deviations from Gaussianity, we
determine the probability of finding a $\chi^2$ value larger than that
of the data for each statistic, as summarized in Table~\ref{tab:chi2}.
The $\chi^2$ value for the data is computed using an estimate of the
covariance matrix between different radial scales determined from the
\cs\ simulations (1000 for each component-separation method), and then
compared to the theoretical $\chi^2$ distribution with 13 degrees of
freedom. The results indicate that the angular profile for the mean is
poorly described by the simulations, of which less than 1\,\% are
found to have a higher $\chi^2$ than the data (when considering the
distribution corresponding to the coldest spot this probability
becomes approximately 2\,\%).  We have checked that this deviation is
not obviously associated with a particular sub-range of angular
scales, implying that the mean profile is anomalous over the full
range considered.  Conversely, the radial profiles of the higher-order
moments are compatible with the Gaussian simulations.  The latter
results are then in contradiction with a similar analysis (using discs
instead of rings) by \cite{Zhao2013} for the WMAP 9-year data.
However, it appears that this may be related to the criteria applied
for the selection of the Gaussian simulations used to define the null
hypothesis. In particular, \citet{Zhao2013} used the coldest pixel in
real space as a means to identify those simulations that should be
retained, as opposed to the existence of cold spots as extreme as the
\cs\ selected in the SMHW coefficient map at $R=300\arcm$. Since it
is not implicit that such a temperature extremum is necessarily
associated with an extended cold region, particularly one defined in
wavelet space, the simulations used by \citet{Zhao2013} did not
contain features comparable to the nature of the \cs. This explains
why the \cs\ seemed to be anomalous when looking at the small-scale
fluctuations.

In conclusion, it appears that only the mean temperature profile of
the \cs\ should be considered anomalous when compared to CMB cold
spots that are as statistically extreme. All other measures of its
internal structure are consistent with expectations.

As a final remark, we note that the high-pass filtering currently
applied to the \Planck\ CMB polarization maps severely limits the
possibility of conducting targeted analyses to discriminate between
different possible origins of the \cs. For example, no polarization
signal would be expected in those models producing secondary
anisotropies due to a gravitational effect, whereas a specific pattern
might be expected in a bubble collision
scenario~\citep{czech2010}. Appropriate tests will be pursued in
future work, once the large-scale polarization data are available.

\section{Dipole modulation and directionality}
\label{sec:dipmodsection}

In this section, we examine isotropy violation related to dipolar
asymmetry, various forms of which have been noted since the early WMAP
releases \citep{eriksen2004b}.  We perform a non-exhaustive series of
tests in an attempt to narrow down the nature of the asymmetry (on the
assumption that it is not simply a statistical fluke).  First, we will
briefly describe some similarities and differences between the tests
that are important for making a proper comparison of the results.

All the tests in this section share in common the fitting of a dipole.
This is done either by
fitting for a dipole explicitly in a map of power on the sky (Sects.~\ref{sec:varasym} and
\ref{sec:power_asymmetry}), by employing Bayesian techniques in pixel
space for a specific model (Sect.~\ref{sec:dipmod}), or by measuring
the coupling of $\ell$ to $\ell \pm 1$ modes in the CMB covariance
matrix (Sects.~\ref{sec:QML}, \ref{sec:biposh}, and
\ref{sec:directionality}).  The differences arise from how the
fitted dipoles are combined, which determines the specific form of
asymmetry that the test is sensitive to.

The tests can be divided into two categories, amplitude-based and
direction-based.  Sections~\ref{sec:varasym} to \ref{sec:biposh} are
all sensitive to the {\em amplitude} of a dipole modulation.
Specifically, Sect.~\ref{sec:varasym} looks for dipole modulation in
the pixel-to-pixel variance of the data, while
Sects.~\ref{sec:dipmod}, \ref{sec:QML}, and \ref{sec:biposh} all
search for dipole modulation of the angular power spectrum.  The
distinction between these two approaches is mainly one of $\ell$
weighting.

Sections~\ref{sec:power_asymmetry} and \ref{sec:directionality} both
examine aspects of {\em directionality} in the data, where the
directions are extracted from dipole fits but combined in different
ways.  Section~\ref{sec:power_asymmetry} fits for dipoles in band
power (with similar results for variance) and only uses the direction
information, while Sect.~\ref{sec:directionality} weights each dipole
equally across all scales and uses the amplitude information as well.

The differences between the approaches of these sections should be
kept in mind when comparing their results.  For example, although
Sects.~\ref{sec:power_asymmetry} and \ref{sec:directionality} both
look for a directional signal in the data, they are optimized for
different forms of deviations from statistical isotropy.  It is
therefore unsurprising that they arrive at different results.
However, the signal found in Sect.~\ref{sec:power_asymmetry}, if not
simply a statistical fluke, is constrained by the results of
Sect.~\ref{sec:directionality}.

\subsection{Variance asymmetry}
\label{sec:varasym}

The study of power asymmetry via the local variance of the CMB
fluctuations was first performed by \citet{Akrami2014} for the
\Planck\ 2013 and WMAP 9-year temperature data.  The approach was
motivated by its conceptual and implementational simplicity, its
directly intuitive interpretation, and by virtue of being defined in
pixel space, a useful complementarity to other mostly harmonic-based
methods.  The statistic was computed over patches of different sizes
and positions on the sky, and compared with the values obtained from
statistically isotropic simulations.  It was found that none of the 1000 available
simulations had a larger variance asymmetry than that
estimated from the data. This suggested the presence of asymmetry at a
statistical significance of at least $3.3\,\sigma$, with a preferred
direction $(l, b)\approx(212\deg, -13\deg)$ in good agreement with
other studies.  In this section, we revisit the variance asymmetry and
report the results of the analysis for the \Planck\ 2015 temperature
maps at full resolution, $N_{\mathrm{side}}=2048$.

The analysis proceeds as follows. We consider a set of discs of
various sizes centred on the pixels of a \healpix\ map defined by a
specific $N_{\mathrm{side}}$ value.  For each sky map, we first remove
the monopole and dipole components from the masked sky and then
compute the variance of the fluctuations on a given disc using only
the unmasked pixels. This yields a local-variance map at the \healpix\
resolution of interest. We also estimate the expected average and
variance of the variance on each disc from the simulations and then
subtract the resulting average variance map from both the observed and
simulated local-variance maps.  Finally, we define the amplitude and
direction of the asymmetry by fitting a dipole to each of the
local-variance maps, where each pixel is weighted by the inverse of
the variance of the variances computed from the simulations at that
pixel. At all stages, we use only the discs for which more than $10\,\%$
of the area is unmasked, although our results are robust against the
choice of this value. The computed local-variance amplitudes are then
used to compare the data with statistically isotropic simulations.  Note
that we use only the dipole amplitudes of the local-variance maps to
measure the significance of the asymmetry; the amplitudes of higher
multipoles were shown by \citet{Akrami2014} to be consistent with
statistically isotropic simulations and we therefore do not consider
them in the present paper.

In \citet{Akrami2014}, the sensitivity of the method to the disc size
was assessed using both statistically isotropic and anisotropic
simulations.  The free parameters, i.e., the number and size of the
discs, were then fixed by these simulations. It was found that for
3072 patches centred on the set of pixels defined at
$N_\mathrm{side}=16$, the simulated asymmetry signals were not
detected when either very small ($r_\mathrm{disc}<4\deg$) or very
large ($r_\mathrm{disc}>16\deg$) discs were used.

The former effect is due to a combination of the low number of pixels
per disc and an insufficient number of discs to cover the entire sky
when $N_{\mathrm{side}}=16$ reference grids are used. However, it has
recently been shown by \citet{adhikari14} that using a larger number
of small discs (by increasing $N_{\mathrm{side}}$ to 32, 64, 128, and
256, depending on the disc size) in order to cover the entire sky
allows the local-variance method to detect the large-scale anomalous
asymmetry as well as the Doppler boost signal from the \Planck\ 2013
data, at a significance of $>3.3\,\sigma$. \citet{fantaye14b} has
demonstrated that the Doppler boost signal can be detected at a
similar level of significance using needlet bandpass filtering of the
data, even with large discs, when simulations are deboosted. Here, in
contrast to the 2013 analysis, we use maps which contain Doppler
boosting, for both simulations and data, and therefore we do not
detect any Doppler boost signal when using a large number of small
discs.

The low observed significance levels when large discs are used is due
to the cosmic variance associated with the largest-scale
modes. Motivated by the analysis of \citet{fantaye14b}, and in order
to address this issue, we also perform analyses using a Butterworth
high-pass filter,
\begin{linenomath*}
\begin{equation}
H(\ell) = \frac{(\ell/\ell_{0})^4}{1+(\ell/\ell_{0})^4} ,
\end{equation}
\end{linenomath*}
centred at multipoles $\ell_{0} =5$, 10, 15, 20, and $30$.  In
addition, the filtering of low multipoles allows us to establish the
contribution of such modes to any detected asymmetry.

Here, based on the analysis of \citet{Akrami2014}, we restrict our
analysis to those disc sizes for which 3072 discs, corresponding to an
$N_{\mathrm{side}}=16$ map, cover the entire sky, i.e., to the range
4\deg--90\deg.  Consistent results can be obtained by choosing other
values of $N_{\mathrm{side}}$ for a given disc size provided that the
entire sky is covered by the discs. Here, for simplicity, we work with
the same $N_{\mathrm{side}}$ ($=16$) for all disc sizes.

\begin{figure}
\begin{center}
\mbox{\epsfig{figure=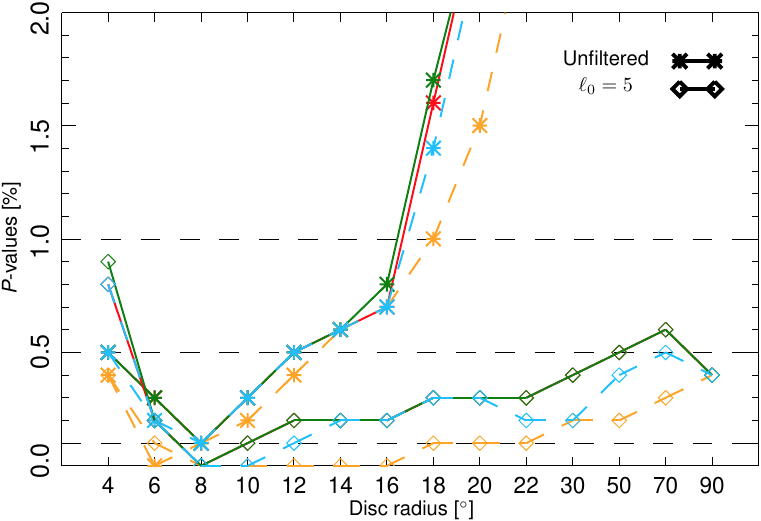,width=1\linewidth}} \\
\vskip 3mm
\mbox{\epsfig{figure=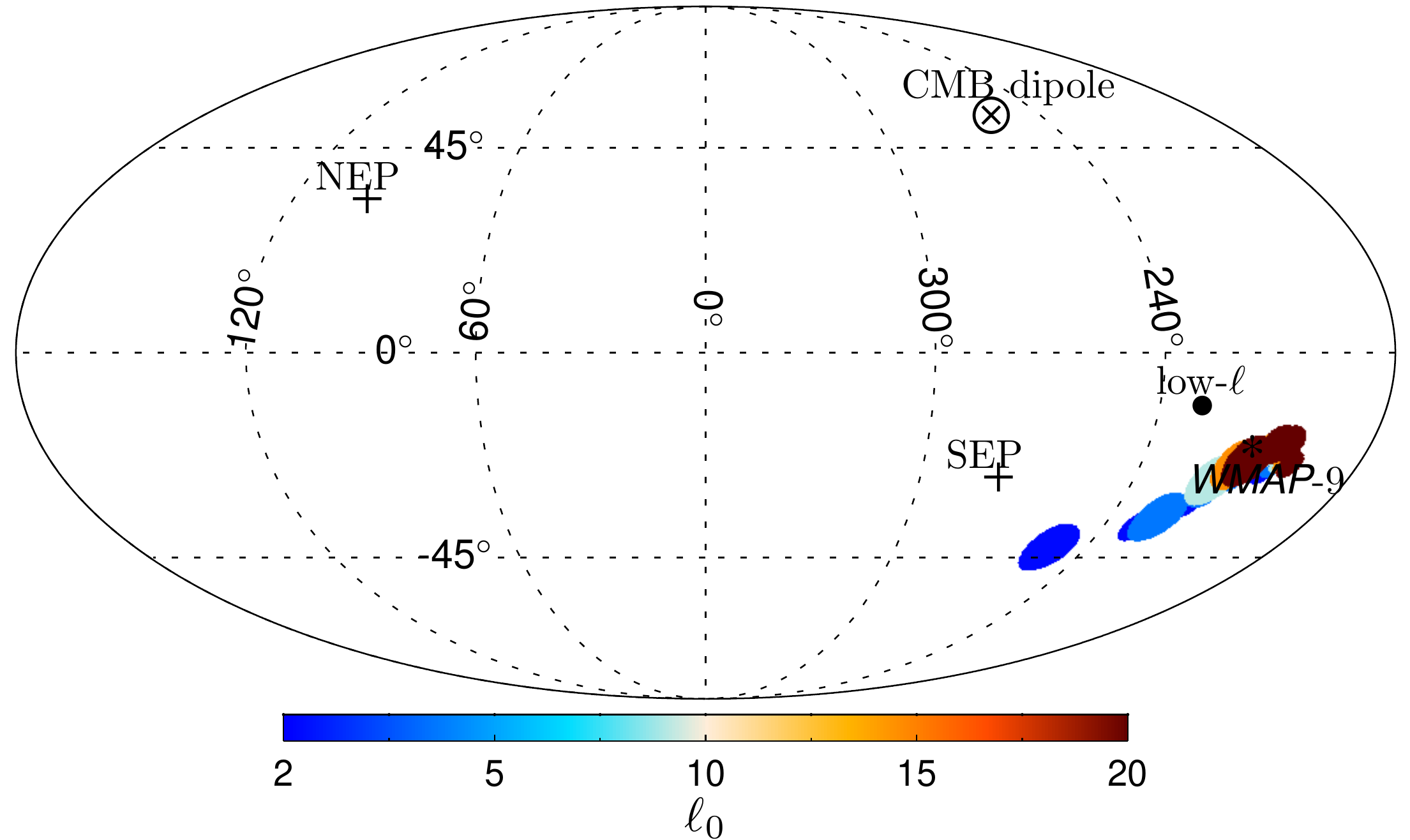,width=1\linewidth}} \\
\end{center}
\caption{{\it Upper panel:} $p$-values for variance asymmetry
  measured as the number of simulations with local-variance dipole
  amplitudes larger than those inferred from the data, as a
  function of disc radius for the four component-separated
  maps, \commander\ (red), \nilc\ (orange), \sevem\ (green),
  and \smica\ (blue), and for unfiltered and high-pass-filtered
  cases. For the filtered case, the \commander\ curve is covered
  by the \smica\ curve for small ($r_\mathrm{disk} \leq 8$) disks, and by
  the \sevem\ curve for large disks  ($r_\mathrm{disk} > 8$). {\it Lower panel:}
  local-variance dipole directions for the \smica\ map. The colours,
  as indicated by the colourbar, correspond to different values of the
  high-pass filter central multipole $\ell_{0}$. The size of a marker
  disc corresponds, from small to large, to the size of the disc
  used in the analysis, namely 4\deg, 12\deg, 20\deg, and 70\deg. The dipole
  directions from the \commander, \nilc, and \sevem\
  component-separation methods are consistent with the case shown
  here.
  The low-$\ell$ and WMAP-9 directions are identical to those in Fig.~\ref{fig:dipoles}.
}  \label{fig:varasym1}
\end{figure}

Our results for the measured amplitude of the variance asymmetry,
compared to the values from the simulations, as well as the
corresponding dipole directions, are shown in
Fig.~\ref{fig:varasym1}. The $p$-values are given for different
disc sizes and in terms of the number of simulations with
local-variance dipole amplitudes greater than the ones measured from
the data. Note that since the discs with different sizes used in our
analysis are correlated, the significance levels are also correlated.
For this reason we choose to show the $p$-values as a function of
disc size instead of combining them into a single number. Moreover,
it should be noted that the significance values we present here do not
incorporate any corrections to account for the choice of parameters
adopted during method calibration, specifically the dipole amplitudes
and directions for the anisotropic simulations that were used to fix
the range of disc sizes and number of patches.

\begin{figure}
\begin{center}
\mbox{\epsfig{figure=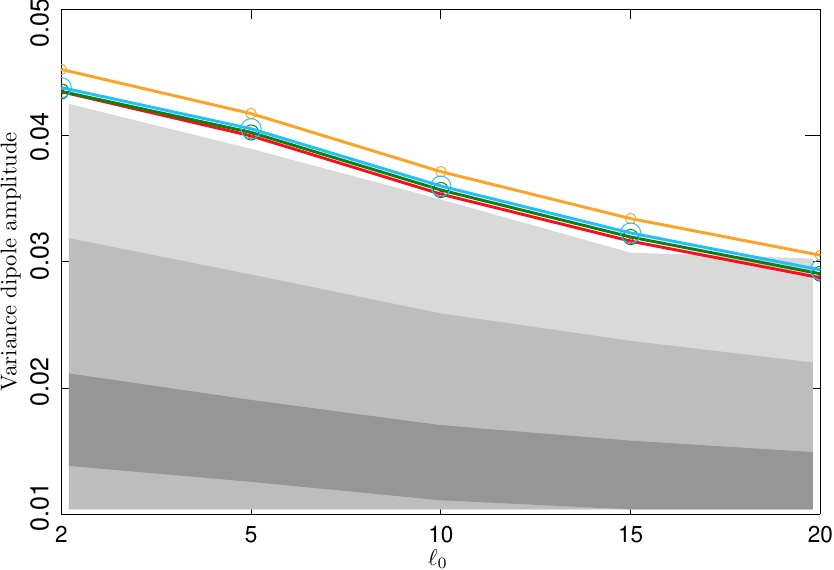,width=1\linewidth}}\\
\vskip 3mm
\mbox{\epsfig{figure=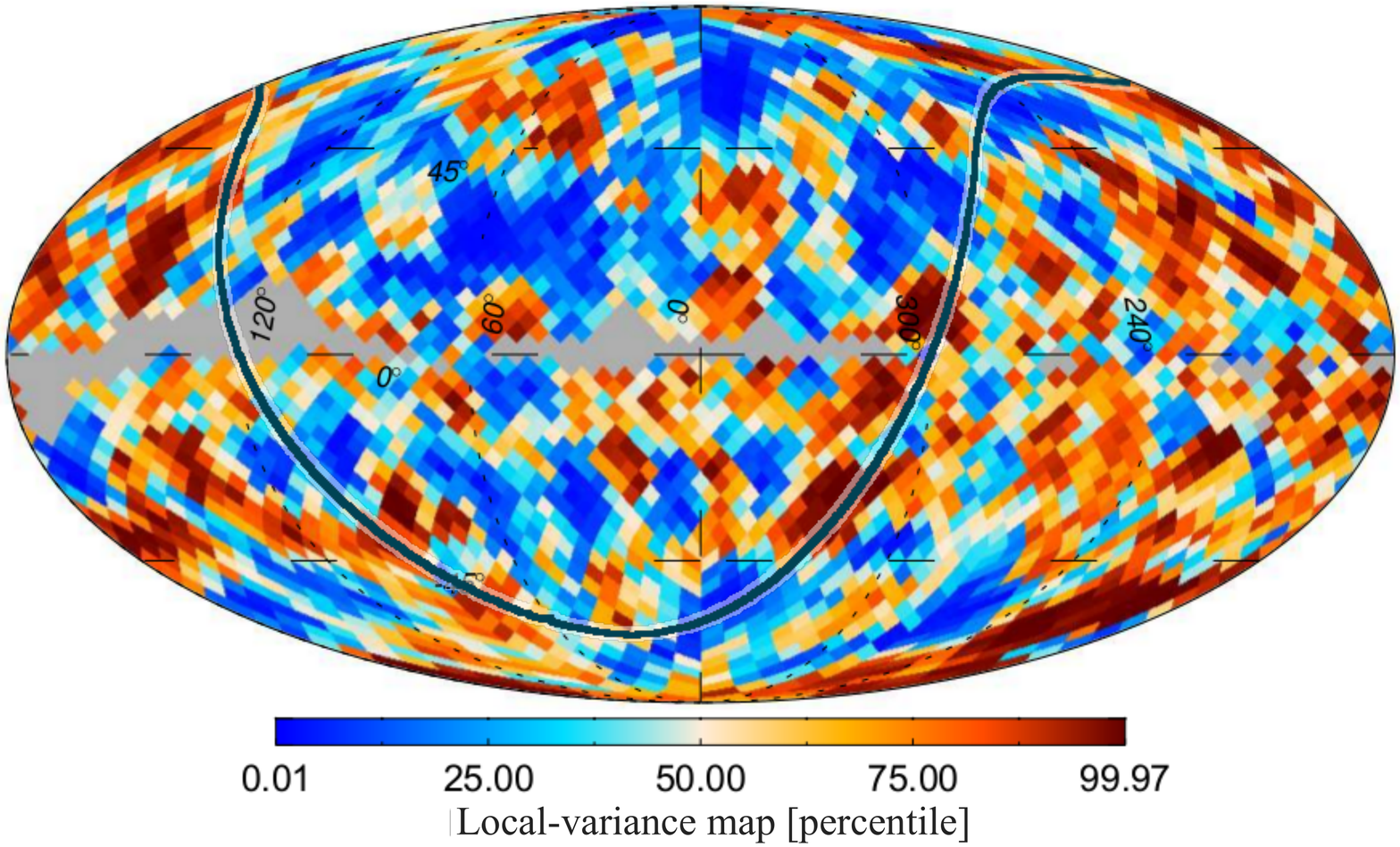,width=1\linewidth}}
\end{center}
\caption{{\it Upper panel:} local-variance dipole amplitude for
  8\deg\ discs as a function of the central multipole of the
  high-pass filter, $\ell_{0}$, for the four
  component-separation methods, \commander\ (red), \nilc\ (orange),
  \sevem\ (green), and \smica\ (blue). The grey regions, from dark to
  light, correspond, respectively, to $1\,\sigma$, $2\,\sigma$, and
  $3\,\sigma$ percentiles from the 1000 FFP8 simulations processed by
  the \commander\ method.  {\it Lower panel:} mean-subtracted and
  inverse-variance-weighted local-variance map for the 8\deg\
  discs and for the \commander\ component-separation method; each
  pixel is given in terms of the lower- and upper-tail probability of
  the measured value on that pixel compared to the values from the
  simulations. The pixels in grey correspond to the centres of the
  8\deg\ discs on which the number of unmasked pixels in the full
  resolution map is lower than our threshold. The black curve
    superposed on the map indicates the boundary of the opposing
    hemispheres along the asymmetry axis. It is clear that the largest fraction
  of $>$95\,\% outliers (red pixels) lie on the positive amplitude hemisphere of the
  local variance dipole, while the $<$5\,\% outliers (blue pixels) are on the
  opposite hemisphere. The corresponding maps for \nilc, \sevem, and
  \smica\ are very similar to the one shown here.}
\label{fig:varasym2}
\end{figure}

It can be seen from the upper panel of Fig.~\ref{fig:varasym1} that
for the unfiltered map the significance of the power asymmetry drops
quickly when we increase the disc size to radii greater than
16\deg. This is no longer the case, however, when the lowest
multipoles are filtered out.  For example, when the filter scale is
set to $\ell_{0}=5$, i.e., when the very low multipoles which are
affected most by cosmic variance are suppressed, the variance
asymmetry is detected at the $3\,\sigma$ level for all disc sizes, as
shown in Fig.~\ref{fig:varasym1}.  Table~\ref{tab:varasym1} presents
the $p$-values of the variance asymmetry using 8\deg\ discs and
for various values of $\ell_{0}$.  Our results show that variance
asymmetry is detected with a remarkable significance for all disc
sizes when very low multipoles are filtered out. In addition, the
variance asymmetry amplitude slowly decreases with increasing
$\ell_{0}$, as seen in the upper panel of Fig.~\ref{fig:varasym2}. For
$\ell_{0}\gtrsim 20$, the dipole amplitude becomes too small and we
find no significant variance asymmetry. It is interesting to note,
however, that the dipole directions found for large $\ell_{0}$ are
closely aligned with those found for $\ell_{0} < 20$.

\begin{table}[h!tb]
\begingroup
\newdimen\tblskip \tblskip=5pt
\caption{$p$-values for the variance asymmetry measured by 8\deg\
  discs for the four component-separated temperature maps and
  different high-pass filter scales. The values represent the
  fraction of simulations with local-variance dipole amplitudes larger than
  those inferred from the data.}
\label{tab:varasym1}
\nointerlineskip
\vskip -3mm
\footnotesize  
\setbox\tablebox=\vbox{
   \newdimen\digitwidth
   \setbox0=\hbox{\rm 0}
   \digitwidth=\wd0
   \catcode`!=\active
   \def!{\kern\digitwidth}
   \newdimen\signwidth
   \setbox0=\hbox{>}
   \signwidth=\wd0
   \catcode`*=\active
   \def*{\kern\signwidth}
\halign{\hbox to 0.8in{#\leaderfil}\tabskip 4pt&
\hfil#\hfil\tabskip 8pt&
\hfil#\hfil\tabskip 8pt&
\hfil#\hfil\tabskip 8pt&
\hfil#\hfil\/\tabskip 0pt\cr
\noalign{\doubleline}
\noalign{\vskip -1pt}
\omit&\multispan4 \hfil $p$-value [\%]\hfil\cr
\noalign{\vskip -4pt}
\omit&\multispan4\hrulefill\cr
\omit \hfil $\ell_{0}$ \hfil&\omit\hfil {\tt Comm.}\hfil&\omit\hfil \nilc \hfil&\omit\hfil \sevem \hfil&\omit\hfil \smica \hfil\cr
\noalign{\vskip 3pt\hrule\vskip 3pt}
Unfiltered & *0.1& *0.1& *0.1& *0.1\cr
5& $<$0.1& $<$0.1& $<$0.1& $<$0.1\cr
10& $<$0.1& $<$0.1& $<$0.1& $<$0.1\cr
15& *0.1& *0.0& *0.1& $<$0.1\cr
20& *0.4& $<$0.1& *0.3& *0.2\cr
30& *1.8& *0.8& *1.8& *1.7\cr
\noalign{\vskip 3pt\hrule\vskip 3pt}}} %
\endPlancktable
\endgroup %
\end{table}

The lower panel of Fig.~\ref{fig:varasym1} shows the dipole directions
we find using different disc sizes and different filter scales for
\smica.  The dipole directions for the \commander, \nilc, and \sevem\
component-separated maps are very similar to those shown.  The
asymmetry directions found here are consistent with those determined
by other analyses in this paper.

In the upper panel of Fig.~\ref{fig:varasym2}, we show the
local-variance dipole amplitudes for the 8\deg\ discs as a function of
the central multipole of the high-pass filter, $\ell_{0}$.  In the
lower panel of the same figure we show, as an example, the
mean-subtracted and inverse-variance-weighted local-variance map using
8\deg\ discs for the \commander\ component-separation method. The
pixels of the map are given in terms of the lower- and upper-tail
probabilities of the values from the data compared to the values from
the simulations. The maps for \nilc, \sevem, and \smica\ are very
similar. The numerical values of the local-variance dipole amplitudes
and directions for the \commander\ method are given in
Table~\ref{tab:varasym2}; the values for \nilc, \sevem, and \smica\
methods are similar.

\begin{table}[h!tb]
\begingroup
\newdimen\tblskip \tblskip=5pt
\caption{Local-variance dipole amplitudes and directions. All values
  quoted here are for 8\deg\ discs. This table is for the
  \commander\ component-separation method, but the results are
  similar for the other methods.
}
 \label{tab:varasym2}
\nointerlineskip
\vskip -3mm
\footnotesize  
\setbox\tablebox=\vbox{
\halign{\hbox to 0.7in{#\leaderfil}\tabskip 6pt&
\hfil#\hfil\tabskip=5pt&\hfil#\hfil\/\tabskip=0pt&\hfil#\hfil \cr
\noalign{\doubleline}
\omit& \omit& Direction \cr
\omit\hfil $\ell_{0}$ \hfil& $A^{\rm a}$& $(l,b)$ [\deg] \cr
\noalign{\vskip 3pt\hrule\vskip 3pt}
Unfiltered& \hfil $0.052 \pm 0.016$ \hfil& \hfil $(210, -26)$\hfil\cr
5& \hfil $0.046 \pm 0.014$ \hfil& \hfil $(208,-24)$\hfil\cr
10& \hfil $0.040 \pm 0.014$ \hfil& \hfil $(199,-16)$\hfil\cr
15& \hfil $0.038 \pm 0.012$ \hfil& \hfil $(206,-16)$\hfil\cr
20& \hfil $0.028 \pm 0.010$ \hfil& \hfil $(202,-18)$\hfil\cr
30& \hfil $0.025 \pm 0.010$  \hfil& \hfil $(199, -19)$\hfil\cr
\noalign{\vskip 3pt\hrule\vskip 3pt}}} %
\endPlancktable
\endgroup %
{\footnotesize $^{\rm a}$ $A = 2(A_{\rm Planck}-\langle A_{\rm FFP8}\rangle )$, where
  $A_{\rm Planck}$ and $A_{\rm FFP8}$ are the local-variance dipole amplitudes of the data
  and the FFP8 simulations, respectively. The quoted errors are the
  dispersion of the simulation amplitudes. Assuming a pure dipole
  modulation model, $A$ to first order would correspond to the
  modulation amplitude.}
\end{table}

\subsection{Dipole modulation: pixel-based likelihood}
\label{sec:dipmod}

\begin{figure}
\begin{center}
\mbox{\epsfig{figure=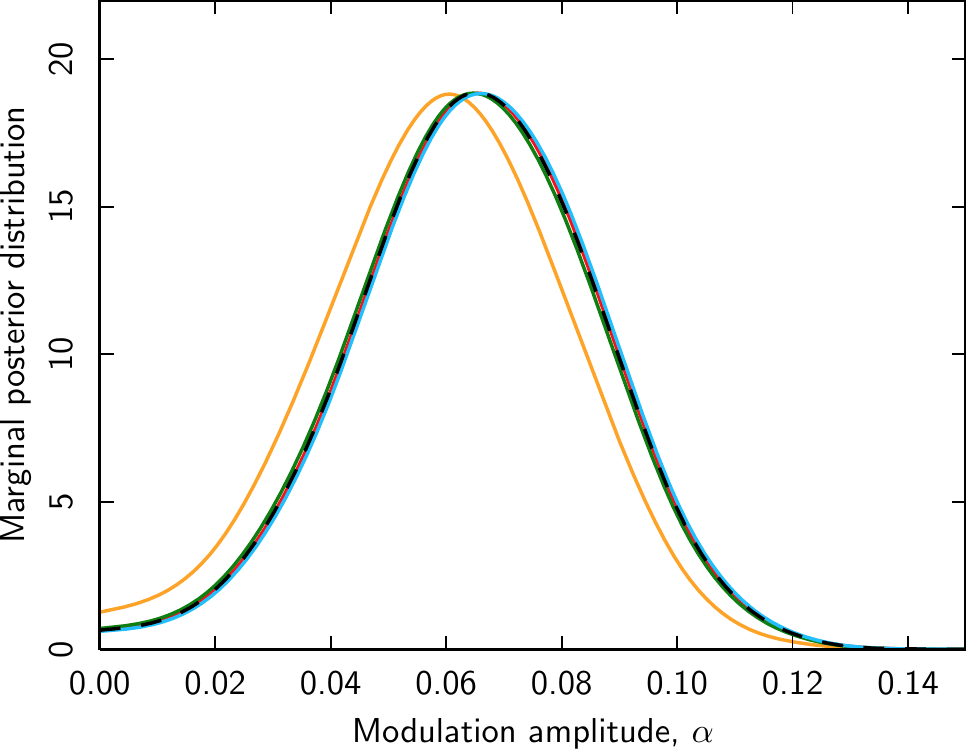,width=\linewidth,clip=}}
\vskip 4pt
\mbox{\epsfig{figure=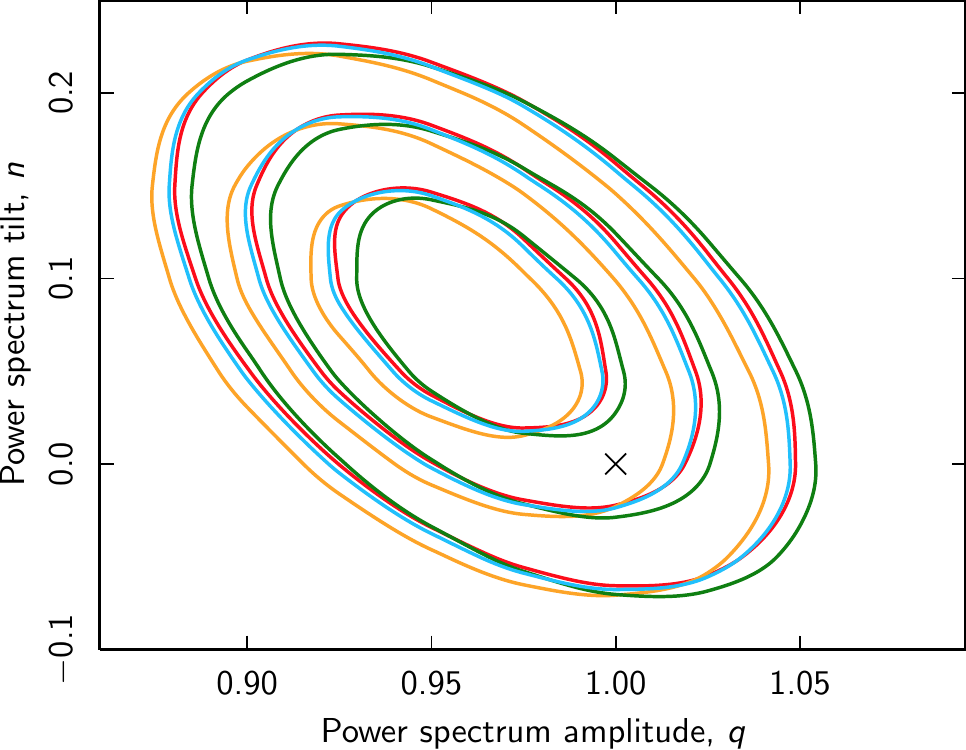,width=\linewidth,clip=}}
\end{center}
\caption{\textit{Top}:  Marginal constraints on the dipole modulation amplitude, as derived
  from \Planck\ 2015 temperature observations at a smoothing scale of
  $5^{\circ}$ FWHM
  for \commander\ (red), \nilc\ (orange), \sevem\ (green), and \smica\ (blue).
  The plot corresponds directly to Fig.~32
  of \citet{planck2013-p09}. The \texttt{Commander}, \texttt{SEVEM},
  and \texttt{SMICA} posteriors coincide almost perfectly both
  internally, and with the corresponding \texttt{SMICA} 2013
  posterior, shown as a dashed black line.
  \textit{Bottom}: Corresponding marginal two-dimensional
  constraints on the low-$\ell$ power spectrum amplitude and tilt,
  $(q,n)$, defined relative to the best-fit \Planck\ 2015 $\Lambda$CDM
  model.}
\label{fig:dipmod}
\end{figure}

\begin{table}
\begingroup
\newdimen\tblskip \tblskip=5pt
\caption{Summary of dipole modulation results at a
  smoothing scale of $5\deg$ for all {\Planck} 2015 CMB temperature
  solutions, as derived by the brute-force likelihood given by
  Eq.~\ref{eq:dipmod_like}.}
\label{tab:dipmod}
\nointerlineskip
\vskip -3mm
\footnotesize
\setbox\tablebox=\vbox{
   \newdimen\digitwidth
   \setbox0=\hbox{\rm 0}
   \digitwidth=\wd0
   \catcode`*=\active
   \def*{\kern\digitwidth}
   \newdimen\signwidth
   \setbox0=\hbox{+}
   \signwidth=\wd0
   \catcode`!=\active
   \def!{\kern\signwidth}
\halign{\hbox to 1.0in{#\leaderfil}\tabskip 2pt&
    \hfil#\hfil\tabskip 8pt&
    \hfil#\hfil\tabskip 0pt\cr
\noalign{\doubleline\vskip -1pt}
\omit \hfil Method\hfil& \omit\hfil 2013\hfil& \omit\hfil 2015\hfil\cr   
\noalign{\vskip 3pt\hrule\vskip 5pt}
\multispan3 \hfil Dipole modulation amplitude, $\alpha$\hfil\cr
\noalign{\vskip 2pt}
{\tt Commander}& $0.078\pm0.021$& $0.066\pm0.021$\cr
{\tt NILC}&      $0.069\pm0.021$& $0.061\pm0.022$\cr
{\tt SEVEM}&     $0.066\pm0.021$& $0.065\pm0.021$\cr
{\tt SMICA}&     $0.065\pm0.021$& $0.066\pm0.021$\cr
\noalign{\vskip 2pt}
\multispan3 \hfil Dipole modulation direction, $(l,b)$ [\deg]\hfil\cr
\noalign{\vskip 2pt}
{\tt Commander}& $(227,-15)\pm19$& $(230,-16)\pm24$\cr
{\tt NILC}&      $(226,-16)\pm22$& $(228,-19)\pm29$\cr
{\tt SEVEM}&     $(227,-16)\pm24$& $(226,-17)\pm25$\cr
{\tt SMICA}&     $(226,-17)\pm24$& $(225,-18)\pm24$\cr
\noalign{\vskip 2pt}
\multispan3{\hfil Power spectrum amplitude, $q$}\hfil\cr
\noalign{\vskip 2pt}
{\tt Commander}& $\cdots$& $0.961\pm0.025$\cr
{\tt NILC}&      $\cdots$& $0.954\pm0.024$\cr
{\tt SEVEM}&     $\cdots$& $0.966\pm0.025$\cr
{\tt SMICA}&     $\cdots$& $0.960\pm0.025$\cr
\noalign{\vskip 2pt}
\multispan3{\hfil Power spectrum tilt, $n$}\hfil\cr
\noalign{\vskip 2pt}
{\tt Commander}& $\cdots$& $0.082\pm0.043$\cr
{\tt NILC}&      $\cdots$& $0.077\pm0.043$\cr
{\tt SEVEM}&     $\cdots$& $0.077\pm0.043$\cr
{\tt SMICA}&     $\cdots$& $0.081\pm0.043$\cr
\noalign{\vskip 2pt\hrule\vskip 3pt}}}
\endPlancktable                    
\endgroup
\end{table}

In \citetalias{planck2013-p09} we presented an analysis of the
apparent anisotropic distribution of large-scale power in the \Planck\
2013 temperature data within the parametric framework defined by
\citet{Gordon2007} and \citet{Hoftuft2009}, who introduced an explicit dipole
modulation field to model potential hemispherical power asymmetry.
The following is a direct update of that analysis using the \Planck\
2015 CMB data at $\nside = 32$, retaining the the 2013 common mask to
explicitly test for consistency with the earlier study. All results are found to be in
excellent agreement. In the following, we therefore
only consider a smoothing scale of $5\deg$ FWHM as a representative
example. This is the highest angular resolution accessible for an
$\nside = 32$ map.

Recall first the basic data model adopted in the dipole modulation
approach: rather than assuming the CMB sky to be a statistically
isotropic Gaussian field, we allow for an additional dipole
modulation, resulting in a data model of the form $\vec{d} =
\tens{B}\tens{M}\vec{s} + \vec{n}$, where $\tens{M_{ij}} =
(1+\alpha\,\vec{\hat{p}}\cdot\vec{\hat{n}_i})\delta_{ij}$ is an offset
dipole field multiplying an intrinsically isotropic signal $\vec{s}$
with a dipole of amplitude $\alpha$ pointing towards some preferred
direction $\vec{\hat{p}}$. $\tens{B}$ denotes convolution with an
instrumental beam, and $\vec{n}$ denotes instrumental
noise. Additionally, we model the power spectrum of the underlying
statistically isotropic field in terms of a two-parameter amplitude--tilt model of
the form $C_{\ell}(q,n) =
q\,\left(\ell/30\right)^{n}\,C_{\ell}^{\Lambda\textrm{CDM}}$, where
$C_{\ell}^{\Lambda\textrm{CDM}}$ is the best-fit \Planck\ 2015
$\Lambda$CDM spectrum \citep{planck2014-a13}. The two parameters $q$
and $n$ can accommodate a deficit in power at low $\ell$ as compared
to the best-fit cosmology that would otherwise create a tension with
the underlying statistically isotropic model and result in the analysis measuring a
combination of both asymmetry and power mismatch.

In the absence of any dipole modulation, $\alpha=0$, the total data
covariance matrix is given by $\tens{C} =
\tens{B}\tens{S}_\mathrm{iso}\tens{B}^\mathrm{T} + \tens{N}$, where
$\tens{S}_\mathrm{iso}$ is the standard statistically isotropic CMB covariance
matrix given by the power spectrum, $C_{\ell}$, $\tens{N}$ is the
noise covariance matrix, and the corresponding likelihood is given by
the usual expression for a multivariate Gaussian distribution. With
dipole modulation, this generalizes straightforwardly to $\tens{C} =
\tens{B}\tens{M}\tens{S}_\mathrm{iso}\tens{M}^\mathrm{T}\tens{B}^\mathrm{T}
+ \tens{N}$, with the likelihood given by
\begin{linenomath*}
\begin{equation}
\mathcal{L}(\alpha,\hat{p},q,n) \propto \frac{\exp{[-\frac{1}{2} \vec{d}^t(\tens{B}\tens{M}\tens{S}\tens{M}^\mathrm{T}\tens{B}^\mathrm{T} + \tens{N})^{-1}\vec{d}]}}   {\sqrt{|\tens{B}\tens{M}\tens{S}\tens{M}^\mathrm{T}\tens{B}^\mathrm{T} + \tens{N}|}}.
\label{eq:dipmod_like}
\end{equation}
\end{linenomath*}

Figure~\ref{fig:dipmod} and Table~\ref{tab:dipmod} summarize this
five-dimensional likelihood in terms of marginal parameters for each
of the four \Planck\ CMB maps, as evaluated over the common mask using
the multi-dimensional grid-based \texttt{Snake} algorithm
\citep{mikkelsen2012}.  All results correspond to a smoothing scale of
$5\deg$ FWHM, the highest resolution supported by an
$N_{\textrm{side}}=32$ {\tt HEALPix} grid, but, as in 2013, we
consider all smoothing scales between $5\deg$ and $10\deg$ FWHM, reaching
similar conclusions in each case: The dipole modulation results
derived from the \Planck\ 2015 temperature maps are essentially
identical to the 2013 results, with improved internal consistency
between the four CMB maps due to better mitigation of systematic errors. The
best-fit dipole modulation amplitude at $5\deg$ FWHM is 6--7\,\% whilst
the low-$\ell$ power spectrum has an approximately 3--5\,\% lower amplitude
compared to the best-fit $\Lambda$CDM prediction.  These results are
fully consistent with expectations given that the \Planck\ 2013 sky
maps were already cosmic-variance-limited on these angular scales, and
the 2015 maps differ from the 2013 maps at the level of only a few
microkelvin \citep{planck2014-a11}.

\subsection{Dipole modulation: QML analysis}
\label{sec:QML}

In this section we use the quadratic maximum likelihood (QML)
estimator introduced in \citet{Moss2011} and described in
Appendix~\ref{sec:mossest} to assess the level of dipole modulation in
our estimates of the CMB sky at $\nside=2048$.  The specific
implementation is essentially identical to that used in
\citet{Hanson2009b}, \citet{planck2013-p12}, and
\citet{planck2013-pipaberration}, and exploits the fact that dipole
modulation of any cosmological parameter is equivalent to coupling of
$\ell$ to $\ell \pm 1$ modes in the CMB covariance matrix to leading
order (see Appendix~\ref{sec:mossest}).  \citet{planck2014-a24}
presents an alternate analysis for a specific isocurvature model.

\begin{table}
\begingroup
\newdimen\tblskip \tblskip=5pt
\caption{Amplitude ($A$) and direction of the low-$\ell$ dipole
  modulation signal determined from the QML analysis for the range
  $\ell \in [2, 64]$. The errors
  are calculated from the cosmic  variance expected for statistically isotropic CMB
  realizations.}
\label{tab:lowellmod}
\nointerlineskip
\vskip -3mm
\footnotesize
\setbox\tablebox=\vbox{
   \newdimen\digitwidth
   \setbox0=\hbox{\rm 0}
   \digitwidth=\wd0
   \catcode`*=\active
   \def*{\kern\digitwidth}
   \newdimen\signwidth
   \setbox0=\hbox{+}
   \signwidth=\wd0
   \catcode`!=\active
   \def!{\kern\signwidth}
\halign{ \hbox to 1.0in{#\leaderfil}\tabskip 4pt&
         \hfil#\hfil\tabskip 8pt&
         \hfil#\hfil\tabskip 0pt\cr                           
\noalign{\doubleline\vskip -4pt}
\omit& \omit& Direction\cr   
\omit\hfil Method\hfil&$A$& $(l, b)$ [\deg]\cr   
\noalign{\vskip 4pt\hrule\vskip 6pt}
{\tt \commander}& $0.063^{+0.025}_{-0.013}$& $(213, -26)\pm28$\cr
\noalign{\vskip 3pt}
{\tt \nilc}& $0.064^{+0.027}_{-0.013}$& $(209, -25)\pm28$\cr
\noalign{\vskip 3pt}
{\tt \sevem}& $0.063^{+0.026}_{-0.013}$& $(211, -25)\pm28$\cr
\noalign{\vskip 3pt}
{\tt \smica}& $0.062^{+0.026}_{-0.013}$& $(213, -26)\pm28$\cr
\noalign{\vskip 3pt\hrule\vskip 4pt}}}
\endPlancktable                    
\endgroup
\end{table}

Since we are interested in dipole modulation there are three
independent estimators. For our particular approach, these are a
real-valued $m=0$ and a complex-valued $m=1$ estimator, and take the form
\begin{linenomath*}
\begin{align}
 \tilde{X}_0 &= \frac{6}{f_{10}}\frac{\sum_{\ell m} \delta C_{\ell \ell+1}A_{\ell m}
(T^*_{\ell m}T_{\ell + 1\,m} - \left< T^*_{\ell m} T_{\ell + 1\,m}
\right>)}{\sum_{\ell} \delta C_{\ell \ell +1}^2(\ell + 1)F_{\ell}F_{\ell+1}},
\label{eq:mossest0}\\
 \tilde{X}_1 &= \frac{6}{f_{11}}\frac{\sum_{\ell m} \delta C_{\ell \ell+1}B_{\ell m} (T^*_{\ell
m}T_{\ell + 1\,m+1} - \left<T^*_{\ell m}T_{\ell + 1\,m+1}\right>) }{\sum_{\ell}
\delta C_{\ell \ell +1}^2(\ell + 1)F_{\ell}F_{\ell+1}}.
\label{eq:mossest1}
\end{align}
\end{linenomath*}
Here $T_{\ell m}$ are $C$-inverse filtered data and $F_\ell \equiv
\left< T_{\ell m} T^*_{\ell m} \right>$. We adopt the inverse-variance
filter from \citet{planck2013-p12}, where the approximate filter
functions are also specified.  We define $\delta C_{\ell\ell+1} \equiv
dC_\ell/dX + dC_{\ell+1}/dX$, where $X$ is the parameter modulated,
and $A_{\ell m}$ and $B_{\ell m}$ are numerical coefficients (details
can be found in Appendix~\ref{sec:mossest}).  The factor $f_{1m}$
corrects the normalization for errors introduced by masking:
\begin{linenomath*}
\begin{align}
 f_{1 m} &\equiv \int \dd\Omega\,Y^*_{1 m}(\Omega)M(\Omega),
\label{eq:flmmask}
\end{align}
\end{linenomath*}
where $M(\Omega)$ is the mask. Finally, we correct the direction for
the effects of inhomogeneous noise which is not accounted for in the
filtering process, by weighting the $\tilde{X}_{m}$ by the inverse of
the variance derived from filtered and mean-field corrected
simulations.

The physics is readily accessible in this estimator: the $\ell$-dependence
 in modulation determined by the parameter $X$ is expressed
in the $\delta C_{\ell\ell+1}$ factor, and the relevant scales appear
directly in the limits of the sum.  We consider the estimator over the
range $\ell_{\mathrm{\min}} = 2 \leq \ell \leq \ell_{\mathrm{\max}}$.
The modulation amplitude and direction are then given by
\begin{linenomath*}
\begin{align}
 \tilde{A} &= \sqrt{\tilde{X}^2_0 + 2| \tilde{X}_1|^2},
\label{eq:amplitude} \\
 \tilde{\theta} &= \cos^{-1}{\left(\frac{\tilde{X}_0}{\tilde{A}}\right)},
\label{eq:thetadir} \\
 \tilde{\phi} &= -\tan^{-1}{\left(\frac{\mathrm{Im}{[ \tilde{X}_1]}}{\mathrm{Re}{[
\tilde{X}_1]}}\right)}.
\label{eq:phidir}
\end{align}
\end{linenomath*}
It is worth re-emphasizing that the quantities $\tilde{A}$,
$\tilde{\theta}$, and $\tilde{\phi}$ are all dependent on the $\ell$
range considered.

As a consequence of the central limit theorem, for sufficiently large
$\ell_{\max}$ the $\tilde{X}$s are Gaussian-distributed with mean
zero, so that the amplitude parameter has a Maxwell-Boltzmann
distribution.  We fit to this distribution for $\ell_{\max} \ge 10$
when computing the \pval, so as not to be influenced by Poisson noise
in the tails of the empirical distribution (and we have determined
that this is a good fit to the simulations by applying a
KS test).  For the case of scalar amplitude
modulation (i.e., $X = A_\mathrm{s}$), and $\ell_{\min} = 2$, the
cosmic-variance-limited expectation for the modulation amplitude from
statistically isotropic skies is
\begin{linenomath*}
\begin{align}
\left\langle\frac{\Delta A_\mathrm{s}}{A_\mathrm{s}}\right\rangle
   &\approx \sqrt{\frac{48}{\pi(\ell_{\max} + 4)(\ell_{\max} - 1)}}.
\label{eq:amplcosvar}
\end{align}
\end{linenomath*}
This is the cosmic variance for a scale-invariant dipole modulation,
and gives a more explicit expression than the $\ell_{\max}^{-1}$
scaling discussed in \citet{Hanson2009b}.

\begin{figure}[h]
\begin{center}
\mbox{\includegraphics[width=\hsize]{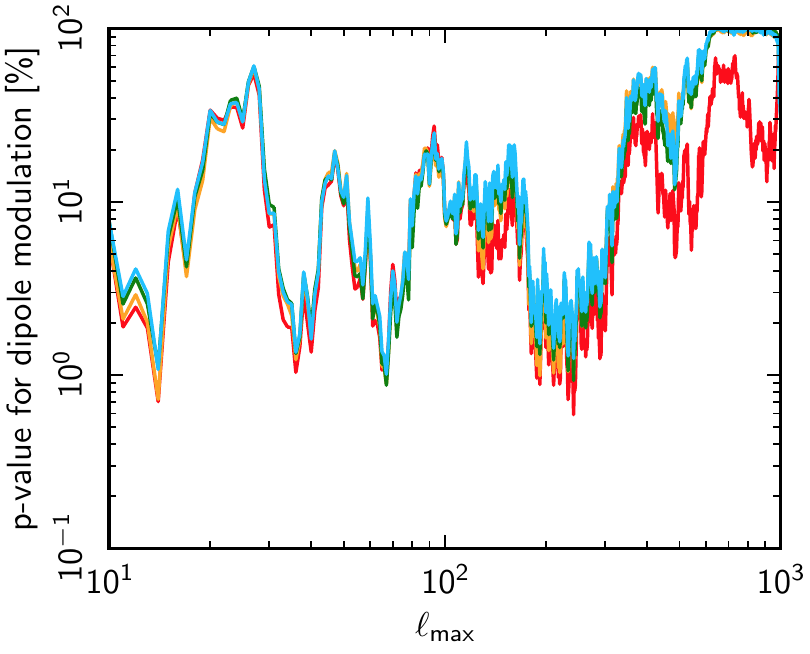}}
\mbox{\includegraphics[width=\hsize]{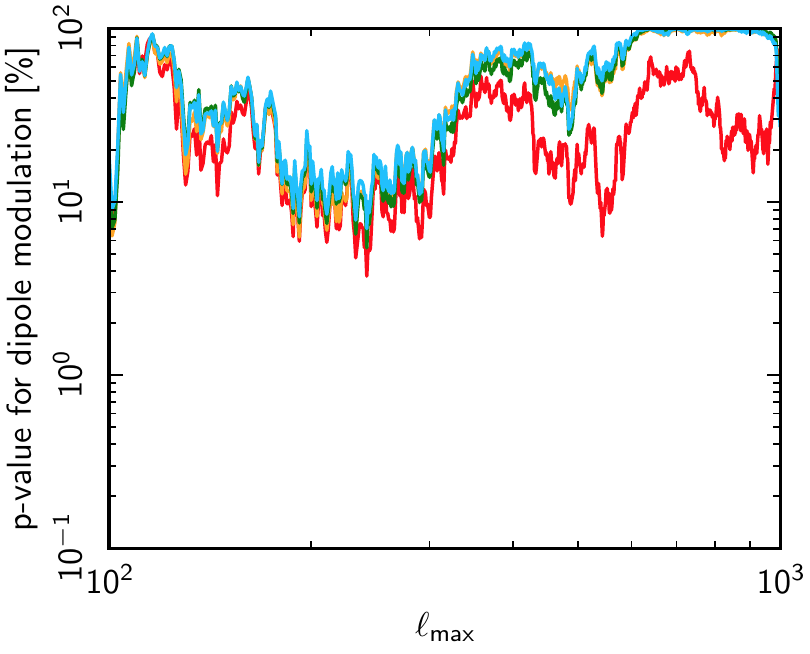}}
\end{center}
\caption{Probability determined from the QML analysis for a Monte
  Carlo simulation to have a larger dipole modulation amplitude than
  the \commander\ (red), \nilc\ (orange), \sevem\ (green), or \smica\
  (blue) data sets, with (top panel) $\ell_{\min} = 2$ or (bottom
  panel) $\ell_{\min} = 100$.  No significant modulation is found once
  the low-$\ell$ signal is removed. We emphasize that the statistic
  here is cumulative and apparent trends in the curves can be
  misleading.}
\label{fig:pvaluesdipmod}
\end{figure}

The top panel of Fig.~\ref{fig:pvaluesdipmod} presents results for the
\pval\ of the fitted modulation amplitude as a function of
$\ell_{\max}$.  Note that there are several peaks, at $\ell \approx 40$
and $\ell \approx 67$ (the focus of most attention in the literature),
and $\ell \approx 240$. The latter peak, while not previously
emphasized, is also present in the WMAP results \citep[see Fig.~15
in][]{bennett2010}.  It is also interesting to note that a modulation
amplitude is observed at $\ell_\mathrm{max} \approx 800$ that is
somewhat lower than what one would typically expect for a statistically isotropic
sky.  However, the significance is not at the level of the excess
dipole modulation at low $\ell$ and will not be discussed further.
The dip at $\ell_\mathrm{max} \approx 67$, with a \pval\ of
0.9--1.0\,\%, corresponds to the well-known low-$\ell$ dipole
modulation.\footnote{Actually only \sevem\ and \smica\ achieve their
  minimum at $\ell_{\max} = 67$, whereas \nilc\ and \commander\
  achieve theirs at $\ell_{\max} = 14$ and 240, respectively. Such
  scatter is expected when searching over a large number of possible
  $\ell$ ranges. The reconstructed amplitudes for each
  component-separation method are well within the error budgets of the
  estimator.}  Table~\ref{tab:lowellmod} presents the corresponding
dipole modulation parameters, which are seen to be consistent with
previous studies.  Note that the mean amplitude expected for a set of
statistically isotropic simulations at this $\ell_\mathrm{max}$ is 2.9\,\% (in close
agreement with the expected value due to cosmic variance,
Eq.~\ref{eq:amplcosvar}).

We have therefore determined a phenomenological signature of
modulation for $\ell = 2$--$67$ with a \pval\ of 0.9--1.0\,\%. If such a
signal had been predicted by a specific model, then we could claim a
significance of about $3\,\sigma$.  However, in the absence of such an a
priori model, we can assess how often we might find a $3\,\sigma$ effect
by chance, given that it could have occurred over any $\ell$
range. Since we are looking for a large-scale phenomenon, we assume
that the analysis should include the corresponding low-$\ell$ modes
and start at $\ell=2$.  In order to correct for a posteriori effects
we then adopt the following scheme.
\begin{enumerate}
\item We calculate the modulation of each simulation on the scales
  2--$\ell$, where $\ell \in [10, \ell_{\max}]$. For each simulation
  we find the modulation that gives the smallest probability, $\eta$
  (in the same way that was done for the data).
\item With the distribution of $\eta$s given by the simulations we
  then compare this to the data.  That is, we calculate the
  probability that one would find oneself in a Hubble patch with a
  modulation amplitude up to $\ell \in [10, \ell_\mathrm{max}]$ that
  is as significant as (or more significant than) the modulation in
  the real data.
\end{enumerate}
\begin{figure}[h]
\includegraphics[width=\hsize]{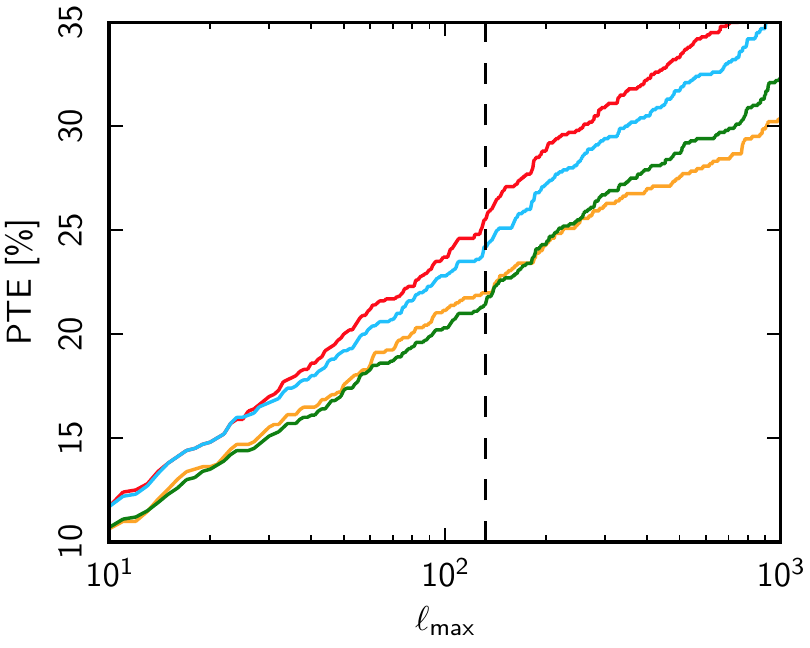}
\caption{Probability determined from the QML analysis for obtaining a dipole
  modulation amplitude at least as anomalous as the \commander\ (red), \nilc\
  (orange), \sevem\ (green), and \smica\ (blue) data sets, for the
  range $\ell \in [10, \ell_\mathrm{max}]$.  The vertical line
  corresponds to $\ell_\mathrm{max} = 132$ which was used as the
  search limit in \citet{bennett2010}. The probability grows
  approximately logarithmically with $\ell_\mathrm{max}$. This means
  that the adopted probability to exceed is fortunately not very
  sensitive to $\ell_\mathrm{max}$, and for any reasonable choice is
  above 10\,\%.}
\label{fig:lookelsewhere}
\end{figure}

If $\ell_\mathrm{max} = 132$ \citep[as chosen by][]{bennett2010}, the
probability of achieving a modulation as large as the \textit{Planck}
data in this range is higher than 10\,\% (see
Fig.~\ref{fig:lookelsewhere}). This is in agreement with the findings
of the WMAP team (which found 10\,\% and 13\,\% in the same $\ell$-range,
using two different masks).  Here, we do not quote a specific PTE for
the dipole modulation since it depends on the choice of both
$\ell_\mathrm{max}$ (albeit not so sensitively) and
$\ell_\mathrm{min}$ (which we have decided not to marginalize over).
However, it appears to be the case that the dipole modulation that we
observe is quite unremarkable.  That is, Gaussian fluctuations in a
statistically isotropic Universe will reasonably often result in a
dipole modulation with a comparable level of significance to that presented
here.

Beyond this, evidence for dipole modulation is found at $\ell \approx
200$--$300$, with a smaller dip at $\ell \approx 500$. Given that the
dipole modulation estimator is a cumulative quantity, it is possible
that these features are statistically enhanced by the usual low-$\ell$
signal. To test this we analyse the dipole modulation as a function of
$\ell_{\max}$ again, with the restriction $\ell_{\min} = 100$ applied
in order to completely remove any low-$\ell$ influence. The outcome is
presented in Fig.~\ref{fig:pvaluesdipmod} (bottom). It is clear that
even before introducing posterior corrections no significant
modulation is found, indicating that the $p$-values of the
features at $\ell > 100$ were indeed exaggerated by the low-$\ell$
modulation.

\subsection{Bipolar spherical harmonics}
\label{sec:biposh}

In the absence of the assumption of statistical isotropy, the CMB
two-point correlation function $C(\hat{n}_1,\hat{n}_2)\not\equiv
C(\hat{n}_1\cdot\hat{n}_2) $ can be most generally expanded in the
bipolar spherical harmonic (BipoSH) basis representation as follows:
\begin{linenomath*}
\begin{equation}
C(\hat{n}_1,\hat{n}_2) = \sum_{LM \ell_1 \ell_2} \tilde{A}^{LM}_{\ell_1 \ell_2}
\{Y_{\ell_1}(\hat{n}_1) \otimes Y_{\ell_2}(\hat{n}_2)\}_{LM} \,.
\end{equation}
\end{linenomath*}
The BipoSH basis functions, $\{Y_{\ell_1}(\hat{n}_1) \otimes
Y_{\ell_2}(\hat{n}_2)\}_{LM} $ are tensor products of ordinary
spherical harmonic functions, and the corresponding expansion
coefficients are termed BipoSH coefficients \citep{AH-TS,AH-TS1}. The
BipoSH basis provides a complete representation of any form of
statistical isotropy violation with the key advantage of
separating the angular scale-dependence of the signal in spherical
harmonic multipoles, $\ell$, from the nature of the violation indexed
in the bipolar multipole space by $L$.  Consequently, it is possible
to simultaneously determine that such a signal is dipolar ($L=1$),
quadrupolar ($L=2$), octopolar ($L=3$), and so on, in nature and that
the power is restricted to specific ranges of angular scales.

The estimation of BipoSH coefficients from CMB maps is a natural
generalization of the more routinely undertaken estimation of the angular
power spectrum $C_l$.  To allow a direct connection to the angular
power, we further introduce a set of BipoSH spectra at every bipolar
harmonic moment, $(L,M)$, labelled by a difference index $d$, defined
as follows:
\begin{linenomath*}
\begin{equation}\label{eqn:evenbiposh}
A^{LM}_{\ell  \ell+d} = \tilde{A}^{LM}_{\ell \ell+d}\frac{\Pi_{L}}{\Pi_{\ell (\ell+d)}\mathcal{C}^{L0}_{\ell 0 (\ell+d) 0}} \,,  \quad ( 0\leq d \leq L )\,,
\end{equation}
\end{linenomath*}
where $\mathcal{C}^{LM}_{\ell_1 m_1 \ell_2 m_2}$ are the
Clebsch-Gordon coefficients and for brevity the notation
$\Pi_{\ell_1\ell_2..\ell_n}= \prod_{i=1}^n \sqrt{(2\ell_i+1)}$.
BipoSH spectra, clearly, are then simply a
generalized set of CMB angular power spectra, with the standard CMB
angular power spectrum $C_\ell= A^{00}_{\ell\ell} $ being one of
them.\footnote{The BipoSH spectra, as defined in
  Eq.~\eqref{eqn:evenbiposh}, restrict us to working with only even-parity
  BipoSH coefficients ($L + d$ is even) due to the vanishing of
  $\mathcal{C}^{L0}_{\ell 0 \ell+d 0}$ otherwise. While most known
  isotropy-violating phenomena like weak lensing, Doppler boost,
  non-circular beams,
  etc., can only produce even-parity BipoSH spectra, measurement of odd-parity
 BipoSH spectra can be used to test for systematic effects, or to search for
  the signatures of exotic effects such as the lensing of CMB photons
  by tensor metric perturbations.}
While $A^{00}_{\ell\ell}$ quantifies the properties of the
statistically isotropic part of the CMB fluctuations, the additional
BipoSH coefficients quantify the statistically anisotropic part of the
CMB two-point correlation function.

Thus BipoSH provides a mathematically complete description of all
possible violations of statistical isotropy in a Gaussian CMB sky
map. It is then always possible to translate any specific model for
such a signal into the language of BipoSH and provide a common
approach for the multiple specialized tests that have been implemented
previously in this paper and elsewhere. However, improving on the
analysis of the 2013 \Planck\ data, a new formalism is developed in
order to reliably analyse a masked sky, 
as concisely described in Appendix~\ref{sec:wvmsf}. \citet{Aluri2015}
provides a more detailed description of the approach and includes an
explicit demonstration of its validity using simulations.

Initially, we revisit the simple phenomenological model of dipole
modulation of the CMB sky from Sect.~\ref{sec:dipmod},
\begin{linenomath*}
\begin{equation}
\label{eq:cmb_mod}
 T(\vec{\hat n})=T_{0}(\vec{\hat n})\left( 1+ \mathcal{M}(\vec{\hat n}) \right) \,,
\end{equation}
\end{linenomath*}
where $T(\vec{\hat n})$ represents the modulated CMB sky, $T_{0
}(\vec{\hat n})$ is the underlying (statistically isotropic) random
CMB sky, and $\mathcal{M}(\vec{\hat n})$ is a dipolar field. The BipoSH
coefficients resulting from such a modulation are given by
\begin{linenomath*}
\begin{eqnarray}
\label{biposh_modulation}
A^{1M}_{\ell \ell+1}&=&\bar{A}^{1M}_{\ell \ell+1}
 + m_{1M}G^{1}_{\ell \ell+1}\, ;
\\
G^{1}_{\ell \ell+1}&=&\frac{C_{\ell}+C_{\ell+1}}{\sqrt{4 \pi}}
\sqrt{\frac{(2\ell+1)(2\ell+3)}{3}} \mathcal{C}^{10}_{\ell 0 (\ell+1) 0}   \,.
\label{modest}
\end{eqnarray}
\end{linenomath*}
Here $\bar{A}^{1M}_{\ell \ell+1}$ corresponds to the BipoSH
coefficients of the unknown, but statistically isotropic, unmodulated
CMB field, $m_{1M}$ are the spherical harmonic coefficients of the
modulation field, and $C_\ell$ is the best-fit CMB angular power
spectrum.

The BipoSH representation further enables an estimate of the
modulation field to be made over specific angular scales by windowing
regions in multipole space in the sum over multipoles $\ell$ in
Eq.~\eqref{modest}.  This additional information is important for
identifying the origin of the isotropy-breaking signal, which could be
either cosmological or due to systematic artefacts.

We perform the analysis for the $N_{\rm side}=2048$ component
separated CMB maps with an apodized version of the common mask at that
resolution and reconstruct the modulation signal in independent bins
of width $\Delta \ell =64$ up to $\ell_\mathrm{max}=512$. The
application of the common mask introduces a mean field bias in the
BipoSH coefficients derived from the data. This bias is estimated from
the FFP8 simulations and subtracted from the derived coefficients.
The process of masking induces a coupling between the modulation field
and the mask that results in a modification of the spectral shape of
the modulation signal by the modified shape function (MSF) (see
Appendix~\ref{sec:wvmsf} for details). Further, the covariance of the
bias-subtracted BipoSH coefficients is not easy to derive analytically
in this case.  To overcome this problem, we consider the diagonal
approximation to the covariance matrix and estimate it from
simulations.

\begin{figure}
\begin{center}
\includegraphics[width=\hsize]{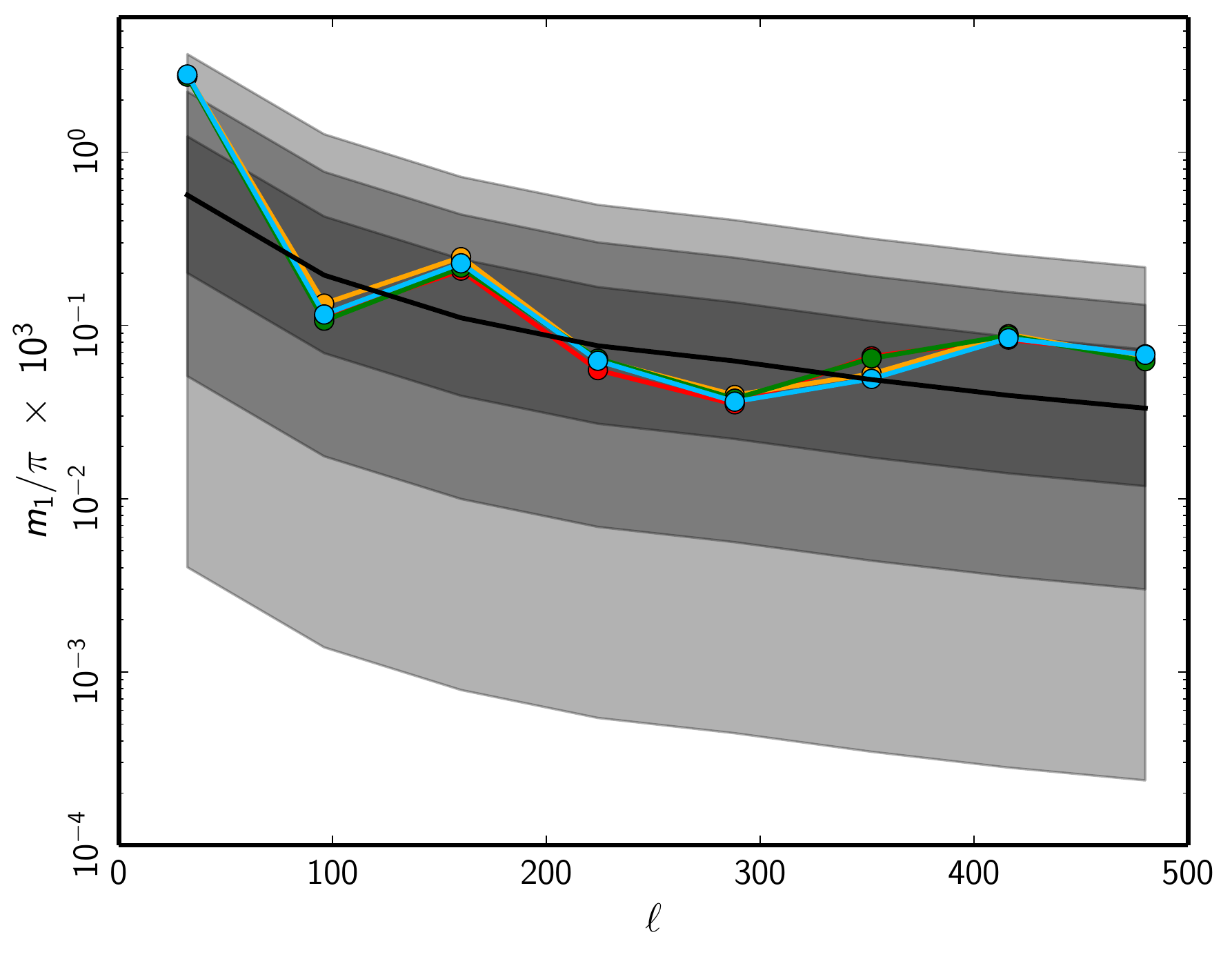}
\includegraphics[width=\hsize]{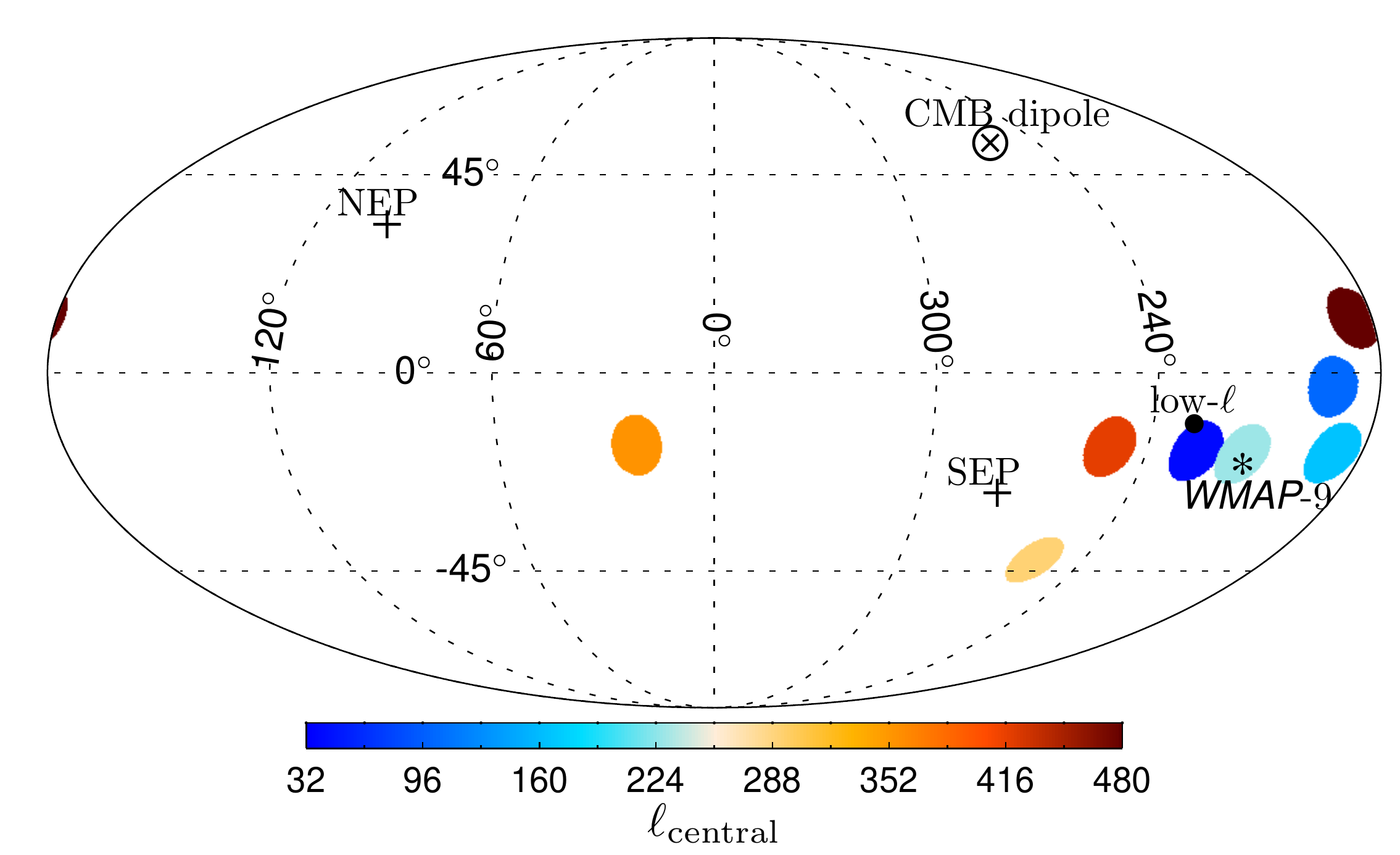}
\end{center}
\caption{{\it Top:} Measured dipole modulation ($L=1$) power in
  non-overlapping CMB multipole bins for \commander\ (red), \nilc\
  (orange), \sevem\ (green), and \smica\ (blue) as determined from a
  BipoSH analysis of the data. The power in the dipole of the
  modulation field is a ${\chi}^2$-distributed variable with 3 degrees
  of freedom. The shaded regions in the plot depict, in dark-grey,
  grey, and light-grey respectively, the 1, 2, and $3\,\sigma$
  equivalent intervals of the distribution function derived from
  simulations, while the solid black line denotes its
  median. Significant power in the dipole modulation is seen to be
  limited to $\ell=2$--$64$ and does not extend to higher multipoles.
  {\it Bottom:} Dipole modulation direction as determined from the
  \smica\ map. The directions found from the other component
  separation maps are consistent with this analysis. The coloured
  circles denote the central value of the multipole bin used in the
  analysis, as specified in the colour bar.
  The low-$\ell$ and WMAP-9 directions are identical to those in Fig.~\ref{fig:dipoles}.
}
\label{fig:dipole_384}
\end{figure}

The results presented in the top panel of Fig.~\ref{fig:dipole_384}
indicate that the dipole modulation signal is most significant in the
lowest multipole window $\ell\in[2,64]$.  Note that the power in the
dipole modulation field
$m_1 = (|m_{11}|^2 + |m_{10}|^2 +|m_{1-1}|^2)/3$ is related to the
dipole amplitude by $A=1.5\sqrt{m_1 / \pi}$.  The best-fit amplitude
($A$) and direction corresponding to the reconstructed dipole
modulation field from this lowest multipole bin is quoted in
Table~\ref{tab:prob_chisq} for each component-separation method.  Also
shown are the corresponding results for the cleaned frequency maps
{\tt SEVEM-100}, {\tt SEVEM-143}, and {\tt SEVEM-217}. As expected for
signals with a cosmological origin, no evidence for frequency
dependence is seen.

\begin{table}[tb]
\begingroup
\newdimen\tblskip \tblskip=5pt
\caption{ Amplitude ($A$) and
  direction of the dipole modulation in Galactic coordinates
  as estimated for the multipole range $\ell \in [2,64]$ using a
  BipoSH analysis.  The measured values of the dipole amplitude and direction are
  consistent for all maps.}
\label{tab:prob_chisq}
\nointerlineskip
\vskip -3mm
\footnotesize
\setbox\tablebox=\vbox{
\halign{\hbox to 0.8in{#\leaderfil}\tabskip 4pt&
\hfil#\hfil\tabskip 8pt&
\hfil#\hfil\/\tabskip=0pt\cr
\noalign{\doubleline}
\omit& \omit& Direction\cr
\omit\hfil Method \hfil& $A$&$(l, b)$ [\deg]\cr
\noalign{\vskip 3pt\hrule\vskip 3pt}
{\tt Commander}& $0.067 \pm 0.023$&  $(230,-18) \pm 31$\cr
\noalign{\vskip 4pt}
{\tt NILC}& $0.069 \pm 0.022$& $(228,-17) \pm 30$\cr
\noalign{\vskip 4pt}
{\tt SEVEM}& $0.067 \pm 0.023$& $(230,-17) \pm 31$\cr
\noalign{\vskip 4pt}
{\tt SMICA}& $0.069 \pm 0.022$& $(228,-18) \pm 30$\cr
\noalign{\vskip 4pt\hrule\vskip 6pt}
{\tt SEVEM-100}& $0.070 \pm 0.023$&  $(231,-19) \pm 30$\cr
\noalign{\vskip 4pt}
{\tt SEVEM-143}& $0.068 \pm 0.023$& $(230,-17) \pm 31$\cr
\noalign{\vskip 4pt}
{\tt SEVEM-217}& $0.069 \pm 0.023$& $(229,-20) \pm 31$\cr
\noalign{\vskip 3pt\hrule\vskip 3pt}}}
\endPlancktablewide                 
\endgroup
\end{table}

Since the amplitude of the dipole modulation field is consistent with
zero within $2\,\sigma$ for all of the higher $\ell$-bins
considered, it is plausible that the simple modulation model in
Eq.~\eqref{eq:cmb_mod} is inadequate to describe the features seen in
the BipoSH spectra and should minimally allow for the amplitude,
$A(\ell)$, of the dipole to depend on CMB multipole, $\ell$.  Although
this may appear to be a more complex model, it does not necessarily
lack motivation.  It is readily conceivable that physical mechanisms
that cause a dipolar modulation of the random CMB sky would be
scale-dependent and possibly significant only at low wavenumbers.  It
is also intriguing to note that, although in most cases the amplitude
of the modulation dipole is seen at low significance, the directions
in the first four bins, $\ell_{32} \in [2,64]$, $\ell_{96}
\in[65,128]$, $\ell_{160} \in [129,192]$, and $\ell_{224} \in
[193,256]$, are seen to be clustered together, as shown in the
bottom panel of Fig.~\ref{fig:dipole_384}. Note that the lower
significance of the modulation for the multipole bins at $\ell > 64$
results in larger errors for their respective directions than the
value quoted for the $\ell \in [2,64]$ bin recorded in
Table~\ref{tab:prob_chisq}.

\begin{figure}
\begin{center}
\includegraphics[width=\hsize]{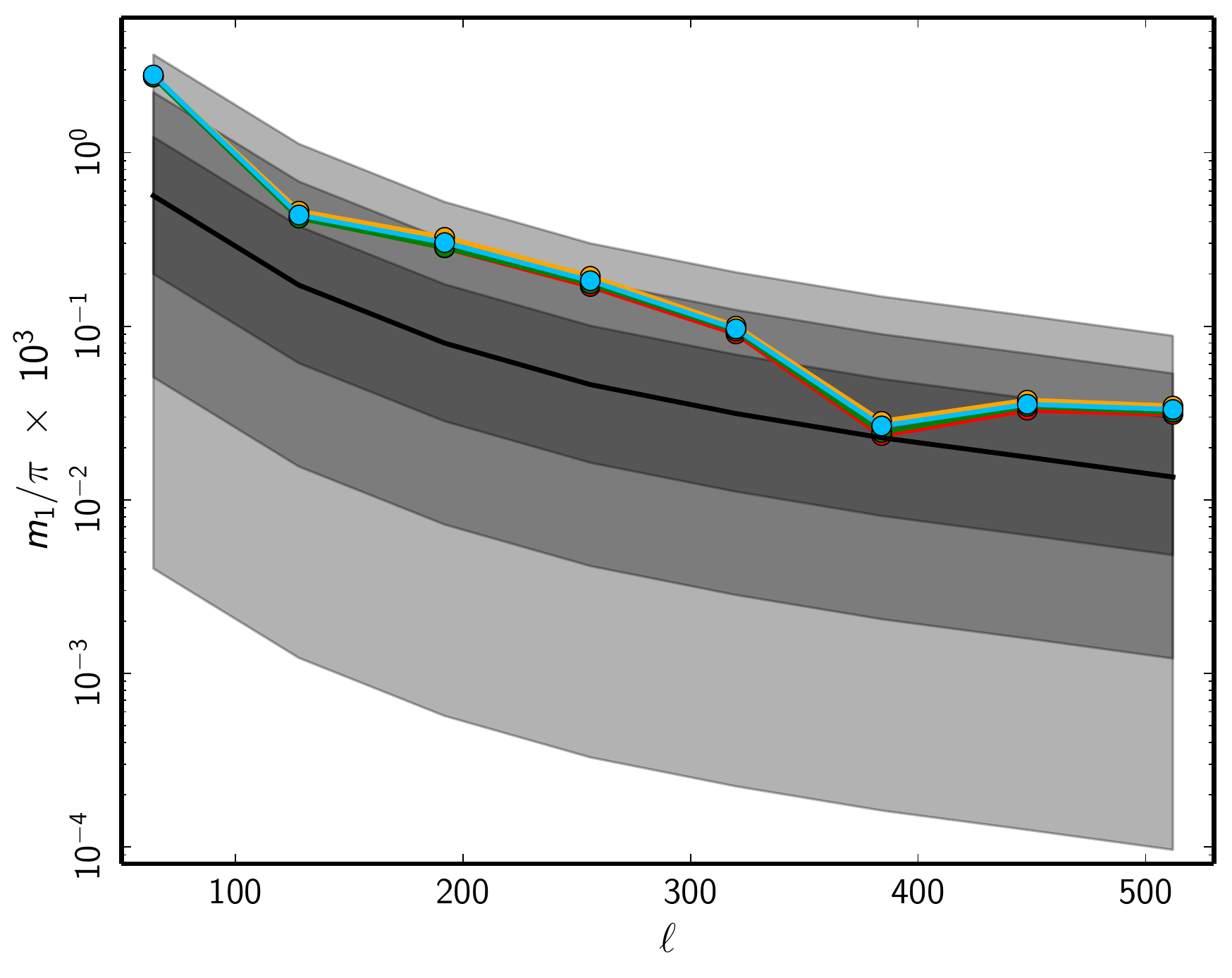}
\includegraphics[width=\hsize]{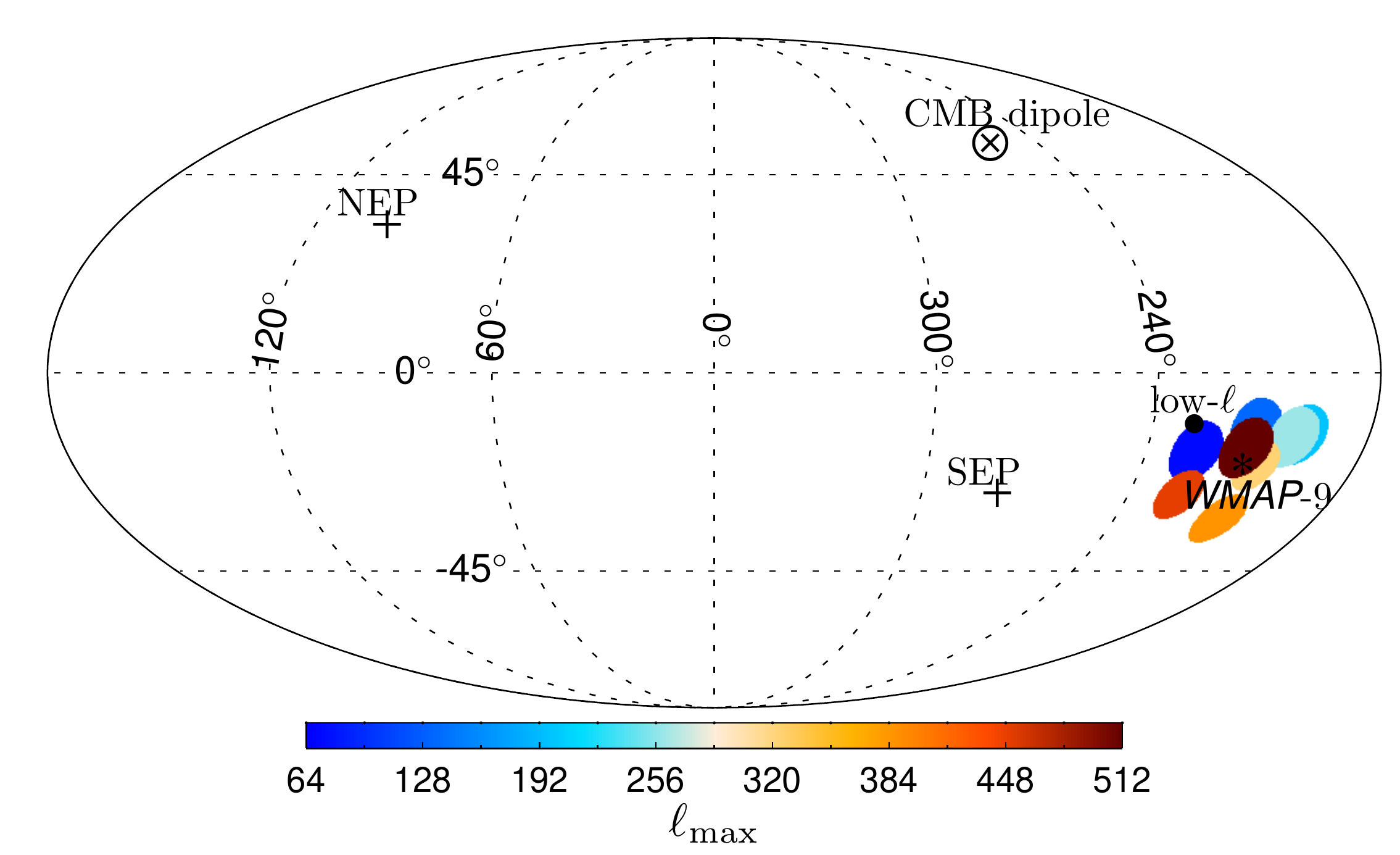}
\end{center}
\caption{{\it Top:} Measured dipole modulation power in cumulative CMB
  multipole bins for \commander\ (red), \nilc\ (orange), \sevem\
  (green), and \smica\ (blue) as determined from a
  BipoSH analysis of the data..  Colour coding as in
  Fig.~\ref{fig:dipole_384}.  Note that the measurements in cumulative
  bins indicate a power in excess of $2\,\sigma$ up to multipole
  $\ell_\mathrm{max} \sim 320$. The value on the horizontal axis denotes the maximum
  multipole used in the analysis, with $\ell_\mathrm{min} = 2$.
  {\it Bottom:} Modulation dipole direction as recovered from the \smica\
  map. The directions found from the other component-separation maps
  are consistent with these directions. The colour-coded points
  represent the directions recovered for the specific $\ell_\mathrm{max}$
  used in the analysis, with $\ell_\mathrm{min}=2$.
  The low-$\ell$ and WMAP-9 directions are identical to those in Fig.~\ref{fig:dipoles}.
}
\label{fig:dipole_cum_512}
\end{figure}

We extend our analysis to carry out the dipole modulation
reconstruction in cumulative bins up to $\ell_\mathrm{max} = 512$,
making cumulative increments in the multipole in steps of $\Delta \ell
= 64$. The results of this analysis are summarized in
Fig.~\ref{fig:dipole_cum_512}.

As noted previously, as a consequence of our motion with respect to
the CMB rest frame, the observed CMB map is expected to be
statistically anisotropic, as has been demonstrated in
\citet{planck2013-pipaberration} and Appendix~\ref{sec:dboost}.
Reassuringly, in \citetalias{planck2013-p09} it was established that
such a signal would not contaminate a dipole modulation signal up to
$\ell_{\rm max}\approx 700$. We now confirm the Doppler boost signal
using the BipoSH methodology.

An equivalent description of the Doppler boost in terms of BipoSH
coefficients is given by
\begin{eqnarray}
\label{biposh_boost}
A^{1M}_{\ell_1 \ell_2}~~~&=&\bar{A}^{1M}_{\ell_1 \ell_2} +
\beta_{1M}G^{1}_{\ell_1 \ell_2} \, ,\\
G^{1}_{\ell_1 \ell_2}~~~&=&\left\{ b_{\nu}[G^{1}_{\ell_1 \ell_2}]^{\mathcal{M}}
- [G^{1}_{\ell_1 \ell_2}]^{\phi} \right\} \times \nonumber \\
&& \sqrt{\frac{(2\ell_1+1)(2\ell_2+1)}{12\pi}} \mathcal{C}^{10}_{\ell_1 0 \ell_2
0} \, , \\
\left[G^{1}_{\ell_1 \ell_2}\right]^{\mathcal{M}}&=&\left[ C_{\ell_1}+C_{\ell_2}
\right] \, , \\
\left[G^{1}_{\ell_1 \ell_2}\right]^{\phi}~&=&\left[ C_{\ell_1}+C_{\ell_2}\right]
\\
& &  +\left[ C_{\ell_1}-C_{\ell_2}\right] \left[
\ell_1(\ell_1+1)-\ell_2(\ell_2+1)\right] /2 \, ,\nonumber
\end{eqnarray}
where $\beta_{1M} = \int \dd n Y_{1M}(\vec{\hat n}) \vec{\beta}\cdot
\vec{\hat n}$, $\vec{\beta}=\vec{v}/c$ denotes the peculiar velocity
of our local rest frame with respect to the CMB, and $b_{\nu}$ is the
frequency-dependent boost factor, as discussed in more detail in
\citet{planck2013-pipaberration}.

Since the Doppler boost signal has a frequency dependence, we perform
our analysis on the {\tt SEVEM-100}, {\tt SEVEM-143}, and {\tt
  SEVEM-217} maps at $\nside = 2048$, and adopt values of
$b_{\nu}=1.51, 1.96$, and $3.07$, respectively.  A minimum variance
estimator for $\beta_{1M}$, as discussed in Appendix~\ref{sec:wvmsf},
is adopted with the shape function $G^L_{\ell_1 \ell_2}$ replaced by
the corresponding Doppler boost term given in
Eq.~\eqref{biposh_boost}.  Corresponding unboosted CMB simulations
were also used, in particular to correct for the mean field bias.
However, we use a set of Doppler-boosted simulations in order to
estimate the error on the reconstructed Doppler boost vector.

\begin{figure}[!htbp]
\includegraphics[width=\hsize]{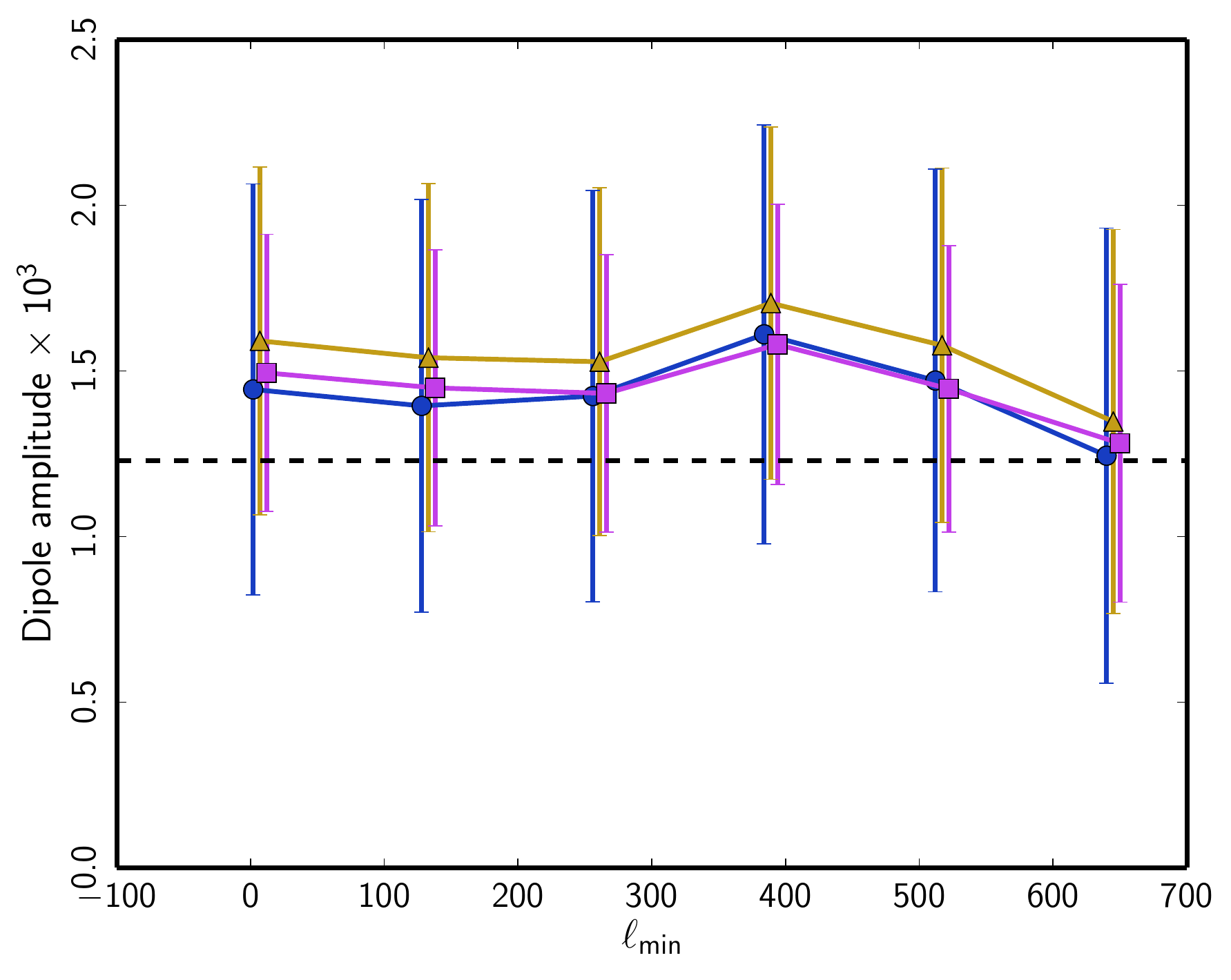}
\includegraphics[width=\hsize]{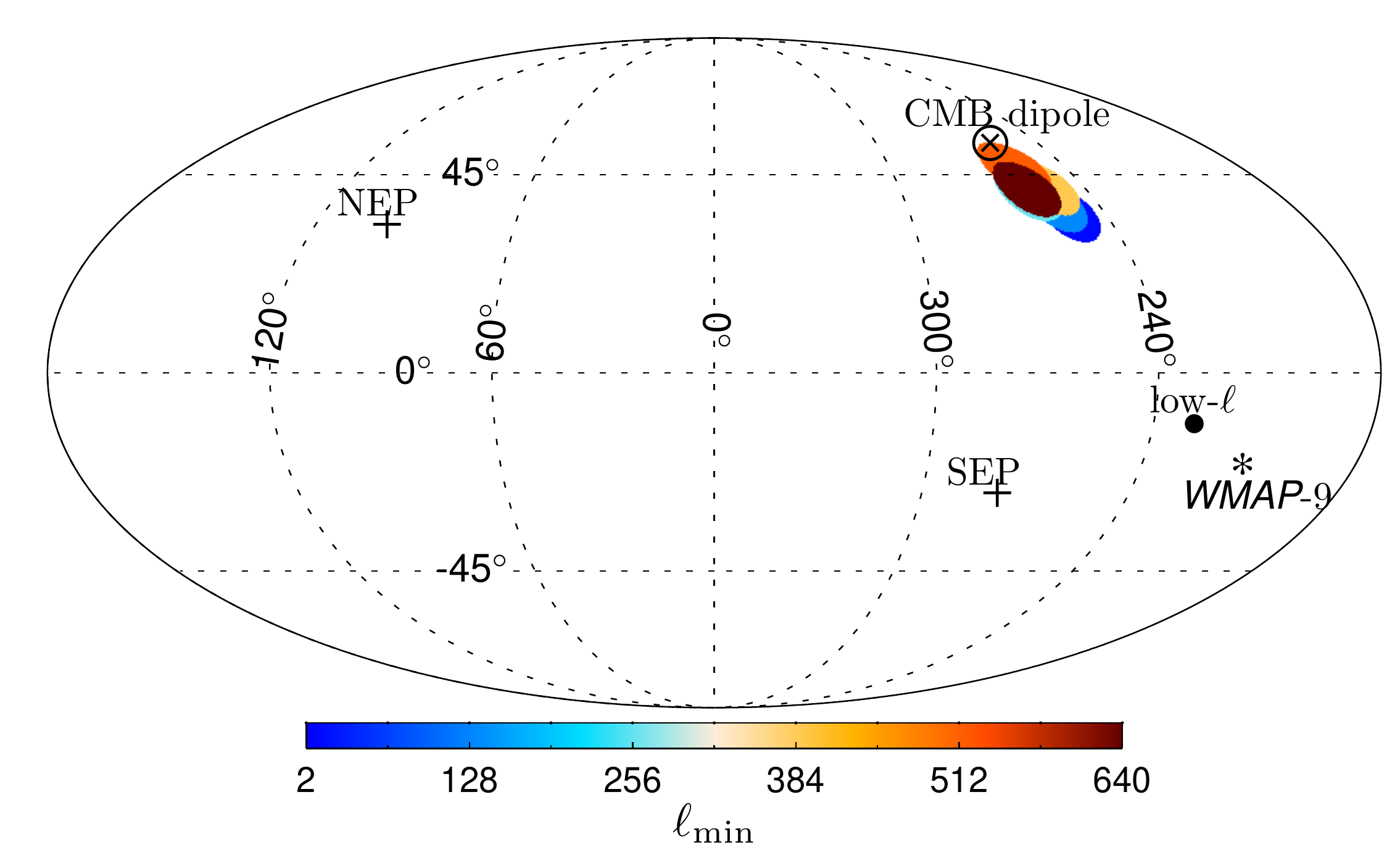}
\caption{{\it Top:} Amplitude $|\beta|$ of the Doppler boost from the
  {\tt SEVEM-100}, {\tt SEVEM-143}, and {\tt SEVEM-217} maps for
  different multipole bins determined using a BipoSH analysis. The
  maximum multipole of each bin is fixed at $\ell_\mathrm{max}=1024$,
  while $\ell_\mathrm{min}$ is incremented from $\ell = 2$ to $\ell =
  640$ in steps of $\Delta \ell = 128$.  The dashed line corresponds
  to the actual dipole boost amplitude, $|\vec{\beta}| = 1.23 \times
  10^{-3}$.  {\it Bottom:} Doppler boost direction $\hat{\beta}$
  measured in Galactic coordinates from {\tt SEVEM-217}. The coloured
  circles denote $\ell_\mathrm{min}$ used in the analysis, while
  $\ell_\mathrm{max} = 1024$ is held fixed.  The low-$\ell$ and WMAP-9
  directions are identical to those in Fig.~\ref{fig:dipoles}.  }
\label{fig:hist-beta-allfrq}
\end{figure}

Since it is expected that the low multipole modes of the
$A^{1M}_{l,l+1}$ spectrum are contaminated by the dipolar signal
reported previously, in order to monitor the impact of this anomalous
signal on the Doppler reconstruction we implement a cumulative
analysis using multipoles with a varying $\ell_\mathrm{min}$ from $2$
to $640$ in increments of $\Delta \ell_\mathrm{min} = 128$ and a fixed
$\ell_\mathrm{max}=1024$.\footnote{We fix $\ell_\mathrm{max}=1024$
  since at higher $\ell$ values the mismatch between the data and
  simulation power spectra becomes more important and is a concern for
  the bias subtraction applied when reconstructing the Doppler boost
  signal.}  The recovered Doppler amplitudes from the three \sevem\
frequency cleaned maps as a function of $\ell_\mathrm{min}$ are shown
in the top panel of Fig.~\ref{fig:hist-beta-allfrq}, while the lower
panel indicates the corresponding direction $\hat \beta$ in Galactic
coordinates determined from the {\tt SEVEM-217} data.
Table~\ref{tab:ampl-angl-pte-allfrq} records the best-fit amplitudes
and directions for $\ell \in [640,1024]$.

\begin{table}[tb]
\begingroup
\newdimen\tblskip \tblskip=5pt
\caption{ The Doppler boost
  amplitude ($|\vec{\beta}|$) and direction in Galactic
  coordinates derived over the multipole range
  $\ell \in [640,1024]$ as evaluated from a BipoSH analysis.
  The errors are estimated from an identical analysis of a set of 1000
  Doppler boosted simulations for each frequency.
}
\label{tab:ampl-angl-pte-allfrq}
\nointerlineskip
\vskip -3mm
\footnotesize
\setbox\tablebox=\vbox{
\halign{\hbox to 0.9in{#\leaderfil}\tabskip 4pt&
\hfil#\hfil\tabskip=5pt&
\hfil#\hfil\/\tabskip=0pt\cr
\noalign{\doubleline}
\omit\hfil& \omit\hfil& Direction\cr
\omit\hfil Method \hfil& $|\vec{\beta}| \times 10^{-3}$& $(l, b)$ [\deg]\cr
\noalign{\vskip 3pt\hrule\vskip 3pt}
{\tt SEVEM-100}& $1.24 \pm 0.66$&  $(277 ,40) \pm 50 $\cr
\noalign{\vskip 4pt}
{\tt SEVEM-143}& $1.35 \pm 0.56$& $(264 , 39) \pm 39 $\cr
\noalign{\vskip 4pt}
{\tt SEVEM-217}& $1.28 \pm 0.45$& $(257 ,42) \pm 32 $\cr
\noalign{\vskip 3pt\hrule\vskip 3pt}}}
\endPlancktablewide                 
\endgroup
\end{table}
%

\subsection{Angular clustering of the power distribution}
\label{sec:power_asymmetry}

In the \Planck\ 2013 data release we reported a possible deviation
from statistical isotropy in the multipole range $\ell= 2$--600, thus
confirming earlier findings based on the WMAP data
\citep{hansen2009,axelsson2013}.  This claim of asymmetry extending to
higher multipoles was made only on the basis of the alignment of
preferred directions as determined from maps of the power distribution
on the sky for specific multipole ranges. In particular, it was found
that the directions of the dipoles fitted to such maps in the
multipole range $\ell=2$--600 were significantly more aligned than in
simulations.  In addition, we showed that the ratio of the power
spectra in the two opposite hemispheres defined by the asymmetry axis
for $\ell=2$--600 was not statistically anomalous \citep[as later
confirmed over the extended multipole range
$\ell=2$--2000 by][]{quartin2014}.

Here, we test for the alignment in the \Planck\ 2015 data set.  We
adopt the  approach for the estimation of the dipole alignment that was
described in detail in \citetalias{planck2013-p09}, a brief summary of
which follows.
\begin{enumerate}
\item Local power spectra are estimated from the data at
  $N_\mathrm{side}=2048$ for 12 patches of the sky corresponding to
  the $N_\mathrm{side}=1$ {\tt HEALPix} base pixels. Only those
  high-resolution pixels surviving the application of the common
  mask are included in the analysis.\footnote{Departing from the analysis in
    \citetalias{planck2013-p09}, we do not use an apodized version of
    the common mask. Simulations indicate that the error on the power
    spectrum for those multipoles in the range 300 to 500 where the
    significance is highest is up to $20\,\%$ larger in this case, with
    the corresponding error on preferred direction being typically
    $8\,\%$ larger.}
 As a consequence of
  this masking, when patches based on {\tt HEALPix} pixels with
  $N_\mathrm{side}>1$ are used, the available sky fraction for those
  patches close to the Galactic plane is too small for power-spectrum
  estimation. For most of the analysis, we use the cross-spectra
  determined from half-mission data sets.\footnote{Note that simulated
    half-mission noise maps were generated by adjusting the properties
    of the existing 1000 (10\,000 in the case of \smica) noise
    simulations appropriately, thus explaining why only 500 (5000)
    simulations are used in this analysis.}  Due to a mismatch between
  the noise level in the data and the simulated maps, the results
  based on auto-spectra are less reliable and also more prone to other
  systematic effects than the cross-spectra.  We therefore do not
  consider such results here.  The spectra are binned over various bin
  sizes between $\Delta\ell=8$ and $\Delta\ell=32$.
\item For each power spectrum multipole bin, an $N_\mathrm{side}=1$
  {\tt HEALPix} map with the local power distribution is constructed.
\item The best-fit dipole amplitude and direction are estimated from
  this map using inverse-variance weighting, where the variance is
  determined from the local spectra computed from the simulations.  We
  do not compute error bars for the direction, but expect this to be
  accounted for in part by the use of equivalently treated simulations
  in the clustering analysis.
\item A measure of the alignment of the different multipole blocks is
  then constructed. In \citetalias{planck2013-p09}, we considered the
  mean angle between all possible pairs of dipole directions up to a
  given $\ell_\mathrm{max}$. Here, for greater consistency with
  Sect~\ref{sec:directionality}, we use the mean of the cosine of the
  angles, rather than of the angles themselves, between all pairs of
  dipoles. This effectively corresponds to the Rayleigh statistic (RS)
  introduced formally in Sect.~\ref{sec:directionality}, and we will refer to it as such,
  although it differs by ignoring all amplitude
  information. Clearly, smaller values of the RS correspond to less
  clustering.
\item The clustering as a function of $\ell_\mathrm{max}$ is then
  assessed using $p$-values determined as follows.  We first
  construct the RS using all multipoles up to $\ell_\mathrm{max}$. The
  $p$-value is then given by the fraction of simulations with a
  higher RS than for the data for this $\ell_\mathrm{max}$. A small
  $p$-value therefore means that there are few simulations that
  exhibit as strong clustering as the data. Note that the $p$-values
  are highly correlated as the RS is a cumulative function of
  $\ell_\mathrm{max}$.
\item We then define two measures of significance. To achieve this, it
  is necessary to reduce the 1499 different $p$-values determined
  for $\ell_\mathrm{max}\in [2,1500]$ to a single measure of
  clustering. We do this in two different ways, using the mean of
  these $p$-values, and by finding the minimum of the {\it
    p}-values, for both the data and for each available simulation.
  We then determine the percentage of simulations with (i) a lower
  mean $p$-value and (ii) a lower minimum $p$-value than the
  data.  Note that these two measures of significance take into
  account different aspects of the data. Note further that since the
  RS is cumulative and the $p$-values therefore correlated,
  different scales are weighted unequally and a detection in the mean
  and/or minimum $p$-value may be difficult to interpret and to
  correct for the multiplicity of tests effect (LEE).
\end{enumerate}

Note that the statistics defined in step 6 above correspond to two
choices of what were referred to as ``global statistics'' in
\citetalias{planck2013-p09} in order to assess the degree to which the
significance of the results depends on a specific choice for
$\ell_\mathrm{max}$.  The mean $p$-value over all available
$\ell_\mathrm{max}$ measures the degree to which clustering is present
over large multipole ranges independently of whether the clustering is
strongly focused in one given direction. Clearly the $p$-values
for different $\ell_\mathrm{max}$ are strongly correlated, but if the
clustering is present only over a small multipole range, the RS will
drop and the corresponding $p$-values will eventually rise. By
comparing this value to simulations, we test not only whether the dipole
alignment in the data is stronger than in statistically isotropic
random simulations, but also whether it is present over larger ranges of
multipoles than expected. The minimum $p$-value will give strong
detections if there is a strong asymmetry over a limited multipole
range or weaker clustering over larger multipole ranges when the
clustering is strongly focused in a given direction.

For \commander, \nilc, and \sevem, only 500 simulations are
available. However, 5000 simulations are available for \smica, which
allows a better estimate of significance to be determined when the
probabilities obtained are very low. In this case, we use half of the
5000 simulations to calibrate the statistic (obtain $p$-values
following step 5 above) and the remaining half to determine
significance levels (compute the mean and minimum over these {\it
  p}-values as a function of $\ell_\mathrm{max}$ following step 6).
When using 500 simulations, it is necessary to use the same set of
simulations to calibrate as well as to obtain probabilities. A related
issue with these results is that this set of simulations
(corresponding to the first 500 out of the 5000 available for
\smica) are observed to yield higher $p$-values for the clustering
angle due to a statistical fluctuation. Another 9 sets of 500
simulations that can be obtained from partitioning the 5000
available \smica\ simulations all result in lower $p$-values. As a
consequence, we observe that results based on the larger number of
simulations often give lower $p$-values than when only 500
simulations are used.

\begin{figure*}[htp!]
\begin{center}
\includegraphics[angle=0,width=1.9\columnwidth]{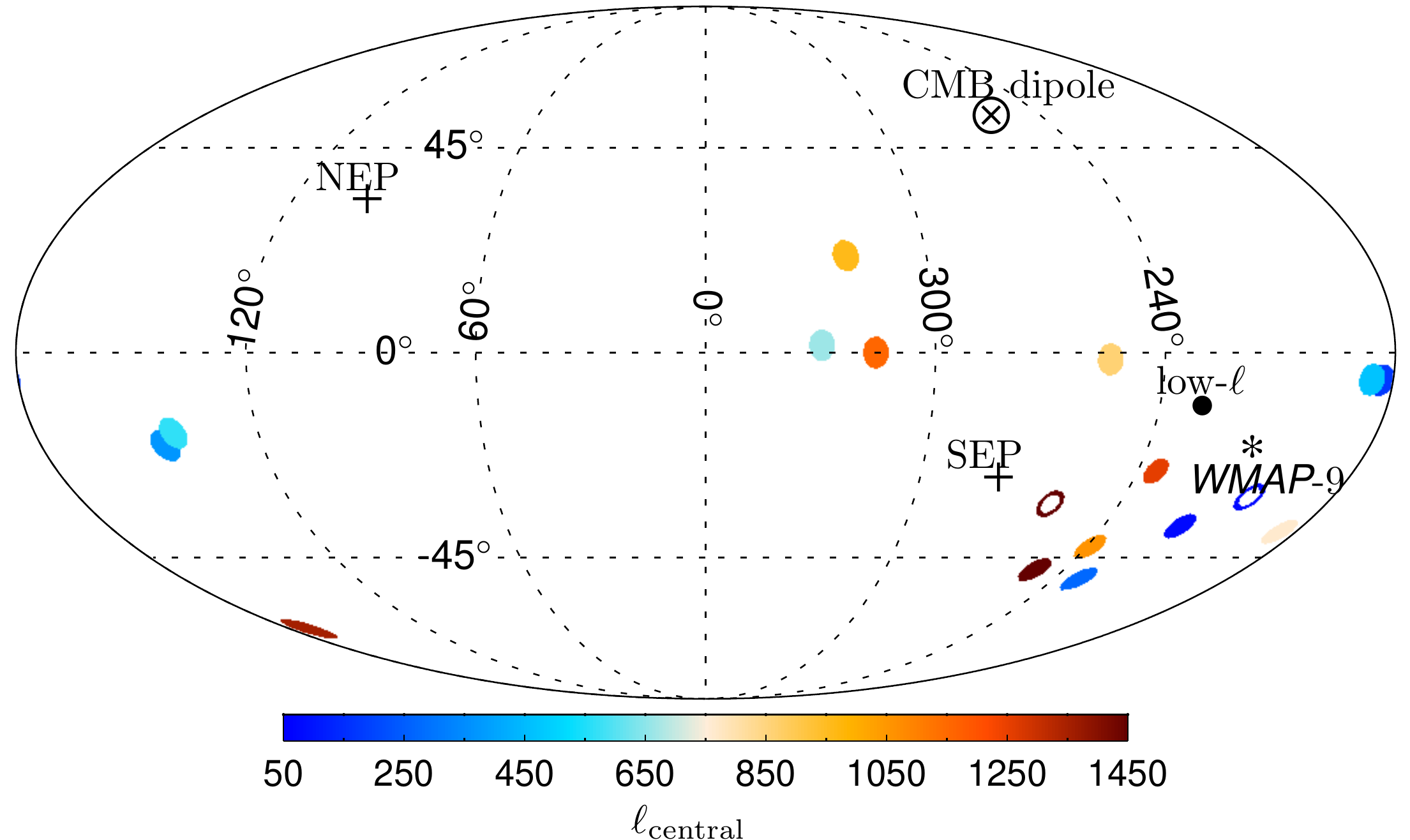}
\caption{Dipole directions for independent 100-multipole bins of the
  local power spectrum distribution from $\ell=2$ to $1500$ in the
  {\tt SMICA} map with the common mask applied.  We also show the
  preferred dipolar modulation axis (labelled as ``low-$\ell$'')
  derived in Sect.~\ref{sec:dipmod}, as well as the total direction
  for $\ell_{\rm max}=600$ determined from WMAP-9
 \citep{axelsson2013}. The average directions determined from the two
  multipole ranges $\ell \in [2,300]$ and $\ell \in [750,1500]$ are
  shown as blue and red rings, respectively. The error on the derived
  direction that results from masking the data is about $60\deg$, with
  only small variations related to bin size.  }
\label{fig:dipoles}
\end{center}
\end{figure*}

In Fig.~\ref{fig:dipoles} we show the dipole directions of the 15
lowest 100-multipole bins for the {\tt SMICA} map.  Here, the binning
has been chosen for visualization purposes; in further analysis of the
\Planck\ data we use finer $\ell$-intervals.  The preferred low-$\ell$
modulation direction determined in Sect.~\ref{sec:dipmod} is also
indicated, along with the WMAP-9 result determined over the range
$\ell=2$ to $600$ \citep{axelsson2013}. The observed clustering of the
dipole directions is similar to that shown in figure~27 of
\citetalias{planck2013-p09}. Note that differences in masking,
foreground subtraction, and residual systematic effects will displace
the direction of a given dipole with respect to the previous
analysis. Similar behaviour is seen for all of the \Planck\
component-separated maps.

\begin{figure}[htbp!]
\begin{center}
\includegraphics[width=\hsize]{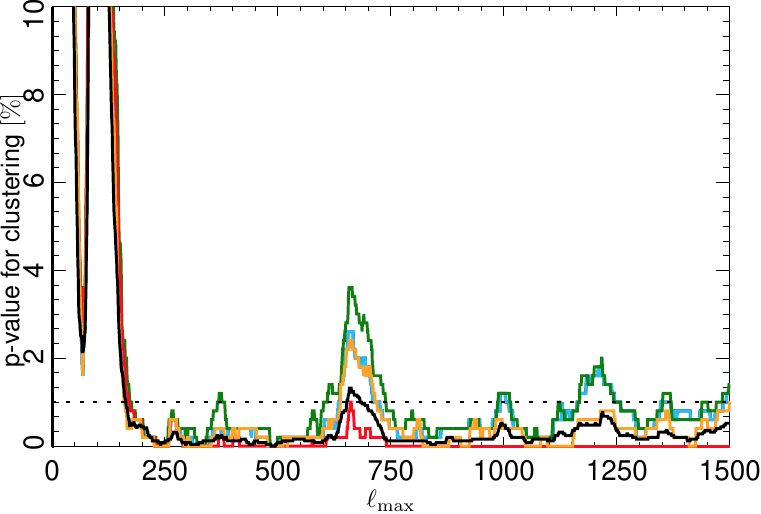}
\caption{Derived $p$-values for the angular clustering of the
  power distribution as a function of $\ell_\mathrm{max}$, determined
  for \commander\ (red), \nilc\ (orange), \sevem\ (green), and \smica\
  (blue), based on 500 simulations. For SMICA, the $p$-values
  based on 2500 simulations are also shown (black). The $p$-values
  are based on the fraction of simulations with a higher RS,
  determined over the $\ell$-range up to the given
  $\ell_\mathrm{max}$, compared to the data. The results shown here
  have been marginalized over bin sizes in the range $\Delta\ell=8$ to
  $\Delta\ell=32$. }
\label{fig:mangle_pvalues}
\end{center}
\end{figure}

In \citetalias{planck2013-p09}, we calculated the mean angle between
all possible pairs of dipole directions determined from maps of the
local power in multipole bins of size $\Delta\ell=16$.  Here we test
the possible bias arising from such a choice by considering bin sizes
between $\Delta\ell=8$ and $\Delta\ell=32$ in steps of 2. The lower
limit avoids significant bin-to-bin coupling in the power spectra for
smaller binnings, whilst the upper limit excludes cases where there
are an insufficient number of derived dipoles from which the mean
angle can be calculated, this leading to poor statistics.  In addition
to showing results for each bin size, we also calculate the variance-weighted
mean of the power spectra over all bin sizes (the $C_\ell$
for a given bin size is weighted by $1/\sqrt{N_{\rm b}}$ where $N_{\rm b}$ is the
bin size). In this way, we marginalize over bin sizes to obtain local
power spectra and thereby the RS for each single multipole.

Figure~\ref{fig:mangle_pvalues} shows the $p$-values for the
different component-separated maps, derived as described in step 5
above. We see that the results based on 500 simulations for \nilc,
\sevem, and \smica\ are in good agreement. The \commander\ results
are less consistent, but this may be related to the fact that
component separation was performed independently for the half-mission
solutions, in contrast to the other methods, where component-separation
solutions were obtained from the full mission data
only. For \smica, we also show $p$-values based on 2500
simulations.  These more accurate results show lower $p$-values,
and may indicate that those determined from only 500 simulations are
not sufficiently stable. Note also that for $\ell<100$ the {\it
  p}-values are not consistent with the detection of a low-$\ell$
asymmetry/modulation, as seen by other methods in this paper. However,
for $\ell<100$, there are very few bins and the variance of the RS
might therefore be too high for this effect to be visible.

In agreement with the conclusions in \citetalias{planck2013-p09}, a
large degree of alignment is seen at least to
$\ell_\mathrm{max}\approx 600$. However, in contrast to the earlier
results where the $p$-values started increasing systematically for
$\ell_\mathrm{max}>1000$, the current $p$-values remain low for
$\ell_\mathrm{max}>750$. The full component-separated maps which have
higher resolution and sensitivity are used for the current analysis,
instead of the single-frequency foreground-cleaned map ({\tt
  SEVEM-143}) used in \citetalias{planck2013-p09}. We note that the
results for the updated {\tt SEVEM-143} map are consistent with the
earlier analysis, both with and without correction for the Doppler
modulation. Note also that the \smica\ results with improved
statistics (based on 2500 simulations) generally show lower {\it
  p}-values than the corresponding results based on 500 simulations.

Table~\ref{tab:alignmentsignificances} presents the fraction of
simulations with a lower mean/minimum $p$-value than in the data
for a number of different cases. The table shows probabilities for
\smica\ with different bin sizes (showing only every second bin size
since these are correlated), as well as for the results marginalized
over bin sizes. We also show results for the different
component-separated maps, results based on half-ring cross-spectra
instead of half-mission cross-spectra, and results using a different
$\ell$-weighting scheme, specifically $(2\ell+1)C_\ell$ instead of
$\ell(\ell+1)C_\ell$, the former being a measure of the variance of
the temperature fluctuations.  The table indicates probabilities of
approximately 0--2\,\% for most of these cases, although results for
the smallest bin size show much less significant results.  This could
be due to the strong anticorrelations between adjacent bins found for
this bin size in those Galactic $N_\mathrm{side}=1$ patches with very
small available sky fraction.  For the other bin sizes, these
correlations are much weaker.  Note that many of the significances
based on minimum $p$-value are only upper limits. This is due to
the fact that the limited number of simulations in some cases results
in the lowest minimum $p$-value being zero. When the minimum {\it
  p}-value in the data is zero, we show the percentage of simulations
which also have zero as the minimum $p$-value. Clearly this
fraction is only an upper limit on the real significance.

\begin{figure}[htbp!]
\begin{center}
\includegraphics[width=\hsize]{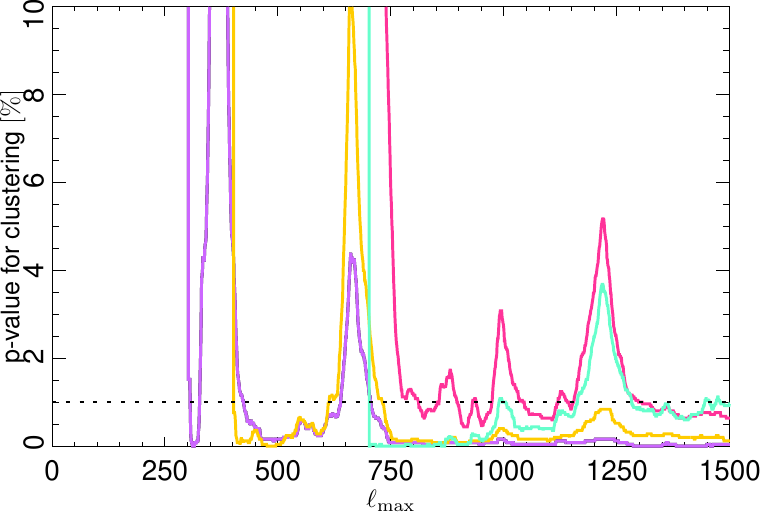}
\caption{Derived $p$-values for the angular clustering analysis as a
  function of $\ell_\mathrm{max}$, determined from \smica\, based on
  2500 simulations. The $p$-values are based on the fraction of
  simulations with a higher Rayleigh statistic up to the given
  $\ell_\mathrm{max}$ than in the data. The RS here is calculated over
  all pairs of dipole directions where one dipole in each pair is
  computed in the range $[\ell_\mathrm{lim},\ell_\mathrm{max}]$, and
  the other is determined in the range $[2,\ell_\mathrm{lim}]$. The
  plot shows $p$-values for $\ell_\mathrm{lim}=300$ (purple),
  $\ell_\mathrm{lim}=400$ (yellow), $\ell_\mathrm{lim}=500$ (pink), and
  $\ell_\mathrm{lim}=700$ (cyan). The results have been
  marginalized over bin sizes in the range $\Delta\ell=8$ to
  $\Delta\ell=32$. }
\label{fig:llim}
\end{center}
\end{figure}

In order to further investigate the $\ell$-dependence of the
asymmetry, we follow two approaches from \citetalias{planck2013-p09}.
Firstly, we restrict the analysis to multipoles above a minimum
multipole $\ell_\mathrm{min}$. Table~\ref{tab:alignmentsignificances}
indicates that clustering at the $<1\,\%$ significance level is still
found when considering only those multipoles with $\ell_\mathrm{min}$
greater than 100. However, when this limit is increased to 200, no
significant clustering is found.  We then calculate the RS between
pairs of dipoles where one dipole is determined from an $\ell$-range
above a certain limiting multipole $\ell_\mathrm{lim}$, and the other
dipole below this limit. Figure~\ref{fig:llim} shows the RS as a
function of $\ell_\mathrm{max}$ for some selected values of
$\ell_\mathrm{lim}$. The $\ell_\mathrm{lim}=300$ curve (purple)
indicates that dipole directions for $\ell>1000$ are significantly
aligned with dipoles for $\ell<300$.  Similarly, the
$\ell_\mathrm{lim}=700$ curve (cyan) indicates that the dipole
directions for $\ell=700$--$1000$ are strongly correlated with the
dipole directions for $\ell<700$.

Combining these results, we note that when using only multipoles with
(i) $\ell>200$, or (ii) $\ell<200$, no significant clustering is
found. The strong clustering significance shown to persist to high
multipoles in Fig.~\ref{fig:mangle_pvalues} must therefore be the
result of clustering of the dipole directions between low and high
multipoles as supported by Fig.~\ref{fig:llim}.  The low {\it
  p}-values can be explained by the alignment of dipole directions for
multipoles extending all the way to $\ell=1500$ correlated with
directions for $\ell<200$. The observed asymmetry is therefore not
consistent with a model based on dipole modulation or power asymmetry
located in one specific multipole range or for one given direction,
but rather as a correlation of the dipole directions between
$\ell<200$ and $\ell>200$. This correlation with lower multipoles is
found to persist all the way to $\ell_\mathrm{max}=1500$.

An advantage of the directional analysis performed here is that it
focuses on a central issue for tests of deviation from isotropy ---
whether there is a preferred direction. Indeed, \citet{Bunn2000} noted
that the CMB may exhibit a pattern that cannot be identified from the
power spectrum, but which would indicate some non-trivial large-scale
structure.  Evidence for the close correlation and alignment of
directions on different angular scales may present a signature of
broken statistical isotropy, since in the standard model, these
directions should all be independent random variables.  In this
context, we do not quote a specific direction for such asymmetry here
since our results indicate a clustering of angles between different
multipoles, but not necessarily that all multipoles are clustered
about one specific direction. However, crucially we have shown that
the measured clustering is driven by the correlations of directions
between higher and lower multipoles.

Some of the analyses in other sections of the paper focus on
dipolar modulation, a specific model for a dipolar power enhancement
of the statistically isotropic CMB field towards a preferred direction of the sky,
and use methods optimized for the detection of such a signal.
While the results of Sect.~\ref{sec:directionality} show no detection
of the clustering of directions, there is no clear contradiction with
the results presented here, since they are based on tests for $a_{\ell
  m}$ correlations between different multipoles as expected in the
dipolar modulation model. The clustering analysis presented here is a
model-independent test for deviations from statistical isotropy which
could induce very different correlation structure. It is therefore
sensitive to other forms of asymmetry, such as the addition of power
in one part of the sky or more general phase correlations.

\begin{table}[tb]
\begingroup
\newdimen\tblskip \tblskip=5pt
  \caption{Significance of the angular clustering of the power distribution.
  We indicate the actual mean/min $p$-value of the data, determined from
  Fig.~\ref{fig:mangle_pvalues} and written as a fraction of the
  number of simulations used to assess the values, together with
  the percentage of simulations with a lower mean/minimum {\it
    p}-value than the data.
  Unless otherwise specified, the numbers are determined from
  half-mission cross spectra $C_\ell\ell(\ell+1)$, for all multipoles
  in the range $\ell=2$--$1500$, and for the common mask.
\label{tab:alignmentsignificances}}
\vskip -3mm
\footnotesize
\setbox\tablebox=\vbox{
   \newdimen\digitwidth
   \setbox0=\hbox{\rm 0}
   \digitwidth=\wd0
   \catcode`*=\active
   \def*{\kern\digitwidth}
   \newdimen\dwidth
   \setbox0=\hbox{<}
   \dwidth=\wd0
   \catcode`!=\active
   \def!{\kern\dwidth}
\halign{\hbox to 0.8in{#\leaderfil}\tabskip 4pt&
\hfil#\hfil&
\hfil#\hfil\tabskip 12pt&
\hfil#\hfil&
\hfil#\hfil&
\hfil#\hfil&
\hfil#\hfil\/\tabskip 0pt\cr
\noalign{\doubleline}
\omit& Bin& Mean& $\%$& Min.& $\%$ \cr
\omit\hfil Method\hfil& size& $p$-value& (mean)& $p$-value& (min)\cr
\noalign{\vskip 3pt\hrule\vskip 3pt}
{\tt SMICA}& *8& 261/2500& !1.60& 35/2500& !16.2\cr
{\tt SMICA}& 10& 51/2500& !0.08& 3/2500& !2.36\cr
{\tt SMICA}& 12& 75/2500& !0.20& 1/2500& !0.96\cr
{\tt SMICA}& 14& 83/2500& !0.16& 2/2500& !1.52\cr
{\tt SMICA}& 16& 78/2500& !0.24& 4/2500& !2.00\cr
{\tt SMICA}& 18& 51/2500& !0.04& 1/2500& !0.68\cr
{\tt SMICA}& 20& 21/2500& <0.04& 1/2500& !0.76\cr
{\tt SMICA}& 22& 60/2500& !0.08& 2/2500& !1.24\cr
{\tt SMICA}& 24& 34/2500& !0.08& 2/2500& !1.00\cr
{\tt SMICA}& 26& 38/2500& !0.08& 1/2500& !0.96\cr
{\tt SMICA}& 28& 42/2500& !0.20& 0/2500& <0.52\cr
{\tt SMICA}& 30& 27/2500& !0.20& 0/2500& <0.60\cr
{\tt SMICA}& 32& 21/2500& !0.04& 0/2500& <0.52\cr
{\tt SMICA}& marg.& 43/2500& <0.04& 0/2500& <1.00\cr
{\tt SMICA$^\mathrm{a}$}& marg.& 48/2500& <0.04& 1/2500& !1.70\cr
{\tt SMICA$^\mathrm{b}$}& marg.& 47/2500& <0.04& 0/2500& <1.16\cr
{\tt SMICA$^\mathrm{c}$}& marg.& 50/2500& <0.04& 0/2500& <0.76\cr
{\tt SMICA$^\mathrm{d}$}& marg.& 254/2500& !1.52& 34/2500& !20.1\cr
\noalign{\vskip 3pt\hrule\vskip 3pt}
{\tt Comm.}& marg.& 9/500& <0.20& 0/500& <2.60\cr
{\tt NILC}& marg.& 10/500& <0.20& 0/500& <3.60\cr
{\tt SEVEM}& marg.& 13/500& <0.20& 0/500& <4.00\cr
{\tt SMICA}& marg.& 11/500& <0.20& 0/500& <3.60\cr
{\tt Comm.$^\mathrm{b}$}& marg.& 11/500& <0.20& 0/500& <3.00\cr
{\tt NILC$^\mathrm{b}$}& marg.& 10/500& <0.20& 0/500& <3.80\cr
{\tt SEVEM$^\mathrm{b}$}& marg.& 12/500& <0.20& 0/500& <3.40\cr
{\tt SMICA$^\mathrm{b}$}& marg.& 11/500& <0.20& 0/500& <3.80\cr
{\tt Comm.$^\mathrm{c}$}& marg.& 8/500& !0.20& 0/500& <4.00\cr
{\tt NILC$^\mathrm{c}$}& marg.& 14/500& !0.20& 1/500& !7.20\cr
{\tt SEVEM$^\mathrm{c}$}& marg.& 17/500& !0.20& 1/500& !8.40\cr
{\tt SMICA$^\mathrm{c}$}& marg.& 15/500& !0.20& 1/500& !7.60\cr
\noalign{\vskip 3pt\hrule\vskip 3pt}}}
\endPlancktablewide                 
\endgroup
 {\footnotesize
 $^\mathrm{a}$ Half-ring maps instead of half-mission maps.\\
 $^\mathrm{b}$ $C_\ell(2\ell+1)$ instead of $C_\ell\ell(\ell+1)$.\\
 $^\mathrm{c}$  Restricted to multipoles $\ell>100$.\\
 $^\mathrm{d}$  Restricted to  multipoles $\ell>200$.}
\end{table}

\subsection{Rayleigh statistic: QML analysis}
\label{sec:directionality}

Results from Sect.~\ref{sec:power_asymmetry} and in
\citetalias{planck2013-p09} suggest that, beyond a dipole modulation
of power on large angular scales, some form of directional asymmetry
continues to small scales.  There are also indications from
Sect.~\ref{sec:power_asymmetry} that the directions of dipolar
asymmetry are correlated between large and small angular scales.
Since the nature of the asymmetry is unknown we use the
RS, a generic test for directionality that makes minimal
assumptions about the nature of the asymmetry. This statistic has been
used both in previous CMB studies \citep{Stannard2005} and other
areas of cosmology \citep{Scott1991}. In our context, for a
statistically isotropic sky this statistic is identical to a three-dimensional
random walk.  The implementation here incorporates all
information pertaining to modulation, not just the direction.  The
approach in this section differs from that of
Sect.~\ref{sec:power_asymmetry} in the method of reconstructing power,
the choice of binning, and the choice of how to weight directions in
each bin.  Another important difference is that
Sect.~\ref{sec:power_asymmetry} only considers the direction of
dipolar asymmetry and does not take into account its amplitude.

The statistic is cumulative and thus narrowing down the specific
scales from which a signal may be originating is a non-trivial task.
However, it is the case that all statistics that measure this form of
asymmetry (dipole modulation or large-scale clustering of power) are
in some way cumulative and so we will not worry about this issue any
further.  Another disadvantage of this approach is that it will
generally be \emph{less} powerful than a test that uses a specific
model for the directionality. Again, this is a distinction shared when
one compares any non-parametric versus parametric statistic.

The construction of the statistic is as follows.
\begin{enumerate}
\item Beginning with the estimator from
  Eqs.~(\ref{eq:mossest0}) and~(\ref{eq:mossest1}), compute the
  following binned quantities for the data and simulation:
\begin{linenomath*}
\begin{align}
 \tilde{X}_{0,~\ell} &= \frac{6}{f_{10}}\frac{\sum_{m} A_{\ell m} (T^*_{\ell
m}T_{\ell + 1\,m} - \left< T^*_{\ell m} T_{\ell + 1\,m} \right>)}{\delta C_{\ell
\ell +1}F_{\ell}F_{\ell+1}(\ell + 1)} , \\
 \tilde{X}_{1,~\ell} &= \frac{6}{f_{11}}\frac{\sum_{m} B_{\ell m} (T^*_{\ell
m}T_{\ell + 1\,m+1} - \left<T^*_{\ell m}T_{\ell + 1\,m+1}\right>)}{\delta
C_{\ell \ell +1}F_{\ell}F_{\ell+1}(\ell + 1)} .
\end{align}
\end{linenomath*}
For each $\ell$ this computes the coupling of $\ell$ to $\ell + 1$.
We emphasize that this is a very natural choice of binning the
estimator, since any parameter that is dipole modulated will lead to
coupling of $\ell$ to $\ell \pm 1$ modes, albeit with different
$\ell$-weightings (below we describe why this is not an important
issue).
\item Construct a three-dimensional vector out of the
  three estimators for both the data and the simulations,%
\footnote{Note that here we have not specified what $\delta C_{\ell \ell +
    1}$ is (it is fully specified by choosing a parameter $X$ to modulate).
    This is because we have decided to weight each $\ell$ equally and thus any
    strictly positive choice for $\delta C_{\ell \ell + 1}$ will be equivalent, since in
    step 3 we force the mean length of the vectors at each $\ell$ to be equal.}
    as defined by Eqs.~(\ref{eq:amplitude}--\ref{eq:phidir}).
  \item Compute the mean amplitude from simulations and divide all
    vectors (data and simulations) by this amplitude. This choice
    ensures that each vector is treated equally, since we have no a
    priori reason to weight some scales more than others.
  \item Add this new vector to the previous vector. If this is the
    first time going through this process the previous vector is the
    zero vector.
  \item Repeat with $\ell \rightarrow \ell + 1$. Note that the
    statistics of this process are identical to a three dimensional
    random walk.
\end{enumerate}

\begin{figure}[h]
\begin{center}
\mbox{\includegraphics[width=\hsize]{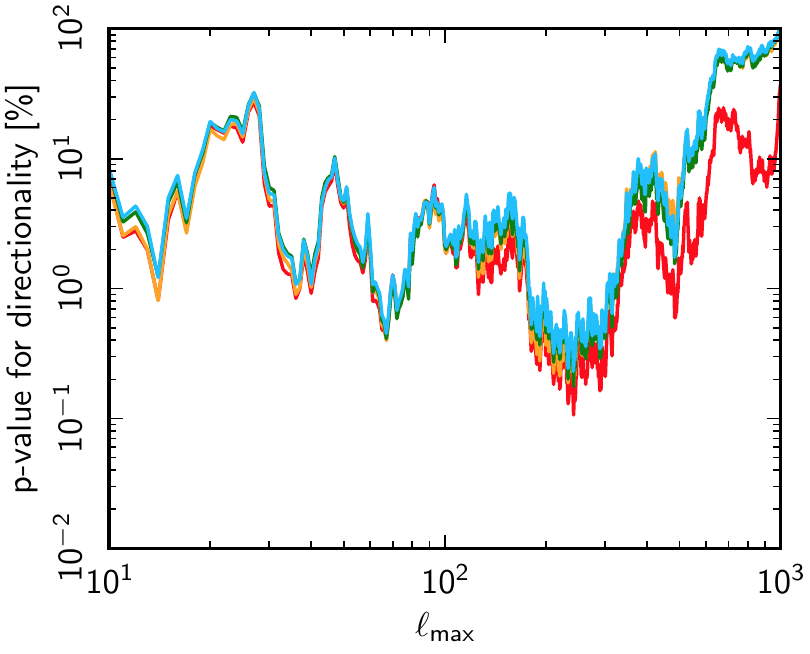}}
\mbox{\includegraphics[width=\hsize]{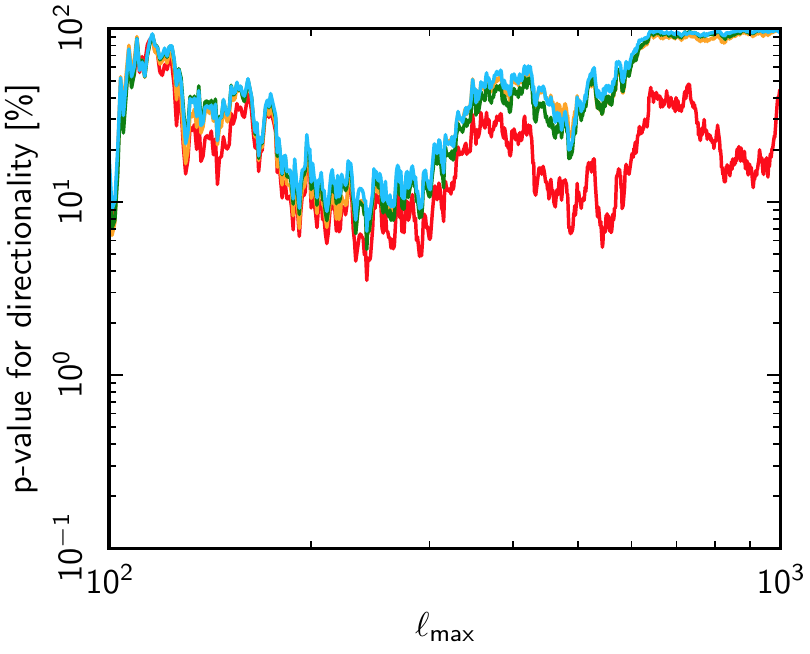}}
\end{center}
\caption{Rayleigh statistic $p$-values determined from the QML analysis as a
  function of $\ell_\mathrm{\max}$ for the \commander\ (red), \nilc\
  (orange), \sevem\ (green), and \smica\ (blue) data sets, with (top
  panel) $\ell_\mathrm{\min} = 2$ and (bottom panel) $\ell_{\min} =
  100$.  The general pattern of peaks is very similar to that in
  Fig.~\ref{fig:pvaluesdipmod}.  We emphasize that the statistic here
  is cumulative and as such trends in the curves can be misleading.}
\label{fig:ray_pvalues}
\end{figure}

Given that a dipole modulation amplitude of roughly $3\,\sigma$
significance is known to exist at low $\ell$ (before a posteriori
correction), one would expect a similar level of detection of
asymmetry to be determined by the RS.  Indeed, we find that asymmetry
is present out to $\ell \approx 240$.  Figure~\ref{fig:ray_pvalues}
(top) presents the $p$-values derived when the RS is
computed as a function of $\ell_\mathrm{\max}$ from $\ell = 2$.  The
minimum \pval\ obtained by the data is 0.1--0.2\,\%, to be compared to
the value of 0.9--1.0\,\% obtained for the dipole modulation amplitude
at $\ell_\mathrm{\max} = 67$. The direction preferred by the data for
$\ell_\mathrm{max} \approx 240$ is $(l, b) = (208\degr, -29\degr)$,
which is approximately $20\degr$ away from the dipole modulation
direction determined to $\ell \approx 64$.

\begin{figure}[h]
\includegraphics[width=\hsize]{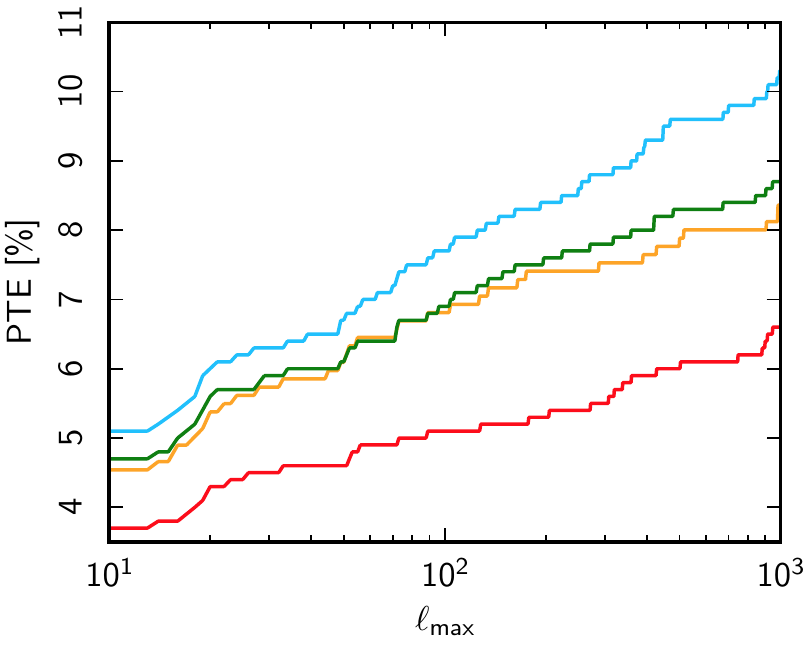}
\caption{Probability to exceed (PTE) the \pval\ of the signal from the
\commander\ (red), \nilc\ (orange), \sevem\ (green), and \smica\ (blue) data at
$\ell = 230$--$240$ (which is the multipole range with the most significant
deviation) when searching over a range of multipoles up to
$\ell_\mathrm{\max}$, for the RS determined from the QML analysis.
Much like the equivalent curve for dipole modulation, the PTE appears to grow
approximately logarithmically with $\ell_\mathrm{\max}$.}
\label{fig:ray_lookelsewhere}
\end{figure}

We correct for a posteriori statistics using the same procedure as in
Sect.~\ref{sec:QML}. Specifically, we count how often simulations find
asymmetry in the range $10 \leq \ell \leq \ell_\mathrm{\max}$ that is
more significant than that found for the data.  From
Fig.~\ref{fig:ray_lookelsewhere} it is clear that generic asymmetry at
the significance level found in our CMB sky occurs about 6\,\% or 8\,\% of
the time (depending on the range of $\ell$ one decides to search
over).

While the PTE here is not very low, it is nevertheless somewhat lower
than for the usual dipole modulation test.  Hence, it seems worth
exploring whether any of this signal comes from higher
multipoles. Therefore we compute the RS starting at
$\ell_\mathrm{\min} = 100$, to avoid the influence of asymmetry at
lower $\ell$. The lower panel of Fig.~\ref{fig:ray_pvalues} presents
the corresponding $p$-values as a function of
$\ell_\mathrm{\max}$.  There is a striking similarity with the lower
panel of Fig.~\ref{fig:pvaluesdipmod}.  It is clear that, even in the
absence of a posteriori correction, we find no significant asymmetry
at larger $\ell$. Hence most of the signal we are seeing in
Fig.~\ref{fig:ray_pvalues} (top) is due to the usual low-$\ell$
asymmetry.

We would like to stress that the results here are very similar to the
results of the previous section. For each of the statistics used we
are simply asking whether there is significant coupling of $\ell$ with
$\ell \pm 1$ modes.  The details of how to optimally combine these
couplings for a given $\ell$ range depends on whether we are talking
about dipole modulation or directionality (or some other related test,
e.g., variance asymmetry). These details will change the range of
scales over which the strongest signal in the data is found.

\section{Sensitivity of anomalies to enhanced sky coverage}
\label{sec:sky_coverage}

\def\lowlcomm{\texttt{Lkl-Commander}}
\def\cscomm{\texttt{Commander}} 
\def\lowlcommabb{\texttt{Lkl-Comm.}}
\def\cscommabb{\texttt{Comm.}}

One of the critical aspects in searching for anomalous features in sky
maps is to ensure that the region being investigated constitutes a
fair and unbiased sample. Since many of the claimed anomalies are on
large angular scales, this implies that minimal masking should be
applied to the data. However, residual foregrounds then become a
significant consideration. The masks applied to the four
component-separated maps studied in the bulk of this paper have been
defined at high resolution, and then conservatively degraded for lower
resolution studies. Such a procedure inevitably reduces the sky
coverage available for analysis, and can be particularly problematic
if significant structures are aligned by chance with the masked
regions. Indeed, the WMAP team \citep{bennett2010} have drawn
attention to several such features in their ILC reconstruction of the
CMB sky, and these are clearly also present in the \Planck\
\commander, \nilc, \sevem,  and \smica\ sky maps. A large cold spot is
seen near to the Galactic centre, a significant fraction of which lies
within the common mask at any resolution. However, despite its
location and visual impression, the feature is neither likely to be
attributable to residual foreground emission, nor is it inconsistent
with the $\Lambda$CDM model \citep{gott2007}. In addition, four
elongated cold fingers stretching from near the Galactic equator to
the south Galactic pole are seen, although no equivalent features are
evident in the northern sky.  \citet{bennett2010} have noted that the
alignment of the $\ell$ = 2 and $\ell$ = 3 multipoles
\citep{Tegmark2003} seems to be intimately connected with these
large-scale cool fingers and the intervening warm regions. One of the
latter also corresponds to the well-known ``Bianchi~VII$_\mathrm{h}$''
main lobe originally found in \citet{jaffe2005}.

Although we would ideally pursue full sky analyses, we prefer to
remain mindful of the influence of residual foregrounds, but still
seek to minimize the extent of any mask applied for analysis. In this
context, and specifically for large-angular-scale studies, we consider
the properties of an additional estimate of the CMB sky, also
generated using the \commander\ component separation methodology. In
particular, we note that the \Planck\ low-$\ell$ likelihood analysis
\citep{planck2014-a13} uses the temperature solution from this study,
degraded to a resolution of $\nside = 16$.  The \lowlcomm\ map, as we
now refer to it, is initially derived from input data sets (32 bands)
at 1\deg\ FWHM resolution and $\nside = 256$.  This includes \Planck\
individual detector and detector set maps from 30--857\,GHz, the
9-year WMAP observations between 23 and 94\,GHz, and the 408\,MHz sky
survey \citep{Haslam408MHz}, whereas the \commander\ map described in
\citet{planck2014-a11} includes \Planck\ data alone.  It is believed
that the 32-band solution is better (on large angular scales) than the
\Planck-only map, because the larger number of input frequencies
allows more detailed foreground modelling, and in particular the
separation of the low-frequency foregrounds into synchrotron,
free-free, and spinning dust components.  An associated confidence mask
(hereafter {\tt LklT$_\mathrm{256}$93}) is then defined based on a
goodness-of-fit measure per pixel, corresponding to a rejection of
7.3\,\% of the pixels on the sky.  A detailed discussion of these
results can be found in \citet{planck2014-a12}.

We now consider the implications of using the \lowlcomm\ map for
studies of several large-angular-scale anomalies observed in previous
sections, in particular since the larger sky coverage permitted by
this data set should constitute a better sample of the Universe. Note
that, at the resolutions of interest for the following analyses, the
noise level is negligible (even accounting for the WMAP contribution)
and should not have significant impact on the results. The exact
details of the noise contribution to simulations is therefore
unimportant.

\subsection{Variance, skewness, and kurtosis}
\label{sec:variance_lowl}

We begin by estimating the variance, skewness, and kurtosis of the
CMB.  We apply the unit variance estimator to the \lowlcomm\ map, and
specifically to the version used in the low-$\ell$ likelihood
analysis, which is smoothed to 440\arcm\ FWHM at a resolution of
$\nside = 16$.  A corresponding low-$\ell$ mask is generated by a
simple degrading of the mask at $\nside = 256$, then setting those
$\nside = 16$ pixels with a value less than 0.5 to zero and all others
to unity. The resultant low-$\ell$ likelihood mask rejects only
6.4\,\% of the sky.  We compare the results for both this mask (also
to be referred to as {\tt LklT$_\mathrm{16}$94}), and the standard
common mask at this resolution ({\tt UT$_\mathrm{16}$58}).  The
results are summarized in Table~\ref{Table:varianceLowEll} and show
that, when using the low-$\ell$ likelihood mask, the lower tail
probability for the variance is 7.0\,\%. This value is higher than the
corresponding values for the component separated maps as shown in
Table~\ref{Table:variance16}. In addition the skewness and kurtosis
are less consistent with Gaussianity than the component separated
maps.  However, when using the standard common mask at $\nside = 16$,
the lower tail probability of the variance, skewness, and kurtosis
become more compatible with those derived earlier.

There are two possible explanations for this behaviour. Either the
variance of the CMB in the region close to the Galactic plane is
intrinsically high, perhaps due to the presence of the various
features noted above, or the presence of residual foregrounds
increases the variance of the map.  In order to attempt to distinguish
between these options, we again apply the unit variance estimator to
the standard component-separated maps\footnote{Note that the \sevem\
  maps used in this section have been inpainted within 3\,\% of the
  sky towards the Galactic centre using a simple diffusive inpainting
  algorithm. This prevents residual foreground contamination from
  propagating to neighbouring regions when downgrading the map. The
  other component-separated maps are not pre-processed in this way
  since some form of inpainting of the most contaminated regions was
  already implemented as part of the component separation
  algorithms.}, but this time utilising the low-$\ell$ mask. Although
the component-separated maps are likely to contain some foreground
contamination in the regions omitted by application of the
UT$_\mathrm{16}$58 mask, it is appropriate to recall that this was
constructed in a conservative way, and may also mask parts of the sky
where the level of residual foregrounds can be considered negligible.
In addition, we investigate the cleaned frequency maps produced by the
\sevem\ algorithm in order to test for the presence of
frequency-dependent residual foregrounds.  The results of the unit
variance estimator analysis are summarized in
Table~\ref{Table:varianceAll}.

\begin{table} \begingroup \newdimen\tblskip \tblskip 4pt
  \caption{Lower-tail probability for the variance,
    skewness, and kurtosis of the \lowlcomm\ map.
}
\label{Table:varianceLowEll}
\nointerlineskip
\vskip -3mm
\footnotesize
\setbox\tablebox=\vbox{
   \newdimen\digitwidth 
   \setbox0=\hbox{\rm 0} 
   \digitwidth=\wd0 
   \catcode`*=\active 
   \def*{\kern\digitwidth}
   \newdimen\signwidth 
   \setbox0=\hbox{+} 
   \signwidth=\wd0 
   \catcode`!=\active 
   \def!{\kern\signwidth}
\halign{\hbox to 1.6in{#\leaderfil}\tabskip 4pt&
         \hfil#\hfil&
         \hfil#\hfil&
         \hfil#\hfil\tabskip 0pt\cr                           
\noalign{\doubleline\vskip -1pt}
\omit&\multispan3 \hfil Probability [\%]\hfil\cr
\noalign{\vskip -4pt}
\omit&\multispan3\hrulefill\cr
\omit\hfil Mask\hfil& Variance& Skewness& Kurtosis\cr   
\noalign{\vskip 3pt\hrule\vskip 5pt}
{\tt LklT$_\mathrm{16}$94}&  *7.0&  *1.5& 94.0\cr
{\tt UT$_\mathrm{16}$58}&  *0.7&  19.9&  82.5\cr
\noalign{\vskip 5pt\hrule\vskip 3pt}}}
\endPlancktable                    
\endgroup
\end{table}

\begin{table} \begingroup \newdimen\tblskip \tblskip 4pt
  \caption{Lower-tail probability for the variance, skewness, and
    kurtosis of the \lowlcomm\ map compared to the component separated
    maps, obtained using the low-$\ell$ likelihood mask {\tt
      LklT$_\mathrm{16}$94}.
  }
\label{Table:varianceAll}
\nointerlineskip
\vskip -3mm
\footnotesize
\setbox\tablebox=\vbox{
   \newdimen\digitwidth 
   \setbox0=\hbox{\rm 0} 
   \digitwidth=\wd0 
   \catcode`*=\active 
   \def*{\kern\digitwidth}
   \newdimen\signwidth 
   \setbox0=\hbox{+} 
   \signwidth=\wd0 
   \catcode`!=\active 
   \def!{\kern\signwidth}
\halign{\hbox to 1.6in{#\leaderfil}\tabskip 4pt&
         \hfil#\hfil&
         \hfil#\hfil&
         \hfil#\hfil\tabskip 0pt\cr                           
\noalign{\doubleline\vskip -1pt}
\omit&\multispan3 \hfil Probability [\%]\hfil\cr
\noalign{\vskip -4pt}
\omit&\multispan3\hrulefill\cr
\omit\hfil Map\hfil& Variance& Skewness& Kurtosis\cr   
\noalign{\vskip 3pt\hrule\vskip 5pt}
\lowlcomm\ & 7.0 & 1.5 & 94.0\cr
\cscomm\ & 7.7 & 1.9 & 96.0\cr
\nilc & 9.6 & 5.0 & 94.4\cr
\sevem & 7.4 & 4.8 & 94.3\cr
\smica & 7.7 & 3.7 & 93.7\cr
\noalign{\vskip 3pt\hrule\vskip 5pt}
{\tt SEVEM-100}& 8.4 & 0.4 & 97.9\cr
{\tt SEVEM-143} & 7.7 & 3.7 & 95.5\cr
{\tt SEVEM-217} & 8.3 & 0.7 & 95.2\cr
\noalign{\vskip 5pt\hrule\vskip 3pt}}}
\endPlancktable                    
\endgroup
\end{table}

All of the component separated maps show an increase in the lower tail
probability from about 0.5\,\% when the {\tt UT$_\mathrm{16}$58} mask
is applied to roughly 7\,\% for the {\tt LklT$_\mathrm{16}$94} mask.
The small variations in results for the different maps may be
attributable to the presence of residual foregrounds close to the
Galactic plane. However, the increased probabilities can also be
explained by the presence of CMB structures with higher variance
within that region which is not rejected by the less conservative mask.
Indeed, since the component-separated maps are affected by different
residual foregrounds, if the source of the changes in probabilities is
due only to the residual foregrounds, then we would expect a
larger dispersion than what is observed.  We also note that when we
apply the low-$\ell$ likelihood mask the skewness and kurtosis values
are shifted towards more extreme values. This implies that the sky
signal is less Gaussian for the larger sky fraction, despite the
results remaining compatible with the $\Lambda$CDM model assumed for
the null tests. Both \commander\ maps are noteworthy in this regard.

An important issue is whether the changes in the statistics can simply
be attributed to differences in the masks. We determine how many
simulations show an increase in variance at least as large as that seen
for the \lowlcomm\ map when comparing the values derived for the
UT$_\mathrm{16}$58 and low-$\ell$ likelihood masks. Similarly, we
determine how many simulations have increased skewness or kurtosis
values with shifts at least as large as observed. When the three
statistics are considered separately, the fraction of simulations that
indicate such changes are 7.6\,\%, 4.3\,\%, and 13.9\,\% for the
variance, skewness, and kurtosis, respectively. Of course, such subsets
of the simulations also include cases where a large shift in the
statistic is observed, but the statistic would not be considered
anomalous for either mask.  If we also impose the requirement that the
simulations have these shifts for all three quantities simultaneously,
then only 2 maps from 1000 are found.  Of course, such a requirement
is rather strong, and at this stage we are likely to be approaching
the limits of what can be said based on model-independent null tests.
Indeed, in order to assess whether these results are sensitive to a
posteriori choices, we repeat the analysis but successively take each
simulation as the reference. Thus, for each simulation the shift in
the variance, skewness, and kurtosis is computed and then we determine
how many times we find a case in which two or less of the remaining
simulations simultaneously show larger shifts for the three moments.
We find that 48 maps from 1000 satisfy these conditions. Given this,
it is difficult to draw strong conclusions about the significance or
otherwise of the mask-related changes in variance.

\subsection{$N$-point correlation functions}
\label{sec:npoint_correlation_lowl}

The connection between sky coverage and the observed structure of the
2-point correlation function for large angular separations has
previously been discussed in the literature, in particular in
connection with the $S_{1/2}$ statistic discussed in
Sect.~\ref{sec:n_point_asymmetry}. \citet{bennett2010} consider that
the use of a Galactic mask when computing these quantities is
sub-optimal, and note that a full-sky computation of the 2-point
correlation function from the 7-year WMAP ILC map lies within the
95\,\% confidence region determined by simulations of their best-fit
$\Lambda$CDM model over all angular separations. However,
\citet{copi2009} suggest that the origin of the inconsistencies
between the full-sky and cut-sky large-scale angular correlations
remains unknown, and that the observed discrepancies may indicate
that the Universe is not statistically isotropic on these scales. We
therefore consider the $N$-point correlation functions, and related statistics, of
the \lowlcomm\ map to contribute to this debate.

We compare results computed for both the \lowlcomm\ and \cscomm\ maps
at $\nside = 64$ after smoothing to a FWHM of 160\arcm. A mask is
constructed for the \lowlcomm\ map by degrading the {\tt
  LklT$_\mathrm{256}$93} mask to $\nside = 64$ and setting all
resulting pixels with a value less than 0.5 to zero, with the
remainder set to unity. The {\tt LklT$_\mathrm{64}$92} mask retains
92\,\% of the sky, to be compared to the 67\,\% usable sky coverage
allowed by the {\tt UT$_\mathrm{64}$67} common mask at this
resolution.

\begin{figure*}[htp!]
\begin{center}
\includegraphics[width=0.47\textwidth]{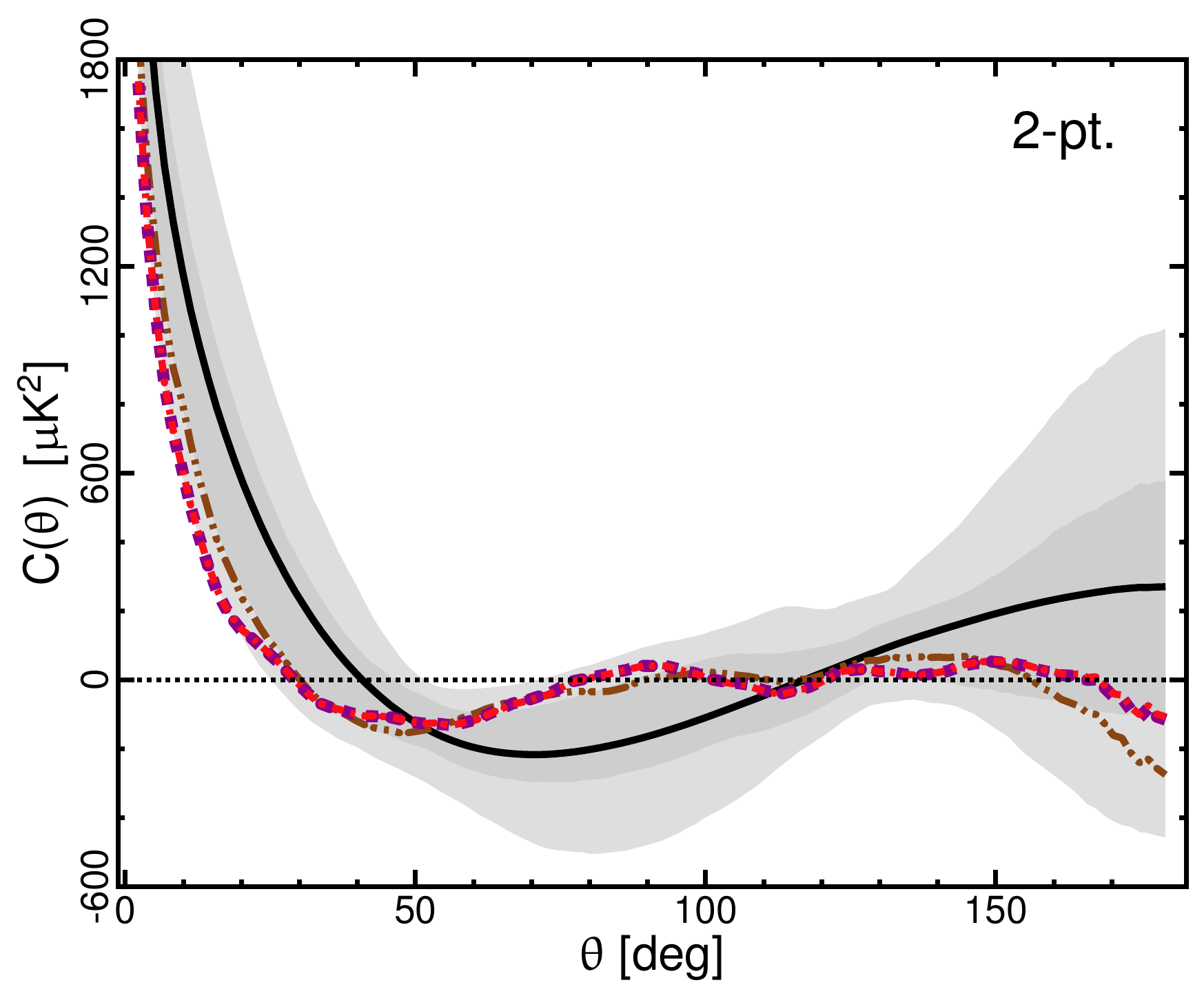}
\includegraphics[width=0.48\textwidth]{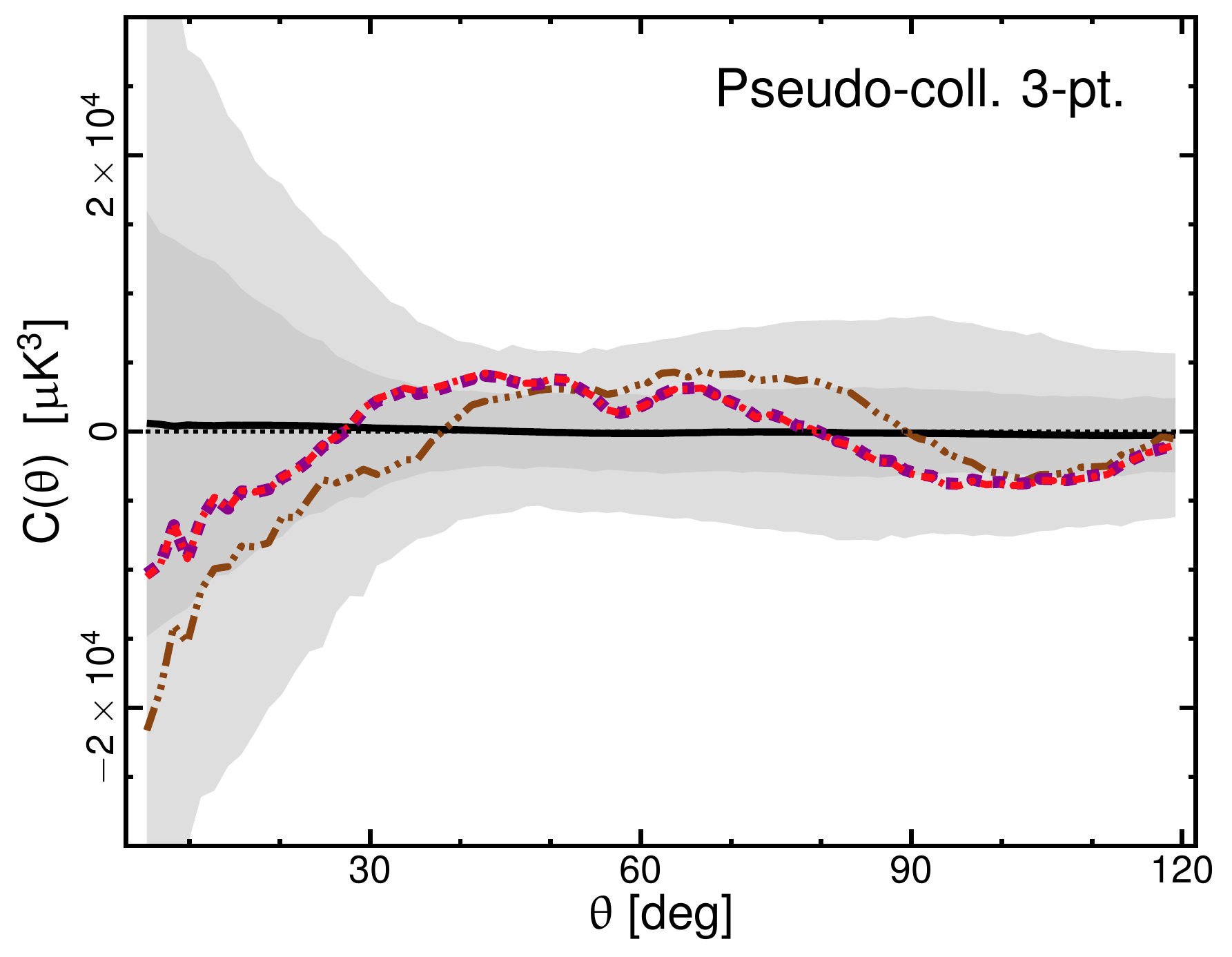}\\
\includegraphics[width=0.48\textwidth]{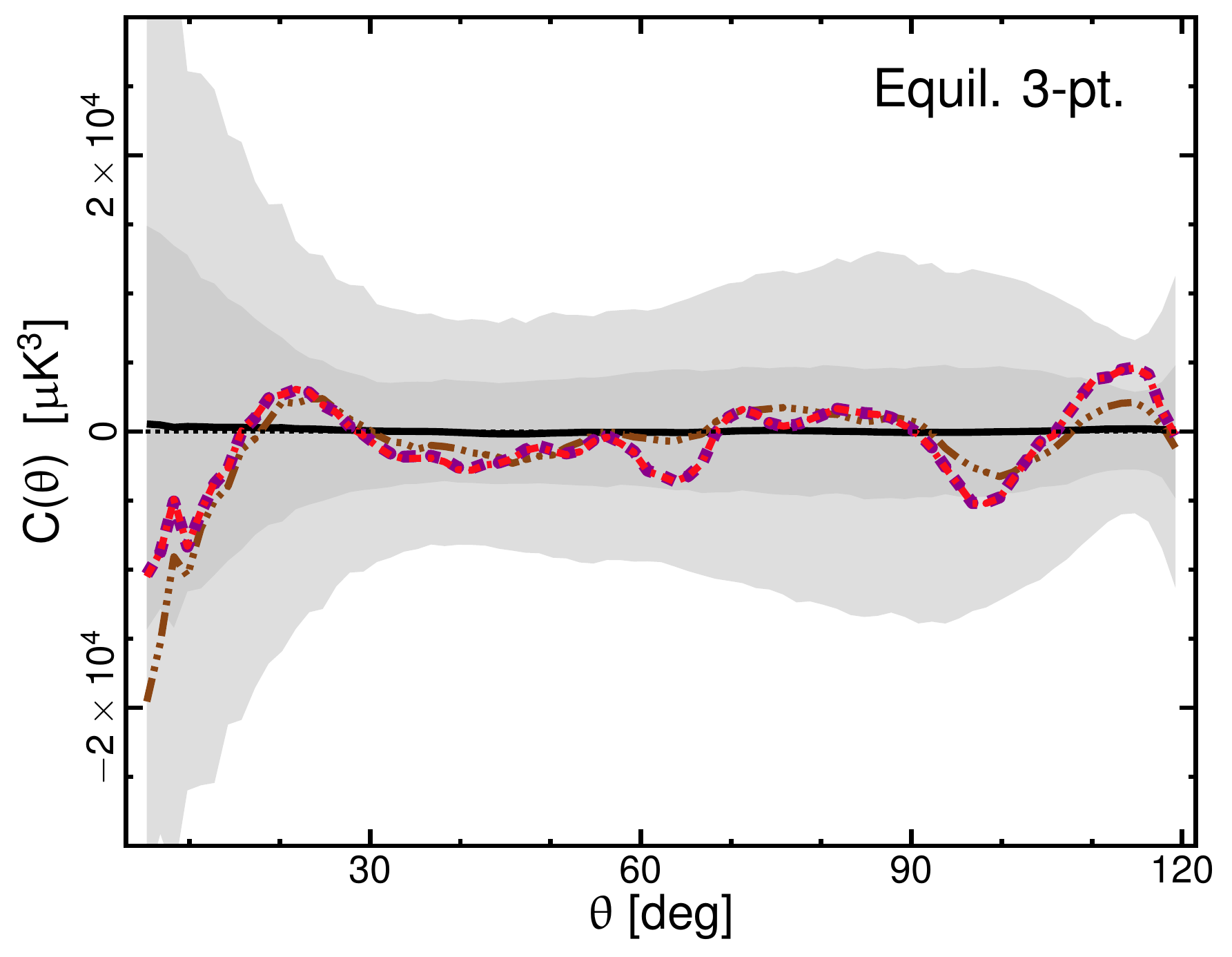}
\includegraphics[width=0.48\textwidth]{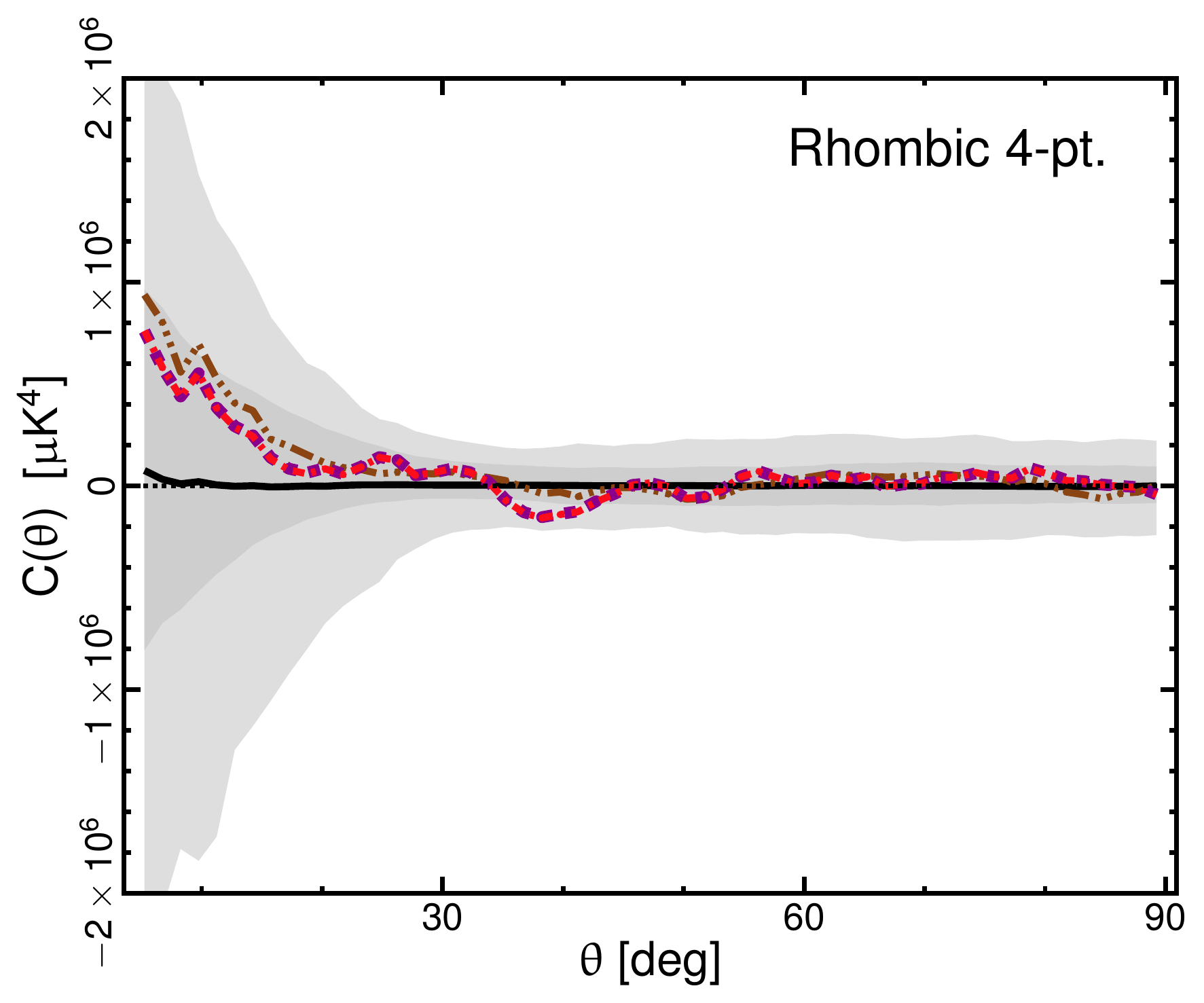}\\
\caption{$N$-point correlation functions determined from the $\nside=64$
  \Planck\ CMB 2015 temperature estimates.  Results are shown for the
  2-point, pseudo-collapsed 3-point (upper left and right panels,
  respectively), equilateral 3-point, and connected rhombic 4-point
  functions (lower left and right panels, respectively). The brown
  three dot-dashed, purple dashed, and red dot-dashed lines correspond
  to the \lowlcomm\ map analysed using the low-$\ell$ and common masks
  and the \cscomm\ map analysed using the common mask, respectively. Note
  that the dashed and dot-dashed lines lie on top of each other. The
  black solid line indicates the mean for 1000 MC simulations. The
  shaded dark and light grey regions indicate the 68\,\% and 95\,\%
  confidence regions, respectively, estimated using 1000 \cscomm\
  simulations. See Sect.~\ref{sec:npoint_correlation} for the
  definition of the separation angle $\theta$.}
\label{fig:npt_data_temp_lowl}
\end{center}
\end{figure*}

The results are presented in Fig.~\ref{fig:npt_data_temp_lowl} where
we compare the $N$-point functions for the data and the mean values
estimated from 1000 \cscomm\ simulations. The probabilities for
obtaining values of the $\chi^2$ statistic for the \Planck\ fiducial
$\Lambda$CDM model at least as large as the observed values are
provided in Table~\ref{tab:prob_chisq_npt_temp_lowl}. For the
estimation of the probabilities, we use the same set of 1000
\cscomm\ simulations for both versions of the \commander\ data. As
noted previously, the details of the simulations for such highly
smoothed data is essentially unimportant. We also provide an analysis
of the \lowlcomm\ map using the common mask to enable a direct
comparison with the analysis of the \cscomm\ map. In this latter case,
the results for both maps are in excellent agreement. However, the
\lowlcomm\ map is more consistent with simulations when the {\tt
  LklT$_\mathrm{64}$92} mask is adopted for the 2-point and
pseudo-collapsed 3-point functions, but less consistent for the
equilateral 3-point and rhombic 4-point function
results. Nevertheless, the results are generally in agreement with
expectations for a Gaussian, statistically isotropic model of the CMB
fluctuations.

\begin{table}[tp] 
\begingroup
\newdimen\tblskip \tblskip=5pt
\caption{Probabilities to exceed the observed values of the $\chi^2$
  statistics for the \lowlcomm\  and \cscomm\ maps at 
  $N_\mathrm{side}=64$.
}
\label{tab:prob_chisq_npt_temp_lowl}
\nointerlineskip
\vskip -3mm
\footnotesize
\setbox\tablebox=\vbox{
   \newdimen\digitwidth 
   \setbox0=\hbox{\rm 0} 
   \digitwidth=\wd0 
   \catcode`*=\active 
   \def*{\kern\digitwidth}
   \newdimen\signwidth 
   \setbox0=\hbox{+} 
   \signwidth=\wd0 
   \catcode`!=\active 
   \def!{\kern\signwidth}
\halign{\hbox to 1.6in{#\leaderfil}\tabskip 4pt&
\hfil#\hfil&\hfil#\hfil&
\hfil#\hfil\tabskip 0pt\cr
\noalign{\doubleline\vskip -1pt}
\omit&\multispan3 \hfil Probability [\%]\hfil\cr
\noalign{\vskip -4pt}
\omit&\multispan3\hrulefill\cr
\omit\hfil Function\hfil& \lowlcommabb$^{\rm a}$& \lowlcommabb$^{\rm b}$& \cscommabb$^{\rm b}$\cr
\noalign{\vskip 3pt\hrule\vskip 3pt}
 2-pt.&  84.3&97.1&97.2\cr
 Pseudo-coll.\ 3-pt.&  76.8&92.1&92.1\cr
 Equil.\ 3-pt.&  96.5&74.0&74.0\cr
 Rhombic 4-pt.&  94.5&65.0&64.6\cr
 \noalign{\vskip 3pt\hrule\vskip 3pt}}}
\endPlancktable                    
\endgroup
 {\footnotesize
 $^{\rm a}$ low-$\ell$ mask, {\tt LklT$_\mathrm{64}$92}\\
 $^{\rm b}$ common mask, {\tt UT$_\mathrm{64}$67}\\
}
\end{table} 

\begin{table}[tp] 
\begingroup
\newdimen\tblskip \tblskip=5pt
\caption{Probabilities for obtaining values of the $S_{1/2}$ and
  $\chi_0^2$ statistics for the simulations at least as
  large as the observed values of the statistic estimated 
  from the \lowlcomm\ and \cscomm\ maps using the 
{\tt LklT$_\mathrm{64}$92} and {\tt UT$_\mathrm{64}$67} masks,
respectively. We also show the corresponding
  estimation of the global {\it p}-value for the $S(x)$ statistic.} 
\label{tab:prob_stat_2pt_temp_lowl}
\nointerlineskip
\vskip -3mm
\footnotesize
\setbox\tablebox=\vbox{
   \newdimen\digitwidth 
   \setbox0=\hbox{\rm 0} 
   \digitwidth=\wd0 
   \catcode`*=\active 
   \def*{\kern\digitwidth}
   \newdimen\signwidth 
   \setbox0=\hbox{+} 
   \signwidth=\wd0 
   \catcode`!=\active 
   \def!{\kern\signwidth}

\halign{\hbox to 1.6in{#\leaderfil}\tabskip 4pt&
\hfil#\hfil&
\hfil#\hfil\tabskip 0pt\cr
\noalign{\doubleline\vskip -1pt}
\omit&\multispan2\hfil Probability [\%]\hfil\cr
\noalign{\vskip -4pt}
\omit&\multispan2\hrulefill\cr
\omit\hfil Statistic \hfil& \lowlcommabb & \cscommabb \cr
\noalign{\vskip 3pt\hrule\vskip 3pt}
$S_{1/2}$&  97.1 &  99.5 \cr
$S(x)$ (global) & 90.9 &   97.7 \cr
$\chi_0^2(\theta > 60^\circ)$ &  95.7 &  98.1 \cr
 \noalign{\vskip 3pt\hrule\vskip 3pt}}}
\endPlancktable                    
\endgroup
\end{table} 

The increased consistency of the 2-point function with simulations
when analysing a larger sky fraction is consistent with the
observations in \citet{copi2009}. We therefore quantify this further
by determining the statistical quantities introduced in
Sect.~\ref{sec:n_point_asymmetry} for the \lowlcomm\ map. In
particular, we reassess the lack of correlation determined previously
for large angular scales. It is evident from
Table~\ref{tab:prob_stat_2pt_temp_lowl} that the results for the $S_{1/2}$
and $\chi_0^2$ statistics are less anomalous when the low-$\ell$ mask
is applied. Moreover, the global {\it p}-value for the $S(x)$
statistic is substantially smaller.

We also repeat the conventional $\chi^2$ analysis but constraining
the computations to the two separate ranges defined by
$\theta<60^\circ$ and $\theta>60^\circ$. The results
of these studies are shown in Table~\ref{tab:prob_chisq_2pt_temp_lowl}.
The analysis for seperation angles $\theta>60^\circ$ indicates that the
unusually good fit of the observed 2-point function to the mean
2-point function determined for the $\Lambda$CDM model is independent
of the mask used in the analysis.  Conversely, the results for the
angles $\theta < 60^\circ$ indicate a strong dependence on the
mask. It appears that the decreased significance of the $\chi^2$
statistic for the 2-point function of the \lowlcomm\ map reported in
Table~\ref{tab:prob_chisq_npt_temp_lowl} is related mainly to
correlations in the data for separation angles smaller than
$60^\circ$. 

\begin{table}[tp] 
\begingroup
\newdimen\tblskip \tblskip=5pt
\caption{Probabilities for obtaining values of the $\chi^2$ statistic
  for the simulations at least as
  large as the observed values of the statistic estimated 
  from the \lowlcomm\ and \cscomm\ maps using the 
{\tt LklT$_\mathrm{64}$92} and {\tt UT$_\mathrm{64}$67} masks,
respectively. } 
\label{tab:prob_chisq_2pt_temp_lowl}
\nointerlineskip
\vskip -3mm
\footnotesize
\setbox\tablebox=\vbox{
   \newdimen\digitwidth 
   \setbox0=\hbox{\rm 0} 
   \digitwidth=\wd0 
   \catcode`*=\active 
   \def*{\kern\digitwidth}
   \newdimen\signwidth 
   \setbox0=\hbox{+} 
   \signwidth=\wd0 
   \catcode`!=\active 
   \def!{\kern\signwidth}
\halign{\hbox to 1.6in{#\leaderfil}\tabskip 4pt&
\hfil#\hfil&
\hfil#\hfil\tabskip 0pt\cr
\noalign{\doubleline\vskip -1pt}
\omit&\multispan2\hfil Probability [\%]\hfil\cr
\noalign{\vskip -4pt}
\omit&\multispan2\hrulefill\cr
\omit\hfil Statistic \hfil& \lowlcommabb & \cscommabb \cr
\noalign{\vskip 3pt\hrule\vskip 3pt}
$\chi^2(\theta < 60^\circ)$& 52.9 &  91.5 \cr
$\chi^2(\theta > 60^\circ)$& 96.5 &  96.8 \cr
 \noalign{\vskip 3pt\hrule\vskip 3pt}}}
\endPlancktable                    
\endgroup
\end{table} 

Our results do appear to indicate that computations made on larger sky
fractions increase the consistency of the 2-point function with
simulations. We therefore also test how the hemispherical asymmetry
observed previously is affected.  The results for the ecliptic frame
are presented in 
Table~\ref{tab:prob_chisq_npt_temp_ecl_lowl}.
We find that the asymmetry is larger for the \lowlcomm\ map than
for the \cscomm\ map in the case of the 2-point function, but
does not change substantially for the 3-point and 4-point functions.

\begin{table}[tp] 
\begingroup
\newdimen\tblskip \tblskip=5pt
\caption{Probabilities for obtaining values of the $\chi^2$ statistic and ratio
  of $\chi^2$ of the $N$-point functions 
  for the \Planck\ fiducial $\Lambda$CDM model at least as
  large as the observed values of the statistic on the northern and
  southern ecliptic hemispheres estimated from the \lowlcomm\ and
  \cscomm\ maps using the {\tt LklT$_\mathrm{64}$92} and {\tt
    UT$_\mathrm{64}$67} masks, respectively. 
}
\label{tab:prob_chisq_npt_temp_ecl_lowl}
\nointerlineskip
\vskip -3mm
\footnotesize
\setbox\tablebox=\vbox{
   \newdimen\digitwidth 
   \setbox0=\hbox{\rm 0} 
   \digitwidth=\wd0 
   \catcode`*=\active 
   \def*{\kern\digitwidth}
   \newdimen\signwidth 
   \setbox0=\hbox{>} 
   \signwidth=\wd0 
   \catcode`!=\active 
   \def!{\kern\signwidth}
\halign{\hbox to 1.6in{#\leaderfil}\tabskip 4pt&
\hfil#\hfil\tabskip 7pt&
\hfil#\hfil\/\tabskip 0pt\cr
\noalign{\doubleline}
\noalign{\vskip -1pt}
\omit&\multispan2 \hfil Probability [\%]\hfil\cr
\noalign{\vskip -4pt}
\omit&\multispan2\hrulefill\cr
\omit \hfil Hemisphere\hfil& \lowlcommabb & \cscommabb \cr
\noalign{\vskip 3pt\hrule\vskip 3pt}
\multispan3 \hfil 2-point function \hfil\cr
 Northern&  98.7&  !89.7\cr
 Southern&  29.4&  !80.5\cr
 $\chi^2$-ratio& 93.3& !22.6\cr
\multispan3 \hfil Pseudo-collapsed 3-point function\hfil\cr
 Northern&  99.4&  >99.9\cr
 Southern&  14.9&  !35.1\cr
$\chi^2$-ratio&  98.6& !98.8\cr
\multispan3 \hfil Equilateral 3-point function\hfil\cr
 Northern&  99.9&  !98.6\cr
 Southern&  47.6&  !45.7\cr
 $\chi^2$-ratio&  94.4& !86.6\cr 
\multispan3 \hfil Rhombic 4-point function\hfil\cr
 Northern&  99.8&  !99.7\cr
 Southern&  24.5&  !22.8\cr
 $\chi^2$-ratio&  97.8& !97.3\cr 
\noalign{\vskip 3pt\hrule\vskip 3pt}}}
\endPlancktable                    
\endgroup
\end{table}

\subsection{Dipole modulation and directionality}

\subsubsection{Variance asymmetry}
\label{sec:variance_asymmetry_lowl}

\begin{table}[h!tb]  
\begingroup 
\newdimen\tblskip \tblskip=5pt 
\caption{{\it p}-values for the variance asymmetry measured by different
  discs from the {\Planck} 2015 \lowlcomm\ and \cscomm\ temperature
  solutions using the 
{\tt LklT$_\mathrm{256}$93} and {\tt UT$_\mathrm{256}$73} masks,
respectively. The values represent the fraction of simulations with
  local-variance dipole amplitudes larger than those inferred from the
  data. No high-pass filtering has been applied to the maps.}
\label{tab:varasym1_lowellcomp}
\nointerlineskip 
\vskip -3mm 
\footnotesize  
\setbox\tablebox=\vbox{ 
   \newdimen\digitwidth 
   \setbox0=\hbox{\rm 0} 
   \digitwidth=\wd0 
   \catcode`!=\active 
   \def!{\kern\digitwidth}
   \newdimen\signwidth 
   \setbox0=\hbox{>} 
   \signwidth=\wd0 
   \catcode`*=\active 
   \def*{\kern\signwidth}
\halign{\hbox to 1.0in{#\leaderfil}\tabskip 4pt&
\hfil#\hfil\tabskip 8pt&
\hfil#\hfil\/\tabskip 0pt\cr
\noalign{\doubleline}
\noalign{\vskip -1pt}
\omit&\multispan2 \hfil {\it p}-value [\%]\hfil\cr
\noalign{\vskip -4pt}
\omit&\multispan2\hrulefill\cr
\omit \hfil Disc radius [\deg] \hfil&\omit\hfil \lowlcommabb
\hfil&\omit\hfil \cscommabb \hfil\cr   
\noalign{\vskip 3pt\hrule\vskip 3pt}
4 & 0.2&  0.7\cr
6&  0.4&  0.3\cr
8&  0.3&  0.1\cr
10& 0.3&  0.7\cr
12& 0.3&  0.8\cr
14& 0.5&  1.5\cr
16& 0.6&  1.9\cr
18& 0.8&  2.7\cr
20& 1.1&  3.7\cr
\noalign{\vskip 3pt\hrule\vskip 3pt}}} %
\endPlancktable 
\endgroup %
\end{table}

\begin{table}[h!tb]  
\begingroup 
\newdimen\tblskip \tblskip=5pt 
\caption{Local-variance dipole directions measured by 8\deg\
  discs for the {\Planck} 2015 \lowlcomm\ and \cscomm\ temperature
  solutions.
}
 \label{tab:varasym2_lowellcomp}
\nointerlineskip 
\vskip -3mm 
\footnotesize  
\setbox\tablebox=\vbox{
\halign{\hbox to 1.4in{#\leaderfil}\tabskip 6pt&
\hfil#\hfil\tabskip=5pt&\hfil#\hfil\/\tabskip=0pt&\hfil#\hfil \cr
\noalign{\doubleline}
\omit\hfil Method \hfil& $(l,b)$ [\deg] \cr
\noalign{\vskip 3pt\hrule\vskip 3pt}
\lowlcommabb& \hfil $(225,-28)$\hfil\cr
\cscomm& \hfil $(214, -24)$\hfil\cr
\noalign{\vskip 3pt\hrule\vskip 3pt}}} %
\endPlancktable 
\endgroup %
\end{table}

Here we apply the local-variance analysis of Sect.~\ref{sec:varasym}
to the \lowlcomm\ map and compare the results with those of the
\cscomm\ map. Contrary to the analysis of Sect.~\ref{sec:varasym},
where full-resolution ($N_{\mathrm{side}}=2048$) maps were used, here
the \cscomm\ map is downgraded to $N_{\mathrm{side}}=256$ in order to
consistently compare the results for both maps. The simulations used
for estimating the significance levels are also downgraded to the same
resolution, and convolved with the corresponding beam
function. Otherwise, the procedure is identical to the one described
in Sect.~\ref{sec:varasym}, e.g., the same number of discs has been
used to construct the local-variance maps. Here we only present the
results when no high-pass filtering has been applied to the maps; this
is to avoid confusion as our objective in this section is only to
compare the general properties of the \lowlcomm\ map to those of the
standard component-separated maps.

Table~\ref{tab:varasym1_lowellcomp} summarizes the significance levels
measured by our variance asymmetry analysis using discs of different
radii, for the {\Planck} 2015 \cscomm\ and \lowlcomm\ temperature
maps. The {\it p}-values represent the fraction of simulations with
local-variance dipole amplitudes larger than those inferred from the
data. We in addition present in Table~\ref{tab:varasym2_lowellcomp}
the preferred variance asymmetry directions for both maps using 8\deg\
discs.
  
Our results show consistency between the two maps. The small
change in the preferred direction is expected from the change in the
mask, and agrees specifically with the directions found by the
analysis of the QML dipole modulation analysis 
in Sect.~\ref{sec:qml_dipmod_lowl}. One interesting
observation is that the large variance asymmetry significance is now
extended to cases where larger discs are used. Note that no high-pass
filtering has been applied in the present analysis, and therefore {\it
  p}-values inferred from the \cscomm\ map increase with the disc
size. As explained in Sect.~\ref{sec:varasym}, the low observed
significance levels for larger discs is due to the cosmic variance
associated with the largest-scale modes. The observed increase in the
significance levels for the \lowlcomm\ map is therefore interestingly
consistent with this picture; the mask in this case is smaller and
therefore a larger fraction of the sky is available. This in turn
provides more data on the largest scales, and therefore lowers the
impact of the cosmic variance.

\subsubsection{Dipole modulation: pixel-based likelihood}
\label{sec:dipmod_lowl}

\begin{table}
\begingroup
\newdimen\tblskip \tblskip=5pt
\caption{Summary of dipole modulation results at a
  smoothing scale of $5\deg$ for the {\Planck} 2015 
  \lowlcomm\ and \cscomm\ temperature solutions, 
  as derived by the brute-force likelihood given by
  Eq.~\ref{eq:dipmod_like}. The former results were derived using the
  {\tt LklT$_\mathrm{32}$93} mask, whereas the latter are those determined
  previously in Sect.~\ref{sec:dipmod}.
}
\label{tab:dipmod_lowl}
\nointerlineskip
\vskip -3mm
\footnotesize
\setbox\tablebox=\vbox{
   \newdimen\digitwidth 
   \setbox0=\hbox{\rm 0} 
   \digitwidth=\wd0 
   \catcode`*=\active 
   \def*{\kern\digitwidth}
   \newdimen\signwidth 
   \setbox0=\hbox{+} 
   \signwidth=\wd0 
   \catcode`!=\active 
   \def!{\kern\signwidth}
\halign{\hbox to 1.0in{#\leaderfil}\tabskip 2pt&
    \hfil#\hfil\tabskip 8pt&
    \hfil#\hfil\tabskip 0pt\cr
\noalign{\doubleline\vskip -1pt}
\omit \hfil Method\hfil& \omit\hfil 2013\hfil& \omit\hfil 2015\hfil\cr   
\noalign{\vskip 3pt\hrule\vskip 5pt}
\multispan3 \hfil Dipole modulation amplitude, $\alpha$\hfil\cr
\noalign{\vskip 2pt}
\lowlcommabb& $\cdots$&    $0.059\pm0.020$\cr
\cscomm& $0.078\pm0.021$& $0.066\pm0.021$\cr
\noalign{\vskip 2pt}
\multispan3 \hfil Dipole modulation direction, $(l, b)$ [\deg]\hfil\cr
\noalign{\vskip 2pt}
\lowlcommabb& $\cdots$& $(223,-17)\pm23$\cr
\cscomm& $(227,-15)\pm19$& $(230,-16)\pm24$\cr
\noalign{\vskip 2pt}
\multispan3{\hfil Power spectrum amplitude, $q$}\hfil\cr
\noalign{\vskip 2pt}
\lowlcommabb&   $\cdots$&       $0.970\pm0.025$\cr
\cscomm& $\cdots$& $0.961\pm0.025$\cr
\noalign{\vskip 2pt}
\multispan3{\hfil Power spectrum tilt, $n$}\hfil\cr
\noalign{\vskip 2pt}
\lowlcommabb&    $\cdots$&      $0.068\pm0.045$\cr
\cscomm& $\cdots$& $0.082\pm0.043$\cr
\noalign{\vskip 2pt\hrule\vskip 3pt}}}
\endPlancktable                    
\endgroup
\end{table}

Table~\ref{tab:dipmod_lowl} presents constraints on the
dipole modulation model as derived from the \lowlcomm\ map and the 
{\tt LklT$_\mathrm{32}$93} mask that includes
93\,\% of the sky, updating the results from Sect.~\ref{sec:dipmod} for
the \cscomm\ map. We find that all previously reported results are
robust with respect to data selection and sky coverage. In particular,
the best-fit dipole modulation amplitude at $5\deg$ FWHM is 5.9\,\% in
the \lowlcomm\ map, and is thus stable to within about $0.3\,\sigma$
when increasing the sky fraction from 78\,\% to 93\,\%. Likewise, the
marginal low-$\ell$ power spectrum amplitude, $q$, shifts upward by
$0.4\,\sigma$, and the power spectrum tilt, $n$, downward by
$0.3\,\sigma$, for the same sky fraction increases.

To assess the statistical significance of these shifts, we compare
with Gaussian statistics, creating two Gaussian random vectors with 78
and 93 elements, respectively, where the first 78 elements of the
latter vector are identical to the first vector. From these, we
compute the difference between the two means, after normalizing each
so that their individual errors in the mean are unity. Repeating this
simple calculation $10^5$ times, we find that 48\,\% of all Gaussian
realizations observe shifts larger than $0.3\,\sigma$, and 34\,\%
observe shifts larger than $0.4\,\sigma$. Thus, the parameter
differences due to the different data selection and sky fractions
reported above are consistent with expectations from random Gaussian
statistics.

\subsubsection{Dipole modulation: QML analysis}
\label{sec:qml_dipmod_lowl}

\begin{table}
\begingroup
\newdimen\tblskip \tblskip=5pt
\caption{Summary of the dipole modulation results for the range $\ell \in [2, 64]$
determined from the {\Planck} 2015 \lowlcomm\ and \cscomm\ temperature solutions, as 
derived by the QML estimator defined in Sect.~\ref{sec:QML} using
the {\tt LklT$_\mathrm{256}$93} and {\tt UT78} masks, respectively.
}
\label{tab:lowellmod_commsol}
\nointerlineskip
\vskip -3mm
\footnotesize
\setbox\tablebox=\vbox{
   \newdimen\digitwidth
   \setbox0=\hbox{\rm 0}
   \digitwidth=\wd0
   \catcode`*=\active
   \def*{\kern\digitwidth}
   \newdimen\signwidth
   \setbox0=\hbox{+}
   \signwidth=\wd0
   \catcode`!=\active
   \def!{\kern\signwidth}
\halign{ \hbox to 1.0in{#\leaderfil}\tabskip 4pt&
         \hfil#\hfil\tabskip 8pt&
         \hfil#\hfil\tabskip 0pt\cr                           
\noalign{\doubleline\vskip -1pt}
\omit& \omit& Direction\cr   
\omit\hfil Method\hfil&$A$& $(l, b)$ [\deg]\cr   
\noalign{\vskip 4pt\hrule\vskip 6pt}
\lowlcommabb& $0.058^{+0.022}_{-0.012}$& $(227, -28)\pm26$\cr
\noalign{\vskip 3pt}
\cscomm& $0.063^{+0.025}_{-0.013}$& $(213, -26)\pm28$\cr
\noalign{\vskip 3pt\hrule\vskip 4pt}}}
\endPlancktable                    
\endgroup
\end{table}

We also repeat the QML dipole modulation analysis of
Sect.~\ref{sec:QML} for the \lowlcomm\ map and corresponding mask.
Table~\ref{tab:lowellmod_commsol} summarizes the results of the
low-$\ell$ dipole modulation for the \lowlcomm\ temperature solution,
compared with the \cscomm\ map.

The best-fit modulation amplitude for \lowlcomm\ is 5.8\,\% and the
small 0.5\,\% shift from the \cscomm\ best-fit amplitude corresponds
to a decrease of approximately $0.4\,\sigma$.  These results mirror
very closely the results found above for the pixel-based likelihood
approach to dipole modulation, as expected, and the observed shifts
are perfectly consistent with those expected from the change in the
mask.

\subsubsection{Bipolar spherical harmonics}
\label{sec:biposh_lowl}

We next perform a dipole modulation analysis on the {\lowlcomm} temperature
map using the BipoSH formalism from Sect.~\ref{sec:biposh}. The dipole
modulation amplitude inferred from the analysis is smaller that that
deduced from analysing the \cscomm\ map as seen in
Table~\ref{tab:biposh-lowell-mod}. However, it should be noted that
the probability for simulations to yield a dipole modulation amplitude
equal to or greater than the amplitude inferred from data is $0.4\,\%$,
which is smaller by a factor of approximately $2.4$ as compared to the {\it
  p}-value inferred from analysis on \cscomm. The reduction in the
dipole amplitude and the enhanced significance can both be attributed
to the reduced power bias which is a result of the increased sky
coverage.

\begin{table}[h!]  
\begingroup
\newdimen\tblskip \tblskip=5pt
\caption{ Amplitude ($A$) and
  direction of the dipole modulation in Galactic coordinates 
  as estimated for the multipole range $\ell \in [2,64]$  using
  the BipoSH analysis on {\lowlcomm}  and {\tt Commander} maps.
The former results were derived using the
  {\tt LklT$_\mathrm{256}$93} mask; the latter are those determined
  previously in Sect.~\ref{sec:biposh}.
}
\label{tab:biposh-lowell-mod}
\nointerlineskip
\vskip -3mm
\footnotesize
\setbox\tablebox=\vbox{
\halign{\hbox to 0.8in{#\leaderfil}\tabskip 4pt&
\hfil#\hfil\tabskip 8pt&
\hfil#\hfil\/\tabskip=0pt\cr
\noalign{\doubleline}
\omit& \omit& Direction\cr 
\omit\hfil Method \hfil& $A$&$(l, b)$ [\deg]\cr
\noalign{\vskip 3pt\hrule\vskip 3pt}
\lowlcommabb& $0.063\pm 0.021$& $(234, -27)\pm31$\cr
\noalign{\vskip 4pt}
\cscomm& $0.067 \pm 0.023$&  $(230,-18) \pm 31$\cr
\noalign{\vskip 3pt\hrule\vskip 3pt}}}
\endPlancktablewide                 
\endgroup
\end{table} 

\subsection{Summary}
\label{sec:summary_lowl}
 
Using a larger sky fraction in our analyses leads to small changes in
the results related to large-angular-scale anomalies, but these are
essentially consistent with expectations from random Gaussian
statistics. In particular, the asymmetry in power on the sky, as
parameterized by a dipole modulation model, is robust to mask
changes. 

\section{Polarization analysis}
\label{sec:polarization_results}

As previously discussed in Sect.~\ref{sec:data}, large angular-scale
CMB fluctuations in the \Planck\ polarization data have been
suppressed by a post-processing high-pass filter to minimize the impact of
systematic artefacts. Therefore, no polarization results concerning
CMB statistical anomalies on such scales are presented in this
paper. In addition, a noise mismatch between simulations and data also
limits our ability to study polarization more generally.
Nevertheless, a local analysis of the polarization data for stacked
patches of the sky can still be performed, in order to test the
statistical properties of the CMB anisotropies. In this case, the
stacking procedure mitigates the impact of the small-scale noise and
potential systematic effects.

Traditionally, the Stokes parameters $Q$ and $U$ are used to describe
the CMB polarization anisotropies \citep[e.g.,][]{Zaldarriaga1997}.  Such
quantities are not rotationally invariant, thus for the stacking
analysis it is convenient to consider a local rotation of the Stokes
parameters, resulting in quantities denoted by $Q_{\mathrm{r}}$ and
$U_{\mathrm{r}}$, as described in Sect.~\ref{sec:stacking}.
Additionally, several other related quantities can be defined.

The polarization amplitude $P \equiv \sqrt{Q^2 + U^2}$ and polarization
angle $\Psi \equiv \tfrac{1}{2}\arctan(U/Q)$, are commonly used
quantities in, for example, Galactic astrophysics. However,
unbiased estimators of these quantities in the presence of anisotropic
and/or correlated noise are hard to define
\citep{Plaszczynski2014}. Of course, a direct comparison of the
observed (noise-biased) quantity to simulations analysed in the same
manner is possible, but we elect here to defer the study of this
representation of the polarization signal, using maps of the
polarization amplitude only to define peaks around which stacking can
be applied.

The rotationally invariant quantities referred to as $E$ and
$B$ modes are commonly used for the global analysis of CMB
data. Although the $E$-mode maps are not analysed in detail here,
they are considered qualitatively, so that it is appropriate to recall
their construction.  Since the quantities $Q\pm iU$, defined relative
to the direction vectors $\vec{\hat n}$, transform as spin-2 variables
under rotations around the $\vec{\hat n}$ axis, they can be expanded as
\begin{linenomath*}
\begin{equation}
(Q \pm iU) (\vec{\hat{n}}) = \sum_{\ell = 2}^{\infty} \sum_{m=-\ell}^{\ell}a_{\ell m}^{(\pm 2)} \,_{\pm 2}Y_{\ell m}(\vec{\hat{n}}),
\end{equation}
\end{linenomath*}
where $\,_{\pm 2}Y_{\ell m}(\vec{\hat{n}})$ are the spin-weighted
spherical harmonics and $a_{\ell m}^{(\pm 2)}$ are the corresponding
harmonic coefficients. If we define
\begin{linenomath*}
\begin{eqnarray}
a^{E}_{\ell m} & = & \frac{1}{2}\left(a_{\ell m}^{(2)} + a_{\ell m}^{(-2)}\right) \,, \\
a^{B}_{\ell m} & = & \frac{-i}{2}\left(a_{\ell m}^{(2)} - a_{\ell m}^{(-2)}\right) \,,
\end{eqnarray}
\end{linenomath*}
then the invariant quantities are given by
\begin{linenomath*}
\begin{eqnarray}
E(\vec{\hat{n}})& = & \sum_{\ell = 2}^{\infty} \sum_{m=-\ell}^{\ell} a^{E}_{\ell m} Y_{\ell m}(\vec{\hat{n}})  \,, \\
B(\vec{\hat{n}}) & = & \sum_{\ell = 2}^{\infty} \sum_{m=-\ell}^{\ell} a^{B}_{\ell m} Y_{\ell m}(\vec{\hat{n}}) \,.
\end{eqnarray}
\end{linenomath*}

\subsection{Stacking around temperature hot and cold spots}
\label{sec:stacking}

The stacking of CMB anisotropies around peaks (hot and cold spots) on
the sky yields characteristic temperature and polarization patterns
that contain valuable information about the physics of recombination
\citep{komatsu:2011}. Statistical analysis of stacked images
differs from the other tests in this paper in several respects. First,
peak-related new physics may be revealed that is difficult to find in
a global analysis, for example, the non-Gaussian CMB cold spots
predicted by a modulated preheating model \citep{BFHK2009}. Secondly,
stacking is a local operation, which naturally avoids mask-induced
complications.  Thus stacking can be used as a transparent and
intuitive method to test the robustness of anomalies found with other
methods. Alternatively, it can be applied as a quality indicator of
the data at the map level.

Our stacking procedure is as follows.  Hot (or cold) peaks are
selected in the temperature map as local extrema with negative (or
positive) second derivatives, and classified relative to a given
threshold $\nu$ (in rms units of the temperature map).  Since the
spinorial components $Q$ and $U$ are expressed in a local coordinate
system, we employ a configuration in which the Stokes parameters
around a peak at the direction $\vec{\hat{n}}_0$ can be superposed
\citep{kamionkowski:1997}.  In particular, we use a locally defined
rotation of the Stokes parameters that is written as:
\begin{linenomath*}
\begin{eqnarray}
Q_\mathrm{r}\left(\vec{\hat{n}};\vec{\hat{n}}_0\right) & = & -Q\left(\vec{\hat{n}}\right)\cos{(2\phi)}-U\left(\vec{\hat{n}}\right)\sin{(2\phi)}  \,, \\
U_\mathrm{r}\left(\vec{\hat{n}};\vec{\hat{n}}_0\right) & = & Q\left(\vec{\hat{n}}\right)\sin{(2\phi)}-U\left(\vec{\hat{n}}\right)\cos{(2\phi)}  \,,
\label{eqn:def_qr}
\end{eqnarray}
\end{linenomath*}
where $\phi$ is the angle between the axis aligned along a meridian
(pointing to the south by convention) in the local coordinate system
centred on a peak at $\vec{\hat{n}}_0$ and the great circle connecting
this peak to a position $\vec{\hat{n}}$. This definition decomposes
the linear polarization into radial ($Q_\mathrm{r} > 0$) and
tangential ($Q_\mathrm{r} < 0$) contributions around the peaks.  This
definition of $Q_\mathrm{r}$ is equivalent to the ``tangential shear''
used in weak lensing studies.

For visualization purposes, a flat patch around each peak is then
extracted, and the average stacked image computed from the subset.  A
position on the sky at an angular distance $\theta$ from the central
peak is labelled with the flat-sky coordinates
\begin{linenomath*}
\begin{equation}
x =  \varpi \cos\phi \,, \quad
y =  \varpi \sin\phi \, . \label{eq:flatxydef}
\end{equation}
\end{linenomath*}
Here $\varpi = 2\sin(\theta/2) \approx \theta $ is the
effective flat-sky radius. For the angular scales of a few degrees considered in the
stacking analyses the difference between $\varpi$ and
$\theta$ is negligible.  We use $\varpi$ for analyses in
the flat-sky approximation, and $\theta$ for
analyses directly on the sphere.

\begin{figure}
\centering
\includegraphics[scale=0.45]{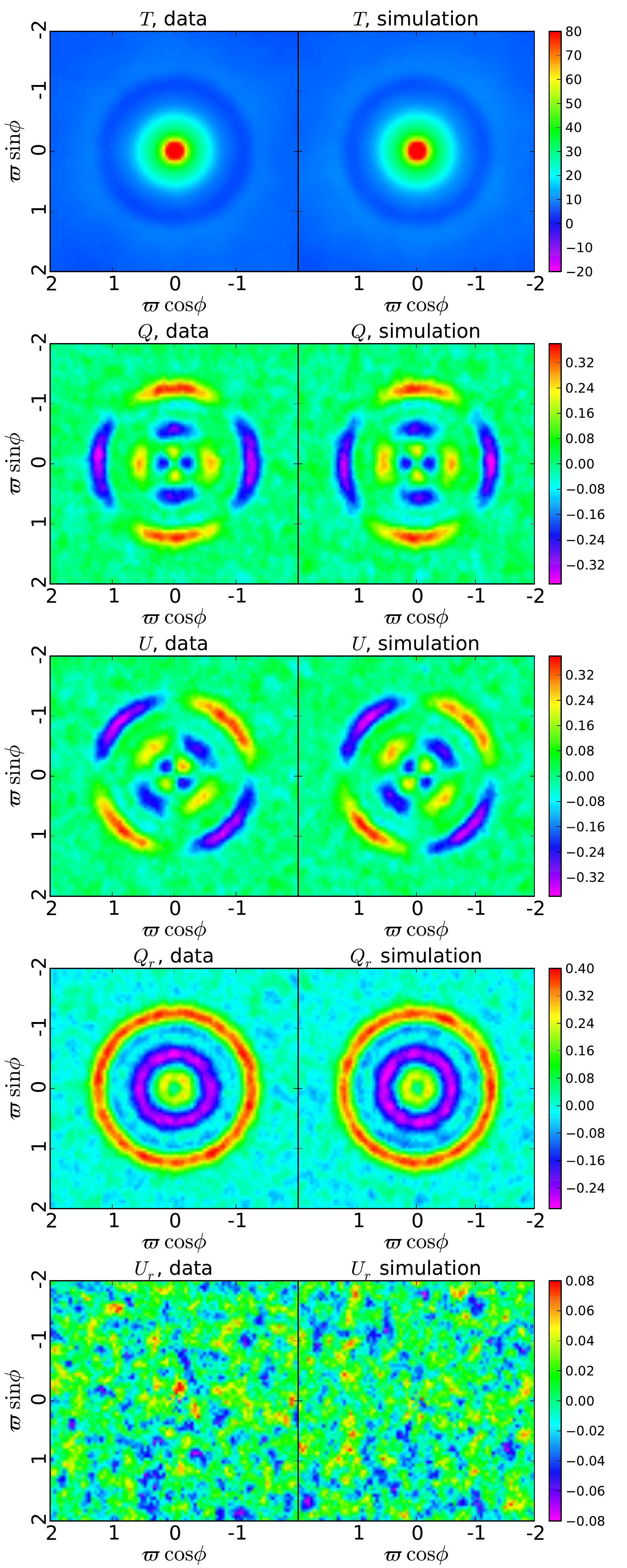}
\caption{From top to bottom, $T$, $Q$, $U$, $Q_\mathrm{r}$, and
  $U_\mathrm{r}$ stacked images (in \microK\ units) extracted around
  temperature hot spots selected above the null threshold ($\nu = 0$)
  in the \commander\ sky map for data (left column) and an equivalent
  simulation (right column). The horizontal and vertical axes of the flat-sky projection are labelled in
    degrees.}
\label{fig:dx11_commander_stacking_patch}
\end{figure}

\begin{figure}
\includegraphics[width=0.48\textwidth]{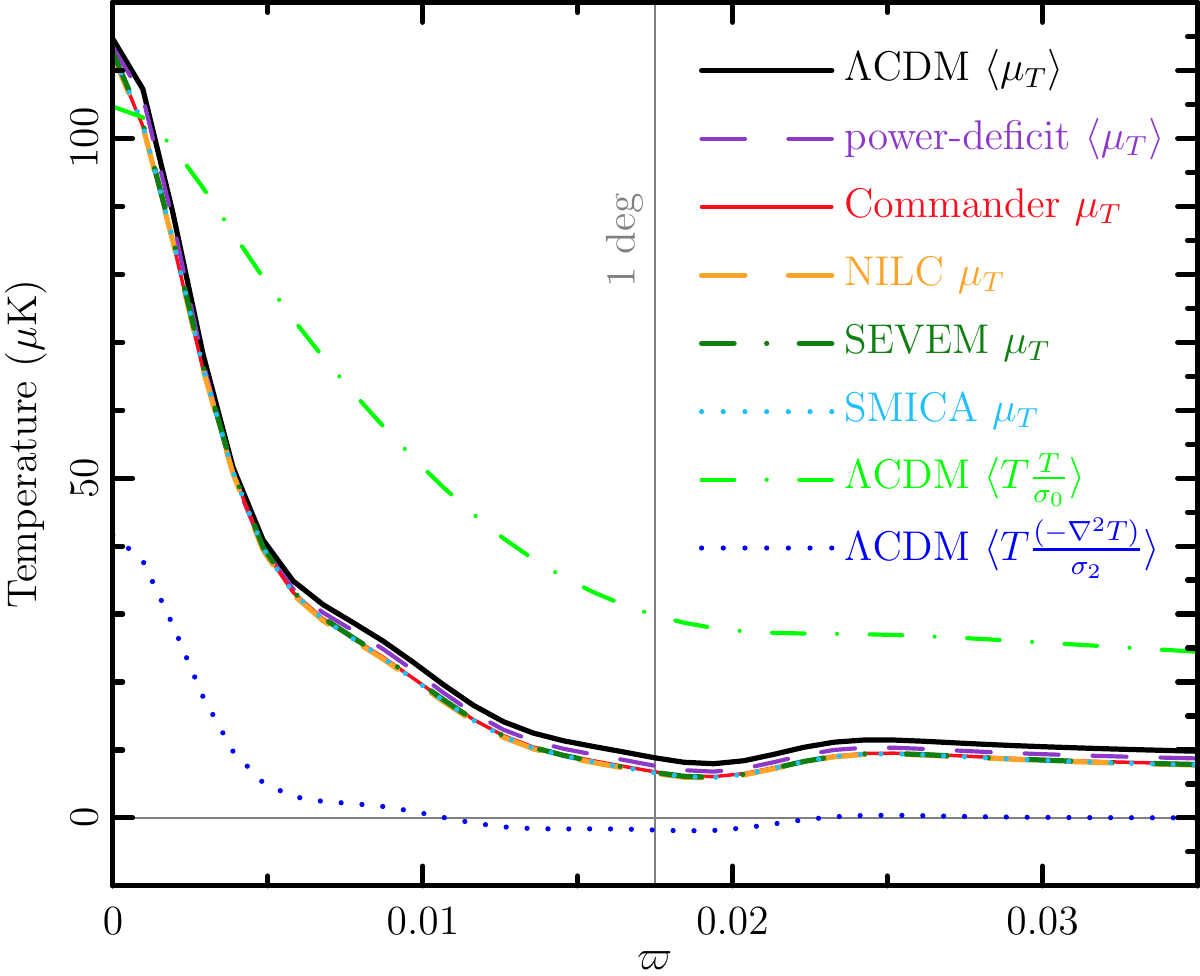}
\caption{Radial profile $\mu_T(\varpi)$ derived from the stacked
 temperature image (see Fig.~\ref{fig:dx11_commander_stacking_patch}
 or \ref{fig:general_stack_T}). The denominators $\sigma_0$ and
 $\sigma_2$ are the theoretical rms values of CMB $T$ and
 $\nabla^2T$, respectively. The theoretical $\langle \mu_T(\varpi)
 \rangle$ is a linear combination of $\langle T (\varpi)
 (T(0)/\sigma_0) \rangle$ (green dash-dotted line) and $\langle
 T(\varpi) (-\nabla^2 T(0))/\sigma_2) \rangle$ (blue dotted line).
 For all four component-separated maps, the deviation of $\mu_T$ from
 the ensemble mean $\langle \mu_T \rangle $ of the fiducial model (here
 the \Planck\ 2015 $\Lambda$CDM best fit) is consistent with cosmic
 variance, and can be related to the low-$\ell$ power deficit. The
 example power-deficit $\langle \mu_T\rangle$ (purple dashed line) is
 the theoretical prediction of $\langle \mu_T\rangle$ if the fiducial
 model $C_\ell$s are reduced by 10\,\% in the range $2\le \ell \le
 50$. \label{fig:stack_T_LapT_corr}}
\end{figure}

The stacking process tends to provide an image with azimuthal symmetry
about its centre, due to the almost uncorrelated orientations of the
temperature peaks. The stacked images of temperature patches around
hot spots selected above the null threshold for both the \commander\
data and a corresponding simulation are shown in the top row of
Fig.~\ref{fig:dx11_commander_stacking_patch}. The observed patterns
are in excellent agreement. Stacking around cold spots yields similar
patterns but with flipped sign. Given the symmetry, it is often useful
to consider the radial profile obtained by averaging the stacked image
over the azimuthal angle $\phi$.  Fig.~\ref{fig:stack_T_LapT_corr}
shows such a profile determined from the stacked temperature image.

At this point, it is useful to consider the underlying physics
represented by the various patterns in the stacked images.  During
recombination, the sound horizon extends an angle $\theta_\mathrm{s} =
r_\mathrm{s}/D_\mathrm{A} \approx 0.011$~(0.61\deg), where
$r_\mathrm{s} \approx 0.15\,\mathrm{Gpc}$ is the size of the sound
horizon at recombination and $D_\mathrm{A} \approx 14\,\mathrm{Gpc}$
is the angular-diameter distance to the last scattering surface. To
understand the ring patterns in the stacking image, projection effects
must be taken into account. Firstly, all 3D modes with wavenumber
$k\ge \ell/D_\mathrm{A}$ contribute to a 2D $\mathbf{\ell}$-mode.
More modes contribute to, and therefore enhance the power at lower
$\ell$. For the first acoustic peak, the net effect is a $\pi/4$ phase
shift towards lower $\ell$, such that $\ell_\mathrm{s} \approx (\pi -
\pi/4) / \theta_\mathrm{s} \approx 220$. The projected acoustic scale on the
temperature map is of order $\theta_\mathrm{s}^\mathrm{2D} =
\pi/\ell_\mathrm{s} = 0.014$~(0.81\deg). Secondly, the stacked 2D
modes around peaks interfere with each other. The first dark ring
appears at $1.22\theta_\mathrm{s}^\mathrm{2D} \approx
0.017$~(1.0\deg). The factor $1.22$ is the ratio
of the first minimum of the projection kernel, the Bessel function
$J_0$, to the first minimum of the unprojected cosine wave.

The dark ring can also be regarded as a consequence of the correlation between
$T$ and $-\nabla^2T$. At the temperature maxima $-\nabla^2
T$ is positive, with an amplitude of order $
T_\mathrm{peak}/(\theta_\mathrm{s}^\mathrm{2D})^2$. Thus, the
quadratic terms in the local expansion of the temperature field have a
negative contribution that grows as $ - T_\mathrm{peak}
(\varpi/\theta_\mathrm{s}^\mathrm{2D})^2$. At $\varpi \gtrsim
\theta_\mathrm{s}^\mathrm{2D}$ the quadratic terms dominate and the
$T$-$(-\nabla^2T)$ correlation becomes negative. Meanwhile, the
$T$-$(-\nabla^2T)$ correlation tends to zero on the scale $\varpi
\gtrsim \theta_\mathrm{s}^\mathrm{2D}$, where the temperature
autocorrelation becomes weak and the local quadratic expansion starts
to fail. As shown in Fig.~\ref{fig:stack_T_LapT_corr}, the dark ring
appears at the critical point where the $T$-$(-\nabla^2T)$ correlation
reaches its minimum and turns back toward zero.

We have discussed the projection effects that make the projected
radial acoustic scale on a stacked $T$ image larger than
$\theta_\mathrm{s}$. For $Q_\mathrm{r}$, the most striking patterns in
the image have more intuitive simple explanations, since the stacking is
essentially the real-space equivalent of the temperature polarization
correlation.  The projection function contains an extra $\ell^2$
factor, which enhances the high-$\ell$ power and reduces the projected
radial acoustic scale, coincidentally, back to $\approx \theta_s$.
The quadrupole responsible for the polarization around peaks is
induced by gravity on angular scales larger than twice the size of the
horizon at decoupling. In the case of an overdensity, this causes a
flow of photons towards the gravitational well on these scales,
inducing a quadrupolar pattern \citep[see, e.g.][]{coulson:1994}. The
spherical symmetry of the gravitational interaction causes a rotation
of the quadrupole in the vicinity of the well, resulting in a radial
configuration in polarization. This radial polarization pattern
implies $Q_\mathrm{r} > 0$ and an overdensity implies $T < 0$ by the
Sachs-Wolfe formulae, which leads to anticorrelation on these
scales. Similarily, an underdensity leads to an outward flow and
induces a tangential polarization pattern, once again leading to
anticorrelation on these scales. At smaller scales, the polarized
contribution is dominated by the dynamics of the photon fluid. The
acoustic oscillations modulate the polarization pattern, leading to the
different rings in the stacked images. The most noticeable rings in
the stacked $Q_\mathrm{r}$ image are approximately at
$\theta_\mathrm{s}$ and $2\theta_\mathrm{s}$. Thanks to the $\ell^2$
enhancement, multiple acoustic peaks in the $TE$ power spectrum may
be captured and projected into ring patterns in the stacked
polarization images. As photons flow towards the
overdensity, they are compressed and the temperature
increases, slowing the fluid descent into the
well. Eventually, the radiation pressure becomes large enough to
reverse the photon flow. This expansion cools the photons until they
fall back towards the well. Note that the resulting inner ring was not
observed in the WMAP analysis \citep{komatsu:2011}, since the
resolution was too low.

Figure~\ref{fig:dx11_commander_stacking_patch} clearly reveals all of
the features described above. The two bright rings at
$\theta_\mathrm{s} \approx 0.011$~(0.6\deg) and $2\theta_\mathrm{s}
\approx 0.021$~(1.2\deg) are the predicted patterns associated with
the first $C_\ell^{TE}$ acoustic peak at $\ell \approx 310$, while the
two faint rings are a striking illustration of the detection of
multiple acoustic peaks in the $TE$ power spectrum. The large-scale
anticorrelation is suppressed due to the scale-dependent bias which
results from the fact that peaks are defined by the second derivatives
of the temperature field \citep[e.g.,][]{desjacques:2008}.

We are now in a position to discuss the consistency of the \Planck\
results with the predictions of a $\Lambda$CDM cosmology.  For
simplicity, further analysis is focused on the angular profiles, and
specifically the mean, $\mu(\theta)$, estimated as the average of the
angular profiles around all hot (cold) peaks above (below) a certain
threshold $\nu$.  This analysis is performed directly on the sphere to
avoid any repixelization error.  Note that the expected value of the
mean temperature angular profile is proportional to $\int \ell\, \dd\ell \,
C_{\ell}^{TT}J_0(\ell\theta)$, whilst the expected values of
the \Qr\ and \Ur\ mean angular profiles are approximately  proportional to
$\int -\ell \, \dd\ell\,  C_{\ell}^{TE}J_2(\ell\theta)$
and $\int - \ell\,  \dd \ell \, C_{\ell}^{TB}J_2(\ell\theta)$,
respectively.  Since $T$ has even parity  and $B$ has odd parity, the
expectation value for $C_{\ell}^{TB}$ is zero, and the \Ur\
mean angular profile is therefore expected to vanish.

\begin{figure*}
\includegraphics[scale=0.38]{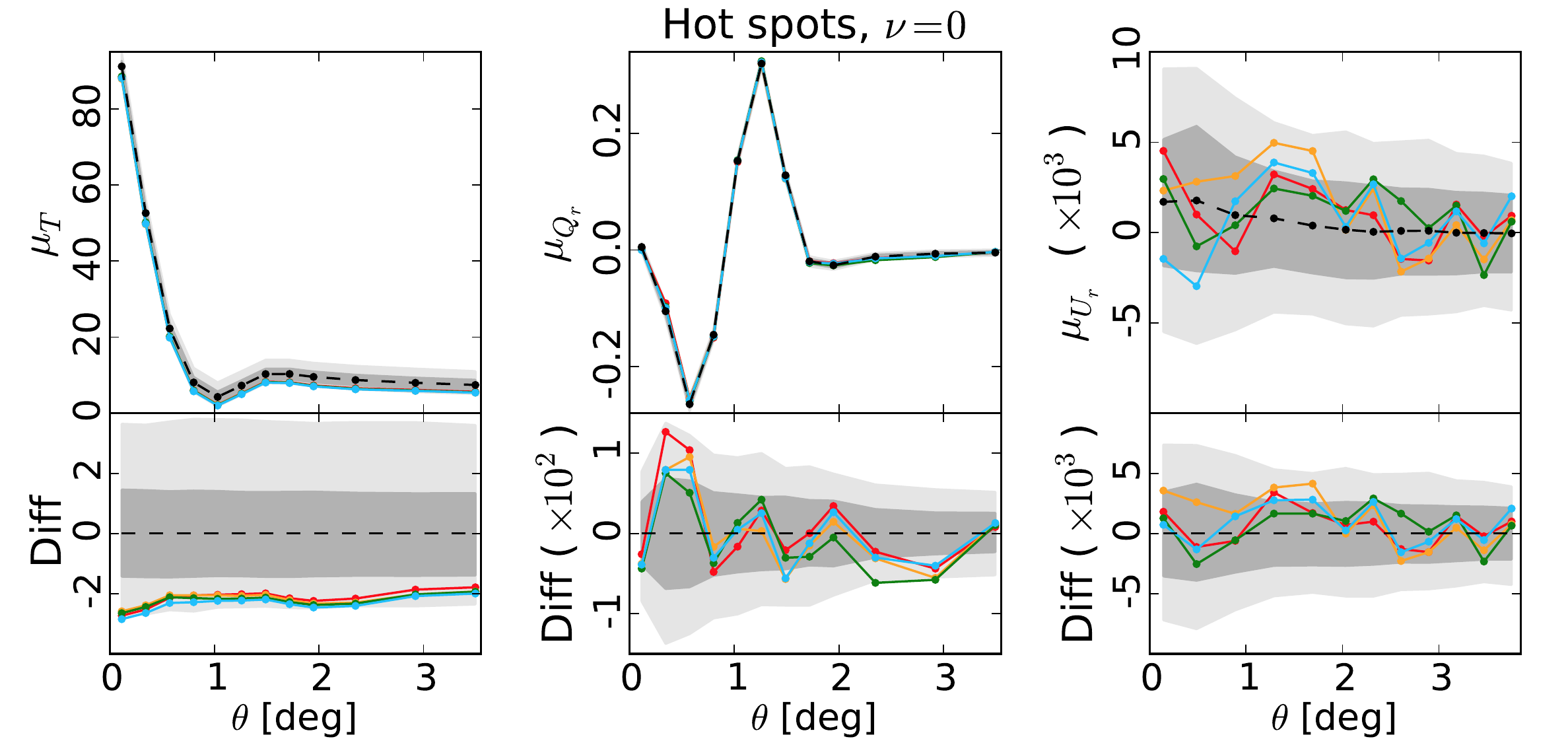}%
\includegraphics[scale=0.38]{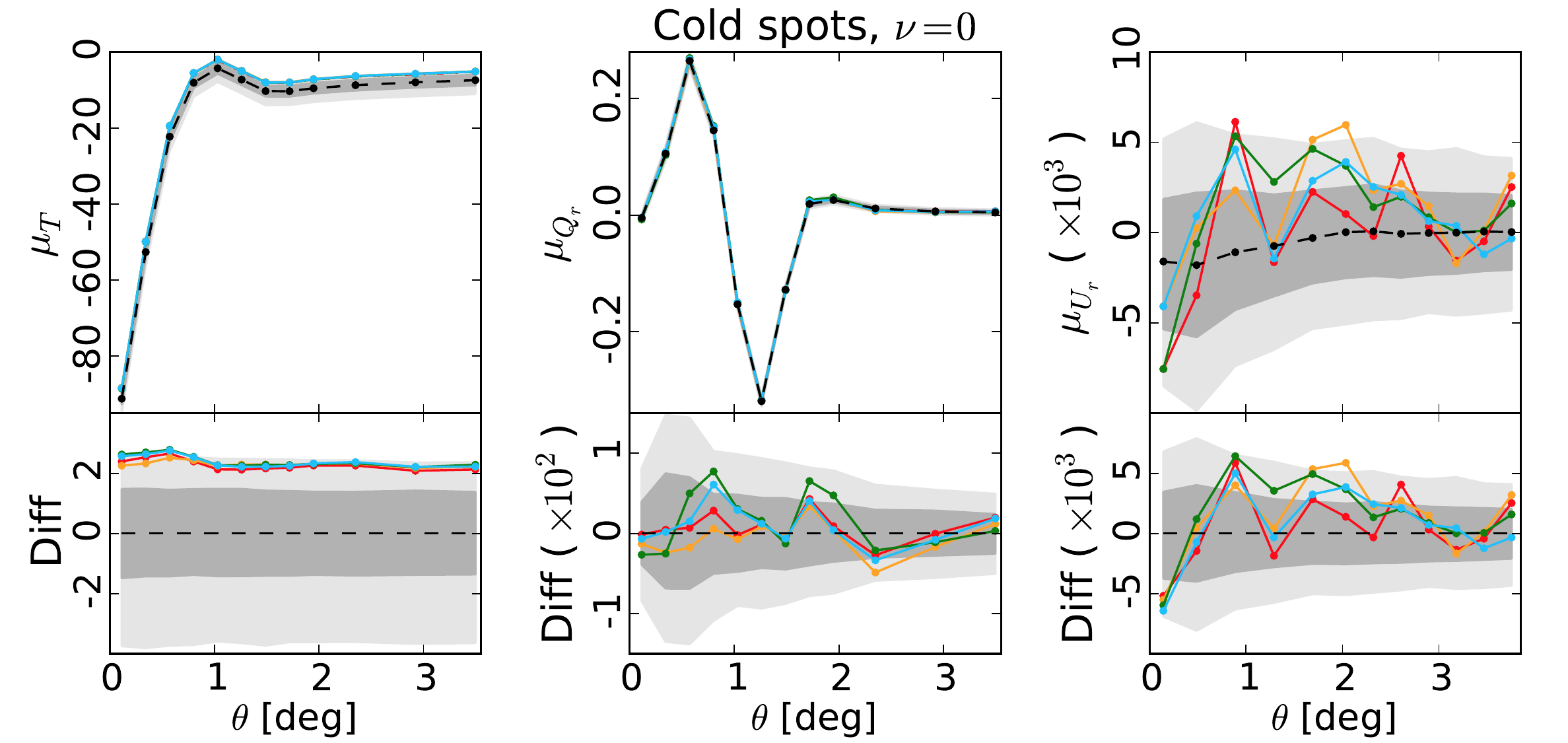}  \\
\includegraphics[scale=0.38]{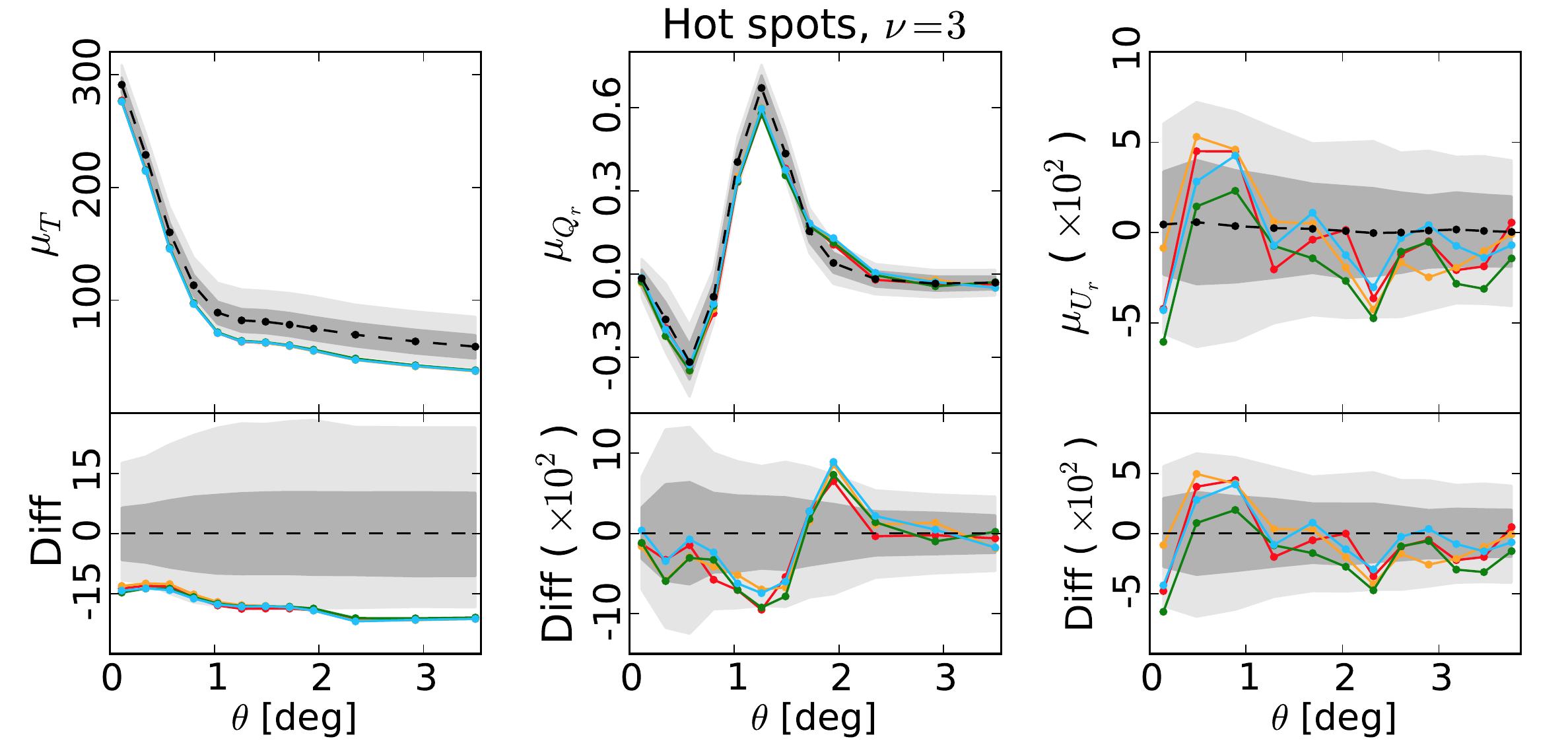}%
\includegraphics[scale=0.38]{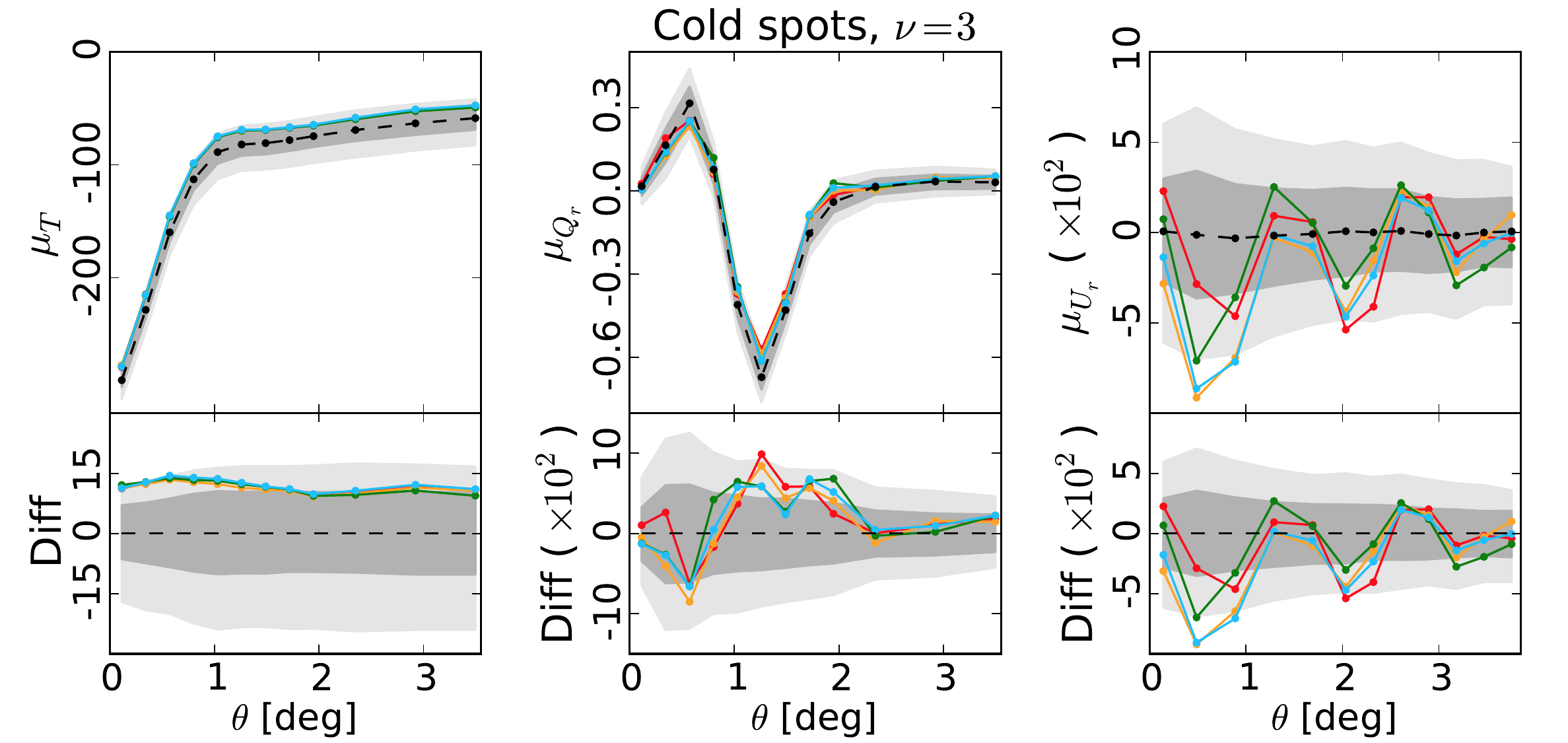}
\caption{Mean radial profiles of $T$, \Qr, and \Ur\ in \microK\
  obtained for \commander\ (red), \nilc\ (orange), \sevem\ (green), and
  \smica\ (blue). Each individual panel contains ({top}) the mean
  radial profiles and ({bottom}) the differences (denoted ``Diff'')
  between the mean profiles of the data and those computed from the
  ensemble mean of the simulations. Results based on stacks around
  temperature hot and cold
  spots are shown in the left and right columns, respectively.
  Upper plots present results for peaks selected above the null threshold, while
  lower plots show the equivalent results for peak amplitudes above
  (hot spots) or below (cold spots) $3$ times the dispersion of the
  temperature map.  The black dots (connected by dashed lines) depict the mean
  profiles and the shaded regions correspond to the $1\,\sigma$ ($68\,\%$) and
  $2\,\sigma$ ($95\,\%$) error bars. The mean profiles and error bars are determined
  from \sevem\ simulations. Note that the Diff curves for each component-separation method
  are computed using the corresponding ensemble average,
  although only the ensemble average from \sevem\ is shown here.}
\label{fig:dx11_commander_stacking_prof}
\end{figure*}

A $\chi^2$ estimator is used to quantify the differences between the
profiles obtained from the data and the expected values estimated with
simulations:
\begin{linenomath*}
\begin{equation}
\chi^2 = \left[\mu(\theta) - \bar{\mu}(\theta)\right] \mathbf{C}^{-1}\left[\mu(\theta) - \bar{\mu}(\theta)\right]^\tens{T},
\end{equation}
\end{linenomath*}
with the covariance matrix defined as
\begin{linenomath*}
\begin{equation}
C(i,j) = \dfrac{1}{N-1}\sum_{k=1}^N{\left[\mu_k(\theta_i)- \bar{\mu}(\theta_i)\right]\left[\mu_k(\theta_j)- \bar{\mu}(\theta_j)\right]},
\end{equation}
\end{linenomath*}
where the sum is over the $N$
simulations used to estimate this matrix and $\bar{\mu}(\theta)$ is
the ensemble average. Note that although the profiles in
Fig.~\ref{fig:dx11_commander_stacking_patch} are derived from data at
a resolution $N_{\mathrm{side}} = 1024$, faster convergence of the
$\chi^2$ statistic is achieved using maps at a lower resolution. We
have verified that the results remain unchanged when adopting data
with $N_{\mathrm{side}} = 512$.

Figure~\ref{fig:dx11_commander_stacking_prof} presents a comparison
between the profiles obtained from the component-separated data and
the mean value estimated from simulations processed through the
\sevem\ pipeline. Note that the error bars for the temperature
profiles are asymmetric due to a bias in the selection of the peaks
above a given threshold. Results for hot and cold spots are shown for two
different thresholds, $\nu = 0$ and $\nu = 3$.  There is
generally excellent agreement between the different component-separation
methods. 
A systematic deviation between the data and the simulations for the hot
peaks in temperature ($\nu = 0$) is seen at a level greater than
$1\,\sigma$. This discrepancy increases at higher thresholds, reaching
values of about $2\,\sigma$ for the $\nu = 3$ case. Similar behaviour
is seen for the cold spots.  For the \Qr\ angular profiles, the most
striking differences appear around $\theta = 2\degr$ in the $\nu = 3$
case for hot peaks, and around $\theta = 1\fdg5$ for the cold
peaks.  For the \Ur\ angular profiles, where a null signal is expected
(i.e., only noise is expected to be present), deviations at similar
levels are seen.

\begin{figure}
\includegraphics[scale=0.35]{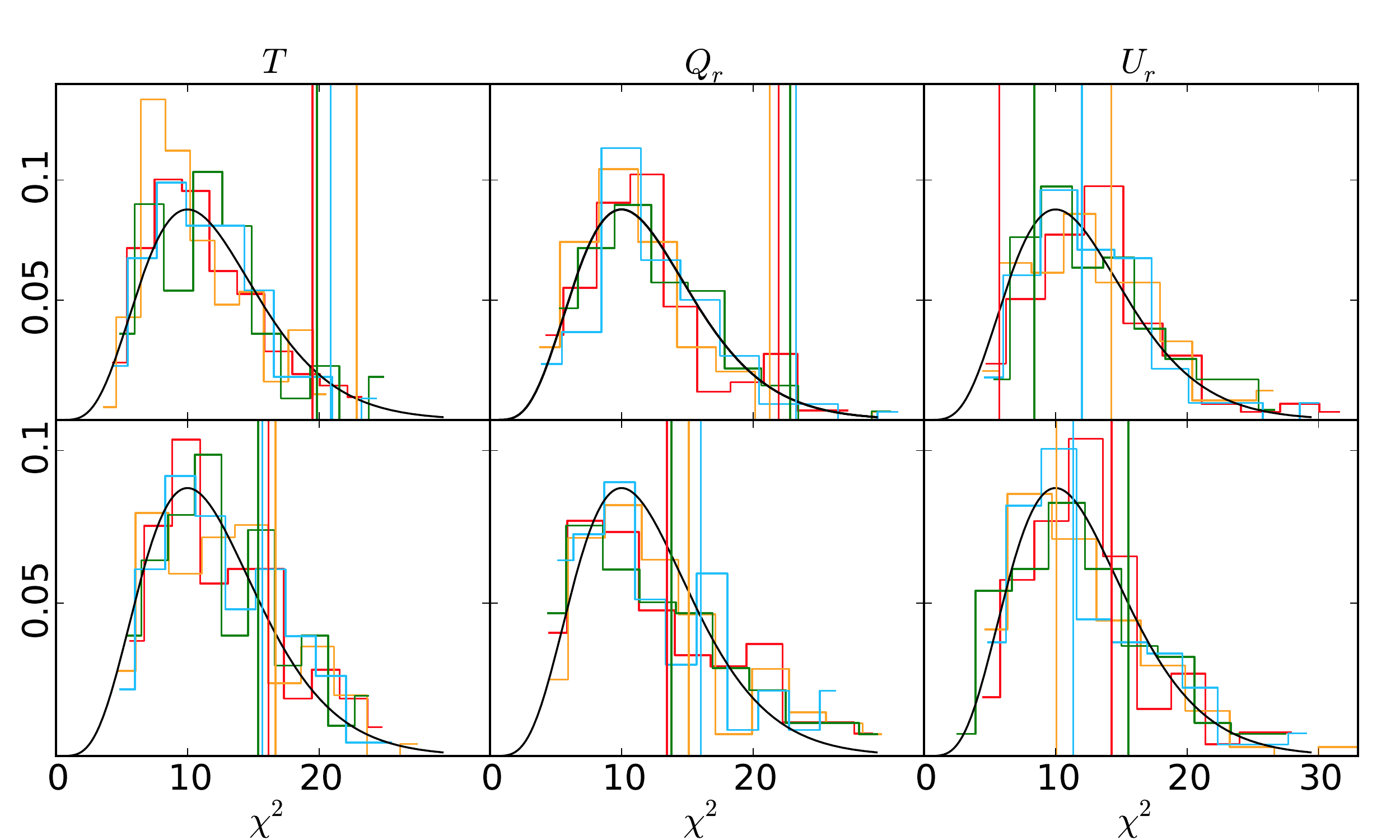}
\caption{$\chi^2$ distributions obtained from the $T$ (left column),
  \Qr\ (middle column), and \Ur\ (right column) mean radial profiles
  centred on temperature hot spots selected above the null threshold (upper row) and three times the
  dispersion of the map (bottom row). The
  black lines correspond to the theoretical $\chi^2$ distribution
  with $12$ degrees of freedom, whilst the histograms show the
  distributions determined from 100 simulations computed through the \commander\ (red), \nilc\ (orange), \sevem\ (green), and \smica\ (blue) pipelines. The vertical lines
  represent the $\chi^2$ values obtained from the data.}
\label{fig:dx11_sevem_stacking_chi2}
\end{figure}

Table~\ref{table:chi2_stacking} presents the corresponding $p$-values
for this comparison.  A theoretical $\chi^2$ distribution
is used to determine the probability that a sky drawn from the
$\Lambda$CDM cosmology has a value larger than that derived from the
data.  We have verified this approach by comparing the empirical
$\chi^2$ distribution estimated from $100$ simulations (in which the
mean value and the covariance matrix are computed from a further $900$
simulations) with the theoretical distribution with the corresponding
degrees of freedom (see Fig.~\ref{fig:dx11_sevem_stacking_chi2}).  The
$\chi^2$ value of the data is then estimated using the mean value and
the covariance matrix determined from simulations.  Although some
differences are found among the component-separation methods, a
general consistency between model and data is found.

Although the $\chi^2$ test has the advantage of being sensitive to
different types of deviations between model and data, does not assume
prior knowledge about possible departures from the model, and can
account for correlations between the various tests from which it  is
constructed, it can nevertheless be suboptimal under certain
conditions.  This appears to be the case when considering the
systematic shift between data and simulations seen in the temperature
profiles $\mu_T$ --- the $\chi^2$ statistic is not particularly sensitive to
systematic deviations of constant sign. We therefore consider an
alternative quantity, the integrated profile deviation, defined as
\begin{linenomath*}
\begin{equation}
\Delta\mu_T(W) =  \int_{0}^{R} \left[\mu_T(\theta) - \bar{\mu}_T(\theta)\right] W(\theta) \, \dd\theta\, , \label{eq:profile_difference}
\end{equation}
\end{linenomath*}
where $R$, the size of stacking patches, is taken to be 3\fdg5  in
this case. The weighting function is chosen to be proportional to the
expected profile, but the results are robust for other choices, e.g.,
$W=1$.  The $p$-values obtained in this case are given in
Table~\ref{table:tau_stacking}. These are consistent with what might
be expected from visual inspection of the plots, i.e., the deviations
are typically close to $2\,\sigma$. These deviations are likely to be
connected to the deficit in the observed power spectrum at low
multipoles, as may be seen in Fig.~\ref{fig:stack_T_LapT_corr}. Here,
the purple dashed line indicates the reduction in $\bar{\mu}_T$ if the
theoretical $C_\ell$ values are reduced by 10\,\% over the range $2\le
\ell \le 50$.

\begin{table}[tp]
\begingroup
\newdimen\tblskip \tblskip=5pt
\caption{$p$-values of the $T$, \Qr, and \Ur\ angular profiles computed
  from the stacking of hot and cold spots selected above the $\nu=0$
  and $\nu=3$ thresholds. }
\label{table:chi2_stacking}
\nointerlineskip
\vskip -3mm
\footnotesize
\setbox\tablebox=\vbox{
\halign{\hbox to 1.0in{#\leaderfil}\tabskip 3pt&
\hfil#\hfil\tabskip 10pt&
\hfil#\hfil\tabskip 8pt&
\hfil#\hfil\tabskip 8pt&
\hfil#\hfil\/\tabskip 0pt\cr
\noalign{\doubleline}
\noalign{\vskip -1pt}
\omit&\multispan4 \hfil Probability [\%]\hfil\cr
\noalign{\vskip -4pt}
\omit&\multispan4\hrulefill\cr
\omit& {\tt Comm.}& {\tt NILC}& {\tt SEVEM}& {\tt SMICA}\cr
\noalign{\vskip 3pt\hrule\vskip 3pt}
\multispan5\hfil $\nu = 0$ (hot spots)\hfil\cr
 $T$ & 8 & 3 & 7 & 5\cr
 $Q_{\mathrm{r}}$ & 4 & 5 & 3 & 3\cr
 $U_{\mathrm{r}}$ & 93 & 28 & 75 & 44\cr
\noalign{\vskip 3pt}
\multispan5\hfil $\nu = 3$ (hot spots)\hfil\cr
 $T$ & 18 & 16 & 22 & 21\cr
 $Q_{\mathrm{r}}$ & 34 & 23 & 31 & 19\cr
 $U_{\mathrm{r}}$ & 28 & 61 & 21 & 50\cr
\noalign{\vskip 3pt}
\multispan5\hfil $\nu = 0$ (cold spots)\hfil\cr
 $T$ & 23 & 38 & 29 & 39\cr
 $Q_{\mathrm{r}}$ & 86 & 85 & 63 & 78\cr
 $U_{\mathrm{r}}$ & 14 & 11 & 39 & 34\cr
\noalign{\vskip 3pt}
\multispan5\hfil $\nu = 3$ (cold spots)\hfil\cr
 $T$ & 24 & 21 & 23 & 20\cr
 $Q_{\mathrm{r}}$ & 21 & 51 & 29 & 52\cr
 $U_{\mathrm{r}}$ & 30 & 13 & 30 & 8\cr
\noalign{\vskip 3pt\hrule\vskip 3pt}
}}
\endPlancktable                    
\endgroup
\end{table}

\begin{table}[tp]
\begingroup
\newdimen\tblskip \tblskip=5pt
\caption{$p$-values of $\Delta \mu_T$ computed from the stacking of hot and cold spots selected above the $\nu=0$ and $\nu=3$ thresholds from the {\tt Commander}, {\tt NILC}, {\tt SEVEM}, and {\tt SMICA} maps.}
\label{table:tau_stacking}
\nointerlineskip
\vskip -3mm
\footnotesize
\setbox\tablebox=\vbox{
\halign{\hbox to 1.0in{#\leaderfil}\tabskip 3pt&
\hfil#\hfil\tabskip 10pt&
\hfil#\hfil\tabskip 8pt&
\hfil#\hfil\tabskip 8pt&
\hfil#\hfil\/\tabskip 0pt\cr
\noalign{\doubleline}
\noalign{\vskip -1pt}
\omit&\multispan4 \hfil Probability [\%]\hfil\cr
\noalign{\vskip -4pt}
\omit&\multispan4\hrulefill\cr
\omit& {\tt Comm.}& {\tt NILC}& {\tt SEVEM}& {\tt SMICA}\cr
\noalign{\vskip 3pt\hrule\vskip 3pt}
\multispan5\hfil Hot spots\hfil\cr
 $T$ ($\nu = 0$) & 96.0 & 95.8 & 96.2 & 97.1\cr
 $T$ ($\nu = 3$) & 98.6 & 98.2 & 98.3 & 98.7\cr
\noalign{\vskip 3pt}
\multispan5\hfil Cold spots\hfil\cr
 $T$ ($\nu = 0$) & 97.1 & 96.9 & 98.1 & 97.9\cr
 $T$ ($\nu = 3$) & 92.0 & 90.6 & 90.6 & 93.0\cr
\noalign{\vskip 3pt\hrule\vskip 3pt}
}}
\endPlancktable                    
\endgroup
\end{table}
\subsection{Generalized stacking}
\label{sec:oriented_stacking}

\def \be{\begin{linenomath*} \begin{equation}}
\def \ee{ \end{equation} \end{linenomath*}}
\def \bea{\begin{linenomath*} \begin{eqnarray}}
\def \eea{ \end{eqnarray} \end{linenomath*} }
\def \vecn{{\mathbf{n}}}
\def \ave#1{{\left\langle #1 \right\rangle}}
\def \bandi#1{{\mathcal{T}_{#1}}}
\def \bandp#1{{\mathcal{P}_{#1}}}
\def \myfigwidth{0.48\textwidth}

In this section, a much wider class of stacking methods is introduced,
with particular emphasis on \emph{oriented stacking}, a novel approach
that has not previoulsy been explored in the literature.
We regard the stacking as \emph{oriented} if the orientation of the
local coordinate frame, and in particular the $\phi = 0$ axis, is
correlated with the map that is being stacked.  Thus, the stacking
methodology in Sect.~\ref{sec:stacking} is considered unoriented,
because the orientation is defined relative to the local meridian
pointing towards the Galactic south, rather than any property of the
data themselves.  Alternative approaches to unoriented stacking can also be
considered. In this subsection, the orientation of each patch is
chosen randomly from a uniform distribution in $[0, 2\pi)$. The
unoriented $T$ and $Q_\mathrm{r}$ images can then be directly compared with
previous sections.

For unoriented stacking, the ensemble average of stacked fields cannot
result in any intrinsic $\phi$-dependence, as this would be averaged
out by the uncorrelated orientation choices. The $\phi$-dependence due
to a specific choice of representation can always be removed via a
local rotation. For example, the ensemble averages of $Q+iU$ around
unoriented temperature peaks are proportional to $e^{2i\phi}$. A local
rotation $(Q, U) \rightarrow (Q_\mathrm{r}, U_\mathrm{r})$ \citep{kamionkowski:1997}
removes the $e^{2i\phi}$ factor and compresses the information into a
single real map $Q_\mathrm{r}$. For oriented stacking, the $\phi$-dependence
can be a mixture of a few Fourier modes ($e^{im\phi}$, for integer $m$).
 Each $m$ mode corresponds to a radial
($\varpi$-dependent) function.

In what follows, we use the $N_{\rm side} = 1024$ component-separated
maps at a resolution of $10\arcmin$ FWHM.  The use of this higher
resolution as compared to the $N_{\rm side} = 512$ data used in
Sect.~\ref{sec:stacking} is motivated by the smaller-scale features
that are expected to result from the oriented stacking.

We also introduce the concept of the noise-free ensemble average
(NFEA), which is defined as the ensemble average of stacked CMB-only
maps for a fiducial cosmology. Recall that the fiducial model for the
simulated sky maps, the \Planck\ 2013 best-fit $\Lambda$CDM
model~\citep{planck2013-p11}, differs from the updated \Planck\ 2015
best-fit $\Lambda$CDM model~\citep{planck2014-a15}.  In previous
sections, this mismatch was partially accommodated by rescaling the
CMB signal by a fixed scale factor. Here, we instead specifically
adopt the 2015 best fit as a fiducial model for the data.  When
comparing the data to the simulations, we subtract the corresponding
NFEA to minimize any bias resulting from cosmology dependence.

In the context of random Gaussian fields, the NFEA can be computed
straightforwardly following \citet{BBKS}:
\begin{linenomath*}
\begin{equation}
\ave{M} = \ave{M \vec{w}^\tens{T}} \ave{\vec{w} \vec{w}^\tens{T}}^{-1} \ave{\vec{w}} \,, \label{eq:stack_mean}
\end{equation}
\end{linenomath*}
where $M$ is the map (around the central peak) to be stacked, and
$\vec{w}$ is the collection of Gaussian variables (on the central
peak) that are related to peak selection and orientation
determination. Eq.~\eqref{eq:stack_mean} is only valid for Gaussian
random variables. If the patch is rotated before stacking, the field
value evaluated at a dynamic coordinate is, in general, not a random
Gaussian variable. However, statistical isotropy guarantees that the
rotation of patches is equivalent to an orientation constraint on the
nonzero-spin field. For example, orienting each patch in the direction
where $U=0$ and $Q>0$ is equivalent to the unoriented stacking case
where only peaks satisfying the additional constraint $-\epsilon/2 <
\arg\left(Q+iU\right)< \epsilon/2$ ($\epsilon \rightarrow
0^+$) are selected.

A further source of statistical bias can arise from noise mismatch
between the simulations and the data. Since the effect of noise is to
introduce random shifts in the peaks and hence suppress patterns in
the stacked images, any noise mismatch can lead to pattern mismatch
between the data and simulations.  For the temperature data, the
contribution due to noise mismatch is estimated to be at the
sub-percent level, lower than the cosmic variance. For stacking on
polarization peaks, the impact of the noise mismatch cannot be safely
ignored. Thus, for quantitative comparisons in this paper, we only
consider stacking on temperature peaks.

\subsubsection{Oriented temperature stacking}
\label{subsec:oriented_stacking_temperature}

The most straightforward way to orient a patch centred on a
temperature peak is to align the horizontal axis with the major
axis defined by a local quadratic expansion of the temperature field
around the peak. The disadvantage of doing so is that the orientation
is dominated by small-scale fluctuations that are noise-sensitive.  A
better choice is to use the major axis of the inverse Laplacian
$\nabla^{-2}T$ that filters out the small-scale power. The inverse
Laplacian is defined as:
\begin{linenomath*}
\begin{equation}
\nabla^{-2}T\left(\vec{\hat{n}}\right) = - \sum_{\ell=2}^{\infty} \sum_{m=-\ell}^\ell \frac{a_{\ell m}^T}{\ell (\ell+1)} Y_{\ell m}\left(\vec{\hat{n}}\right) \, ,
\end{equation}
\end{linenomath*}
where $a_{\ell m}^T$ are the harmonic coefficients of the masked
temperature map.  Spin-2 maps $Q_T$, $U_T$ are then defined by:
\begin{linenomath*}
\begin{equation}
\left(Q_T \pm i U_T\right) \left(\vec{\hat{n}}\right) = \sum_{\ell=2}^{\infty}\sum_{m=-\ell}^{\ell} a_{\ell m}^T \left[\,_{\pm 2}Y_{\ell m}(\vec{\hat{n}})\right] \, .
\end{equation}
\end{linenomath*}
In the flat-sky limit, $Q_T \approx (\partial_x^2
- \partial_y^2)(\nabla^{-2}T)$ and $U_T \approx -
2 \partial_x\partial_y \nabla^{-2}T$. To align the
$\nabla^{-2}T$ axes of the patches, we rotate each patch
so that $U_T$ vanishes and $Q_T\ge 0$ for the central peak.

\begin{figure}
\includegraphics[width=\myfigwidth]{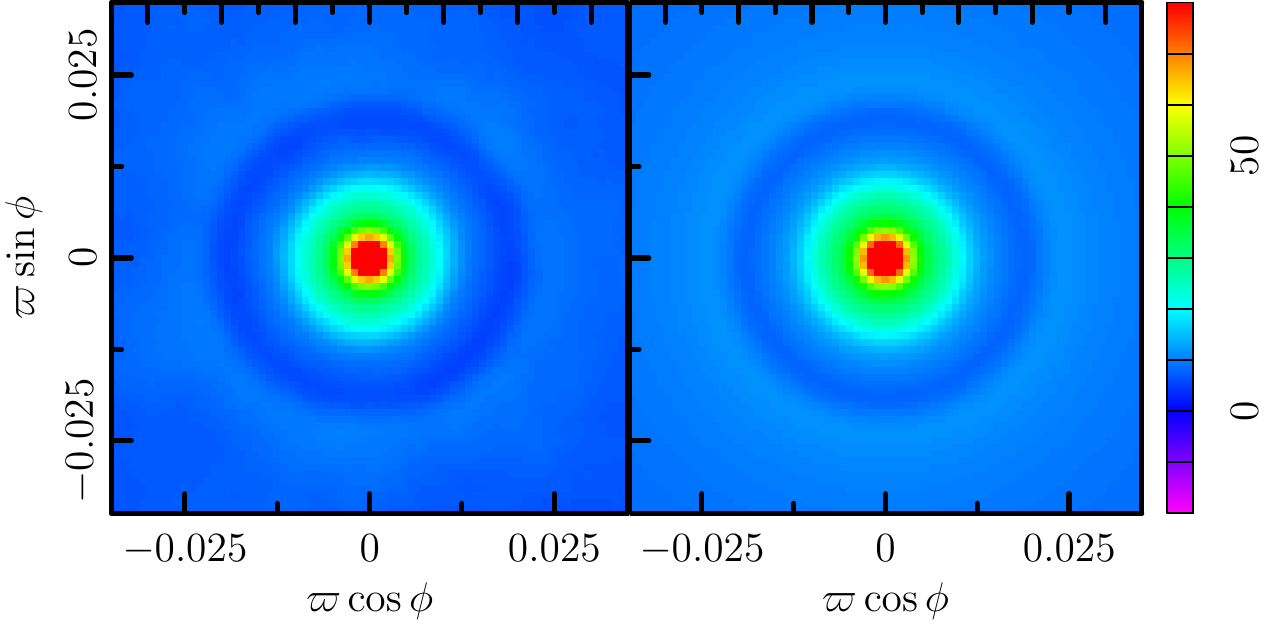}
\includegraphics[width=\myfigwidth]{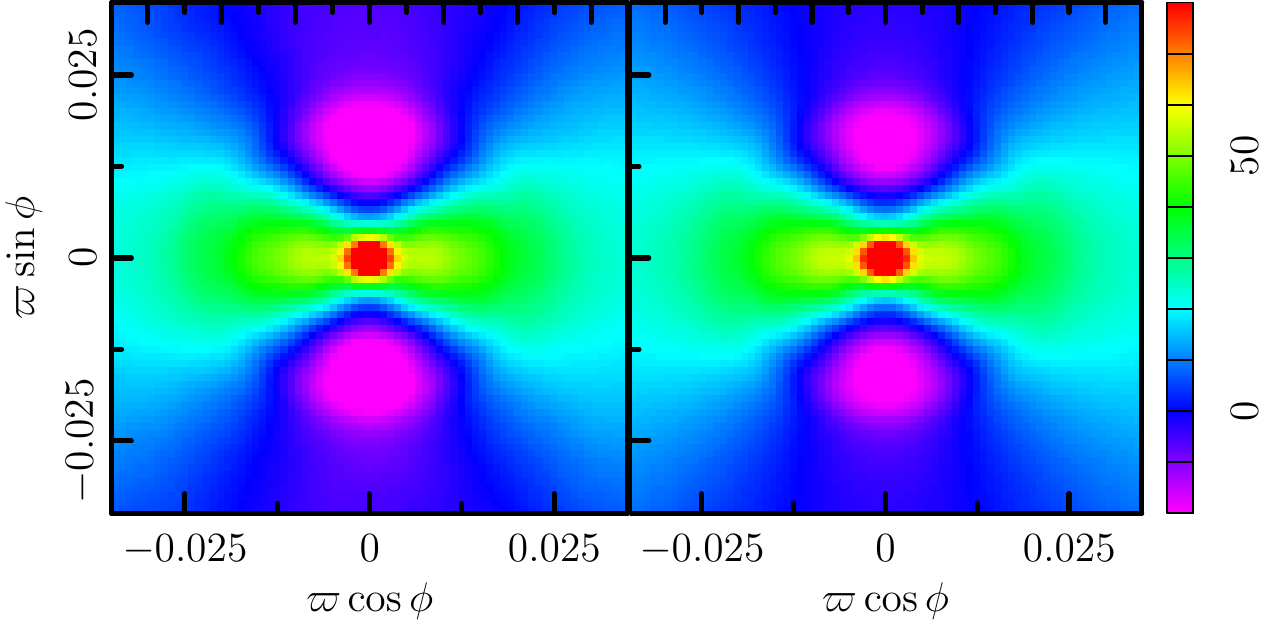}
\caption{Comparison between unoriented stacking (upper panels) and
  oriented stacking (lower panels) of temperature patches around
  temperature hot spots selected above the null threshold ($\nu =
  0$). The left panels are the stacked \smica\ maps, and the right
  panels their corresponding NFEAs. The image units  are \microK.}
  \label{fig:general_stack_T}
\end{figure}

Figure~\ref{fig:general_stack_T} presents the stacked images of
\smica\ temperature patches centred on temperature hot spots selected
above the threshold $\nu=0$, in both unoriented and oriented
forms. These are seen to be in excellent agreement with their
accompanying NFEAs, and, in the case of the unoriented stacks, with
the results shown in Fig.~\ref{fig:dx11_commander_stacking_patch},
despite the different stacking methodologies adopted (and component
separation method selected for visualization purposes).

The oriented $T$ image is notably different from the unoriented
one. The alignment between the major axis (of $\nabla^{-2}T$) and
the horizontal axis results in an ellipse elongated along the horizontal axis,
rather than a central disc. Moreover, the quadratic-term contribution
is suppressed along the horizontal axis where the temperature profile is
smoother, and enhanced along the vertical axis where the temperature
profile is sharper. As a consequence, the dark ring visible in the
upper panel at $1\degr$ splits into two cold blobs along the vertical axis.

To proceed with a quantitative analysis, we extract Fourier modes
$T_m(\varpi)$ from the stacked map $T_{\rm stack}(\varpi,\phi)$ as
follows:
\begin{linenomath*}
\begin{equation}
T_m(\varpi) =  \frac{1}{(1+\delta_{m0})\pi}\int_0^{2\pi} T_{\rm stack}\left(\varpi, \phi\right) \cos{m\phi} \,\dd\phi\,,
\end{equation}
\end{linenomath*}
where $\delta_{m0}$ is the Kronecker delta function.  For odd $m$, the
NFEA $\ave{T_m}$ vanishes due to statistical isotropy.  For even $m$,
a straightforward calculation shows that only $T_0(\varpi)$, which is
equivalent to $\mu_T(\varpi)$, and $T_2(\varpi)$ have nonzero NFEAs.

As discussed previously in Sect.~\ref{sec:stacking}, there are some
shortcomings of the standard $\chi^2$ procedure that is generally used
to assess the consistency of the data with simulations.  The problem
is simplified by studying the statistics of an integrated profile
deviation:
\begin{linenomath*}
\begin{equation}
\bandi{m}(W) =  \int_{0}^{R} \left[T_m(\varpi) -\ave{T_m(\varpi)}\right]W(\varpi) \, \dd\varpi \, , \label{eq:oriented_stacking_Sm}
\end{equation}
\end{linenomath*}
where $R$, the size of the stacking patches, is taken to be 2\deg\ in
our examples. The purpose of removing the NFEA, $\ave{T_m(\varpi)}$,
which differs for the data and the simulations, is to minimize the
impact of the cosmology dependence. A natural choice for the filter is
$\ave{T_m(\varpi)}$ itself with a proper normalization:
\begin{linenomath*}
\begin{equation}
W(\varpi) = \frac{ \ave{T_m(\varpi) }}{\int_0^R\ave{T_m(\varpi) }^2\dd\varpi} \,  \label{eq:oriented_stacking_WI}.
\end{equation}
\end{linenomath*}
For the filter given by Eq.~\eqref{eq:oriented_stacking_WI}, the
integrated profile deviation $\bandi{m}$ describes the relative
deviation from the NFEA. If $\Lambda$CDM is the correct model, the
deviation is due to cosmic variance and noise. The distribution of
$\bandi{m}$ is obtained from simulations.

Table~\ref{tbl:oriented_stacking_S_comparison} presents a comparison
of the $\bandi{m}$ values derived from the \Planck\ data and the FFP8
simulations.  No inconsistencies in excess of the $3\,\sigma$ level have
been found, although tensions around $2\,\sigma$ are seen.

The $m=0$ projection kernel $J_0[(\ell+1/2) \varpi]$ peaks at low
$\ell$. Thus $\bandi{0}$ is cosmic-variance sensitive and the apparent
discrepancy in it could be related to a low-$\ell$ power deficit. An
example is shown in Fig.~\ref{fig:stack_T_LapT_corr} for
illustration. To test the robustness of this result, we have tried
three additional filters: a top-hat filter $W = 1$, a linear filter $W
= \varpi$, and a Gaussian filter $W = \exp(-\varpi^2/\sigma_\mathrm{g}^2)$ with $\sigma_\mathrm{g}
= 1\deg$. In all cases, the power deficit remains at about the
2$\,\sigma$ level.

\begin{table*} \begingroup \newdimen\tblskip \tblskip=5pt
  \caption{$\bandi{m}$, as defined in
    Eqs.~\eqref{eq:oriented_stacking_Sm} and
    \eqref{eq:oriented_stacking_WI}, for different thresholds
    $\nu$. The expected values, together with the $1\,\sigma$ (68\,\% CL) and $2\,\sigma$
    (95\,\% CL) ranges determined from simulations are given in brackets.}
\label{tbl:oriented_stacking_S_comparison}
\nointerlineskip
\vskip -3mm
\footnotesize
\setbox\tablebox=\vbox{
   \newdimen\digitwidth
   \setbox0=\hbox{\rm 0}
   \digitwidth=\wd0
   \catcode`*=\active
   \def*{\kern\digitwidth}
   \newdimen\signwidth
   \setbox0=\hbox{+}
   \signwidth=\wd0
   \catcode`!=\active
   \def!{\kern\signwidth}
\halign{\hbox to 1.0in{#\leaderfil}\tabskip 4pt&
         \hfil#\hfil&
         \hfil#\hfil&
         \hfil#\hfil&
         \hfil#\hfil\tabskip 0pt\cr                           
\noalign{\doubleline\vskip -1pt}
\omit& \multispan2\hfil $\bandi{0}$\hfil&\multispan2\hfil $\bandi{2}$\hfil\cr
\noalign{\vskip -4pt}
\omit&\multispan2\hrulefill&\multispan2\hrulefill\cr
\omit\hfil Method\hfil& Hot spots & Cold spots & Hot spots & Cold spots \cr 
\noalign{\vskip 3pt\hrule\vskip 3pt}
\omit & \multispan4\hfil threshold $\nu = 0$ \hfil \cr
\noalign{\vskip 3pt}
{\tt Commander} & $-0.03(!0.04^{+0.06+0.15}_{-0.04-0.07})$
& $-0.04(!0.04^{+0.06+0.16}_{-0.04-0.07})$
& $0.07(0.04^{+0.01+0.03}_{-0.02-0.04})$
& $0.06(0.04^{+0.01+0.03}_{-0.02-0.04})$\cr
\noalign{\vskip 2pt}
{\tt NILC} & $-0.04(!0.04^{+0.06+0.15}_{-0.04-0.07})$
& $-0.05(!0.04^{+0.06+0.16}_{-0.04-0.07})$
& $0.06(0.03^{+0.01+0.02}_{-0.02-0.04})$
& $0.05(0.03^{+0.01+0.03}_{-0.02-0.04})$\cr
\noalign{\vskip 2pt}
{\tt SEVEM} & $-0.03(!0.04^{+0.06+0.16}_{-0.04-0.07})$
& $-0.04(!0.04^{+0.06+0.16}_{-0.04-0.07})$
& $0.06(0.04^{+0.01+0.03}_{-0.02-0.04})$
& $0.06(0.04^{+0.01+0.03}_{-0.02-0.04})$\cr
\noalign{\vskip 2pt}
{\tt SMICA} & $-0.03(-0.01^{+0.03+0.07}_{-0.02-0.04})$
& $-0.05(-0.00^{+0.03+0.07}_{-0.02-0.04})$
& $0.06(0.04^{+0.01+0.03}_{-0.02-0.04})$
& $0.06(0.04^{+0.01+0.03}_{-0.02-0.04})$\cr
\noalign{\vskip 3pt}
\omit &\multispan4 \hfil threshold $\nu=1$ \hfil \cr
\noalign{\vskip 3pt}
{\tt Commander} & $-0.06(!0.05^{+0.09+0.22}_{-0.05-0.10})$
& $-0.06(!0.05^{+0.09+0.21}_{-0.06-0.09})$
& $0.06(0.03^{+0.02+0.03}_{-0.02-0.04})$
& $0.04(0.03^{+0.02+0.03}_{-0.02-0.04})$\cr
\noalign{\vskip 2pt}
{\tt NILC} & $-0.06(!0.05^{+0.09+0.22}_{-0.05-0.10})$
& $-0.07(!0.05^{+0.09+0.21}_{-0.05-0.09})$
& $0.06(0.02^{+0.02+0.03}_{-0.02-0.04})$
& $0.04(0.02^{+0.02+0.03}_{-0.02-0.04})$\cr
\noalign{\vskip 2pt}
{\tt SEVEM} & $-0.06(!0.06^{+0.09+0.22}_{-0.05-0.10})$
& $-0.06(!0.06^{+0.09+0.22}_{-0.05-0.10})$
& $0.06(0.03^{+0.02+0.03}_{-0.02-0.04})$
& $0.04(0.03^{+0.02+0.03}_{-0.02-0.04})$\cr
\noalign{\vskip 2pt}
{\tt SMICA} & $-0.06(-0.01^{+0.04+0.10}_{-0.03-0.06})$
& $-0.07(-0.01^{+0.04+0.10}_{-0.03-0.06})$
& $0.06(0.03^{+0.02+0.03}_{-0.02-0.04})$
& $0.04(0.03^{+0.02+0.03}_{-0.02-0.04})$\cr
\noalign{\vskip 4pt\hrule\vskip 3pt}}}
\endPlancktable                    
\endgroup
\end{table*}

Although the $\bandi{0}$ deficit is not significant enough to falsify
the $\Lambda$CDM model, further investigation of its properties may
still be interesting and help us understand the other anomalies
discussed in this paper. We consider two possibilities.  Firstly the
amplitude of the $\bandi{0}$ deficit is of order 5--10\,\%, which
coincides with the level of hemispherical power asymmetry discussed in
Sect.~\ref{sec:varasym}. To test whether the $\bandi{0}$ deficit is
localized on one hemisphere, we define the ``north'' direction to be
aligned with the power asymmetry direction at $(l, b) = (212\deg,
-13\deg)$ \citep{Akrami2014} and compute $\bandi{0}$ on the northern
and southern hemispheres separately. The results are presented in
Table~\ref{tbl:oriented_stacking_ha}. Although the $\bandi{0}$ deficit
is more significant for the southern hemisphere, it remains consistent
with the $\Lambda$CDM prediction.  Secondly, it is of interest to
determine whether the $\bandi{0}$ deficit is related to the \cs\
discussed in Sect.~\ref{sec:coldspot}.  We therefore mask out the \cs\
using a disc of radius 6\deg\ and repeat the calculation. The impact
of this region on the $\bandi{0}$ deficit is insignificant.

\begin{table*} \begingroup \newdimen\tblskip \tblskip=5pt
  \caption{$\bandi{0}$, as defined in
    Eqs.~\eqref{eq:oriented_stacking_Sm} and
    \eqref{eq:oriented_stacking_WI}, for different thresholds
    $\nu$ and hemispheres. The ``north'' hemisphere is centred on the Galactic
    coordinate $(l, b) = (212\deg, -13\deg)$ and the
    ``south'' hemisphere in the opposite direction. The expected
    values, together with the  $1\,\sigma$ (68\,\% CL) and $2\,\sigma$
    (95\,\% CL) ranges determined from simulations are given in
    brackets.}
\label{tbl:oriented_stacking_ha}
\nointerlineskip
\vskip -3mm
\footnotesize
\setbox\tablebox=\vbox{
   \newdimen\digitwidth
   \setbox0=\hbox{\rm 0}
   \digitwidth=\wd0
   \catcode`*=\active
   \def*{\kern\digitwidth}
   \newdimen\signwidth
   \setbox0=\hbox{+}
   \signwidth=\wd0
   \catcode`!=\active
   \def!{\kern\signwidth}
\halign{\hbox to 1.0in{#\leaderfil}\tabskip 4pt&
         \hfil#\hfil&
         \hfil#\hfil&
         \hfil#\hfil&
         \hfil#\hfil\tabskip 0pt\cr                           
\noalign{\doubleline\vskip -1pt}
\omit& \multispan2\hfil ``North'' $\bandi{0}$  \hfil&\multispan2\hfil  ``South'' $\bandi{0}$\hfil\cr
\noalign{\vskip -4pt}
\omit&\multispan2\hrulefill&\multispan2\hrulefill\cr
\omit\hfil Method\hfil& Hot spots & Cold spots & Hot spots & Cold spots \cr 
\noalign{\vskip 3pt\hrule\vskip 3pt}
\omit & \multispan4\hfil threshold $\nu = 0$\hfil\cr
\noalign{\vskip 3pt}
{\tt Commander}&   $-0.02(!0.03^{+0.07+0.16}_{-0.04-0.07})$
& $-0.03(!0.03^{+0.07+0.18}_{-0.04-0.07})$
& $-0.05(!0.03^{+0.07+0.18}_{-0.05-0.07})$
& $-0.06(!0.03^{+0.07+0.18}_{-0.04-0.07})$\cr
\noalign{\vskip 2pt}
{\tt NILC}     &  $-0.02(!0.02^{+0.07+0.16}_{-0.04-0.07})$
& $-0.03(!0.02^{+0.07+0.17}_{-0.04-0.07})$
& $-0.05(!0.02^{+0.07+0.18}_{-0.04-0.07})$
& $-0.06(!0.02^{+0.07+0.18}_{-0.04-0.07})$\cr
\noalign{\vskip 2pt}
{\tt SEVEM}  & $-0.02(!0.03^{+0.07+0.17}_{-0.04-0.07})$
& $-0.03(!0.03^{+0.07+0.18}_{-0.04-0.07})$
& $-0.05(!0.03^{+0.07+0.18}_{-0.05-0.07})$
& $-0.06(!0.03^{+0.07+0.18}_{-0.04-0.07})$\cr
\noalign{\vskip 2pt}
{\tt SMICA}   & $-0.02(-0.01^{+0.04+0.09}_{-0.03-0.05})$
& $-0.03(-0.01^{+0.04+0.09}_{-0.03-0.05})$
& $-0.05(-0.01^{+0.04+0.08}_{-0.03-0.05})$
& $-0.07(-0.01^{+0.04+0.08}_{-0.03-0.05})$\cr
\noalign{\vskip 3pt}
\omit &\multispan4 \hfil threshold $\nu=1$ \hfil\cr
\noalign{\vskip 3pt}
{\tt Commander} & $-0.04(!0.03^{+0.09+0.22}_{-0.06-0.10})$
& $-0.05(!0.03^{+0.09+0.23}_{-0.06-0.10})$
& $-0.08(!0.04^{+0.09+0.25}_{-0.06-0.11})$
& $-0.08(!0.04^{+0.09+0.24}_{-0.06-0.10})$\cr
\noalign{\vskip 2pt}
{\tt NILC}  & $-0.05(!0.03^{+0.10+0.23}_{-0.06-0.10})$
& $-0.06(!0.02^{+0.09+0.23}_{-0.06-0.10})$
& $-0.08(!0.03^{+0.09+0.25}_{-0.06-0.11})$
& $-0.08(!0.03^{+0.09+0.24}_{-0.06-0.10})$\cr
\noalign{\vskip 2pt}
{\tt SEVEM}      & $-0.04(!0.04^{+0.10+0.23}_{-0.06-0.10})$
& $-0.05(!0.03^{+0.10+0.23}_{-0.06-0.10})$
& $-0.08(!0.04^{+0.09+0.25}_{-0.07-0.11})$
& $-0.08(!0.04^{+0.09+0.24}_{-0.06-0.11})$\cr
\noalign{\vskip 2pt}
{\tt SMICA}  & $-0.04(-0.02^{+0.05+0.13}_{-0.04-0.07})$
& $-0.05(-0.02^{+0.05+0.13}_{-0.04-0.07})$
& $-0.08(-0.02^{+0.05+0.11}_{-0.04-0.07})$
& $-0.09(-0.02^{+0.05+0.12}_{-0.04-0.07})$\cr
\noalign{\vskip 4pt\hrule\vskip 3pt}}}
\endPlancktable                    
\endgroup
\end{table*}

Tensions at the $2\,\sigma$ level are also seen for $\bandi{2}$.
However, due to the additional $\ell^2$ factor in the projection
kernel, the oriented ($m=2$) component $\bandi{2}$ is more sensitive
to high-$\ell$ power where the cosmic variance is small, and an
understanding of the noise properties of the data is more
important. The former implies that the related uncertainty in
$\bandi{2}$ is, in general, smaller than that in $\bandi{0}$. However,
a mismatched cosmology, perhaps arising from a different primordial
power amplitude $A_\mathrm{s}$, can then lead to significant tension between
the data and the simulations.  Indeed, we find that without
application of our cosmology calibration (i.e., the subtraction of the
NFEA in Eq.~\ref{eq:oriented_stacking_Sm}) the $\bandi{2}$-tension
between the data and simulations increases by about $0.5\,\sigma$,
whereas the variation of the $\bandi{0}$-tension is $\lesssim
0.2\,\sigma$.  The high-$\ell$ sensitivity of $\bandi{2}$ also requires
the use of an accurate noise model, and it is possible that the
1--2$\,\sigma$ tension in $\bandi{2}$ may be alleviated once improved
noise simulations are available.

\subsubsection{Oriented polarization stacking}
\label{subsec:oriented_stacking_polarization}

\begin{figure}
\includegraphics[width=\myfigwidth]{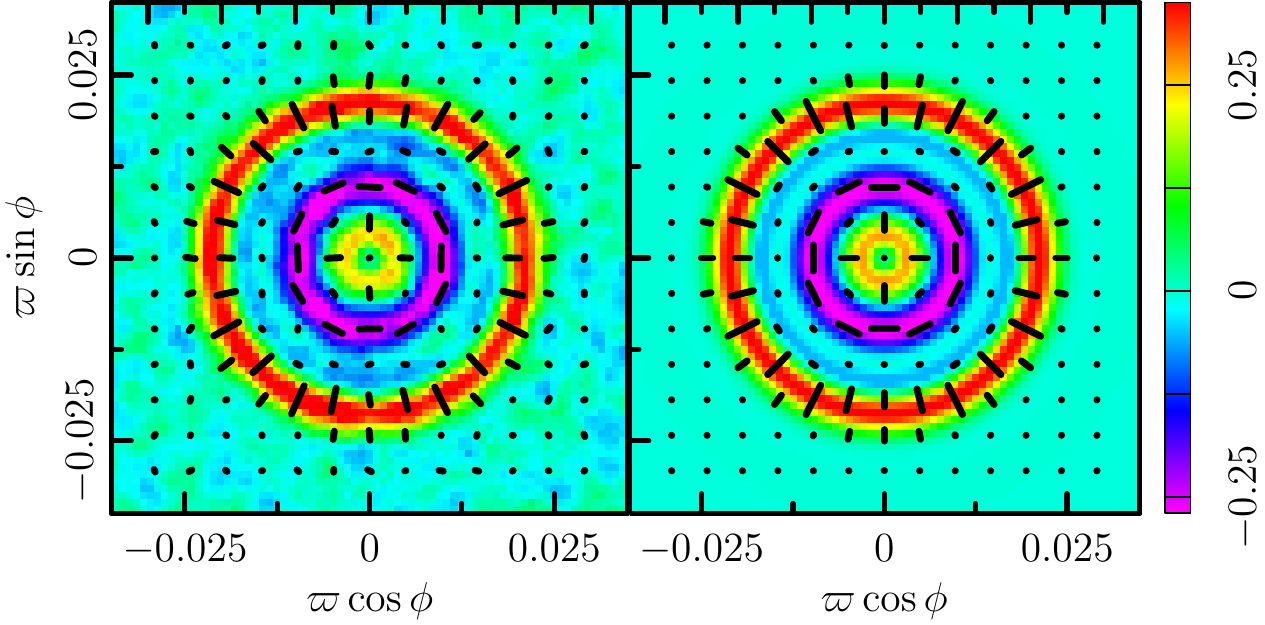}
\caption{Stacked $Q_\mathrm{r}$ image around temperature hot spots selected
  above the null threshold ($\nu = 0$) in the \smica\ sky map.
  The left panel corresponds to the observed data and the right panel
  shows the NFEA. The image units are \microK. The headless
  vectors (black solid lines) are the polarization directions for
  stacked $Q_{\rm stack}$, $U_{\rm stack}$. The lengths of the
  headless vectors are proportional to the polarization amplitude
  $P_\mathrm{stack}$.}
  \label{fig:stacking_Qr_high_resol}
\end{figure}

The stacked $Q$ and $U$ images can be decomposed into Fourier modes,
$Q+iU = \sum_{m=-\infty}^{\infty} P_m(\varpi) e^{im\phi}$. For
unoriented $Q + iU$ stacking on temperature peaks, only $P_2(\varpi)$
has a non-zero NFEA, and it can be linked to the conventional $Q_\mathrm{r}$
stacking via $P_2 = - Q_\mathrm{r}$.  Figure~\ref{fig:stacking_Qr_high_resol}
shows that the stacked $Q_\mathrm{r}$ image is in excellent agreement with its
NFEA and the corresponding stacked image (fourth panel) in
Fig.~\ref{fig:dx11_commander_stacking_patch}, despite the different
stacking methodologies adopted (and component-separation method
selected for visualization purposes). The length and orientation of
the headless vectors represent the polarization amplitude,
$P_\mathrm{stack}\equiv\sqrt{Q_\mathrm{stack}^2+U_\mathrm{stack}^2}$,
and direction.

\begin{figure}
\includegraphics[width=\myfigwidth]{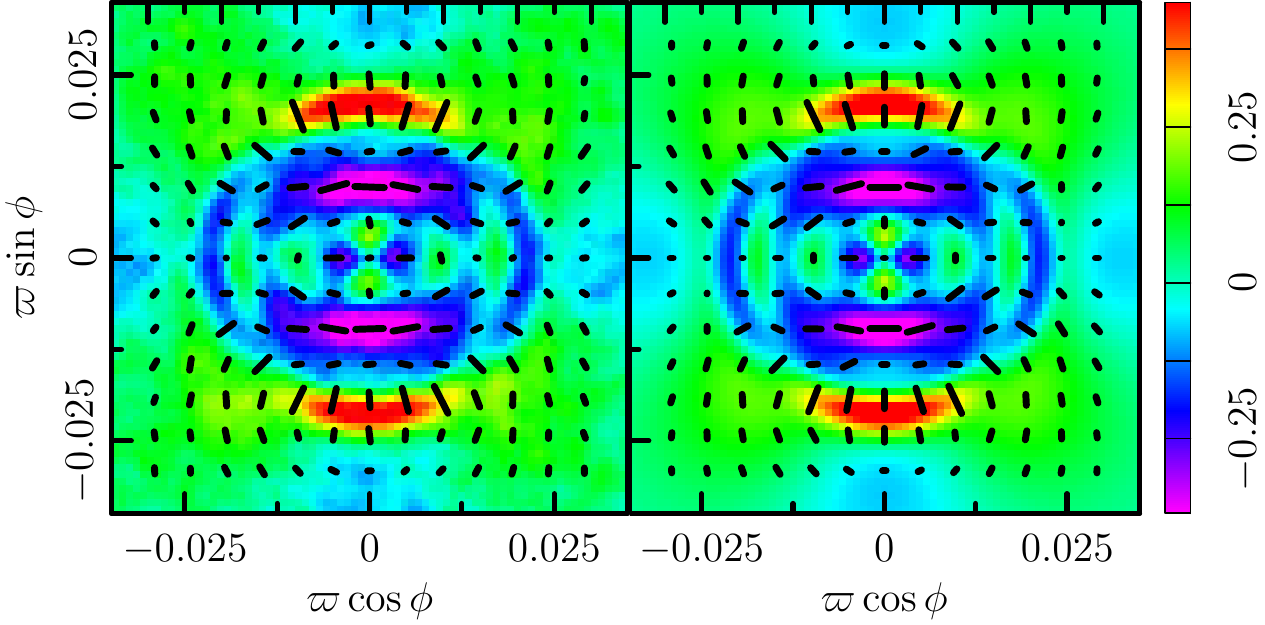}
\includegraphics[width=\myfigwidth]{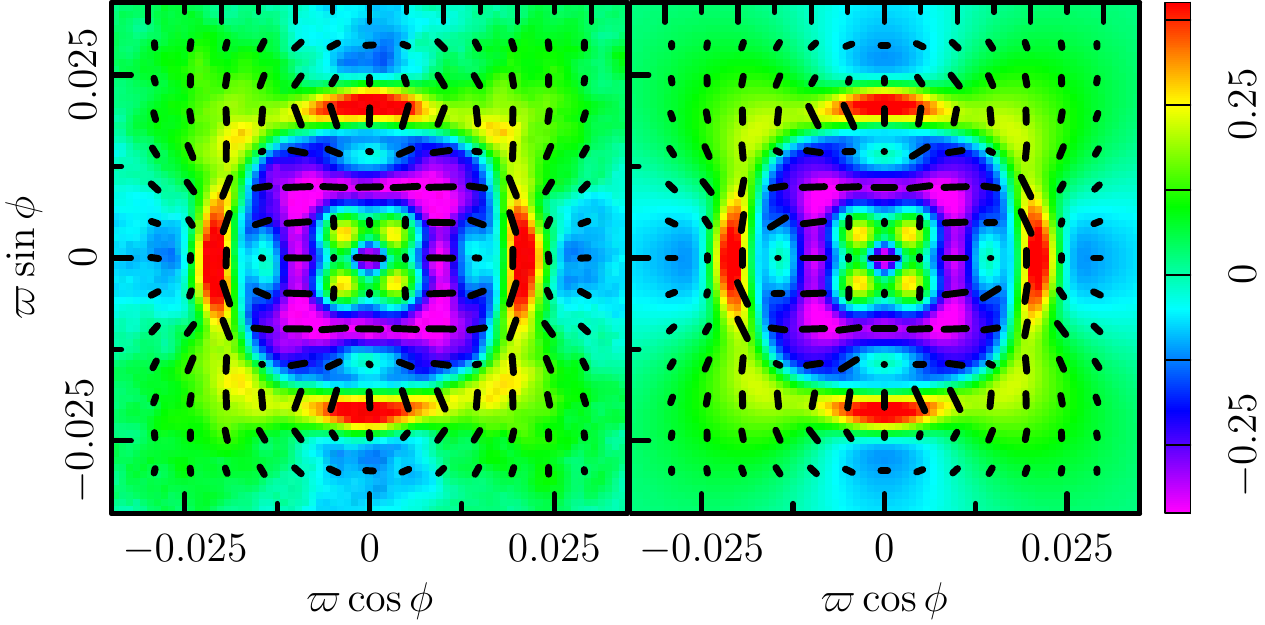}
\caption{Oriented stacking of polarization fields ($Q$, $U$) on
  temperature maxima (upper panels) and $P_T$
  maxima (lower panels). In both cases the threshold $\nu = 0$ is used and the
  orientation is chosen such that $U_T=0$ and $Q_T\ge 0$ on the
  central peak. The image units are \microK. The left panels are the stacked \smica\ maps, and
  the right panels their NFEAs. See Fig.~\ref{fig:stacking_Qr_high_resol} for
  the meaning of the headless vectors (black dashed lines).
\label{fig:general_stack_QU}}
\end{figure}

We next consider oriented stacking of the polarization maps, again
using $Q_T$, $U_T$ to define the orientation of the patches. The
stacked polarization images around temperature peaks have $m=0, 2, 4$
Fourier components. We can also choose to stack the polarization maps
on $P_T$ peaks, where $P_T=\sqrt{Q_T^2+U_T^2}$. This picks up $m=0, 4$
Fourier modes with no circularly symmetric ($Q_\mathrm{r}$, $m=2$) mode. In
Fig.~\ref{fig:general_stack_QU} we compare the $(Q,U)$ images stacked
centred either on $T$ peaks  (top panel) or on $P_T$ peaks (bottom panel)
with their corresponding NFEAs, and find excellent agreement.

For a quantitative comparison, we only consider stacking on
temperature peaks and define the polarization integrated profile
deviation
\begin{linenomath*}
\begin{equation}
\bandp{m}(W) = \int_0^R \left(P_m(\varpi) - \langle P_m(\varpi) \rangle \right)W(\varpi) \, \dd\varpi \, , \label{eq:oriented_stacking_Pm}
\end{equation}
\end{linenomath*}
where by default the filter is
\begin{linenomath*}
\begin{equation}
W(\varpi) = \frac{\ave{P_m(\varpi)}}{\int_0^R \ave{P_m(\varpi)}^2 \dd\varpi} \, . \label{eq:oriented_stacking_WP}
\end{equation}
\end{linenomath*}
The comparison of $\bandp{m}$ ($m=0,2,4$) between the data and the
simulations is shown in
Table~\ref{tbl:oriented_stacking_Pm_comparison}, where the results are
seen to be in excellent agreement.

\begin{table*} \begingroup \newdimen\tblskip \tblskip=5pt
  \caption{$\bandp{m}$, as defined in
  Eqs.~\eqref{eq:oriented_stacking_Pm} and
  \eqref{eq:oriented_stacking_WP}, for different thresholds $\nu$.
  The expected values, together with $1\,\sigma$ (68\,\% CL) and $2\,\sigma$
    (95\,\% CL) ranges determined from simulations are given in brackets.}
\label{tbl:oriented_stacking_Pm_comparison}
\nointerlineskip
\vskip -3mm
\footnotesize
\setbox\tablebox=\vbox{
   \newdimen\digitwidth
   \setbox0=\hbox{\rm 0}
   \digitwidth=\wd0
   \catcode`*=\active
   \def*{\kern\digitwidth}
   \newdimen\signwidth
   \setbox0=\hbox{+}
   \signwidth=\wd0
   \catcode`!=\active
   \def!{\kern\signwidth}
\halign{\hbox to 1.0in{#\leaderfil}\tabskip 4pt&
         \hfil#\hfil&
         \hfil#\hfil&
         \hfil#\hfil\tabskip 0pt\cr                           
\noalign{\doubleline\vskip -1pt}
\omit\hfil Method\hfil& \hfil  $\bandp{0}$  & \hfil  $\bandp{2}$ & \hfil $\bandp{4}$ \cr
\noalign{\vskip 3pt\hrule\vskip 3pt}
\omit & \multispan3\hfil Hot spots, threshold $\nu = 0$ \hfil\cr
\noalign{\vskip 3pt}
{\tt Commander}  & $0.06(0.02^{+0.03+0.05}_{-0.03-0.06})$
& $-0.01(!0.01^{+0.01+0.02}_{-0.01-0.02})$
& $0.04(0.01^{+0.02+0.05}_{-0.02-0.05})$\cr
\noalign{\vskip 2pt}
{\tt NILC}  & $0.05(0.02^{+0.03+0.05}_{-0.03-0.06})$
& $-0.02(!0.00^{+0.01+0.02}_{-0.01-0.02})$
& $0.03(0.01^{+0.02+0.05}_{-0.02-0.05})$\cr
\noalign{\vskip 2pt}
{\tt SEVEM} & $0.05(0.02^{+0.03+0.06}_{-0.03-0.06})$
& $!0.01(!0.01^{+0.01+0.02}_{-0.01-0.02})$
& $0.04(0.01^{+0.02+0.05}_{-0.02-0.05})$\cr
\noalign{\vskip 2pt}
{\tt SMICA} & $0.05(0.03^{+0.03+0.05}_{-0.03-0.05})$
& $-0.02(!0.00^{+0.01+0.02}_{-0.01-0.02})$
& $0.03(0.01^{+0.02+0.05}_{-0.02-0.05})$\cr
\noalign{\vskip 3pt}
\omit &\multispan3 \hfil Cold spots, threshold $\nu=0$ \hfil\cr
\noalign{\vskip 3pt}
{\tt Commander} & $0.06(0.02^{+0.03+0.05}_{-0.03-0.06})$
& $-0.01(!0.01^{+0.01+0.02}_{-0.01-0.02})$
& $0.03(0.01^{+0.03+0.05}_{-0.02-0.05})$\cr
\noalign{\vskip 2pt}
{\tt NILC}   & $0.06(0.02^{+0.03+0.05}_{-0.03-0.06})$
& $-0.01(!0.00^{+0.01+0.02}_{-0.01-0.02})$
& $0.04(0.01^{+0.02+0.04}_{-0.03-0.06})$\cr
\noalign{\vskip 2pt}
{\tt SEVEM} & $0.06(0.03^{+0.03+0.05}_{-0.03-0.06})$
& $!0.01(!0.01^{+0.01+0.02}_{-0.01-0.02})$
& $0.03(0.01^{+0.02+0.05}_{-0.03-0.05})$\cr
\noalign{\vskip 2pt}
{\tt SMICA} & $0.05(0.03^{+0.03+0.05}_{-0.03-0.06})$
& $-0.02(!0.00^{+0.01+0.02}_{-0.01-0.02})$
& $0.03(0.02^{+0.02+0.04}_{-0.02-0.05})$\cr
\noalign{\vskip 3pt}
\omit & \multispan3\hfil Hot spots, threshold $\nu = 1$ \hfil\cr
\noalign{\vskip 2pt}
{\tt Commander} & $0.04(0.02^{+0.03+0.06}_{-0.04-0.07})$
& $-0.02(-0.00^{+0.02+0.03}_{-0.02-0.03})$
& $0.05(0.01^{+0.03+0.06}_{-0.03-0.06})$\cr
\noalign{\vskip 2pt}
{\tt NILC} & $0.06(0.02^{+0.03+0.07}_{-0.04-0.07})$
& $-0.02(-0.01^{+0.01+0.03}_{-0.02-0.03})$
& $0.05(0.01^{+0.03+0.06}_{-0.03-0.06})$\cr
\noalign{\vskip 2pt}
{\tt SEVEM} & $0.05(0.02^{+0.04+0.07}_{-0.04-0.07})$
& $-0.01(-0.00^{+0.02+0.03}_{-0.01-0.03})$
& $0.05(0.01^{+0.03+0.06}_{-0.03-0.06})$\cr
\noalign{\vskip 2pt}
{\tt SMICA} &  $0.04(0.03^{+0.03+0.07}_{-0.03-0.07})$
& $-0.02(!0.00^{+0.01+0.02}_{-0.01-0.03})$
& $0.06(0.01^{+0.03+0.05}_{-0.03-0.06})$\cr
\noalign{\vskip 3pt}
\omit &\multispan3 \hfil Cold spots, threshold $\nu=1$ \hfil\cr
\noalign{\vskip 2pt}
{\tt Commander} & $0.07(0.02^{+0.03+0.06}_{-0.03-0.07})$
& $-0.00(-0.01^{+0.02+0.03}_{-0.02-0.03})$
& $0.01(0.01^{+0.03+0.06}_{-0.03-0.07})$\cr
\noalign{\vskip 2pt}
{\tt NILC} & $0.08(0.02^{+0.03+0.06}_{-0.04-0.07})$
& $-0.01(-0.01^{+0.01+0.03}_{-0.02-0.03})$
& $0.01(0.01^{+0.03+0.06}_{-0.03-0.07})$\cr
\noalign{\vskip 2pt}
{\tt SEVEM} & $0.09(0.02^{+0.03+0.07}_{-0.03-0.07})$
& $-0.00(-0.00^{+0.02+0.03}_{-0.02-0.03})$
& $0.02(0.01^{+0.03+0.06}_{-0.03-0.06})$\cr
\noalign{\vskip 2pt}
{\tt SMICA} & $0.06(0.03^{+0.03+0.06}_{-0.03-0.07})$
& $-0.01(!0.00^{+0.01+0.02}_{-0.01-0.03})$
& $0.02(0.01^{+0.03+0.06}_{-0.03-0.06})$\cr
\noalign{\vskip 4pt\hrule\vskip 3pt}}}
\endPlancktable                    
\endgroup
\end{table*}

\begin{figure}
\includegraphics[width=\myfigwidth]{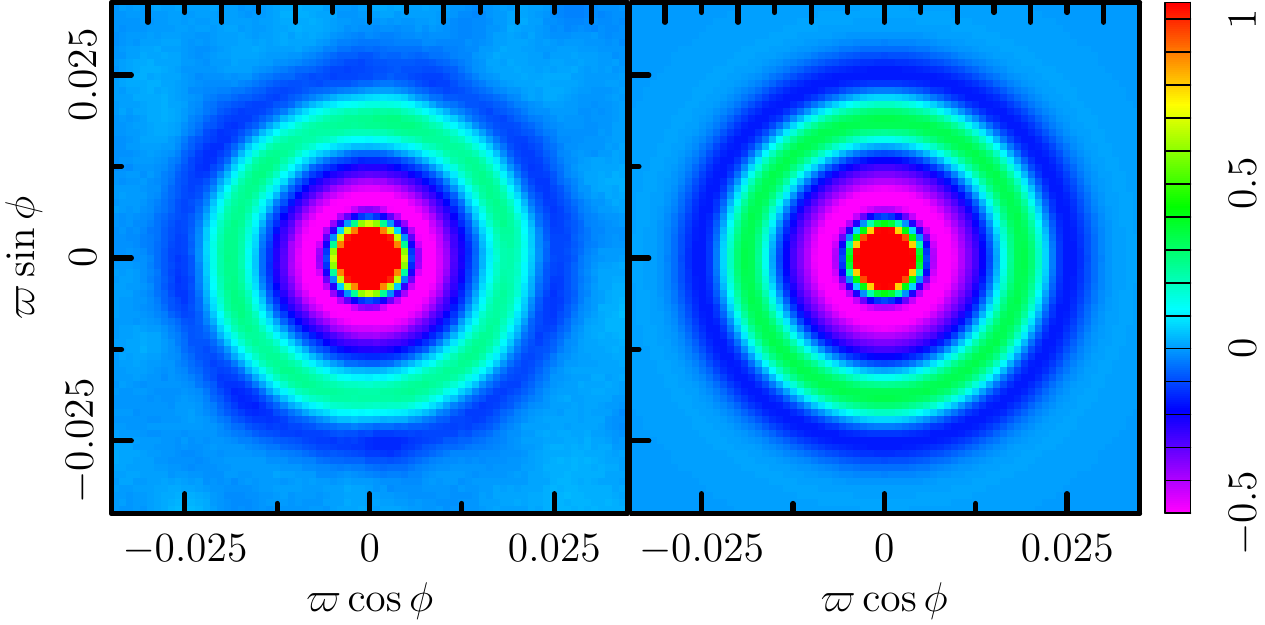}
\includegraphics[width=\myfigwidth]{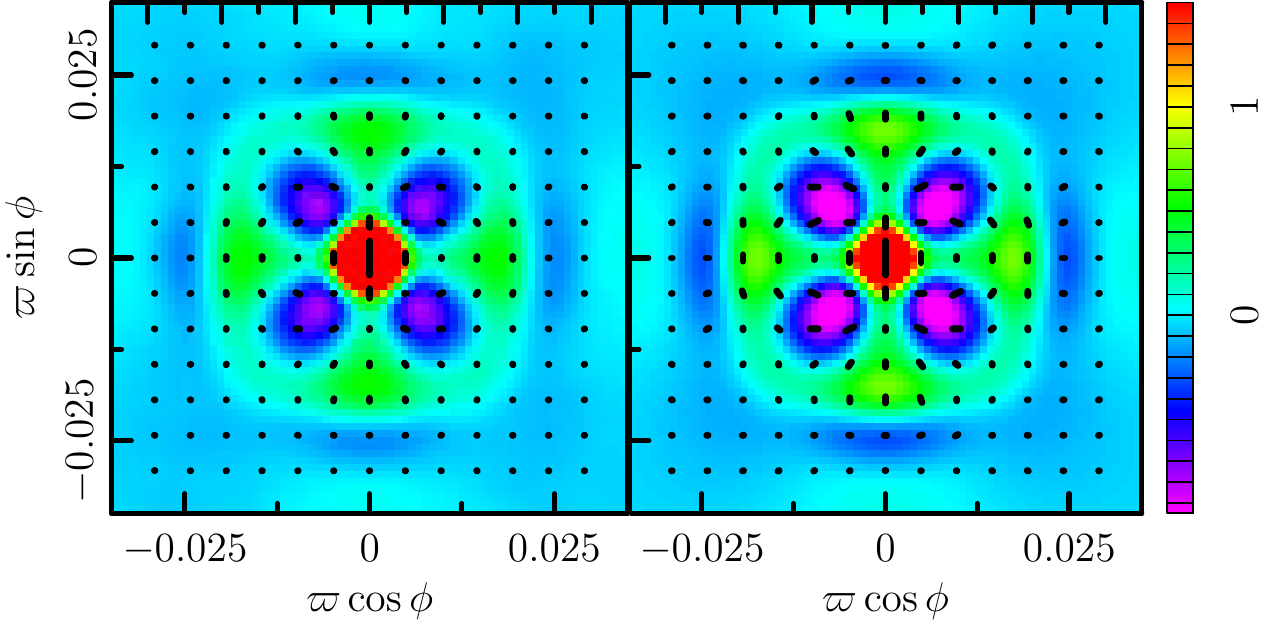}
\caption{\textit{Top}: $E$-mode maps stacked on the unoriented $E$-mode maxima
  computed above the null threshold $\nu =0$. \textit{Bottom}: $Q$ stacked around oriented
  polarization amplitude ($P$) maxima. In this case, no
  threshold is used and the orientation is chosen such that $U=0$ and
  $Q\ge 0$ on the central peak. The left panels are the stacked \smica\
  maps, and the right panels their corresponding NFEAs. See
  Fig.~\ref{fig:stacking_Qr_high_resol} for the meaning of the
  headless vectors (black dashed lines). The image units are \microK.}
  \label{fig:stack_on_pol_peak}
\end{figure}

Finally, we note that the peak selection does not have to be made from
the temperature map. In Fig.~\ref{fig:stack_on_pol_peak} we show a few
examples of stacking on polarization peaks using the $N_{\rm side} =
512$ maps. The higher-resolution polarization data are too noisy for
peak selection. In the upper panels, we compare stacked images of the
$E$-mode map centred around $E$-mode peaks with the corresponding
NFEA. We find that the noise impact is relatively minor for
$\mathrm{FWHM}=20\arcmin$ maps and the plots are in qualitatively good
agreement. Another possibility, shown in the lower panels, is to stack
polarization maps centred on peaks determined from the corresponding
polarization amplitude map.  In this case the peaks are strongly
biased by the quadratic noise contribution and quite visible deviation
from the NFEA is observed in the stacked image.

\section{Conclusions}
\label{sec:conclusions}

  In this paper, we have presented a study of the statistical isotropy
and Gaussianity of the CMB using the \Planck\ 2015 data, including the
full mission for temperature.  We do not claim that our results support or
refute any particular physical model. Rather, we focus on
null-hypothesis testing: a number of tests are performed, then
  $p$-values are calculated and reported. It is in the very nature of
such a model-independent approach to leave the detailed interpretation
to the reader. However, we do address the important subject of a
posteriori correction where possible.

  The statistical tests are performed on maps of the CMB anisotropy that
result from the application of the four component-separation methods
described in \citet{planck2014-a11}.  All of the results presented
here are robust with respect to the choice of component-separated CMB
map. This is important since it demonstrates the high quality and
equivalence of the \Planck\ component-separated data products rendered
by different methodologies under varying assumptions.

We find that the CMB is largely consistent with statistical isotropy,
although there are a few indications of anomalies with respect to the
expectations of \LCDM.  Some of the tests we have performed are the
same as those in \citetalias{planck2013-p09}, in which case the
results are consistent.  Since many of these anomalies were also
observed in the WMAP temperature data, we re-emphasize explicitly the
statement we made in 2013 --- that the agreement between the two
independent experiments effectively rules out the possibility that the
origin of these features can be found in residual systematic artefacts
present in either data set. We have also performed a number of new
tests, in order to try to narrow down the nature of the apparent
violations of statistical isotropy.  In addition, although the
component-separated polarization maps contained in the \Planck\ 2015
release are high-pass filtered, we have performed a stacking analysis
that tests some aspects of the polarized sky while mitigating the
impacts of noise and systematic effects.

  In Sect.~\ref{sec:non_gaussianity}, we examined aspects of the
Gaussianity of the CMB fluctuations.  Tests of skewness, kurtosis,
multi-normality, $N$-point functions, and Minkowski functionals
yielded no indications of significant departures from Gaussianity,
while the variance of the CMB map was found to be low, in agreement
with previous studies \citepalias{planck2013-p09}.  First-order
moments of filtered maps also exhibit the low-variance anomaly, as
well as a kurtosis excess on certain scales associated with the Cold
Spot.  A new study of peak statistics finds results consistent with
the expectations for a Gaussian random field, although the Cold Spot
is again detected.

  Section~\ref{sec:anomalies} provides an updated study of several
previously known peculiarities.  We study in detail the low variance
anomaly, which appears to be associated with the known low-$\ell$
deficit in the angular power spectrum.  We confirm the lack of
large-scale angular correlations, relatively featureless northern
ecliptic hemisphere 3- and 4-point functions, and indications of
violations of point- and mirror-parity symmetry, although we make
little or no attempt to correct these for a posteriori effects.  We
place tight constraints on a quadrupolar power modulation.  The Cold
Spot is examined further, and, while we find variance, skewness, and
kurtosis angular profiles consistent with the expectations of
statistically isotropic simulations, the mean temperature profile is
anomalous at roughly the 1\,\% level, apparently due to the surrounding hot
ring --- the feature that deviates most from the Gaussian model.

  In Sect.~\ref{sec:dipmodsection} we perform a series of tests probing
the well-known large-scale dipolar power asymmetry.  We detect the
asymmetry via pixel-to-pixel variance, as well as by measuring power
explicitly or indirectly via $\ell$ to $\ell \pm 1$ mode coupling.
The latter approach lends itself to a posteriori correction, which
reduces the significance of the asymmetry substantially when no model
for the anomaly is assumed.  In addition, we perform two independent
but related tests of directionality.  One finds suggestions of
anomalous clustering of directions out to relatively small scales
while the other does not, evidently due to being optimized for
slightly different forms of directionality.

Section~\ref{sec:sky_coverage} demonstrates that the significances of
several large-angular-scale anomalies are robust to the use of larger
sky coverage, with the observed small changes being consistent with
expectations from random Gaussian statistics. 

  Finally, Sect.~\ref{sec:polarization_results} presents the results of
the stacking of temperature and polarization peaks.  We find results
that are largely consistent with statistically isotropic simulations,
both for oriented and unoriented stacking.  The exception is a low
unoriented temperature profile, which seems to be yet another
reflection of the large-scale power deficit.

  With the \Planck\ 2015 release, we are probably near the limit of our
ability to probe the CMB anomalies with temperature fluctuations
alone.  The use of large-angular-scale polarization, expected for the
final \Planck\ release, should enable {\em independent\/} tests of
these peculiar features.  Importantly, this will reduce or eliminate
the subjectivity and ambiguity in interpreting their statistical
significance.  It is a tantalizing possibility that some of the
anomalies described in this paper will take us beyond the standard
model of cosmology.

\begin{acknowledgements}
\label{sec:acknowledgements}

The \Planck\ Collaboration acknowledges the support of: ESA; CNES and
CNRS/INSU-IN2P3-INP (France); ASI, CNR, and INAF (Italy); NASA and DoE
(USA); STFC and UKSA (UK); CSIC, MINECO, JA, and RES (Spain); Tekes,
AoF, and CSC (Finland); DLR and MPG (Germany); CSA (Canada); DTU Space
(Denmark); SER/SSO (Switzerland); RCN (Norway); SFI (Ireland);
FCT/MCTES (Portugal); ERC and PRACE (EU). A description of the Planck
Collaboration and a list of its members, indicating which technical or
scientific activities they have been involved in, can be found at
\href{http://www.cosmos.esa.int/web/planck/planck-collaboration}{\texttt{http://www.cosmos.esa.int/web/planck/planck-collaboration}}.
Some of the results in this paper have been derived using the {\tt
  HEALPix} package.

\end{acknowledgements}


\bibliographystyle{aat}
\bibliography{Planck_bib,IandS}

\begin{appendix}
\section{Generalized Savitzky-Golay polynomials}
\label{asec:gsgp}

In the construction of optimal linear filters, one needs to combine
information about the (statistically isotropic) CMB signal,
anisotropic instrumental noise, masking to be applied for the
elimination of foreground contributions, and a model for any
non-Gaussian signal for matched filtering. These can be combined in a
general framework of normalized convolutions \citep{knutsson1993},
where the filtered field is defined as
\begin{linenomath*}
\begin{equation}
  U = \frac{ a\vec{B} \star w\vec{T} }{a\vec{B} \star \vec{B}^\dagger w},
\end{equation}
\end{linenomath*}
where $\vec B$ is the (multiscale) filtering beam function, $\vec T$ is the
temperature, $a$ and $w$ their respective weights, and $\star$ denotes
the usual convolution operation
\begin{linenomath*}
\begin{equation}
  \{ a\vec{B} \star w\vec{T} \}(\vec{\xi}) = \sum\limits_{\vec{x}}
    a(\vec{x}) \vec{B}(\vec{x}) \cdot w(\vec{\xi} - \vec{x}) \vec{T}(\vec{\xi} - \vec{x}).
\end{equation}
\end{linenomath*}
In the absence of a specific model for the non-Gaussian signal, the
beam functions can be taken to be orthogonal polynomials on a disc,
weighted by some smoothing function, while the weights applied to the
temperature maps are determined by the CMB and noise covariance.

In a simple approach, the information about the CMB signal can be
utilized by pre-whitening the map by convolving it with an isotropic
beam function $w_\ell = C_\ell^{-1/2}$ derived from the isotropic
best-fit CMB power spectrum combined with a diagonal approximation to
the instrumental noise covariance. After the component-separated CMB
maps are pre-whitened, and the corresponding mask is applied to the
resulting map, the multiscale filtering kernel $b_\ell$ is applied at
various scales.

\begin{figure*}
  \centering
  \begin{tabular}{cc}
  \includegraphics[width=80mm]{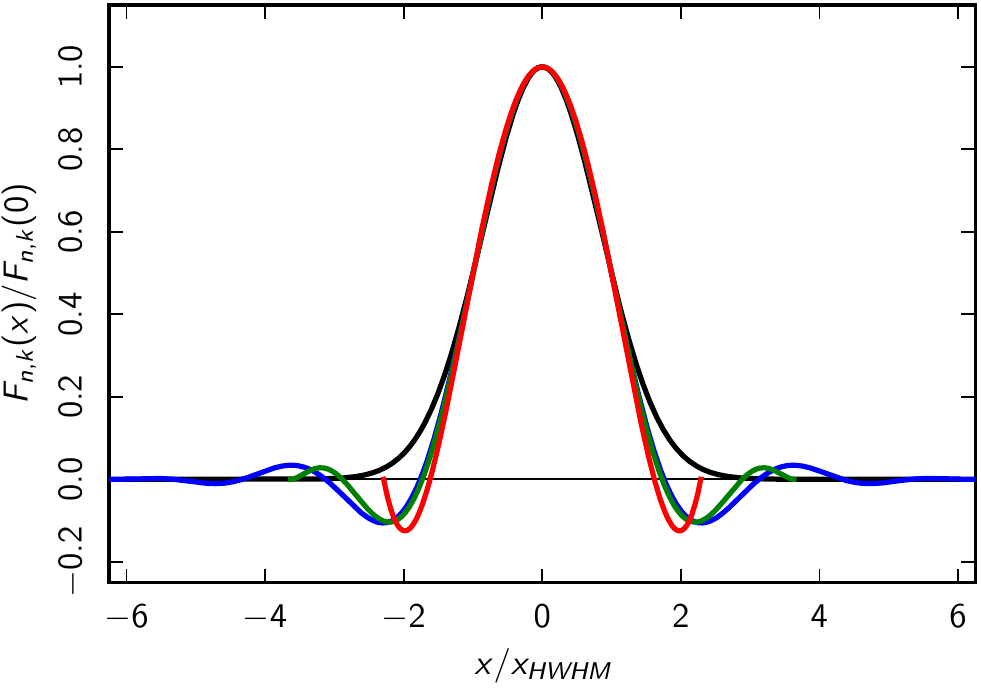} &
  \includegraphics[width=80mm]{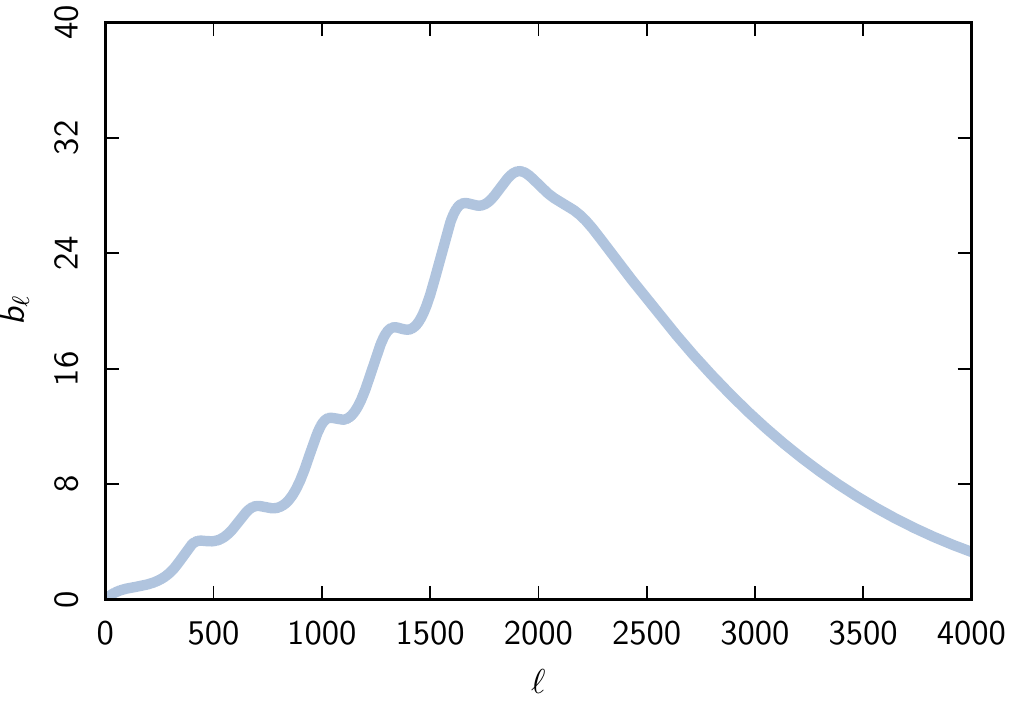} \\
  \includegraphics[width=80mm]{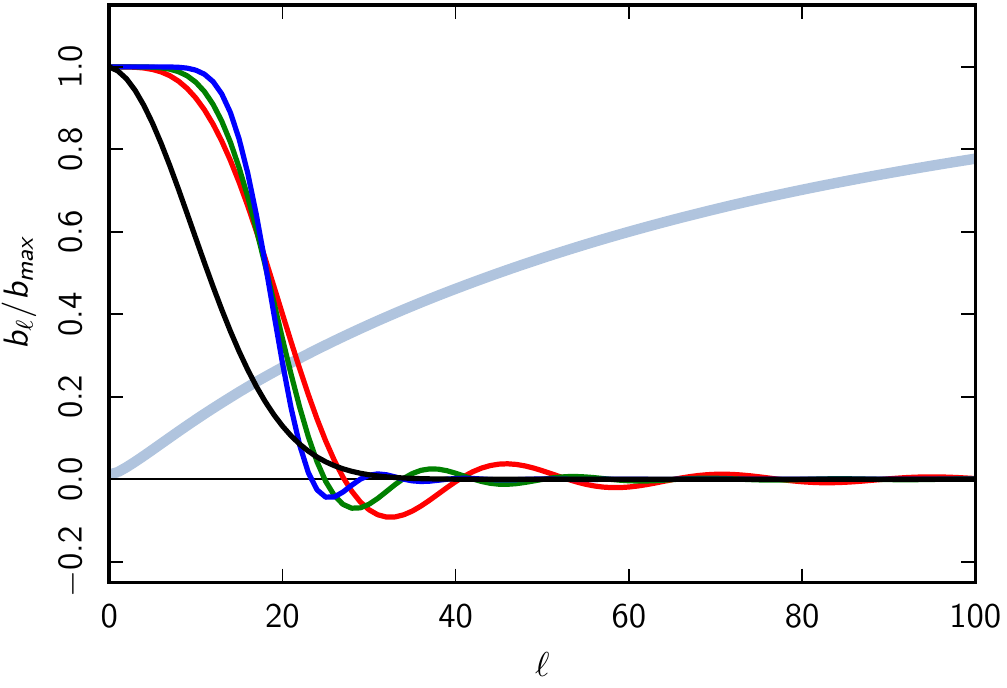} &
  \includegraphics[width=80mm]{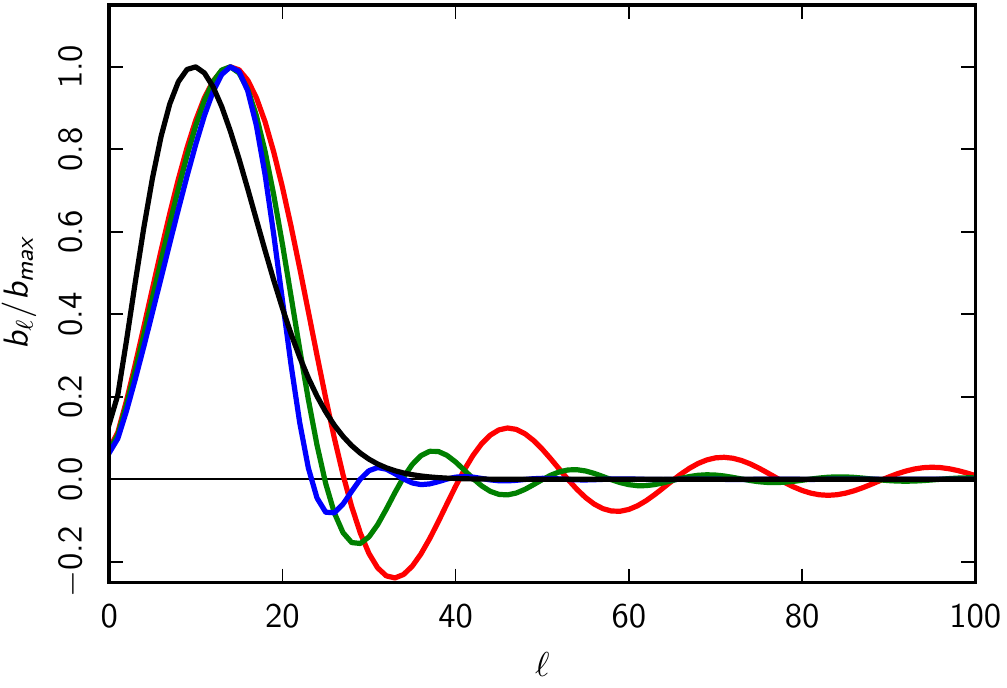}\\
  \end{tabular}
  \caption{Generalized Savitzky-Golay polynomials are orthogonal to
    polynomials up to degree $n$ on a disc, with smoothing weight
    applied.  Upper left panel shows a few representative polynomial
    kernels (\texttt{SSG21} in red, \texttt{SSG42} in dark green,
    \texttt{SSG84} in blue) and Gaussian (in black) as a function of
    radius (scaled to the same FWHM of 800\arcm), lower left shows their harmonic
    space representation. Right column shows pre-whitening kernel for
    \smica\ temperature map on the top (in light blue), and the
    corresponding composite kernels (\texttt{WHITE*SSG21}, etc)
    on the bottom (in the same colors).}
  \label{fig:peaks:ssg}
  \label{fig:peaks:beams}
\end{figure*}

In this paper, the maps are pre-whitened with the 2013 best-fit
cosmological parameter CMB spectrum \citep{planck2013-p08}, co-added
to an isotropic noise power spectrum derived from the half-mission,
half-difference noise maps appropriate for each component-separation
method.  No adjustment is made either for the recalibration of the
2015 data relative to the nominal results that the cosmological
spectrum is derived from, or for the mismatch in noise level between
the half-mission, half-difference and full-mission maps. This implies
that the filtering is sub-optimal, but the data and simulations are
treated consistently so there should be no significant impact on the
results.  The resulting pre-whitening beam function $w_\ell$ for the
\smica\ temperature map is shown in Fig.~\ref{fig:peaks:beams}.

The peak detector wavelets are taken to be Savitzky-Golay polynomials
\citep{Savitzky1964}, generalized to be defined on a disc with a
polynomial smoothing weight function applied, as shown in
Fig.~\ref{fig:peaks:ssg}. A generalized spherical Savitzky-Golay
kernel of order $n$ and smoothing weight $k$ (referred to as {\tt
  SSGnk} in the text) is defined by a polynomial function of a radial
variable $x = \sin(\theta/2)/\sin(\theta_{\text{max}}/2)$,
\begin{linenomath*}
\begin{equation}
  F_{n,k}(x) = \left(\sum\limits_{i=0}^{n/2} a_i x^{2i}\right) (1-x^2)^k,
\end{equation}
\end{linenomath*}
which is normalized to have unit mean on a disc and is orthogonal to
all non-constant polynomials up to order $n$,
\begin{linenomath*}
\begin{equation}
  \int\limits_{0}^{1} x F_{n,k}(x)\, \dd x = 1, \quad
  \int\limits_{0}^{1} x^{i+1} F_{n,k}(x)\, \dd x = 0.
\end{equation}
\end{linenomath*}
These are essentially high-order low-pass filters in harmonic space,
but have compact support on the sphere. A few representative
Savitzky-Golay polynomials are compared to a Gaussian kernel in
Fig.~\ref{fig:peaks:ssg}.  Combined with pre-whitening, the total
effect of the filters applied is described by the composite beam
functions shown in Fig.~\ref{fig:peaks:beams}.

One should note a slight $\ell$-space bandwidth mismatch between
differently shaped kernels with the same FWHM
value in real space, which is clear from the lower left panel of
Fig.~\ref{fig:peaks:ssg}. While not a problem in general, some care
should be exercised when directly comparing results for different
shape kernels.  In particular, the $\ell$ value at which the filter
kernel coefficient reaches $b_\ell = b_{\text{max}}/2$ differs by
a factor of $1.58$ between the {\tt GAUSS} and {\tt SSG84} kernels of
the same FWHM.

\section{Doppler boosting}
\label{sec:dboost}

The main effect of our relative motion with respect to the CMB rest
frame is a dominant contribution to the CMB dipole ($C_1$); this is
boosting of the monopole and has been detected previously
\citep[][]{kogut2003,Fixen:dipole,hinshaw2009}. A subtler
consequence of our motion is the boosting of all other multipoles. In
fact, there are really two effects at work.  The first is a modulation
effect which increases power by approximately $0.25\,\%$ in the
direction of our motion and decreases it by the same amount in the
opposite direction. This can equivalently be thought of as coupling
between the multipoles $\ell$ and $\ell \pm 1$.  The second is an
aberration effect which shifts the apparent direction in which CMB
photons arrive at our detectors toward the velocity direction.  

\citet{planck2013-pipaberration} reported a detection of this Doppler
boosting, and an associated measurement of its velocity signature of
$384\pm 78$ (statistical) ${}\pm
115$ (systematic) km\,s$^{-1}$  in the known dipole direction, $(l,
b) = (264\degr, 48\degr)$. Here, we demonstrate that the \Planck\
2015 data release remains in agreement with this result, by
considering the angular scales $500 \leq \ell \leq 2000$. However,
since the simulations employed in the analysis contain the effects of
Doppler boosting, we report a consistency check rather than a
detection.

It is useful to perform a harmonic transform on the peculiar velocity
vector,
\begin{linenomath*}
\begin{align}
 \beta_{LM} &= \int \dd{\hat{\vec n}} \, Y^*_{LM}(\hat{\vec n}) {\vec \beta}\cdot {\hat{\vec n}},
\label{eq:velocity_vector}
\end{align}
\end{linenomath*}
where only the ${L} = 1$ modes are non-zero. Following the
convention in \citet{planck2013-pipaberration}, we rotate to an
orthonormal basis, labelled ${\vec \beta}_{||}$ (along the expected
velocity direction), ${\vec \beta}_\times$ (parallel to the Galactic
plane), and ${\vec \beta}_\perp$ (the remaining vector).

\begin{table}
\begingroup
\newdimen\tblskip \tblskip=5pt
\caption{Significance measures for the $\vec\beta$ estimates for the $143\times217$
  data set. $\chi^2$ is formed from the three modes of $\vec\beta$ using the
  covariance matrix measured from Doppler boosted simulations.}
\label{tab:betachisquared}
\nointerlineskip
\vskip -3mm
\footnotesize
\setbox\tablebox=\vbox{
   \newdimen\digitwidth
   \setbox0=\hbox{\rm 0}
   \digitwidth=\wd0
   \catcode`*=\active
   \def*{\kern\digitwidth}
   \newdimen\signwidth
   \setbox0=\hbox{+}
   \signwidth=\wd0
   \catcode`!=\active
   \def!{\kern\signwidth}
\halign{ \hbox to 1.0in{$#$\leaderfil}\tabskip 2em&
\hfil$#$\hfil&
\hfil$#$\hfil\tabskip 0pt\cr                           
\noalign{\doubleline\vskip -4pt}
\omit\hfil Estimator \hfil&\omit $\chi^2$ \hfil&\omit PTE [\%]\cr   
\noalign{\vskip 3pt\hrule\vskip 3pt}
\hat{\vec\beta}_{\parallel}& 3.28& 7.01\cr
\hat{\vec\beta}_{\perp}& 0.21& 64.39\cr
\hat{\vec\beta}_{\times}& 0.08& 77.53\cr
\hat{\vec\beta}& 3.38& 33.70\cr
\noalign{\vskip 3pt\hrule\vskip 3pt}}}
\endPlancktable                    
\endgroup
\end{table}

The peculiar velocity is detected using estimators that pick out the
off-diagonal components of the CMB covariance matrix
\begin{linenomath*}
\begin{align}
 \left\langle T_{\ell_1 m_1} T_{\ell_2 m_2} \right\rangle_\mathrm{CMB} = \sum_{LM}
{(-1)}^{M}
\left(
\begin{array}{ccc}
\ell_1 & \ell_2 & {L} \\
m_1 & m_2 & {M}
\end{array}
\right) \nonumber\\
\times\sqrt{\frac{(2\ell_1 + 1) (2\ell_2 + 1) (2{L} +
1)}{4\pi}}W^{\beta_v}_{\ell_1\ell_2{L}}\beta_{LM}.
\end{align}
\end{linenomath*}
The weight function $W^{\beta_{v}}$ is a sum of the modulation
($b_{v}W^\tau$) and aberration ($W^\phi$) effects. We quote results
based on orthogonalized weight matrices,
\begin{linenomath*}
\begin{align}
 W^{\hat{\phi}} &= W^\phi - W^\tau
\mathcal{R}^{\phi\tau}/\mathcal{R^{\tau\tau}} \\
 W^{\hat{\tau}} &= W^\tau - W^\phi
\mathcal{R}^{\tau\phi}/\mathcal{R}^{\phi\phi}.
\end{align}
\end{linenomath*}
Due to the clear connection between the velocity estimators and those
used for the lensing analysis, we adopt the same data (143\,GHz and 217\,GHz sky
maps, with dust foregrounds removed using the 857\,GHz data as a
template) and mask as used in \citet{planck2014-a17}.
The results are summarized in Table~\ref{tab:betachisquared}, and show
good consistency with previous results.

\section{Generalized modulation estimator}
\label{sec:mossest}

Consider a parameter $X$ that the (primary) CMB power spectrum is
dependent on.  Let $X$ have a dipolar dependence of the form
$X(\hat{\vec n}) = X_0 + \Delta X\hat{\vec n}\cdot\hat{\vec m}$ (this could
correspond to a gradient in $X$ across our observable volume), where
$X_0$ is the average value, $\hat{\vec n}$ is the direction to the last
scattering surface, and $\hat{\vec m}$ is the gradient direction. To linear
order in $\Delta X/X$, the measured spherical harmonics coefficients
are given by
\begin{linenomath*}
\begin{align}
 a_{\ell m} &= a_{\ell m}^{\mathrm{iso}} + \sum_M \Delta X_M \sum_{\ell'm'}
\frac{d a_{\ell'm'}^{\mathrm{iso}}}{dX}\xi^M_{\ell m\ell'm'},
\label{eq:modulatedmodes}
\end{align}
\end{linenomath*}
where the $a_{\ell m}^{\mathrm{iso}}$ are the unmodulated
statistically isotropic modes. The $\xi^M_{\ell m\ell'm'}$ are
coupling coefficients given by
\begin{linenomath*}
\begin{align}
 \xi^0_{\ell m\ell'm'} &= \delta_{m'm}\left(\delta_{\ell'\ell-1}A_{\ell-1\,m}
+ \delta_{\ell'\ell+1}A_{\ell m}\right), \\
 \xi^{\pm 1}_{\ell m\ell'm'} &= \delta_{m'm\mp 1}\left(
\delta_{\ell'\ell-1}B_{\ell-1\,\pm m-1} - \delta_{\ell'\ell+1}B_{\ell\,\mp m}
\right),
\label{eq:couplingcoeffs}
\end{align}
\end{linenomath*}
where
\begin{linenomath*}
\begin{align}
 A_{\ell m} &= \sqrt{\frac{(\ell+1)^2 - m^2}{(2\ell+1)(2\ell+3)}}, \\
 B_{\ell m} &= \sqrt{\frac{(\ell+m+1)(\ell+m+2)}{2(2\ell+1)(2\ell+3)}}.
\end{align}
\end{linenomath*}
From Eq.~\eqref{eq:modulatedmodes} we can find the covariance matrix
to first order in the components $\Delta X_M$:
\begin{linenomath*}
\begin{align}
 C_{\ell m\ell'm'} &= C_\ell\delta_{\ell\ell'}\delta_{mm'} +
\frac{\delta C_{\ell\ell'}}{2}\sum_M \Delta X_M\xi^M_{\ell m\ell'm'},
\label{eq:cmbcovariance}
\end{align}
\end{linenomath*}
where $\delta C_{\ell\ell+1} = dC_{\ell}/dX + dC_{\ell+1}/dX$. To
determine the best-fit parameters, we proceed by maximizing the CMB
likelihood function
\begin{linenomath*}
\begin{align}
 \mathcal{L} &= \frac{1}{\sqrt{2\pi|C|}}\exp(-{\vec d}^\dagger C^{-1}{\vec d}/2),
\label{eq:cmblikelihood}
\end{align}
\end{linenomath*}
where $\vec d$ is the CMB temperature
data. Equation~\eqref{eq:cmblikelihood} is maximized for the ${\Delta
  X_M}$ that satisfy
\begin{linenomath*}
\begin{align}
 {\vec d}^\dagger C^{-1} \frac{\dd C}{\dd\Delta X_M}C^{-1} {\vec d} &= \Tr{\left( C^{-1}\frac{\dd
C}{\dd\Delta X_M}\right)}.
\end{align}
\end{linenomath*}
From Eq.~\eqref{eq:cmbcovariance} it is clear that the CMB covariance
can be decomposed into an isotropic part ($C_\ell$) and a small
anisotropic part proportional to $\Delta X_M$. By inverting
Eq.~\eqref{eq:cmbcovariance} and using the orthogonality of the
$\xi^M_{\ell m\ell'm'}$, we can determine the best-fit parameters
\begin{linenomath*}
\begin{align}
 \Delta X_0 &= \frac{6\sum_{\ell m}\frac{\delta
C_{\ell\ell+1}}{C_{\ell}C_{\ell+1}}A_{\ell m}a^*_{\ell
m}a_{\ell+1\,m}}{\sum_{\ell}\frac{\delta C_{\ell\ell+1}^2}{C_\ell
C_{\ell+1}}(\ell + 1)}, \\
 \Delta X_{+1} &= \frac{6\sum_{\ell m}\frac{\delta
C_{\ell\ell+1}}{C_{\ell}C_{\ell+1}}B_{\ell m}a^*_{\ell
m}a_{\ell+1\,m+1}}{\sum_{\ell}\frac{\delta C_{\ell\ell+1}^2}{C_\ell
C_{\ell+1}}(\ell + 1)},
\end{align}
\end{linenomath*}
and $\Delta X_{-1} = -\Delta X^*_{+1}$, to first order in the
anisotropy.  These estimators are the full-sky, no-noise versions of
Eqs.~\eqref{eq:mossest0} and~\eqref{eq:mossest1}.

Errors can easily be found by expanding the log-likelihood about the
best-fit parameters. The Fisher matrix is defined as
\begin{linenomath*}
\begin{align}
 F_{MM'} &\equiv \frac{1}{2}\Tr{\left( \frac{\partial C}{\partial \Delta
X_M}C^{-1}\frac{\partial C}{\partial \Delta X_{M'}} C^{-1} \right)}.
\end{align}
\end{linenomath*}
Upon switching bases, we find
\begin{linenomath*}
\begin{align}
 F_{0,0} &= \frac{1}{4}\sum_{\ell m}\frac{\delta C^2_{\ell\ell+1}}{C_\ell
C_{\ell+1}}A^2_{\ell m}, \\
 F_{\Re{(\Delta X_{+1})}, \Re{(\Delta X_{+1})}} &= \frac{1}{2}\sum_{\ell m}
\frac{\delta C^2_{\ell\ell+1}}{C_\ell C_{\ell+1}}B^2_{\ell m}.
\end{align}
\end{linenomath*}
We can then assign the standard errors, $\sigma = \sqrt{F^{-1}}$.

\section{Weighted-variance modified shape function estimator}
\label{sec:wvmsf}

The BipoSH representation characterizes the off-diagonal elements in
the covariance matrix and is a generalization of the angular power
spectrum, $C_{\ell}$,
\begin{linenomath*}
\begin{eqnarray}
A^{LM}_{\ell_1 \ell_2}=\sum_{m_1 m_2}\langle a_{\ell_1 m_1}a_{\ell_2 m_2}\rangle C^{LM}_{\ell_1 m_1 \ell_2 m_2}.
\end{eqnarray}
\end{linenomath*}
In general, it is not possible to analyse the full sky even for
component-separated maps, due to the presence of residual
contributions from diffuse Galactic emission and point
sources. However, the application of a mask leads to coupling between
the spherical harmonic modes. Hence, the correlation function is no
longer described only by $C(\theta)$ or the power spectrum $C_{\ell}$,
and other quantities are required to completely quantify the
statistical field.

We obtain an analytic expression for the observed BipoSH coefficients
after the application of a mask in terms of the corresponding
coefficients of the unmasked sky, and those of the mask itself,
\begin{linenomath*}
\begin{eqnarray}\label{BipoSH-mask}
\tilde A^{LM}_{\ell_1 \ell_2}=\sum_{\ell_3 \ell_4 }\frac{\Pi_{\ell_{3}\ell_{4}}}{\sqrt{4\pi}}\sum_{\ell_5 \ell_6}
\frac{\Pi_{\ell_{5}\ell_{6}}}{\sqrt{4\pi}}C^{\ell_1 0}_{\ell_3 0 \ell_5 0}C^{\ell_2 0}_{\ell_4 0 \ell_6 0}\times\nonumber\\ \sum_{L_1 M_1 J K} \Bigg\{ \begin{array}{ccc}
L & \ell_1 & \ell_2 \\
L_1 & \ell_3 & \ell_4 \\
J & \ell_5 & \ell_6 \end{array} \Bigg\}
  \Pi_{L_1}\Pi_{J}A^{L_1 M_1}_{\ell_3 \ell_4}W^{JK}_{\ell_5 \ell_6}C^{LM}_{L_1 M_1 J K}
\end{eqnarray}
\end{linenomath*}
where $\Pi_{\ell}=\sqrt{2\ell+1}$, $\tilde{A}^{LM}_{\ell_1 \ell_2}$
are the BipoSH coefficients of the masked sky map, $A^{LM}_{\ell_1
  \ell_2}$ correspond to the BipoSH coefficients of the unmasked sky,
$W^{LM}_{\ell_1 \ell_2}$ are the BipoSH coefficient of the mask
itself, $C^{LM}_{lml'm'}$ are the Clebsch-Gordon coefficients, and the
term $\{\,\}$ in Eq.~\eqref{BipoSH-mask} is the $9j-$symbol.  This
quantifies the coupling between the BipoSH coefficients of the CMB sky
map and those of the mask itself.

The underlying CMB sky may have deviations from statistical isotropy,
as discussed in Sect.~\ref{sec:biposh}, due either to a dipole
modulation ($L=1$) of unknown origin, or to Doppler boosting ($L=1$) of
the temperature field. The BipoSH coefficients of such statistical
isotropy-violating fields can be given by
\begin{linenomath*}
\begin{eqnarray}
A^{LM}_{\ell_1 \ell_2}=\bar A^{LM}_{\ell_1 \ell_2}+\phi_{LM}G^{L}_{\ell_1 \ell_2} \,.
\end{eqnarray}
\end{linenomath*}
Here $\bar A^{LM}_{\ell_1 \ell_2}$ corresponds to the BipoSH
coefficients of the unknown but statistically isotropic CMB
field. This couples with BipoSH coefficients of the mask to introduce
a mean field linear bias $\langle \mathcal{A}^{LM}_{\ell_1
  \ell_2}\rangle_\mathrm{mask}$, which is estimated from simulations
and subtracted from the BipoSH coefficients obtained from the masked
sky.  The $\phi_{LM}$ are the spherical harmonic coefficients of the
field that breaks statistical isotropy, and $G^{L}_{\ell_1 \ell_2}$ is
the shape function.  Shape functions for dipole modulation and Doppler
boosting are given in Eqs.~\eqref{biposh_modulation} and
\eqref{biposh_boost}, respectively.

Due to symmetries of the mask, which is largely defined by foreground
residuals towards the Galactic plane, the dominant BipoSH modes of the
mask correspond to $J=\{0,2\}, K=0$.  Hence, for all practical
purposes, signal is retained in the $L=1$ mode itself, although
masking modifies the shape function, now defined as the modified shape
funtion in the rest of the text. A weighted variance modified
shape function is defined as
\begin{linenomath*}
\begin{equation}
\hat{\phi}_{LM} = \sum_{\ell_1 \ell_2} {w}^{LM}_{\ell_1 \ell_2} \frac{\hat{A}^{LM}_{\ell_1 \ell_2}}{K^{LM}_{\ell_1\,\ell_2}} \,,
\end{equation}
\end{linenomath*}
where $\hat{A}^{LM}_{\ell_1 \ell_2} = \tilde{A}^{LM}_{\ell_1 \ell_2} -
\langle\mathcal{A}^{LM}_{\ell_1 \ell_2}\rangle_\mathrm{mask}$ and the
weights are chosen such that $\sum_{\ell_1 \ell_2}{w}^{LM}_{\ell_1
  \ell_2}=1$.

Here ${K^{LM}_{\ell_1\,\ell_2}}$ is the MSF, which can be evaluated as
\begin{linenomath*}
\begin{eqnarray}
{K^{LM}_{\ell_1\,\ell_2}}&=&\sum_{\ell_3 \ell_4 }\Pi_{L}G^{L}_{\ell_3 \ell_4}\frac{\Pi_{\ell_{3}\ell_{4}}}{\sqrt{4\pi}}\sum_{\ell_5 \ell_6}
\frac{\Pi_{\ell_{5}\ell_{6}}}{\sqrt{4\pi}}C^{\ell_1 0}_{\ell_3 0 \ell_5 0}C^{\ell_2 0}_{\ell_4 0 \ell_6 0}\times\nonumber\\ && \sum_{J K}  \Bigg\{ \begin{array}{ccc}
L & \ell_1 & \ell_2 \\
L & \ell_3 & \ell_4 \\
J & \ell_5 & \ell_6 \end{array} \Bigg\}
 \Pi_{J}W^{JK}_{\ell_5 \ell_6}C^{LM}_{L M J K}.
\end{eqnarray}
\end{linenomath*}
The weights are then given by
\begin{linenomath*}
\begin{equation}
w^{LM}_{\ell_1 \ell_2} = \frac{1}{\sum_{M}  \left({\sigma_{\mathcal{A}^{LM}_{\ell_1 \ell_2}}}/{K^{LM}_{\ell_1 \ell_2}}\right)^2}
\left[ \sum_{\ell'_1 \ell'_2} \frac{1}{\sum_{M} \left({\sigma_{\mathcal{A}^{LM}_{\ell'_1 \ell'_2}}}/{K^{LM}_{\ell'_1 \ell'_2}}\right)^2} \right]^{-1} \,.
\end{equation}
\end{linenomath*}

\end{appendix}

\raggedright

\end{document}